\documentclass[preprint]{aastex}

\shorttitle{Superluminal Motion in Gamma-Ray Blazars}
\shortauthors{Jorstad et al.}

\begin{document}

\title{Multi-Epoch VLBA Observations of EGRET-Detected Quasars and
BL Lac Objects: Superluminal Motion of Gamma-Ray Bright Blazars} 
    
\author{Svetlana G. Jorstad\altaffilmark{1,2,3}, Alan P. Marscher\altaffilmark{1}, John R. Mattox\altaffilmark{1,4}}
\author{Ann E. Wehrle\altaffilmark{5}, Steven D. Bloom\altaffilmark{6}, \and
Alexei V. Yurchenko\altaffilmark{2}}
\altaffiltext{1}{Institute for Astrophysical Research, Boston University, 
      725 Commonwealth Ave., Boston, MA, 02215}
\altaffiltext{2}{Astronomical Institute, St. Petersburg State University,
       Bibliotechnaya pl. 2, Petrodvorets, St. Petersburg, 198904,
       Russia}
\altaffiltext{3}{Formerly S. G. Marchenko}
\altaffiltext{4}{Current address: Department of Chemistry, Physics and Astronomy,
Francis Marion University, P.O. Box 100547, Florence, SC 29501-0547}
\altaffiltext{5}{Jet Propulsion Laboratory, MS 301-486, 4800 Oak Grove Dr.,
Pasadena, CA 91109}
\altaffiltext{6}{Hampden-Sydney College, Box 821, Hampden-Sydney, VA 23943}

\slugcomment{\apjs, 12/31/00}
\begin{abstract}
We present the results of a program to monitor the structure of the radio emission
in 42 $\gamma$-ray bright blazars
(31 quasars and 11 BL~Lac objects) with the VLBA at 43, 22, and occasionally 15
and 8.4~GHz, over the period from November 1993 to July 1997. We determine proper
motions in 33 sources and find that the apparent superluminal motions in $\gamma$-ray 
sources are much faster than for the  general population of bright compact radio sources.
This follows the strong dependence of the $\gamma$-ray flux on the level of
relativistic beaming for both external-radiation Compton and 
synchrotron self-Compton emission. There is a positive correlation (correlation
coefficient $r$=0.45) between the
flux density of the VLBI core and the $\gamma$-ray flux and a moderate correlation 
(partial correlation coefficient $r$=0.31)
between $\gamma$-ray apparent luminosity and superluminal velocities of jet components,
as expected if the $\gamma$-ray emission originates in a very compact region of the 
relativistic jet and is highly beamed. In 43\% of the sources the jet bends by more than
20$^\circ$ on parsec scales, which is consistent with amplification by projection effects
of modest actual changes in position angle.

In 27 of the sources in the sample
there is at least one non-core component that appears to be stationary 
during our observations. Different characteristics of stationary features close to
and farther from the core lead us to suggest two different classes of stationary components: 
those within
about 2~milliarcseconds (mas) of the core, probably associated with standing hydrodynamical
compressions, and those farther down the jet, which tend to be associated with bends
in the jet.

\end{abstract} 

             
\section{Introduction}

Among the most significant accomplishments of the Energetic Gamma Ray Experiment
Telescope (EGRET) aboard the {\it Compton} Gamma Ray Observatory was the
detection of nearly 70 blazars at photon energies of 0.1--3 GeV
(Hartman et al. 1999; Mattox et al. 1997a). (Blazars here are defined as
quasars and BL Lac objects containing compact, variable radio sources with flat
cm-wave spectra.) One of the primary characteristics of blazars is strong
radio emission on scales less than 1 mas, usually in the form of a jet that
is observed with very long baseline interferometry (VLBI) on scales down to
$\sim 0.1$ mas.  It is commonly thought that the $\gamma$-ray emission occurs
in the jet, but closer to the ``central engine'' than the radio emission
(based on timescales of variability; see, e.g., Dermer \& Schlickeiser 1994).
An important question, therefore, is whether there is a relation between the
radio and $\gamma$-ray emission. For example, one might expect that a temporary
increase in the velocity and/or energy flux of the plasma injected into the jet
might first cause an outburst of $\gamma$ rays, followed by the appearance of a
knot propagating down the jet at a superluminal apparent speed. It could be
possible, on the other hand, that any such perturbation is damped by radiative
losses or dissipation before it reaches the radio-emitting portion of the jet.

A number of studies attempting to relate the $\gamma$-ray flux and emission at other
bands have established a significant correlation between
$\gamma$-ray flux and radio flux, and an even better correlation 
between $\gamma$-ray flux and radio flux contained in the compact jet
(e.g., Dondi \& Ghesellini 1995; Valtaoja \& Ter\"asranta 1996; Mattox et al. 1997a;
Zhou et al. 1997). Therefore, one might
expect the VLBI properties of $\gamma$-ray blazars to be distinct from those of
other subclasses of flat radio-spectrum, non-$\gamma$-ray AGNs,
especially in the innermost sub-milliarcsecond region of the jet close to the
central engine, where the $\gamma$-ray emission most likely originates.    

In an effort to establish whether correlations indeed exist between the properties
of the compact radio jets and the $\gamma$-ray emission, we have undertaken
an extensive study of changes in the structure of the compact radio jets of
EGRET-detected blazars with the Very Long Baseline Array (VLBA)\footnote{The VLBA
is an instrument of the National Radio Astronomy Observatory, a facility of the
National Science Foundation operated under cooperative agreement by Associated
Universities, Inc.}. This multi-epoch survey was carried out mostly at the highest
frequencies available with the VLBA (22 and 43 GHz) in order to obtain the highest
possible resolution. This allows us to probe the jets at sites as close as possible
to the central engine by imaging the radio emission on scales ranging from
0.1--0.3 to $\sim 10$ mas. Our sample contains 42 sources, roughly 60\% of the known
$\gamma$-ray blazars (Hartman et al. 1999) and covers 14 epochs over 3.5 years between
November 1993 and July 1997 (with each source observed only at some of these
epochs). Our sample size was increased as more blazars were reported to have
been detected by EGRET (von Montigny et al. 1995a; Mukherjee et al. 1997; Hartman et
al. 1999). Known $\gamma$-ray blazars were excluded from our sample based on
declinations that are too far south for the VLBA, positions not known with
sufficient accuracy for high-frequency VLBI, and a few cases of inadvertent
omission.

EGRET-detected active galactic nuclei represent $\sim 15$\% of known blazars. Since
the $\gamma$-ray emission is known to be variable in many of these sources (Hartman
et al. 1999), as many as $\sim 25$\% of all blazars would be expected to be
bright $\gamma$-ray sources (i.e., detectable by EGRET) some of the time
(cf. Mattox et al. 1997a). However, not all blazars are $\gamma$-ray bright.
The high $\gamma$-ray flux and short timescale of variability, combined with
the requirement that the pair-production optical depth be less than unity, suggest
that the $\gamma$-ray emission must be relativistically beamed (Mattox et al. 1993;
Wehrle et al. 1998). If the $\gamma$ rays arise from inverse Compton scattering
of either synchrotron photons from the jet or photons external to the jet, 
the EGRET-detected objects should be those with the highest Doppler beaming
factors. This implies high Lorentz factors of the jet flows and small
angles between the jet axes and the lines of sight (Lister \& Marscher 1999).
These are the conditions required for superluminal apparent speeds. Lister \& Marscher
therefore predicted that the distribution of apparent velocities of bright
features in the jets of $\gamma$-ray blazars should peak at high values. By
contrast, the motions observed in a radio flux-limited
sample of compact, flat-spectrum radio sources are heavily weighted toward
low apparent velocities (Pearson et al. 1998).
                        
This paper is a compendium of our multi-epoch VLBA observations of the sample of
42 $\gamma$-ray blazars. Our emphasis is on the presentation of the images
and model fits from this study, along with some analysis of the results.
Separate papers will present single-epoch polarization observations of most of the
sample (Marscher et al. 2001) and comparison of epochs of component
ejection with times of high $\gamma$-ray flux (Jorstad et al. 2001).

In what follows, we calculate the apparent transverse velocity, $\beta_{\rm app}$,
from standard Friedmann cosmology 
with deceleration parameter q$_o$=0.1, cosmological constant
$\Lambda$=0, and Hubble constant  $H_o$=100h~km~s$^{-1}$~Mpc$^{-1}$.
[Over the redshift range of the blazars in our sample, the result differs
by no more than 10\% from that obtained for a value of $\Lambda$ such that
$\Omega{\rm (matter)} = 0.3$ and $\Omega(\Lambda) = 0.7$.]
The formula (e.g., Pearson \& Zensus 1987) used is:

$$\beta_{app} = \mu\frac{z}{H_o(1+z)}\left[\frac{1+(1+2q_oz)^{1/2}+z} 
{1+(1+2q_oz)^{1/2}+q_oz}\right].$$

\section{Sample Selection}

The first session of VLBA observations, carried out at 8.4 and 22~GHz 
on 1993 November 8, included two of the brightest blazars, 3C~279 and 3C~273,
known previously as superluminal radio sources at lower
frequencies and identified as variable $\gamma$-ray sources (Thompson et al. 1993,
Fichtel et al. 1993). Starting in 1994 the sample contained
21 objects from the list  of $\gamma$-ray sources detected during the first EGRET
cycle of observation and
identified with AGNs (Michelson et al. 1993) and the 2~Jy sample of compact 
extragalactic radio sources suitable for millimeter VLBI 
observations (Valtaoja et al. 1992).  In 1995 the sample was expanded
to 38 sources and increased further to 43 objects in 1996. The selection criteria
were: (1) detection by EGRET (E$>$100~MeV; note that the flux limit of EGRET varied
greatly with position on the sky depending on the exposure time and the distance of
the object from the center of the field); (2) flux density at 37~GHz $\gtrsim$ 1~Jy
(allowing for variability); and (3) declination (J2000) $\ge$ $-$30$^\circ$.
The quasar 0954+556 was detected by the VLBA at 8.4~GHz (single epoch)
but not at 22 GHz; in addition, the source did not show structure smaller than 1~mas 
(Marscher et al. 2001). We therefore do not consider it to be a true blazar (it
appears similar to compact symmetric objects) and do not include it in our analysis.

Table 1 presents the entire sample of 42 $\gamma$-ray
blazars, which we observed at 22 and 43~GHz with the VLBA. (Note: only a
fraction of the sources were observed at both frequencies.) The contents of the 
table are as follows: (1) IAU source designation; (2) common name; (3) source type
(Q - quasar, B - BL Lac object); (4) redshift $z$, taken from the NED database;
(5) maximum $\gamma$-ray flux (E$>$100~MeV) in units of
10$^{-8}$~phot~cm$^{-2}$~s$^{-1}$, from the third EGRET catalog (Hartman
et al 1999); (6) $\gamma$-ray spectral index $\alpha$ [flux density 
$S_\nu\propto\nu^\alpha$] at
E$>$100~MeV, also from the third EGRET catalog; (7) maximum observed radio flux
density at 37~GHz, from the radio light curves obtained at the Mets\"ahovi
Radio Observatory from 1991 to 1997 (Ter\"asranta, private communication);
(8) radio spectral index at 22--37~GHz,
calculated at times of maximum flux in the Mets\"ahovi light curves at 22 and 37~GHz
from 1991 to 1997.
                                   
\section{Observations and Data Analysis}

We monitored the milliarcsecond scale structure of the
42 blazars with the VLBA at frequencies of
22, 43, and, at some epochs, 8.4 and 15.4~GHz during the period from
8 November 1993 to 31 July 1997 (14 epochs); only a subset of the sample 
was observed at each epoch and at each frequency.  
Prior to 1995, the partially-completed VLBA consisted of 5 to 7 operating
antennas. The early observations (November 1993 to the end of 1994) used the
Mk 2 VLBI recording system, which utilized four 2-MHz wide channels (IFs).
Starting in January 1995 the observations were carried out with the full
VLBA using four (1995--1996) or eight (1997)
8~MHz wide channels in both right and left circular polarization. Each source
was observed in ``snapshot'' mode with 3--6 scans of 5--15 minute duration.
Not all antennas were available at all epochs
owing to equipment failures or very poor weather.
Fig. 1 shows the typical {\it uv} coverage for
four sources with different declinations at one of the epochs, 
while Fig. 2 shows the {\it uv}-coverage for one object at four
different epochs.

\subsection{VLBA Data Analysis}

Initial correlation was carried out at the California Institute of
Technology (for all data prior to 1995) and the National Radio Astronomy
Observatory (NRAO) Array Operations Center in Socorro, NM.
The subsequent calibration was performed at Boston University with
the Astronomical Image Processing System (AIPS) software supplied
by NRAO, while imaging was performed at Boston University and St. Petersburg
State University using the Caltech software Difmap (Shepherd et al. 1994). 

The calibration included the following steps: (1) initial editing; (2)
nominal amplitude calibration using the system temperatures and gain
curves supplied by NRAO; (3) opacity corrections using
total flux densities from  Mets\"ahovi Radio Observatory 
at 22 and 37~GHz and the University of Michigan Radio Astronomy
Observatory at 8 and 14.5~GHz; (4) an initial fringe fit on a scan of
a bright source to apply a first-order correction to the residual instrumental
delay (frequency dependence of the phase); and (5) a global fringe fit
to all the data at that epoch to determine and correct for the residual
instrumental delays and rates (time dependence of the phase, which is also
affected by the Earth's atmosphere and ionosphere), with subsequent smoothing
of the solutions. The data were then split into single-source files while
averaging over frequency, and imported into Difmap, which was
used to perform the final editing and imaging.

Initial images were made in Difmap using the CLEAN algorithm (H\"ogbom 1974)
and self-calibration of the visibility phases and amplitudes in an iterative
procedure. These images
were then transferred to AIPS and used to self-calibrate the phases of the
pre-Difmap data files. (The phases of each polarization were therefore
self-calibrated separately, which is not possible in Difmap; this step was
unnecessary for the pre-1995 single-polarization observations.) The use of the
initial image rather than a point-source model in the first phase
self-calibration guards against the self-calibration routine in AIPS
(CALIB) setting noisy phases to an average value of zero. This effect was
further avoided by deleting data before self-calibration when the phases
appeared to be random (which implies nondetection of fringes).

After this procedure, the data were once again transferred to Difmap for
final imaging and self-calibration through an iterative procedure that results
in the production of a ``hybrid map'' (Cornwell \& Wilkinson 1981).
Uniform weighting of the {\it uv} data was usually used to obtain the highest 
resolution, although Gaussian tapering (downweighting of the longer-baseline
{\it uv} data) was employed during some of the iterations in order to
image better more extended emission. Since the nominal amplitude calibration was
inaccurate at a level of tens of percent, we applied a posteriori corrections
to the intensities by requiring that the total flux densities on images of
nearly unresolved sources match the single-dish values. Table 2 gives the
values of the intensity correction coefficients at 22 and 43 GHz.

The dynamic range (ratio of peak intensity to noise level) of the final images
depended on the {\it uv} coverage as well as on the quality of the data, which
can be affected by weather and problems with the receivers. In no case was the
theoretical thermal noise level reached.

\subsection{Model fitting}

Model fitting of the emission structure of each image was performed with Difmap
using the self-calibrated data files. We used the task `modelfit' to create
a model source using components with circular Gaussian brightness distributions.
The procedure started by initially fitting the flux density of the core (the
bright, compact component at one end of the structure in most sources, 
at most slightly resolved and usually the brightest feature on the image)
while setting the FWHM to 0.01 mas. Additional components were introduced one
at a time at the position of the next brightest feature on the residual image.
The program then iteratively adjusted the flux density and position of only the
newly added component until the value of the $\chi^2$ statistic was minimized.
The iterations were ended when the remaining features on the residual image
were at the noise level. A final set of 100 iterations was executed with all
parameters of all components allowed to vary.
For some sources with diffuse jet emission we fit a singular extended circular
Gaussian component to a feature whose structure on the hybrid map was complex.
In this case, we tapered the data (see above) before introducing this component
and fit its parameters while holding constant those of the other components.
The final iterations were always performed with no tapering and with the
parameters of the diffuse components held fixed.
For a number of sources the final values of $\chi^2$ characterizing
the goodness of fit of the model to the {\it uv} data are less than unity.
The majority of models have reduced $\chi^2$ values in the range from 1 to 1.5.
The models derived by the above procedure are listed in Table 4.

We used the model-fitting program SLIME, which runs in the AIPS
environment, to determine the errors in the parameters of the components
from the Difmap models. This program gives 1$\sigma$ errors for the 
parameter values under the assumption that the errors
have a Gaussian distribution and that the 
parameters are independent of each other (Flatters 1996).
To estimate the errors, a fully calibrated {\it uv} data file and
corresponding model were imported into  AIPS and loaded into SLIME.
The command  `FIT DATA' with the `LEAST-SQUARES TOOL' option was employed
with a maximum of 100 iterations to determine the uncertainties.
Usually we could not get a good result for the parameters of those
components of a Difmap-generated model that were insufficiently
constrained by the data owing to their faintness. (In some cases,
the component can be removed without resulting in an unacceptably
high value of the reduced $\chi^2$.) For this reason, for some
components in Table 4 uncertainties in the parameters are
not listed. However, our experience based on repeatability of faint
features on the hybrid maps from different epochs suggests that a
reasonable estimate of the uncertainties is on the order 15$-$20\% of 
the parameter values.

\subsection{Presentation Of Results}

The hybrid-map images of all sources at all epochs are presented in
Figures  3--38 in the form of time sequences of images. The parameters of
the images are indicated in Table 3. The 100\% contour level
corresponds to the maximum brightness on the images across all epochs.
The hybrid-map model, which is a grid of $\delta$-functions, is convolved 
with a beam corresponding to the average CLEAN beam over all epochs.
In general, deviations from the avarage do not exceed 1$-$2\% for
the major and minor axes and 20\% for the position angle of the beam;
however, in cases in which the position angle of the beam at a single epoch
was significantly different from the average, then its value was not
included in the averaging. Images of the same source at
different frequencies are presented separately with the exception 
of the sources 0716$+$714 and 0827$+$243, for which 
images at 22 and 43~GHz are presented together to illustrate
better the proper motions that we derive.
The contents of the columns of Table 3 are as follows: (1) IAU source 
designation; (2) frequency of observation in GHz; (3) number epochs of 
observation at the given frequency; (4) epoch of maximum intensity
of the map peak; (5) intensity of the map peak at this epoch
in Jy beam$^{-1}$; (6--8) major axis, minor axis, and position
angle of the average CLEAN beam in mas; (9) lowest contour level on
the image; contours are in factors of 2 up to 64\% of the peak
intensity, plus a contour at the 90\% level; (10) rms intensity of
the highest-intensity residuals of the hybrid map (i.e., the noise peak)
in mJy beam$^{-1}$; (11) figure number of the images.

Table 3 contains 36 sources of the 42 blazars in our sample. The images 
and parameters of model fitting for the remaining 6 sources
are presented in separate papers:   
0202$+$149 (Pyatunina et al. 2000), 0336$-$019 (CTA~26)
(Mattox et al. 2001), 0440$-$003 (NRAO~190) (Yurchenko, Marchenko-Jorstad, \& Marscher 2000),
1101$+$384 (Mkn~421) and 1652$+$398 (Mkn~501) (Marscher 1999), 
and 2200$+$420 (BL~Lac) (Denn, Mutel, \& Marscher 2000). However, for completeness,
Tables 5, 6, 7, and 8 (see below) contain the proper motions,
parameters of stationary knots, and jet geometry for the entire sample.  

The parameters of the components found in our model fitting procedure
are given in Table 4, with the same designations as the features
marked in Figures 3--38. The columns of Table 4 correspond to:
(1) IAU source designation; (2) epoch of observation; (3) frequency of 
observation in GHz; (4) designation of component; for all
sources the core is designated by the letter ``A''; (5) flux density of 
the component in Jy; (6) distance of the component from the core
in mas; (7) position angle of the component relative to the core
in degrees, measured from north through east; (8) size of the component in 
mas.

It is often difficult to identify components across epochs with confidence.
We base our identifications on the minimization of changes in the observed
parameters of a given feature: flux density, position angle, and position.
(The last of these is perhaps less important, since motions away from the 
core are expected.)
Nevertheless, the identifications are not always unique and the reader
can form his or her own opinion concerning their reliability.
The components are designated with letters
and numbers: ``A'' corresponds to the core in all cases; ``B'' followed
by a number designates a moving jet component at the main observed frequency,
with higher numbers correspond to later epochs of ejection or lower distance
from the core if the component was present on the earliest image; ``C'' designates
a stationary component, with lower numbers corresponding to closer distances
from the core; other letters (D, E, F, G) refer to special cases, such as multiple
frequencies or components that appear at one (non-initial) epoch at a
considerable distance from the core.

Comparison of images and model fits of each source at different epochs
allows us to classify components as either moving or stationary features
over our period of observation. The results of the classification are
summarized  in Tables 5 and 6. In addition, for each source in which we
detect proper motions, we present a plot giving the positions of the
components as a function of time relative to the presumed stationary
core. (Hereafter we refer to  these as ``PT'' figures: position vs. time).
Some weak, difficult to follow, components are not plotted on the PT
figures in order to prevent the graphs from appearing too messy.
Table 5 lists the proper motions $\mu$ derived from the results
of the model fitting (Table 4). In all cases these are
least-squares linear fits and are shown as solid lines in the PT figures.
Linear back extrapolation gives the time $T_o$ of zero separation of each
component from the core. 
                  
For some entries in Table 5, a sum of components is indicated
(for example, $D1+B3$); this means that there are observations
at two different frequencies (see Table 3), with the same (according
to our interpretation) component given a different label at each
frequency. In such cases the model-fitting results at both
frequencies are used to determine its proper motion. 
 
Table 6 lists the parameters of stationary features in the jets of many
of the sources.  The mean distance and position angle with respect to
the core, $\bar{R}$ and $\bar{\Theta}$, are arithmetic averages over
all epochs.   Positions of stationary knots
are given by dashed lines in the PT figures.

For some sources with multiband observations
the spectral indices $\alpha$ of the core (``A''), stationary (``S'') and moving 
(``M'') components are listed in Table 7, under the assumptions that the flux density
S$(\nu)\propto\nu^\alpha$ and that the positions are approximately
independent of frequency.

The geometric parameters of the jets, based on the  
hybrid maps and model fitting, are summarized in Table 8.
Here, $\theta$ is the mean position angle of
the jet within 1~mas of the core, $\Delta\theta_{\rm max}$ is
the maximum local bend observed, and  $R_{\Delta\theta}$ is the distance
from the core at which the maximum bend occurs. Contours below the
1\% level are not considered in the determination of the bending
so that sources with different image dynamic ranges can be compared.
The quantity $S_{\rm jet}/S_{\rm core}$ is the ratio of the flux density
summed over all jet components (as indicated by the model fitting)
to that of the core, calculated at the
epoch of minimum core flux density at the highest observed
frequency. To facilitate comparison of sources at different redshifts,
and therefore observed at different rest frequencies, the flux
densities of the jet components are
K-corrected under the assumption that the jet spectrum is a power law
with spectral index $\alpha$=$-$0.7. No K-correction was applied to the
cores, since their spectra tend to be flat.

\section{Notes On Individual Sources}

In order to relate the findings of our high-frequency VLBA observations
to the observed properties of the blazars in our sample, we summarize here
some relevant characteristics of each source as found by previous studies.
This review of the literature is by no means exhaustive, rather it is
meant to provide a guide for placing our results for each object
into a broader context.

{\bf 0219$+$428 (3C~66A):}
This BL Lac object has been observed extensively at optical wavelengths.
Analysis of the optical light curve suggests a periodicity of 65 days
in the bright state (Lainela et al. 1999), and microvariability
was detected by Xie et al. (1994). Price et al. (1993) have presented
VLBI maps at 6 and 20~cm, which show
an extended jet out to $6''$, with gradual bending from
PA$\sim $150$^\circ$ to 180$^\circ$.   
 
Our 43 and 22~GHz images (Fig. 3) reveal a jet
of complex structure out to 3~mas to the south of the core,
with a gradual change of PA from $-$150$^\circ$ to 180$^\circ$.
We model the jet with multiple knots, 
identification of which across epochs might not be unique. We favor
the interpretation shown in Fig. 3 (c) that gives a high superluminal
speed (up to 19$h^{-1}c$) of four components ($B2$, $B3$, $B4$, and $B5$). 
This is the first detection
of superluminal motion in this object. A proper motion of $\sim$0.68~mas~yr$^{-1}$
is tentantively found for component $B5$, but not indicated in Table 5.   
The moving components appear to pass through a stationary feature
(component $C$/$D3$ at 43/22~GHz) and
appear to be blended with the stationary feature at epochs 1996.34,
1996.60, and 1997.58. The model fits indicate that a new component,
$B6$, appeared in 1997.58. If we speculate that this component 
has a proper motion similar to the average value of the four
previously ejected components (0.8$\pm$0.2~mas~yr$^{-1}$), the ejection
of components would appear to have a quasi-period of about 0.6$\pm$0.2~yr. 
      
{\bf 0234$+$285 (CTD 20)}:
This quasar displays a prominent jet extending to the north 
out to 3.5~mas from the core on our 22~GHz images (Fig. 4). 
We model the jet with three main components,
all stationary during the two years of monitoring. 
The model parameters of component $C3$ were obtained from a tapered
image (see above); however, the component has complex structure
both along the jet and in the transverse direction.
The morphology is similar to that 
seen in the March 1997 15~GHz image of Kellermann et al. (1998) and the
extent of the mas-scale jet is no longer than that found by Wehrle et al.
(1992) at 5 GHz.
Components 2 and 3 in the July 1994 8.55~GHz image of Fey et al. (1996) 
have similar parameters to components $C2$ and $C3$, respectively. 
The model fitting indicates the presence of an unresolved component,
$C1$, near the core (at $R\sim$0.3~mas) at all epochs, the
reality of which needs to be verified by higher frequency
observations. Although no motion is detected in the 22 GHz images
during our observations, superluminal motion with a speed as high as
12~$h^{-1}c$ has been reported by Vermeulen \& Cohen (1994).

{\bf 0235$+$164}:
This BL Lac object exhibits violent variations at optical 
and radio frequencies (e.g., Chu et al. 1996).
Antonucchi \& Ulvestad (1985) detected no extended emission
above 0.1\% of the peak flux density of the core at 20~cm, although
a faint symmetric halo has been observed at lower frequencies
(0.3--0.4 GHz; Gopal-Krishna 1977; Stannard \& McIlwrath 1982).
The 5 and 3.6~GHz images of Gabuzda et al. (1992) and Gabuzda \& Cawthorne (1996)
did not reveal any jet. However, 22 GHz VLBI observations
by Jones et al. (1984) showed a component  0.7 mas from the core
at PA=$7^\circ$. Chu et al. (1996) interpreted differences in position
angle of components relative to the core at three epochs (ranging from
$\sim 47^\circ$ to $7^\circ$) as
evidence for superluminal motion with $\beta_{app} \sim 30$. Our 43~GHz
images (Fig. 5) confirm the presence of weak emission north of the core.
The jet is either strongly bent or has a wide opening
angle ($\sim 50^\circ$) within about 0.5~mas
of the core. We trace the proper motions of two components ($B1$ and $B2$)
at 4 epochs. These have bent trajectories and superluminal speeds
up to 30~$h^{-1}c$. The Lorentz factor would need to be at least 45, which
allows Doppler factors of 90 or more in places where the bent jet points
directly along the line of sight. This conforms with the
expectations of Doppler factors as high as 100 from radio variability
(Kraus et al. 1999).                                                                         

{\bf 0420$-$014 (OA~129)}:
Wagner et al. (1995) found that this quasar exhibits pronounced flares at optical
frequencies. They noted that the optical light curve is
consistent with the assumption of repeated outbursts at 13-month intervals and suggested 
that the flares are caused by knots of enhanced particle density propagating along a helical
trajectory in a precessing jet.  
At kiloparsec scales the source has structure directly to the south out to
25 arcsec and a weak secondary component $\sim 20''$ northeast of the core
(Antonucci \& Ulvestad, 1985). Britzen et al.(1998) investigated the VLBI structure 
at 3.6~cm  between 1989.32 and 1992.48 and found four jet components that 
move at apparent superluminal speeds of 7.9, 5.3, 3.9, and 2.1$ h^{-1}$~c,
decreasing as a function of distance from the core. Hong et al. (1999) reported two
superluminal components with apparent velocities of 1.9 $\pm$0.6 and
3.4$\pm 0.9 h^{-1}c$ at 5~GHz.

At 43~GHz the source consists of a broad jet to the south, with a flare to
the southwest between 1.5 and 2 mas of the core, similar to the structure found
at 15 GHz by Kellermann et al. (1998).
A weak feature (component $C2$, Fig. 6a) to the southwest appears to be a stationary
knot. Another, more prominent stationary component ($C1$) is closer to the core.
During 1996 the model fits reveal motion of component $B$ at $6.2\pm 0.6h^{-1}c$
toward $C1$, after which the two merge in 1997.58. Fig. 6b shows the images at the
last four epochs on a scale that accentuates the region near the core where 
component $B$ appears.

{\bf 0446$+$112}: We could find no previous published images of this quasar, in which
we detect a stationary component ($C$) 1.28$\pm$0.02 mas southeast of the core
(Fig.~7). At the first epoch of observation the modeling reveals
the presence of a component ($B$) close to the core. Given the sparse time
coverage between epochs of our observations, we cannot identify the component after
this. 

{\bf 0454$-$234}:
The 2.3 and 8.5~GHz images of Fey et al. (1996) show that this quasar
has a core-dominated structure with a weak jet out to 5~mas along PA$\sim -130^\circ$.
Our images at 43~GHz (Fig. 8) show an unresolved core, plus
a weak component ($B$) detected only in May 1996
at a similar PA as at the lower frequencies.
The faint structure to the east in 1995.47 is probably an artifact of the
noisy data and limited {\it uv} coverage.                                                                                                     

{\bf 0458$-$020}:
In the 15~GHz image of Kellermann et al. (1998; see also the 5 GHz image of Wehrle
et al. 1992), this quasar shows structure to the
northwest starting about 1.5 mas north of the core. Our higher resolution
43 GHz images reveal structure north of the core in the inner 1.5 mas plus a weak jet
to the northwest (Fig. 9a). Taken together, the 15 and 43 GHz images indicate that
the jet bends strongly, by about 60$^\circ$, between about 1 and 2 mas from the core.   
We derive an apparent speed of $9.1\pm 0.6h^{-1}c$ for component $B2$
over four epochs. 
 
{\bf 0528$+$134}:
The broadband $\nu S_{\nu}$ spectrum of PKS 0528+134 is dominated by the  
$\gamma$-ray emission, even when the quasar was at its lowest $\gamma$-ray state observed
by EGRET (Mukherjee et al. 1999). The VLBI structure of the source has been extensively
studied by Zhang et al. (1994), Pohl et al. (1995),  Britzen et al. (1998), and
Krichbaum et al. (1998).  Pohl et al.(1995) described the source structure at 22~GHz
as a one-sided bent jet of 5~mas length and find that the PA
of the jet ranges from $\sim$60$^\circ$ near the core to $\sim 5^\circ$--$20^\circ$
beyond 5~mas at 8~GHz. They derived steeper spectral indices at 8--22~GHz
for components that were farther from the core. 
Britzen et al. (1998) measured apparent speeds for four components 
(5.5$\pm$1.4, 4.7$\pm$0.4, 5.2$\pm$0.4, and 5.2$\pm 0.6~h^{-1}c$)
between 1986.25 and 1994.07 on the basis of geodetic VLBI observations.
Krichbaum et al. (1998) noted that new jet components seem to appear at
times of local minima in the 90~GHz light curve. Circular polarization 
in the parsec-scale radio jet of 0528$+$134 was detected at 15~GHz by
Homan \& Wardle (1999).

We have a thorough set of images at 43~GHz covering a period of 3.6~years. 
Five moving components with superluminal speeds as high as 23~$h^{-1}c$
are detected.  Our estimates of apparent speeds are significantly
higher than those reported by Britzen et al. (1998) at lower frequencies
but similar to that reported by Pohl et al. (1995)
at 22~GHz ($\beta_{app}$ up to $20h^{-1}c$). In addition, we find that the proper
motions of the components increase with distance from the core. 
The trajectories of all components show that the 
jet has a sharp bend of $\sim$ 70$^\circ$ within 
$\sim$1~mas of the core.  Comparison of the model fits at 22 and 43~GHz 
at epochs 1995.01 and 1995.15 leads to the identification of
components $F3$ and $F4$ at 22~GHz with
components $B1$ and $B2$ at 43~GHz, respectively.  
The average (over two epochs) values of
spectral indices  are $0.6\pm0.2$ for the core ($A$) and
$-0.3\pm0.3$~ and $-0.95\pm0.05$ for components $B1/F3$ and $B2/F4$, respectively.
The core spectrum is therefore inverted, as found also by Pohl et al. (1995).
It is interesting to note that component $B1$ has the same spectral index
as the similarly located (relative to the core) component $C1$ on the Pohl
et al. (1995) images. However, component $B2$, which is
closer to the core than $B1$, has a spectrum that is significantly steeper.
This is even more curious when one considers that the flatter-spectrum
component was ejected at about the time of the highest $\gamma$-ray flux
detected for this object, in March 1993 (Mukherjee et al 1999), when
radiative energy losses of the electrons might be expected to have been most severe.

{\bf 0716$+$714:}
This BL Lac object is one of the best studied examples of intraday variability.
Simultaneous radio, optical, UV, and X-ray monitoring yielded a short
duty cycle of variability at all frequencies, plus correlation
between the rapid variatons at different frequency regimes (Wagner et al. 1996).
The 1.6 GHz map of Polatidis et al. (1995) reveals a jet out to 10~mas from
the core at PA $\sim 17^\circ$. 
VLBA images at 5 and 8.4/22 GHz show
a compact, one-sided core-jet at PA$\sim 13^\circ$ (Gabuzda et al. 1998).
These authors also detected high linear polarization $\sim$50\% in the innermost jet 
component and estimated the apparent velocity to be $> 0.82\pm 0.11h^{-1}c$
(under the assumption that $z > 0.3$; Wagner et al. 1996),
which is similar to the result obtained by Witzel et al. (1988)
at 5 GHz on the basis of 2 epochs (1979.93, 1983.25) with a proper motion
of 0.09$\pm 0.11$~mas yr$^{-1}$.

Our images at 22 and 43~GHz (Fig. 11a) show a weak jet
at PA$\sim 10^\circ$. Although the cross-epoch identification of components
is not unique, we obtain consistent results (i.e., identification of all
components and small deviations from constant proper motions for each
component) if we adopt the interpretation shown in Table 4 and Figure 11b.
In this case, we trace components $B2$ and $B3$ over 4 epochs and identify
component $B2$ at 22~GHz with component $B2^*$ at 43~GHz. The proper motions
are then 0.9--1.2 mas yr$^{-1}$, or $> 11-15~h^{-1}$c. The lower proper
motions derived by earlier authors (see above) could then be explained
by inadequate time sampling leading to misidentification of short-lived components
across epochs. If we speculate that component $B5$, which appeared at the
last epoch, has a similar proper motion ($\sim$1~mas~yr$^{-1}$) as that of components
$B2$, $B3$, and $B4$, then the ejection of components appears to occur
with a quasi-period of $\sim 0.7$ yr (see Fig. 11b).

{\bf 0804$+$449 (OJ~508):}
Aller et al. (1999) found that this quasar exhibits variability on the shortest
timescale (in the source frame) --- about 22~days ---
of any of the active galaxies which they have monitored at radio frequencies
over more than 30 years.
A 5~GHz VLBI image by Pearson \& Readhead (1988) contains a very weak
jet out to 3~mas, with a sharp bend near the core
from PA$\sim$85$^\circ$ to PA$\sim 140^\circ$. The 8.4 and 22 GHz
maps of Bloom et al. (1999) show only weak emission east of the core.
The 15 GHz image of Kellermann et al. (1998) reveals a broad jet
ranging from east to southeast from about 1 to 3 mas from the core.
Our 22 GHz images suggest that the structure between
about 0.5 and 1 mas from the core corresponds to a twisted jet.
We detect no significant change
in the structure of the jet over a time span of 1.24 yr (Fig. 12).

{\bf 0827$+$243:}
VLA images at 20 and 6~cm (Price et al. 1993) show a bright core and a
secondary component  $\sim 8''$ from the core with a peak at PA$\sim -165^\circ$.
This component has a jet-like structure to the east-northeast. 

Our parsec-scale images at 22 and 43~GHz reveal a weak jet 
to the southeast out to 4~mas (Fig. 13a). Component 
$C$ seems to have a stationary position relative to the core
and appears to be blended at 1995.31 with a brighter 
component ($B1$) that was passing through the stationary feature.
Two detected moving components $B1$ (near the core) and $D2$ (farther out) have
similar high apperent speeds. The overall structure can be interpreted as a jet
that starts out to the southeast on mas scales
and bends on arcsec scales to the west-southwest, terminating in a hotspot.

{\bf 0829$+$046:}
This BL~Lac object exhibits strong intraday variability of polarization and 
total flux density at optical and near-IR wavelengths (Sitko et al. 1985, Smith et al. 1987).
A VLA image at 20~cm (Antonucci \& Ulvestad, 1985) 
shows a sharply curved jet extending to the south out to $30''$.    
A 43~GHz polarized intensity VLBA image (Lister et al. 1998) 
displays a well-defined jet extending to the east-northeast, starting at PA$\sim$62$^\circ$
and bending to PA$\sim 81^\circ$ at a distance of 1~mas from the core.
High fractional polarization, up to 17\%, is detected about 
0.6~mas from the core.

Our 22~GHz images show a prominent jet out to 6~mas to the east-northeast (Fig. 14a).
Components $C1$ and $C2$, the knots closest to the core, do not
appear to move over a one-year period. Components $B2$ and $B3$ have
similar apparent speeds of about 6$h^{-1}c$, although overall the proper
motions increase with distance from the core.

{\bf 0836$+$710:}
This high redshift quasar contains prominent extended structure, with
a secondary diffuse component located at a distance of $\sim 150$~mas
to the southwest of the core at 1.7 GHz (combined MERLIN and VLBI images
presented by  Hummel et al. 1992). The line between the core and this component
points toward diffuse structure 50~arcsec from the core.
On the basis of 5~GHz VLBI images, Krichbaum et al. (1990) 
found two moving components with significantly different proper motions
(0.23$\pm$0.05 and 0.14$\pm$0.05), plus a stationary feature 3~mas from the core
that did not show any motion between 1979 and 1983.
Otterbein et al. (1998), using multi-frequency VLBI observations
from 1993 to 1996, found a moving component with an extrapolated time of ejection 
close to the epoch of the strong optical flare and $\gamma$-ray detection.
  
Our 22~GHz images (Fig. 15b) show a curved jet extending to the south-southwest. 
We follow the evolution of moving component $B$, which consists
of two subcomponents ($B1$ and $B2$) at 43~GHz (Fig. 15a),
over 5 epochs. The proper motion (0.24$\pm$0.02) agrees
with that found by Otterbein et al. (1998) at earlier epochs.
Components $C2$ and $C3$ at 22~GHz are most likely
associated with components $C2^*$ and $C3^*$ at 43~GHz;
this appears to be a complex of emission that is stationary relative to the
core. It probably corresponds to the stationary feature
found by Krichbaum et al. (1990); if so, it has persisted for 17~years. 
A weak component ($C1$) near the core also appears to be a stationary
feature of the jet. Its double structure in 1996.90 at 43~GHz corresponds
to the appearance of new moving component ($B3$), which is blended
with component $C1$ at 22~GHz in 1997.58.
In this case, its proper motion ($\sim 0.28$~mas~yr$^{-1}$) is similar 
to the proper motion of component $B$.
A comparison of flux densities of the core and components $B$ and $C2+C3$
at 22 and 43~GHz at two epochs reveals that the core has 
a flat spectrum ($\alpha = 0.05\pm 0.15$), while the
spectral indices of the jet components are
steep ($-0.77\pm 0.25$ for $B$ and $-1.5\pm 0.4$ for $C2+C3$), with
a slope that increases with distance from the core.

{\bf 0851$+$202 (OJ~287):}
This well-studied BL Lac object is noted for the reported 12-year periodicity
in its optical light curve, with global outbursts described by
twin-peak structure (Sillanp\"a\"a et al. 1996).
A VLA image at 5~GHz shows a weak jet extending toward the west
(Kollgaard et al. 1992).  A 3.6~cm polarized intensity VLBI map by Gabuzda \& Cawthorne
(1996) reveals the presence of a parsec-scale jet out to 3~mas from the
core at PA$\sim -100^\circ$, with a component polarized at 11\% located
0.7~mas from the core. Gabuzda et al. (1989) detected apparent superluminal
motion ($0.20\pm 0.03$ and $0.27\pm 0.03$~mas~yr$^{-1}$).
Tateyama et al. (1999) report higher proper motions, ranging from
0.74 to 0.40~mas~yr$^{-1}$ (6 components) on the basis of 8.5 and 2.3~GHz
geodetic VLBI observations between October 1990 and December 1996.

Our 43, 22 and 15~GHz images show a jet extending to
1.5~mas west of the core, with a stationary component at 1.2~mas.
The model-fit parameters of this stationary component, designated at the
different frequencies as $C2$ (43~GHz), 
$E2$ (22~GHz), and $G2$ (15~GHz), are consistent and give 
a spectral index of $-1.3\pm 0.3$, averaged 
over two epochs. The core has an inverted spectrum
with $\alpha = 0.32\pm 0.02$. The 43~GHz images show
a bright component ($C1$) near the core, which appears to
be another stationary knot in the jet. This component
can be seen on the 22~GHz images and as a slight extension to the core
on the 15~GHz images. Component $C1$ has an optically thick spectrum,
with $\alpha = 0.59\pm 0.09$ between 22 and 43~GHz. 
Three moving components ($B1$, $B2$, and $B3$) with  
proper motion from 0.43 to 0.67~mas~yr$^{-1}$ are seen to travel between the
two  stationary components. The superluminal speeds of the components 
are similar to those obtained by Tateyama et al. (1999) but considerably 
higher than those found by Gabuzda et al. (1989); the latter may have resulted from
blending of moving and stationary components at the lower resolution
of their images.

{\bf 0917$+$449:}
The image from the first Caltech-Jodrell Bank VLBI survey at 18~cm
(Polatidis et al. 1995) shows that this quasar has a prominent bent jet to the southwest
at PA$\sim -160^\circ$ near the core and PA$\sim -140^\circ$ 
at a distance of 40~mas. The 5~GHz map from the same survey (Xu et al. 1995)
contains multi-component, curved structure out to 10~mas, with a 
weak component at 20~mas. The parsec-scale structure of the source is revealed
in the 15~GHz image of Kellermann et al. (1998), where the source has a
well-defined, straight 
jet to the south out to 3~mas and very weak structure at 7~mas to the southwest.

At 22~GHz (Fig. 17a) the source has a prominent jet to the south
out to 1.5~mas with what appears to be a stationary feature $C2$ at the end
of the jet. Due to the high resolution in the direction transverse to the jet,
we can distinguish through model fitting a stationary knot ($C1$) to the
southeast.  The superluminal motion of component $B$ is directed toward 
stationary feature $C2$.

{\bf 0954$+$658:}
The 5~GHz VLA map of Kollgaard et al. (1992) shows a curved jet to the
southwest with a bright feature at the end, about 4~mas from the core.
The polarized intensity 3.6~cm VLBI image of Gabuzda \& Cawthorne (1996)
indicates that the jet bends from
PA$\sim -14^\circ$ near the core to directly west out to 3~mas.
The maximum polarized flux is detected in a knot emerging from the core.
They identified two superluminal components with apparent speeds of 7.4$\pm 0.7$
and 4.4$\pm 0.7 h^{-1}c$. 
   
At 22~GHz (Fig. 18) the jet emerges from the core at PA$\sim -20^\circ$ and 
appears to execute a sharp bend  to the west
between 1 and 2~mas.  Despite the fact that during three epochs of
observation the structure of the jet is modeled by three stationary
features ($C1$, $C2$, and $C3$) relative to the core, it is possible
to assume that component $D1$ seen in the 1995.31 image was blended
with the stationary knot $C2$ in 1995.15, since the model fitting indicates
a significantly larger size of component $C2$ at this epoch.
In this case $D1$ has 
a superluminal motion $\sim 19h^{-1}c$ that is much higher than   
that measured by Gabuzda \& Cawthorne (1996).
However, we do not list component $D1$ as a moving knot in Table 5 owing to the
uncertainty of identification across widely-spaced epochs.    

{\bf 1127$-$145:}
A VLBI image of this quasar at 5~GHz (Wehrle et al. 1992) shows that the jet
emerges from the core at PA$\sim 81^\circ$ and bends at a distance
of about 3~mas into the direction of the kiloparsec-scale radio structure at
PA$\sim 52^\circ$ (Bondi et al. 1996). The 15~GHz map of
Kellermann et al. (1998) reveals a prominent knot at the position of the
bend, which appears to be rather sharp. Vermeulen \& Cohen (1994) did not find
any motion of this knot over two epochs at 5 GHz.

The 22~GHz images (Fig. 19a) show a well-defined jet to the east with 
a bright stationary feature ($C2$) at 4~mas from the core where the jet flares
to the northeast, in agreement with the image of Kellermann et al. (1998).
A weak component ($C1$) between the core and stationary knot $C2$ appears to be
stationary as well. Knot $B1$ is observed to move superluminally    
between stationary knots $C1$ and $C2$. Component $B2$, which is closer to the core,
has a somewhat slower proper motion. A new bright component ($B3$) appears to have 
been ejected
shortly before the last epoch of our observations.

{\bf 1156$+$295 (4C 29.45):}
McHardy et al. (1990) presented VLBI maps at 18, 6, 2.8, and 1.35~cm,
which show a jet emerging from the core along PA$\sim 20^\circ$ and then
bending to the east farther out. On their VLA and Merlin images the jet
extends out to $\sim 2"$ along PA $-19^\circ$, at which point it abrubtly
bends by almost $90^\circ$ to the northeast. McHardy et al. tentatively (based on
images at different frequencies) derived the
proper motion of the outermost VLBI component to be 1.15 mas yr$^{-1}$, which translates
to $30~h^{-1}c$.  Piner \& Kingham (1997b) identified four moving components at 8 and 2~GHz
with superluminal speeds of 7.7$\pm 1.8$, 4.4$\pm 1.6$, 8.7$\pm 1.5$, and
13.2$\pm 2.6h^{-1}c$ (q$_\circ$=0.1).

Our data at 22~GHz (Fig. 20a) reveal that
the jet either has a wide opening angle or a changing flow direction,
since two moving components ($B2$ and $B3$) are observed at different
position angles with $\Delta$PA$\sim 40^\circ$ at the same distance
from the core. Component $B3$ seems to be merging with stationary 
knot $C$ at epoch 1997.58, based on the significant
increase in brightness of component $C$ at that epoch.
The results of model fitting give the motion of the component ($D1$) farthest 
from the core as $\beta = 15\pm 4h^{-1}$, although
we do not list it in Table 5 since it is very weak, with size comparable to the
change in distance from the core. We detect motion in two other components at
superluminal speeds similar to those found by Piner \& Kingham (1997b). 
Piner \& Kingham suggested that the jet oscillates in position angle from 0$^\circ$ 
near the core to about 20$^\circ$ 1 mas  downstream and then back again
to $\sim -20^\circ$ at 3 mas.
Our data suggest that the trajectories of components can fall anywhere in
the range of position angles from $-25^\circ$ to $+25^\circ$ out to a
distance of $\sim$1~mas from the core.

{\bf 1219$+$285 (ON~231):} 
The 5~GHz VLA map of this BL Lac object by Kollgaard et al. (1992) shows very faint
extended emission to southwest out to 10 arcsec, while
the 5~GHz VLBA image of Gabuzda et al. (1994) reveals 
a complex and heavily resolved jet toward the east
at PA$\sim 100^\circ$. The 15~GHz VLBA image of Kellermann et al. (1998)
contains a long, thin jet to the east with a prominent extended 
knot at the end at a distance of about 10~mas from the core. Based on two epochs
of observation, Gabuzda et al. (1994) find a series of superluminal components at 5~GHz,
with proper motions of 0.14, 0.18, 0.28, 0.37, and 0.55~mas yr$^{-1}$ in order of
increasing distance from the core. 

At 22~GHz (Fig. 21a) the source has a multi-knot jet along PA $\sim 105^\circ$,
which shows a bend toward the south at a distance of $\sim 2.5$~mas from the core.
A comparison of the nearest epochs (1996.90 and 1997.58)
indicates a significant change in the jet, but with such limited data we
cannot identify components unambiguously across epochs.
Fig. 21b  shows two different possible interpretations.
One minimizes the proper motions (solid lines in Fig. 21b; these 
results are given in Table 5); the estimated proper motions are then similar
to those obtained by Gabuzda et al.(1994) at 5~GHz
and also show an increase of proper motion with distance from the core
(0.13, 0.32, 0.60, 0.47, and 0.50~mas yr$^{-1}$, respectively).
However, it is also possible that the proper
motions are higher ($\sim$1.5~mas yr$^{-1}$, dashed lines in Fig. 21b). In this case,
the component identified with B3 (1995.31), B5 (1996.90), and B4 (1997.58)
would have an extrapolated time of zero separation from the core at
(1995.1$\pm$0.2), close to the epoch of
a $\gamma$-ray flare in 1995.334 (Hartmann et al. 1999). 

{\bf 1222$+$216 (4C~21.35):}
Price et al. (1993) observed this quasar with the VLA at 5~GHz,
finding a jet that is bent to the northeast with a secondary extended component
at a distance of about $12"$ from the core.  Based on two-epoch (1986.42 and
1988.24) VLBI observations at 5~GHz, Hooimeyer et al. (1992) derive an apparent speed
of $1.6\pm 0.6h^{-1}c$ for a component located at PA$\sim -14^\circ$.
 
At 22~GHz (Fig. 22a) the source has a prominent jet toward the north
at PA $\sim -10^\circ$. Although we only observed 1222$+$216 at two epochs,
it is clear that two bright knots ($B1$ and $B2$)
separated from the core at different speeds, which are significantly higher
than the velocity reported by Hooimeyer et al. (1992).    

{\bf 1226$+$023 (3C~273)}:
This relatively nearby object is one of the brightest flat-spectrum radio sources
and the brightest X-ray quasar. As such, it has been extensively studied at a number
of wavelengths (e.g., McHardy et al. 1999; von Montigny et al. 1997).
It is well known to have superluminal motion (see Porcas 1987), which
extends out to at least 120~pc at 1.7 GHz (Davis et al. 1991).
According to Homan \& Wardle (1999), the core of 3C~273 contains significant
circular polarization.
                   
In our program the source was observed at 3 frequencies, which allows us 
to trace the jet out to 16~mas at 8.4~GHz and 8~mas at 22~GHz.
The structure of the jet is similar to the description by
Mantovani et al. (1999), who observed 3C~273 at 22 and 43~GHz
during 42~days in December 1992 and January 1993 to search for short-term variability. 
In our images we identify 7 moving components from $\sim$0.5~mas to 8~mas
from the core. To determine the proper motions
of these components, we combine the results of our model fitting at different
frequencies with that of Mantovani et al. (1999), whose model fits
appear to be compatible with ours.  The results are shown in Fig. 23(d,e), 
where Fig. 23d presents the separation of components
from the core in the innermost part of the jet out to 4~mas, while
Fig. 23e presents the same for the segment of the jet between 4 and 10~mas. 
The values of the proper motions cover a wide range
from 0.3~mas~yr$^{-1}$ to 1.6~mas~yr$^{-1}$, with a tendency toward faster
motions farther from the core. 

{\bf 1253$-$055 (3C~279):}
This quasar, considered the first superluminal source (Whitney et al. 1971),
is often nearly as bright as 3C~273 at high frequencies. Since it is also a bright
$\gamma$-ray source (Hartman et al. 1999), 3C~279 has been the target of a number
of campaigns of contemporaneous monitoring from $\gamma$-ray to radio
frequencies (e.g., Wehrle et al. 1998, Grandi et al. 1996). In addition, it was
the first active galactic nucleus with circular polarization in
the radio core measured by VLBI (Wardle et al. 1998).
 
We performed observations at 3 frequencies (43, 22, and 8.4~GHz)
from November 1993 to July 1997 (Fig. 24a, b, and c, respectively) .
A comparison of model-fit parameters at the different
frequencies at the same epochs shows that at 8.4$-$43~GHz
the core spectrum rises toward higher frequencies
($\alpha$=0.24$\pm$0.02), and that long-lived
western component $D$ has the spectrum of an optically thin
synchrotron source ($\alpha = -0.87 \pm 0.15$), which is also
the case for another moving knot labeled as $E2$ at 22~GHz and
as $B1$ at 43~GHz ($\alpha = -0.77\pm 0.05$). 
From the model fitting we derive a higher proper motion for
component $D$ ($6.4h^{-1}c$) than was found at 22~GHz ($4.3h^{-1}c$)
by Unwin et al. (1998). Wehrle et al. (2001), on the other hand obtain an
intermediate value of
$5.25\pm 0.3h^{-1}c$. From the long-term separation vs. time plot of Wehrle
et al., it is apparent that our observations covered a period when the proper
motion of the centroid of this component was temporarily faster than average
after a slower-than-average period. Comparison of the proper motions of
four components suggests a possible increase of apparent
speed with distance from the core. A weak, stationary component 
($C$) at 43~CHz can be identified with component $F$
at 22~GHz and, probably, with component $G$ at 8.4~GHz.     

{\bf 1406$-$076:}
Ours are the first published images of this source.
At 22~GHz (Fig 25a) this quasar has a weak bent jet to the west 
out to $\sim$1~mas. The two moving knots show significantly different 
proper motions. The component with higher proper motion gradually turns
from west-northwest to west-southwest respect to the core, with a total change in
PA of about 35$^\circ$. 

{\bf 1510-089:}
This is a highly polarized quasar at optical wavelengths, with rather bright and flat-spectrum
X-ray emission (e.g., Singh et al. 1997). 
VLA images at 20 and 6~cm (Price et al. 1993) reveal an unresolved core
and a secondary component about $8''$ to the southeast.
The 15~GHz image of Kellermann et al. (1998) reveals a short jet
extended to the north and then bending to the northwest within 2 mas of the
core. In the VLBI image at 1.67 GHz presented by Bondi et al. (1996) the
dominant component lies to the north, which suggests that the core is
faint at this frequency. That map also contains faint secondary components
to the north-northwest and south-southeast that are probably artifacts caused by
small calibration errors. 
 
At 43~GHz (Fig. 26a) the jet has complex diffuse structure out to 1.5~mas.
This seems to be in the form of a
broad jet that appears to bend from PA $\sim -40^\circ$
near the core to PA $\sim -10^\circ$ beyond 1~mas, consistent with the
lower resolution 15~GHz image of Kellermann et al. (1998).
Component $C$ at a distance $\sim$0.6~mas from the core
is probably a stationary feature; its model-fit
parameters are similar to those of component 2 in the
8.55~GHz map of Fey \& Charlot (1997) obtained in April 1995.
In the last two epochs, moving component $B1$ appears to merge with
component $C$. The model fit for epoch 1997.58 indicates the presence of
a new component ($B2$) near the core. Our model fitting suggests a double
structure for the farthest component, $D1$ and $D2$, with different proper
motions and therefore different times of zero-separation. However, this
region is heavily resolved, leading us to consider these to belong to
the same, complex feature.

{\bf 1606$+$106:}
The VLBI structure of this quasar was investigated at 2, 8, and 15~GHz from 1992
to 1997 by Piner \& Kingham (1998).
Their images at 8~GHz show a jet with a sharp bend at a distance of 
1~mas from the core from PA$\sim -80^\circ$ to PA$\sim -40^\circ$.
These authors found four moving components ($C1$, $C2$, $C3$, and $C4$) 
with apparent speeds of $8.0\pm 5.9$, $6.1\pm 6.0$, $3.4\pm 1.7$,
and $2.4\pm 1.1~h^{-1}c$, respectively, plus stationary component $C5$.

At 22~GHz (Fig. 27) the jet has a sharp bend from the west to 
the north at about 1~mas from the core. 
Model fitting indicates a weak knot ($C2$) located at the position where the jet
bends, whose distance from the core is constant within the uncertainties.
However, its position angle differs significantly in 1996.90 from that of the other three
epochs. Our data also do not indicate any motion for components $C1$
and $C3$. It seems likely that component $C1$ is the same as
component $C5$ of Piner \& Kingham (1998),
and that their component $C4$ is a superposition of our components
$C2$ and $C3$.  The proper motion measured by Piner \& Kingman for $C4$
is determined mainly from the first two epochs, without which
it would be difficult to draw a conclusion about its motion. Furthermore,
according to these authors the later observations have better $uv$-coverage.
No motion is detected in our images. 

{\bf 1611$+$343 (DA~406):}
At 1.7 GHz, VLBI maps of this quasar by Bondi et al. (1996) show only a single
compact component. No flux density or structural variations were detected over
three epochs (1980.1, 1981.8, and 1987.9), although this source is a well-known
variable at lower frequencies (Cotton \& Spangler 1979; Altschuler et al. 1984).
Altschuler et al. (1984) detected only modest variability at 1.4 GHz, however.
The 2 and 8~GHz VLBI images of Piner \& Kingham (1997a) reveal
a complex jet extending south from the core.
These authors used model fits to measure the apparent speeds of four components to
be $11.5\pm 2.3$, $7.6\pm 1.3$, $3.8\pm 1.4$, and $6.7\pm 1.6~h^{-1}c$,   
(from the innermost component outward).

We present a sequence of 9 images of the object at 22~GHz (Fig. 28b)
and two images each at 43 and 15~GHz (Fig. 29a,c). These show a thin jet to the
south, ending in a bright
diffuse component 2.9~mas from the core at the site of a sharp bend to the east
of as much as 90$^\circ$. Our model fitting of the 22~GHz data finds two stationary
features: $C1$ near the core and $C2$ at the site of the sharp bend of the jet.
Two components ($B4$ and $B3$) move between these stationary knots,
while two other moving components ($B2$ and $B1$) lie beyond $C2$. 
The model fitting indicates that  components
$B2$ and $B1$ are blended with component $C2$
at the first epoch, 1994.76, and resolved from it at the next epoch.
Although these components move at different proper motions
(and, therefore, have  different times of zero separation
from the core if we extrapolate linearly) according to the model fits,
the sequence of images gives the impression that they are features of the same
disturbance in the jet. The proper motions of the innermost components are significantly 
higher than those of the outer knots, which agrees with the results obtained
by Piner \& Kingham (1997a), although our values of apparent velocities
are somewhat higher.    
The bright knot $C2$ at 22~GHz is identified with components $E5$ and $F3$
at 43 and 15~GHz, respectively. This allows us to estimate the spectral index 
of this stationary knot between 15 and 43~GHz to be $\alpha = -1.5\pm 0.3$ , which is
quite steep.  
The model fitting  at 15~GHz shows a large and bright component ($G1$),
which can be identified as the combination of components $B1$ and $B2$, 
while at 43~GHz component $D1$ appears to correspond to component $B2$.
Component $B1$ seems to be too weak to detect at 43~GHz.
According to this cross-frequency identification, component $B2$ has 
a spectral index $\alpha = -1.32\pm 0.03$ between 22 and 43~GHz. 
 
{\bf 1622$-$253:}
According to Perley (1982), this quasar has a weak secondary component at
PA $-57^\circ$ at $2''$ and ``diffuse secondary'' emission at PA $90^\circ$
at $6''$ at 1.4 and 5 GHz. VLBI maps at 2 and 8~GHz presented by Fey et
al. (1996) show very weak jets to the northwest and north, respectively.
The VLBA image at 8.3~GHz obtained in February 1997 by Tingay et al. (1998)
displays a core-dominated compact source with a weak jet-like
structure extending out to 2~mas from the core along PA$\sim -54^\circ$.
 
At 22~GHz (Fig. 29a) our three epochs of VLBA observations over 1 yr reveals
that the flux density of the core varied by a factor of 5.
The images contain a moving component, $B$, whose PA 
rotates from west to northwest with greater separation
from the core as it moves at a high apperant speed $\sim 14~h^{-1}c$.
This is the first detection of superluminal motion in this object.
The location of component $B$ in 1997.58 appears to agree with
the position of a feature in the 8~GHz image of Tingay et al. (1998).
With tapering of the data and model fitting, we detect a diffuse 
component ($D1$, $D2$, and $D3$, with the numbers indicating epoch) to the
north at a distance $\sim$1.5~mas from the core.  
It appears that component $B$ was turning
toward the direction of this diffuse component during our observations. 

{\bf 1622$-$297:} 
This source is the brightest, most luminous $\gamma$-ray blazar ever 
detected (Mattox et al. 1997b). During the period of observation
by EGRET from 6 June 1995 to 25 July 1995
the source showed the most rapid GeV $\gamma$-ray flux change yet seen 
for any blazar, with a doubling time $< 3.6$~hr.
Perley (1982) found weak secondary structure (at PA $22^\circ$, $14''$ from
the main emission) at 1.4 GHz. A VLBA image of the source at 1.65~GHz by
Tingay et al. (1998) contains a jet along 
PA$\sim -69^\circ$ consisting of a prominent component approximately
15~mas from the core and diffuse structure out to 30~mas. The VLBA image at
5~GHz obtained by the same authors reveals a strong component $\sim$4~mas from
the core.

Our 15~GHz images (Fig. 30c) show a weak, stationary component ($G$)
located 15.4$\pm$0.1~mas from the core at PA$\sim -69^\circ$. 
We find components ($F2$ and $F1$) that move with apparent speeds of
$\sim$10$h^{-1}c$ (Fig. 30e) at distances of $\sim$ 3 and 5~mas from the core.
Inside 1.5~mas from the core, the jet consists of three components
at 15~GHz ($E3$, $E2$, and $E1$), and 2 components at 22 GHz ($D2$, $D1$)
and 43~GHz ($B1$, $B2$). Comparison of the model-fit parameters of the
different features allows us to identify component $B2$ (43 GHz) to be
the same as $D2$ (22 GHz) and $E3$ (15 GHz).  We further identify component
$B1$ (43 GHz) with $D1$ (22 GHz) and $E2$ (15 GHz) (see Fig. 30d).
Component $E1$ and feature $B1$/$D1$/$E2$ have proper motions similar to
those of components $F1$ and $F2$ farther down the jet.
The model-fit parameters of the innermost components
reveal significantly different superluminal speeds
($\sim 2$ times less) than for the downstream components.
Our simultaneous observations at three epochs (1996.34, 1996.76,
and 1997.19) at 15, 22, and 43~GHz allow us to estimate the 
spectral indices of the core and jet components.
The core has an inverted spectrum 
($\alpha = 0.67\pm 0.03$ and $0.25\pm 0.04$ in 1996.76 and 1997.19,
respectively) with the flatter spectrum at the higher 
flux density. In 1996.34 component $E3$ appears to be
blended with the core at 15~GHz. Component $B1$ has a steep spectral index 
($-0.91\pm 0.05$); this is also the case for component $B2$
($\alpha = -0.86\pm 0.08$).

{\bf 1633$+$382:}
A 1.7 GHz VLBI image of this quasar by Polatidis et al. (1995)
shows a prominent jet to the west out to 50~mas,
where a sharp bend to the south is observed.
The 2.3~GHz VLBI map by Fey \& Charlot (1997) contains a weak 
jet out to 20~mas. The 8~GHz image by the same
authors reveals curvature of the jet within 5~mas 
of the core from PA$\sim -85^\circ$ to $\sim -62^\circ$.
A proper motion of $0.16\pm 0.03$~mas yr$^{-1}$ was determined
by Barthel et al. (1995) on the basis of three epochs
(1979.25, 1984.40, and 1986.89) of VLBI observations at 5~GHz.
 
Our 22~GHz images (Fig. 31a) show a prominent jet
out to 1.5~mas along PA$\sim -87^\circ$ and reveal
motion of bright component $B1$ throughout 6 epochs of observation.
Model fitting indicates the presence of a component ($B2$) upstream
of component $B1$ with significantly lower apparent speed. Component $B1$
has somewhat higher proper motion than detected previously for this
source. However, a new component ($B3$), which is seen during the last
three epochs, has a proper motion
corresponding to the estimate of Barthel et al. (1995).   
 
{\bf 1730$-$130 (NRAO~530):}
The 1.7~GHz VLBI map of this quasar by Bondi et al. (1996) shows a strong secondary
component 25~mas from the core at PA$\sim 0^\circ$, while the structure seen on 
the 8.4~GHz image by Tingay et al. (1998) consists of a diffuse jet out to 5~mas 
along PA$\sim 15^\circ$. In 1995 the source underwent a dramatic radio 
outburst with amplitude higher than any found over 30~years of monitoring
(Bower et al. 1997).
According to these authors, their VLBI observations at 86 GHz reveal
the creation of new components in the jet that move to the southwest at
apparent velocities of 7.9 ($C2$, their notation) and $7.4~h^{-1}c$ ($C1$). 
They associate the creation of component $C1$ with a radio flare
observed in mid-1994 and the creation of their component $C2$
with the 1995 outburst.

Our 43~GHz images (Fig. 32a) were obtained at epochs during and after the 1995 flare.
Based on better time sampling, we interpret the evolution of the source in a
different manner from Bower et al. (1997). The time development is simpler if
it we assume that the component ejected during the big flare in 1995 is the
brightest feature on the VLBI images over at least the next two years.  In this case,
the structure and motion are both roughly to the north. The Bower et al. 
interpretation (which assumes that the brightest feature is always the core), on
the other hand, requires
trajectories that are bent by nearly $180^\circ$ to connect with the structure we
find north-northeast of the core on our images.
According to our interpretation, a very strong component ($B2$) was ejected along
PA$\sim 30^\circ$ at a superluminal speed similar 
to the velocity indicated by Bower et al. (1997) for their component $C2$. 
The innermost jet appears to be bent by about 40$^\circ$ and a second moving
component ($B1$, which can be identified with component $D3$ at 22~GHz)
is seen at PA$\sim -10^\circ$. The extrapolated time of zero separation from the core
of component $B1$ associates it with the radio flare in the middle of 1994.
Our 22~GHz images (Fig. 32b) contain a stationary feature ($C$) to the north at a
distance $\sim 1.4$~mas from the core, beyond which the jet appears to wiggle through
one more turn of about 25$^\circ$ toward the direction of a weak, diffuse component
($D1$). The diffuse component is absent in the lower dynamic-range images of 1994.48 and
1994.59, and appears at a distance $\sim$3.5~mas from the core in the 1995.15 image,
just before the flare at mm-wavelengths.  The proper motion of this feature from the
model fits is quite high over the three epochs of observation,
with an apparent speed of $29\pm 9~h^{-1}c$, which corresponds 
to an extrapolated time of ejection 1991.4$\pm$1.2.
Bower et al. (1997) also detected an expanding ``halo''
with an apparent velocity of about 26 $h^{-1}c$ between April 1994 and
April 1995. In addition, they found a correlation between the rise in $\gamma$-ray
flux and increase in radio flux from 8.0 to 90 GHz that implies an association 
of $\gamma$-ray activity with the creation of the superluminal VLBI component.
Our 43~GHz data show that 
the epoch of zero-separation of component $B2$ coincides with
the epoch of the $\gamma$-ray light curve peak within $\sim$2~weeks.
     
{\bf 1739$+$522:}
The 1.7 GHz VLBI map of this quasar by Polatidis et al. (1995) contains  very
weak extended structure to the east up to 10~mas and, perhaps, a faint secondary
component to the northwest 20~mas from the core.
A 2.3~GHz VLBI image reveals a weak, thin structure out to 15~mas
along PA$\sim 28^\circ$ (Fey et al. 1996). The 8.55~GHz map of these authors
at the same epoch shows a jet-like structure curved from PA$\sim 36^\circ$ near
the bright core to PA$\sim -10^\circ$
beyond 1~mas. The 5~GHz image by Pearson \& Readhead (1988)
contains a dominant core plus a weak component at PA$\sim 3^\circ$ 
1.7~mas from the core.

Our images at 22~GHz (Fig. 33) during the last two epochs of our program
are also dominated by the unresolved core.
The lower dynamic-range 1993.86 image contains a short jet along PA$\sim 77^\circ$
plus the suggestion (near the noise level) of a sharply curved extension
to the north.  Our images indicate strong variability of the intensity of
the jet, but the absence of features that can be followed negates any
possibility of estimating the jet velocity.
However, a comparison of the results of model fitting of our 
1993.86 observations with the Gaussian model of the 8.55~GHz image at epoch
July 1994 presented by Fey et al. (1996) suggests that our component $B1$ in 1993.86
might correspond to their component 2 at 8.55~GHz. If we make this identification
and assume no frequency-dependent gradients in separation from the core, we can
make a very tentative estimate of the proper motion of $\sim$0.56~mas yr$^{-1}$,
which corresponds to a high apparent speed, $\sim 24~h^{-1}c$ (not included
in Table 5 or in our subsequent analysis).

{\bf 1741$-$038:}
Perley (1982) found no arcsec-scale structure ($< 0.3$\% of the brightness peak)
in this quasar at 1.4 and 5 GHz. 
The source has been imaged at 2.3 and 8.55~GHz at two epochs:
July 1994 (Fey et al. 1996) and October 1995 (Fey \& Charlot 1997).
These images indicate the presence of a component (labeled by the authors as ``2'') 
to the southwest at a distance of 0.3 and 0.4~mas from the core, respectively.

Our 22~GHz images (Fig. 34) show a weak southern component ($B1$), which is present 
at both epochs. Our first epoch is very close to the date of the observation by 
Fey \& Charlot (1997); $B1$ lies at the same position angle as their component $2$.  
A two-epoch determination of the proper motion of $B1$ gives the value of
$\sim 0.24$~mas yr$^{-1}$, which corresponds to an apparent speed of $\sim 8.3~h^{-1}c$.
However, with only two epochs of observation, the equatorial position of the source
(which causes calibration errors to create artifacts that appear to be significant
components), and a very weak jet, we do not consider this result sufficiently
sound to list in Table 5.  On the other hand, if we adopt this value and
assume that the viewing angle is that required to maximize the apparent superluminal
motion, we obtain a viewing angle ($4.5^\circ$) and Doppler factor ($\sim 13$) similar
to those obtained by Wajima et al. (2000) from two-epoch VSOP observations at
1.65 and 5~GHz.

{\bf 1908$-$201:}
The source is a BL Lac object with unknown redshift (but $z \ge 0.2$ based on the
absence of a detectable host galaxy).
Perley (1982) finds no arcsec-scale structure at 1.4 and 5 GHz
at brightness levels less than 0.8 and 0.4\%, respectively.
A VLBI image of the source by Tingay et al. (1998), obtained
in February 1997 at 8.4~GHz, shows jet-like structure
out to 3~mas from the core along PA$\sim 52^\circ$. 

Our 22~GHz images (Fig. 35a) show  a sharp bend in the jet by $\sim 60^\circ$ 
at $\sim 1$~mas, plus rather 
messy looking structure beyond 2~mas from the core. Our model fitting
indicates a possible stationary feature ($C$) at a distance $\sim 0.5$~mas
from the core and a component ($B2$) with changing PA, which separates from
the core at a high apparent speed. 
This is the first detection of superluminal motion for this source.

{\bf 2209$+$236:}
This object is characterized by the NED database
as a quasar with unknown redshift. The first VLBI
image of the source was published by Kellermann et al. (1998), whose
15~GHz image contains a weak bent jet extending northeast from the core.
Our 22~GHz images (Fig. 36) show a bright core with a weak jet along
PA$\sim 16^\circ$. Three epochs of observation reveal a stationary feature 
1.4~mas from the core that underwent a slight increase in brightness as the
core faded. 

{\bf 2230$+$114 (CTA102):}
Perley (1982) find a single secondary (3.6\% of peak intensity) component
1.6 arcsec from the main emission at PA $140^\circ$ at 1.4 GHz in this quasar.   
Antonucci \& Ulvestad (1985) confirmed the presence of secondary
structure to the southeast.
A 32~cm VLBI image by Rantakyr\"o et al. (1996) contains three components:
a bright double knot containing the core and
a jet component as well as a faint feature to the north, which also was
detected by Antonucci \& Ulvestad (1985).
A VLBI image at 8.55~GHz by Fey \& Charlot (1997) shows a prominent
jet to the southeast out to 20~mas, with a sharp bend by up to 90$^\circ$
at a distance of about 10~mas. Miltiple twists of the jet are apparent 
in the 15~GHz map of Kellermann et al. (1998). A very high proper motion
(0.65$\pm$0.15~mas yr$^{-1}$) was found at 932~MHz by B{\aa}{\aa}th (1987),
although Wehrle \& Cohen (1989) found no significant motion near the core 
at 5~GHz, with an upper limit of 0.5~mas yr$^{-1}$. Rantakyr\"o et al. (1996)
detected 4 moving components with proper motions from $0.3\pm 0.6$
to $0.6\pm 0.2$~mas yr$^{-1}$; this suggests a possible increase in
apparent motion as a function of distance from the core. 
 
Our tapered 43~GHz images (Fig. 37b) allow us to follow the jet out
to 8~mas from the core. Our model fits indicate that
a new component emerged from the core at PA $\sim 130^\circ$ and 
moved toward the direction of component $C$ (PA$\sim 140^\circ$),
which has complex structure (Fig. 37a) and seems
to be stationary over all our epochs of observation, as well as compared 
with previous observations (Rantakyr\"o et al. 1996; Wehrle \& Cohen 1989).
Component $C$ is located at a site where the jet bends sharply by about 55$^\circ$
into the direction of a weak component ($D1$). At $D1$, the jet turns to the east
through a bend of $\sim 35^\circ$ toward the direction of component $D2$, which
is located at a distance of $\sim$7.3~mas. Components $D1$ and $D2$ do not
show any motion during our observations. Superluminal motion is detected for three
components near the core: $B1$, $B2$, and $B3$ (see Fig. 37c).
Components $B1$ and $B2$ have a similar apparent speed while 
component $B3$, which is the nearest to the core, has a lower 
proper motion, which agrees with the results of Rantakyr\"o et al. (1996). 
Note that a linear back-extrapolation of motion of components $B1$, $B2$, and $B3$
suggests that knots are ejected quasi-periodically at intervals of $\sim$0.9~yr.  

{\bf 2251$+$158 (3C~454.3):}
This quasar is a bright, violently variable source whose mas-scale structure has
been observed extensively with VLBI
(Pauliny-Toth et al. 1987; Cawthorne \& Gabuzda 1996;
Kemball et al. 1996; Marscher 1998;
Pauliny-Toth 1998; Kellermann et al. 1998; G\'omez et al. 1999). 
The $\sim 5$ yr VLBI monitoring program of Pauliny-Toth et al.
(1987) at 10.7 GHz found superluminal components with 
proper motions between 0.21 and 0.35~mas yr$^{-1}$, as well as 
two stationary features about 0.6 and 1.0~mas respect from
the core. The VLBI maps obtained at 3 epochs at 1.7 GHz by Bondi et al.
(1996) contain a bright core and a secondary component at about 6.6~mas 
at PA$\sim -62^\circ$ embedded in a lower surface brightness bridge;
there was no significant change in the separation between the bridge and
the core over 7.8 yr. Recent results at 5 and ~8.4~GHz presented by
Pauliny-Toth (1998) give mean proper motions of $0.68\pm 0.02$~mas yr$^{-1}$.
G\'omez et al. (1999) found a moving component at 22 and 43~GHz
with much slower proper motion of $0.14\pm 0.02$~mas yr$^{-1}$.
A 7~mm polarized intensity image of the source by Kemball et al. (1996)
contains three-component structure, consisting
of the core, a stationary component at $\sim$0.6~mas and a new component
located between the two. The stationary component has a magnetic field
perpendicular to the jet axis, as previously observed at 5~GHz
by Cawthorne \& Gabuzda (1996). The 22~GHz polarized intensity images
of G\'omez et al. (1999) revealed a sudden change in the polarization structure of 
the core region, which they ascribed to a newly ejected component.
 
Our well-sampled 43~GHz images (Fig. 38a) show that the highly variable 
(more than by a factor of 5) core is located at the eastern end of the jet. We have
detected three moving components: $B1$, $B2$, and $B3$. The moving components
emerge from the core at PA$\sim -80^\circ$ and move toward
stationary component $C$. Feature $C$ is located at PA$-66^\circ\pm 2^\circ$
at a distance $0.66\pm 0.04$~mas, as is the case in the image of
Kemball et al. (1996). The same stationary
component was first detected in 1983 by Pauliny-Toth et al. (1987) and
extensively analyzed by G\'omez et al. (1999). Despite the similar trajectories
of the moving components, their proper motions differ significantly, ranging
from 0.14 to 0.53~mas yr$^{-1}$ (Fig. 38b).
The component with the slowest proper motion ($B3$) is detected at
five epochs and corresponds to component $A$ in the 22 and 43~GHz images  
of G\'omez et al. (1999). In addition, G\'omez et al. (1999) find a complex
jet beyond stationary feature $C$ that extends to the north
as well as to the south. In our images this structure is modeled by two components,
$D1$ (southern) and $D2$ (northern).

If we register the images according to the position of the centroid of component
$C$, then either the position of the core appears to shift upstream after the
ejection of a component, or the core nearly disappears from the images from
time to time. This effect is most apparent at epochs 1995.47 and 1996.34.
Although in 1996.34 this could be caused by a moving component blending
with $C$, hence translating its centroid first upstream and then downstream,
such an explanation for epoch 1995.47 would require  component $B2$
suddenly to have deviated sharply from its constant proper motion. As is discussed
by Marscher (1998), the change in position or disappearance of the core is
consistent with the model in which the core represents a system of standing
shocks whose position and perhaps strength change as the Mach number of the
flow varies (Daly \& Marscher 1988; G\'omez et al. 1995).
 
\section {Discussion}

Previous studies based on VLBI observations of more limited samples of
EGRET sources at lower frequencies have led to conflicting conclusions regarding
possible differences or similarities in VLBI properties between $\gamma$-ray 
blazars and the general population of compact radio sources. [We use this term
to represent a flux-limited sample of flat-spectrum radio sources, such as the
CJF sample defined by Pearson et al. 1998.]
Vermeulen \& Cohen (1994) list proper motions of 13 EGRET sources, of which 
four show no motion, four have low apparent speeds --- from 1.5 to $3~h^{-1}c$ ---
and the remainder have high superluminal velocities, up to $30~h^{-1}c$.
In their studies of EGRET blazars observed with geodetic VLBI, Piner \& Kingham
(1997a; 1997b; 1998) measured proper motions for five sources, finding high
superluminal velocities in two. However, they saw no indication that EGRET blazars
are more  strongly beamed than non-EGRET blazars. In a VLBI investigation of southern
EGRET sources, Tingay et al. (1996; 1998) found a tendency for less pronounced jet
bending relative to similar compact radio sources not detected by EGRET. Kellermann
et al. (1995) observed 26 $\gamma$-ray AGNs along with
23 radio sources not detected by EGRET, and identified no obvious difference 
in radio morphology between these two groups. Dondi \& Ghisellini (1995) studied
a sample of 46 $\gamma$-ray radio sources to conclude that $\gamma$-ray luminosity 
correlates better with radio luminosity than with either optical or X-ray luminosity.
These authors also found a strong correlation between radio--optical and
radio--$\gamma$-ray spectral indices. Zhou et al. (1997) determined that a significant 
correlation exists between compact radio flux density (i.e., that measured
with VLBI) and the  $\gamma$-radio flux of EGRET-detected AGNs, along with a
significant correlation between the 
luminosities: L$_{100 Mev}\propto$ L$_{5 GHz}^{1.05\pm0.06}$. However, M\"ucke et al.
(1997) applied a Spearman correlation analysis to simultaneously observed radio
(cm-wave) and $\gamma$-ray fluxes, which indicated that there is no significant
flux-flux correlation. Mattox et al. (1997a) established that EGRET did not detect
a significant number of any type of AGN other than blazars and found that the radio
sources identified with EGRET detections have greater correlated VLBI flux densities
than the parent population of sources with flat radio spectra.    

Our program of multi-epoch VLBA observations of 42 $\gamma$-ray blazars allows us 
to examine the radio morphology of parsec-scale jets in a more uniform manner than
has been possible with these previous, less comprehensive or lower resolution
observations. Using our data set, we can establish the salient
features of the jet as close to the central engine as possible
and relate the VLBI properties of $\gamma$-ray blazars to the $\gamma$-ray emission.
   
Fig. 39 plots the maximum (over the epochs of our observations) VLBI-core flux
densities vs. the maximum $\gamma$-ray fluxes for all the sources in our sample.
There is a positive correlation between these quantities, with a correlation coefficient
of 0.45. This decreases slightly to 0.39 when each point is assigned a weight
corresponding to 1/$\sigma$, where $\sigma$ is the statistical uncertainty in the flux
measurement. According to the t-test, both values of the correlation coefficient
correspond to a probability of 95\% that the  fluxes are related.  The $\gamma$-ray
luminosity -- radio luminosity dependence is displayed in Fig. 40. These are
apparent (i.e., relativistically beamed) luminosities, calculated from
the observed fluxes listed in Table 1 for the case of isotropic emission. 
K-corrections of the observed fluxes 
have been performed with spectral indices indicated in the same table.
In the cases in which there is no value for the radio flux, this was estimated from
VLBI data at the highest observed frequency (Table 4). Unmeasured spectral indices
are taken equal to 0.0 in the radio band and $-$1.0 at $\gamma$-ray energies.  
The $\gamma$-ray luminosities correspond to frequency 10$^{23}$~Hz.
The measured flux density at 37 GHz (or estimated from VLBI-data at 43/22~GHz) 
is used to calculate the radio luminosity.
Since  luminosities are strongly correlated with redshift
we examine the correlation between luminosities excluding the dependence on redshift
using a partial correlation analysis (Padovani 1992). Let
$r_{xy}$ be the correlation 
coefficient between variables $x$ and $y$. In the case of three variables the 
correlation coefficient between two of them, excluding the effect of the third one ($z$), is 
$$ r_{xy,z}=\frac{r_{xy}-r_{xz}r_{yz}}{\sqrt{1-r_{xz}^2}\sqrt{1-r_{yz}^2}}. $$
This method gives a partial coefficient of correlation between $\gamma$-ray and radio luminosities
equal to 0.82. The solid line in Fig. 40 corresponds to the relation  
$L_\gamma \propto L_{radio}^{1.10\pm 0.04}$ that best fits the data. 
If all the fluxes were equal, then the
slope of this line would be 1.0 (the dotted line in Fig. 40) owing 
to the multiplication of both radio and
$\gamma$-ray fluxes by the square of the distance to obtain the luminosities. The
deviation from a slope of 1.0 --- significant at the 2.5$\sigma$ level ---
indicates that a positive correlation between L$_\gamma$/L$_{radio}$ and L$_{radio}$
exists. This can be explained if the apparent $\gamma$-ray luminosity depends more
strongly on the Doppler factor than does the radio luminosity, as expected
(see Lister \& Marscher 1999). However, we stress that a found correlation
between these luminosities is rather marginal and should be confirmed 
by further observations, especially of low luminosity blazars.
   
\subsection{Superluminal motion}

In our sample of 31 quasars and 11 BL~Lac objects we have obtained
velocities for jet components in 23 quasars and 10 BL~Lac objects.
Of these, we have observed 16 quasars and 8 BL~Lac objects to have multiple moving knots.
The BL~Lac objects with more than one moving knot can be separated into 
two groups: those with similar component speeds (0716$+$741,
MKN~421, and ON~231) and those with significantly different velocities (3C~66A,
0235$+$164, 0829$+$046, OJ~287, and BL~Lac). 
The second group has characteristics similar to those of quasars with multiple moving knots.
The proper motions in the majority of such quasars (CTA~26, 0528$+$134, 0829$+$046, 3C279,
1622$-$297, and CTA~102) are higher
farther from the core. (We do not, however, have sufficient time coverage to determine
whether the motions of individual components accelerate.)
The apparent velocities in some sources (3C~66A, 1222$+$216, 1633$+$382, and
3C~454.3) vary from one component to another at similar distances from the core. In some
cases, the knots ejected later propagate along the same trajectory   
as the previous one, while in others (0235$+$164, 1156$+$295, and BL~Lac) they follow
somewhat different trajectories. Of the 33 objects with proper motions detected, 21 
contain stationary features in the jet.
Some of these standing components were observed at similar positions in
previous VLBI studies (0234$+$285, 0836$+$710, 1127$-$145, CTA~102, and
3C~454.3), which confirms their stationary nature. We do not include stationary
components in our statistical analysis of moving components, i.e., we do not treat them
as features similar to superluminal knots but with proper motions equal to zero. Rather,
we consider the standing components to be a distinct phenomenon.

Fig. 41 shows the dependence of the apparent speeds on redshift $z$, with each
component having a separate entry. (Note that only three quasars are represented in the
highest and lowest redshift bins.) The solid horizontal lines
denote the arithmetically averaged velocity for quasars over redshift intervals of 0.5,
with the standard deviation of these averages indicated by dashed error bars.
The solid diagonal line fits the dependence of apparent speed on $z$ for BL~Lac objects. 
We also plot stationary components (triangles) on this diagram to display their
redshift distribution. The number of stationary features per redshift interval
is, to first order, proportional to the number of sources and therefore
independent of $z$.  The average value ($\pm$ standard deviation) of apparent velocities 
in quasars (57 knots)
is $10.6\pm 6.2~h^{-1}c$, twice as high as the average value for 
BL~Lac objects ($4.2\pm 3.1~h^{-1}c$ for 28 knots). Both values are significantly faster
than the average values of apparent velocities of quasars and the BL~Lac objects
in the Caltech-Jodrell Bank flat-spectrum sample (CJF) observed at 5~GHz (Pearson et al.
1998, Britzen et al. 1999).  The mean apparent velocities of quasars tend to
increase with redshift from a value of $6.8\pm 3.5~h^{-1}c$ in the range $0 < z \leq 0.5$
to $15.9\pm 6.6~h^{-1}c$ over the interval of highest redshifts, $2 \leq z \leq 2.5$.
The mean apparent velocities of BL~Lac objects in our sample increases dramatically 
with distance at a rate of $0.025\pm 0.009~h^{-1}c$~Mpc$^{-1}$.
A slight increase in apparent velocity with redshift for the CJF sample has been reported 
previously for both quasars and BL~Lac objects (Britzen et al. 1999).

There are no sources with apparent velocities between $\sim 2$ and $8.5~h^{-1}c$ at
redshifts $z~>~1$.
There is no obvious reason based on population statistics as to why this should be the case
(cf. Lister \& Marscher 1997), since a sample of highly beamed jets should contain some
sources pointing almost directly toward the line of sight and therefore having low
proper motions. 
  
In Fig. 42 we plot apparent velocities vs. both radio and $\gamma$-ray apparent
(i.e., under the assumption of isotropic emisson and no Doppler boosting) luminosity.
Table 9 presents the results of a partial correlation analysis, where
$r_{gz}$ ($r_{rz}$) is the coefficient of correlation between $\gamma$-ray (radio) luminosity 
and redshift,
$r_{\beta z}$ is the coefficient of correlation between apparent speed and redshift,
$r_{\beta g}$ ($r_{\beta r}$) is the coefficient of correlation between apparent speed
and $\gamma$-ray (radio) luminosity, and  $r_{\beta g,z}$ ($r_{\beta r,z}$) is the coefficient
of correlation between apparent speed and $\gamma$-ray (radio) luminosity excluding the
dependence on redshift.
A sufficiently strong correlation is apparent for the BL~Lac objects, with a correlation coefficient of
0.66 for $\gamma$-ray luminosity and 0.47 for radio luminosity. For the quasars there
is no correlation with radio luminosity (correlation coefficient of $-$0.13), while apparent  
velocities and $\gamma$-ray luminosity are weakly connected (correlation coefficient of 0.17).
In general, a moderate correlation (0.31) is observed between apparent velocities and $\gamma$-ray
luminosity of $\gamma$-ray blazars and it is twice as high as for the radio luminosity.
Previously, Pearson et al. (1998) found a clear correlation between apparent velocity
and apparent radio luminosity in the CJF sample at 5~GHz. An increase in 
apparent speed with luminosity could be explained by a positive correlation
between luminosity and  Lorentz factor, while a partial correlation analysis 
for $\gamma$-ray blazars
indicates that the $\gamma$-ray luminosity more strongly depends on Lorentz factor
than on radio luminosity.
However, the Monte-Carlo simulations of Lister \& Marscher (1997) showed 
that, in flux-limited samples, such a correlation is also expected from selection
effects (owing to Doppler beaming and the
Malmquist bias) even if there is no relationship between bulk Lorentz factor and
luminosity.

Fig. 43 presents the distributions of the apparent velocities of the jet 
components in quasars and BL~Lac objects. The distribution for quasars peaks
at 8--9~$h^{-1}c$, while the distribution for the BL~Lac objects
has a global maximum at low apparent speeds, 2--3~$h^{-1}c$. 
The Kolmogorov-Smirnov test
gives a probability of 99.9\% that the $\beta_{app}$ values
for the quasars and BL Lac objects belong to different distributions. 
The result that typical VLBI component speeds in BL Lac objects are systematically
lower than those in quasars was found previously by Gabuzda et al. (1994) and Wehrle et al.
(1992) at 3.6 and 6~cm. Although our distributions of apparent speeds show the same tendency,
it is noteworthy that more than 30\% of the BL Lac objects in our sample
have very high apparent velocities, $\beta_{app} > 10~h^{-1}c$.       
Our sample of BL Lac objects partially overlaps with that of Gabuzda et al. (1994)
(0716$+$714, OJ~287, Mkn~421, ON~231, Mkn~501, BL~Lac). Of these, only the first two have
apparent speeds substantially higher in our study than found by Gabuzda et al. at lower
frequencies.
Our apparent velocity distribution for BL~Lacs objects
contains a high-velocity tail populated by components in sources
in which we report the first detection of superluminal motion (3C~66A, 0235$+$164,
0829$+$046, and 1908$-$201). This  demonstrates that  jet components 
in $\gamma$-ray bright BL~Lac objects might have similar high apparent velocities as 
those in quasars. In addition, the source 0829$+$046 shows
proper motion over a wide range of apparent speeds covering the ranges typical for both BL~Lac objects
and quasars. The shift of the maximum of the distribution for quasars
into higher apparent speeds and the high-velocity tail for BL~Lac objects in our sample
compared with the results
of Caltech-Jodrell flat-spectrum radio source (CJF) survey (see Pearson et al. 1998)
demonstrate that the apparent speeds of the $\gamma$-ray
bright blazars are considerably faster than in general population of
bright compact radio sources.   

The highest apparent speeds that we measure exceed $40c$ (for $h = 0.65$). This is
about half an
order of magnitude higher than the theoretical limit for the bulk Lorentz factor
of the jet flow, $\Gamma_{\rm max} \sim 10$ (e.g., Begelman et al. 1994).
These high speeds correspond in many cases to components that are quite weak,
hence they should be confirmed by further observations. Nevertheless, a number
of the apparent speeds in excess of $20c$ are quite well established. This
suggests that the Lorentz factor of the jet may only reach its maximum value
somewhere in the radio emitting portion of the jet rather than farther upstream
where the intense radiation environment would cause Compton drag.

We detect no proper motions in six objects with sufficient time coverage to do so:
0234+285, 0804+499, 0954+658, 1606+106, 1739+522, and 2209+236. Several of these
have been reported previously to have superluminal knots (see \S4); therefore, our
inability to detect motion over the course of our observations does not imply that
a source never produces moving components.

\subsection{Stationary Features in the Jets}    

In 27 of the 42 $\gamma$-ray blazars in our sample, we find a total of 45 components
that show no motion respect to the core over the duration of our VLBA observations (see Table 6). 
Out of these
27 sources, 21 are observed to contain moving components as well, some of which
propagate at very high apparent speeds, $> 10~h^{-1}c$. Hence, the existence of
standing features does not correspond to very slow flow speeds in the jet.
Rather, stationary hotspots are a common characteristic of compact jets.

Figure 44 presents the distribution of the distances of stationary knots from their
respective cores, in projection on the sky plane. The distribution shows a significant 
global maximum in the range of projected distances of 1--$2~h^{-1}$ pc from the core,
which corresponds to $< 0.5$~mas for 30\% of the sources in the sample. 
In the few cases of simultaneous observations at different frequencies 
we can define the spectral indices of stationary and moving
components (Table~7). Fig. 45 shows the dependence of spectral indices on distance
from the core for stationary and moving knots. For moving knots the average spectral
index ($\pm$ standard deviation) $\langle \alpha \rangle = -0.7\pm 0.3$ is flatter 
than for stationary features 
($\langle \alpha \rangle = -1.2\pm 0.3$). In general, the spectral index of a component 
is steeper at greater distance from the core. The core itself usually has an inverted spectrum,
with average spectral index $\langle \alpha \rangle = 0.3\pm 0.4$.

Models for stationary components in supersonic jets fall into three categories:
(1) standing recollimation shocks caused by pressure imbalances at the boundary between
the jet fluid and the external medium; (2) sites of maximized Doppler beaming where a
curved jet points most closely to the line of sight; and (3) stationary shocks where
the jet bends abruptly, presumably as a result of striking an obstacle (dense cloud)
that deflects it.  Stationary knots near the core and, perhaps, the core itself
most likely are caused by process (1) (Daly \& Marscher 1988).
Numerical simulations of the relativistic hydrodynamics 
and emission of jets (G\'omez et al. 1995, 1997) show that when a moving component passes 
through a stationary feature produced by a standing shock, both components appear to
be blended into a single feature and the position of the merged components appears to shift 
downstream with respect to the pre-disturbance location of the stationary knot.
After the collision,
the two components appear to split up, with the centroid of the quasi-stationary feature 
returning upstream. Within this scenario a superluminal component would not show
any significant changes in its proper motion. This situation was observed
by G\'omez et al. (1999) for moving component $A$ and stationary component $St$ in
3C~454.3. Component $St$ corresponds to component $C$ in our designation, and
we observed a similar behavior when moving components $B2$ and, probably, $B1$ 
approached $C$ (see Fig. 38b). In our sample there are a number of examples 
of this scenario: merging of component $B4$ with stationary feature $C$ in 3C~66A;
approach of component $B$ to stationary knot $C$ in
0420$-$014; blending of component $B1$ with stationary feature $C$ in
0827$+$243; blending of component $B$ with component $C$ in 0836$+$710;
motion of knot $B2$ through stationary feature $C1$ in OJ~287;
blending of component $B3$ with knot $C$ in 1156$+$295; and motion of component B1
through stationary feature $C$ in 1510$-$089. Although the uncertainties in the
positions of individual stationary features are usually greater than the shifts in their 
locations, the mean shift relative to the average position of stationary knots over all such 
events is significant: $0.08\pm 0.05$~mas.
Furthermore, there is no systematic change in proper motion of the moving components
involved in these events. 

In the cases in which a stationary feature is produced at a bend
in the jet, the proper motion of a moving knot should change and the component
should either become brighter or fainter as it curves around the bend. This has
been observed in the best-observed event of this type, in 4C~39.25 (Alberdi et al. 1993). 
In our sample there are a few suitable
examples of bright stationary features located at a bend in the jet:
component $C2$ in 0836$+$710, component $C2$ in
1127$-$145, component $C2$ in 1611$+$343, and component $C$ in CTA~102.
Only in two of these cases was a moving component observed near the stationary feature.
In 1127$-$145 a moving component $B1$ approached stationary knot $C2$.
Although component $B1$ is very weak and observed only at three
epochs, its proper motion determined from the two last epochs was
three times less than the proper motion between the first two epochs (see Fig. 19b). 
Since component $B1$ was very faint, the change in its flux is highly uncertain; however, 
the stationary knot ($C$) underwent a significant increase in flux. In the case of quasar 
1611$+$343, moving component $B3$ decelerated and increased somewhat in flux
density as it approached stationary feature $C2$ (Fig. 28d).
The lower proper motion of components observed beyond 
knot $C2$ compared with that of components near the core might be explained
by this scenario. 
 
\subsection{Jet Bending}                                
 
Von Montigny et al. (1995b) suggested two ways to explain that 
not all radio loud AGNs with flat spectrum are detectable $\gamma$-ray
sources. One possible reason is that the $\gamma$-ray emission may be more narrowly
beamed than the radio emission. The second hypothesis was based on the
possibility that EGRET-identified sources 
have preferentially straight jets while $\gamma$-ray-quiet AGNs have bent
jets. In this case, if a jet bends downstream of the $\gamma$-ray
emitting region before the radio emission is produced, the $\gamma$-ray
emission could be beamed away from our line of sight but the
radio emission could still be beamed toward us. This seems an improbable
scenario, since it is much less likely for a jet to bend into a narrow cone
about the line of sight than to bend away.
Nevertheless, Tingay et al. (1998) found evidence supporting the second suggestion
using the sample of 42 $\gamma$-ray blazars listed by Mattox et al. (1997a)
and a list of 26 $\gamma$-ray-quiet radio sources from the
sample of Pearson \& Readhead (1988). However, Tingay et al. note
that the VLBI images which they used for the EGRET-identified radio sources
are less representative of their actual structure than are 
the VLBI images of the $\gamma$-ray-quiet radio sources due to
the wide variety of VLBI arrays and observing frequencies employed.

Our sample of $\gamma$-ray blazars is sufficiently uniform to check whether
bending is more or less prevalent in $\gamma$-ray bright blazars than in
the general population. Indeed, as our images show, there is considerable
bending of the jets of EGRET-detected blazars on parcec scales.
The jet bending is quite similar to that found 
by Kellermann et al. (1998) in their 15~GHz survey: 
a number of jets have gradual curvature
(3C~66A, 0836$+$710, 1219$+$285, 3C~273, 3C~279), some jets
display sharp bends up to 90$^\circ$ or more (0954$+$658, 1611$+$343, MKN~501,
1908$-$201), in other sources the jet appears to wiggle
through several turns (0202$+$149, 0458$-$020, CTA~102),
and some jets are quite broad within a few mas of the core
(0235$+$164, 1156$+$296). 

Table~8 lists the parameters related to bending for each source (see the description
in the section on ``Presentation of Results'' above). (Note that bending near
the core can be detected only over a size scale larger than the resolving beam.)
Fig.~46 presents the distribution of jet position angles, demonstrating that
there is no preferential direction of projected jet position angle
for $\gamma$-ray blazars. The distribution of jet bending (Fig. 47) indicates that 
46\% of the sources in the sample have jets that curve by more than 20$^\circ$, with
63\% of these (29\% of the total) containing bends sharper than 50$^\circ$.
In addition, we note that many sources that show no significant curvature within the
1\% contour level isophote bend at lower contour levels (e.g., 0440$-$003, 0716$+$714,
0829$+$046, and 1510$-$089). According to Fig.~48, the first bend of the jet 
takes place in the innermost part of the jet, within a projected distance
$\sim 7~h^{-1}$ pc of the core. Our results demonstrate that EGRET
sources contain jets with levels of bending similar those of compact
radio sources not detected by EGRET (Tingay et al. 1998; Kellermann et al. 1998).

\section{Summary}

We have completed an extensive VLBA monitoring program of the majority of
$\gamma$-ray bright blazars. Our findings can be summarized as follows:

\begin{enumerate}
\item In our sample of 42 $\gamma$-ray bright blazars apparent
superluminal jet velocities are measured in 33 sources. The results strongly support
the thesis that $\gamma$-ray emission originates in a highly
relativistic jet. A positive correlation between VLBI core flux and
$\gamma$-ray flux suggests that the production of the $\gamma$-ray 
emission takes place in the most compact region of the relativistic jet,
close to the VLBI-core.     

\item The distribution of apparent velocities of jet components peaks
at $8-9~h^{-1}c$, which is significantly higher than the average superluminal
speed of jet components in the general population of strong compact radio sources. This 
follows the expectations of inverse Compton models in which the $\gamma$-rays
are more highly beamed than is the radio synchrotron radiation.
  
\item The maximum in the distribution of apparent speeds of jet components in
BL~Lac objects occurs at lower apparent speeds than is the case
for quasars. However, the BL Lac distribution has a long high-velocity tail,
corresponding to a substantial number of $\gamma$-ray bright 
BL~Lac objects that have high apparent speeds similar to those in quasars.

\item A higher level of correlation between apparent speeds of jet components and apparent
$\gamma$-ray luminosity than between apparent speeds and radio luminosity further supports
the conclusion that the
$\gamma$-ray emission depends more strongly on the Lorentz factor of the jet flow than 
does the radio emission.  This is also consistent with both
synchrotron self-Compton and external inverse Compton models. It is possible
that non-inverse Compton models could produce similar correlations between
$\gamma$-ray flux and Lorentz factor. The $\gamma$-ray emission in such models generally
arises in the high radiation environment close to the central engine. However,
in a subsequent paper (Jorstad et al. 2001) we conclude from the relative timing
of superluminal ejections and $\gamma$-ray flares that the $\gamma$-ray emission
originates downstream of the radio core.  

\item The existence of a number of sources with multiple moving knots
with different proper motions in the same source indicates that either the Lorentz
factor of the jet flow or the pattern speed of bright features (or both) is variable.

\item The majority of jets of $\gamma$-ray blazars bend significantly on 
parsec scales, although no more so than do jets of blazars not detected by EGRET.

\item In 27 of the sources in the sample there is at least one non-core 
component that
appears to be stationary during our observations. Stationary knots can be
separated into two groups: those nearest to the core ($1-3~h^{-1}$ pc) and those
that are farther downstream. The characteristics of these groups are different,
which leads us to suggest that the first group corresponds to stationary shocks
and the second to bends in the jet.

\item  Multi-frequency observations allow us to determined the spectral indices 
of jet components in 9 sources for 12 moving and 5 stationary knots.
The spectra of jet components are those  expected for 
optically thin synchrotron radio sources. The spectral index is steeper for components
farther from the core. There is a possible difference between
spectral indices of moving and stationary components in the sense that moving
components have flatter spectra.  

\end{enumerate}

The population of bright $\gamma$-ray blazars detected by EGRET can therefore be
categorized as highly superluminal, with apparent speeds as high as $\sim 40c$ for
a Hubble constant of 65 km s$^{-1}$ Mpc$^{-1}$ ($h=65$). Furthermore, the jets tend to
be bent, a sign of orientation close to the line of sight. We suggest that
the jets of EGRET-detected sources might be almost directly aligned with the line of
sight at the section where $\gamma$-ray emission originates. Such jets, if they bend 
(as most jets seem to do), have a much higher probability of curving away from the line
of sight than toward it (i.e., almost all directions lead away from the line of sight).
This causes a slight misalignment with the line of sight beyond
the core, an orientation that is favorable to the detection of apparent
superluminal motion.

The correlations presented here should be confirmed and perhaps strengthened through
better time sampling of the $\gamma$-ray light curves of more blazars, coupled with
regularly and closely spaced VLBA observations. The authors hope that a concerted
effort can be carried out with the VLBA when the planned GLAST $\gamma$-ray mission
provides well-sampled monitoring of the $\gamma$-ray fluxes of hundreds or even
thousands of blazars.

\begin{acknowledgments}

This work was supported in part by NASA through CGRO Guest Investigator
grants NAG5-7323 and NAG5-2508, and by U.S. National Science Foundation
grant AST-9802941. A.E.W. was supported in part by the NASA Long Term Space Astrophysics
program. This research has made use of data from the University of
Michigan Radio Astronomy Observatory, which is supported by the National
Science Foundation and by funds from the University of Michigan. This research has
also made use of the NASA/IPAC Extragalactic Database (NED), which is operated by
the Jet Propulsion Laboratory, California Institute of Technology, under contract
with the National Aeronautics and Space Administration. We thank
Dr. H. Ter\"asranta for providing flux densities measured at the Mets\"ahovi
Radio Research Station in Finland.
\end{acknowledgments}
\clearpage
\begin{deluxetable}{llccllll}
\singlespace
\tablenum 1
\tablecolumns{7}
\tablecaption{\small\bf Source List}
\tabletypesize{\footnotesize}
\tablehead{
\colhead{Source}&\colhead{Name}&\colhead{z}&\colhead{Source}&
\colhead{S$_\gamma$}&\colhead{$\alpha_\gamma$}&\colhead{S$_{37}$ }&\colhead{$\alpha_{22-37}$}\\
\colhead{}&\colhead{}&\colhead{}&\colhead{Type}&\colhead{10$^{-8}$cm$^{-2}$s$^{-1}$}&\colhead{}
&\colhead{ Jy}&\colhead{}
}
\startdata
0202$+$149&&0.833?&Q&52.8$\pm$26.4&$-$1.23$\pm$0.28&3.93$\pm$0.10&$+$0.49$\pm$0.06   \\
0219$+$428&3C 66A&0.444&B&25.3$\pm$5.8&$-$1.01$\pm$0.14&1.56$\pm$0.06&$-$0.10$\pm$0.05   \\
0234$+$285&CTD 20&1.213&Q&31.4$\pm$11.6&$-$1.53$\pm$0.22&4.19$\pm$0.26&$+$0.05$\pm$0.03      \\
0235$+$164&OD 160&$\ge$0.94&B&65.1$\pm$8.8&$-$0.85$\pm$0.1&6.84$\pm$0.12&$+$0.93$\pm$0.15     \\
0336$-$019&CTA 26&0.852&Q&177.6$\pm$36.6&$-$0.84$\pm$0.25&3.56$\pm$0.31&$-$0.19$\pm$0.08     \\
0420$-$014&OA 129&0.915&Q&64.2$\pm$34.2&$-$1.44$\pm$0.19&8.55$\pm$0.16&$+$0.27$\pm$0.05     \\
0440$-$003&NRA0 190&0.844&Q&85.9$\pm$12.0&$-$1.37$\pm$0.18&1.42$\pm$0.19&$-$0.17$\pm$0.04    \\
0446$+$112&&1.207&Q&32.1$\pm$17.7&$-$1.27$\pm$0.16&2.11$\pm$0.09&$-$0.06$\pm$0.04     \\
\enddata
\end{deluxetable}
\clearpage
\begin{deluxetable}{ccll}
\singlespace
\tablenum 2
\tablecolumns{4}
\tablecaption{\small\bf Calibration Coefficients}
\tabletypesize{\footnotesize}
\bf
\tablehead{
\colhead{Epoch}&\colhead{Freq.~[GHz]}&\colhead{Calibration Sources}&\colhead{Coeff.} }
\startdata
1993.86&22.2&1253$-$055&1.00$\pm$0.10 \\
1994.45&22.2&1253$-$055&1.05$\pm$0.10 \\
1994.65&22.2&0528$+$134&1.10$\pm$0.10 \\
1994.76&22.2&1606$+$106&1.20$\pm$0.10 \\
1995.01&22.2&0528$+$134&1.25$\pm$0.10 \\
1995.01&43.2&0528$+$134,~2251$+$158&1.57$\pm$0.20 \\
1995.15&22.2&0528$+$134&1.00$\pm$0.10 \\
1995.15&43.2&0528$+$134,~2251$+$158&1.20$\pm$0.20 \\
1995.31&22.2&0827$+$243,~1156$+$295,~1611$+$343&1.18$\pm$0.10 \\
1995.31&43.2&0528$+$134,~2251$+$158&1.30$\pm$0.20 \\
1995.47&22.2&0716$+$714&1.00$\pm$0.20 \\
1995.47&43.2&2251$+$158&1.57$\pm$0.20 \\
1995.59&22.2&0234$+$285&1.06$\pm$0.10 \\
1995.59&43.2&2251$+$158&1.60$\pm$0.20 \\
1995.79&22.2&1406$-$076&1.11$\pm$0.10 \\
1995.79&43.2&02251$+$158&1.52$\pm$0.20 \\
1996.34&22.2&0234$+$285,~1156$+$295,~1606$+$106,&1.18$\pm$0.10 \\
&&1611$+$343,~1739$+$522& \\
1996.34&43.2&0528$+$134,~2251$+$158&1.29$\pm$0.20 \\
1996.60&22.2&0234$+$285,~1611$+$343,~1156$+$295&1.02$\pm$0.10 \\
1996.60&43.2&0528$+$134,~2251$+$158&1.12$\pm$0.20 \\
1996.90&22.2&0827$+$243,~1156$+$295,&1.39$\pm$0.10 \\
&&1606$+$106,~1611$+$343& \\
1996.90&43.2&0528$+$134,~2251$+$158&1.40$\pm$0.20 \\
1997.58&22.2&1156$+$295,~1606$+$106,&1.16$\pm$0.10 \\
&&1611$+$343,~1739$+$522& \\
1997.58&43.2&0528$+$134,~0420$-$014,&1.23$\pm$0.20 \\
&&0446$+$114& \\
\enddata
\end{deluxetable}
\clearpage
\begin{deluxetable}{cccccllrlcl}
\singlespace
\tablenum 3
\tablecolumns{11}
\tablecaption{\small\bf Parameters of maps}
\tabletypesize{\footnotesize}
\bf
\tablehead{
\colhead{Source}&\colhead{Freq.}&\colhead{NE}&\colhead{Epoch of}&
\colhead{$I_{peak}$}&\colhead{$\Theta_{maj}$}&
\colhead{$\Theta_{min}$}&\colhead{P.A.}&\colhead{Cont.}&
\colhead{RMS}&\colhead{ Fig.} \\
\colhead{}&\colhead{[GHz]}&\colhead{}&\colhead{$I_{peak}$}&
\colhead{[Jy/Beam]}&\colhead{[mas]}&\colhead{[mas]}&\colhead{[deg.]}&
\colhead{[\%]}&\colhead{[mJy/Beam]}&\colhead{}
}
\startdata
0219$+$428&43.2&2&1995.79&0.95& 0.35&0.20&$+$15&0.50&1.83& 3a   \\
0219$+$428&22.2&2&1994.65&0.90& 0.75&0.40&$+$10&1.00&10.8& 3b  \\
0234$+$285&22.2&4&1996.60&1.47& 0.60&0.30&$-$4&0.25&1.10&  4   \\
0235$+$164&43.2&6&1995.47&0.96& 0.30&0.15&$+$0&0.25&2.05&  5a  \\
0420$-$014&43.2&7&1995.31&3.53& 0.40&0.15&$-$7&0.25&2.32&  6a   \\
0446$+$112&43.2&4&1995.47&0.82&0.40&0.15&$-$2&0.25&2.03&  7 \\
0454$-$234&43.2&2&1996.34&0.83&0.40&0.15&$-$4&0.50&1.98&  8 \\
0458$-$020&43.2&5&1995.31&1.53&0.45&0.17&$-$3&0.25&3.20&  9a \\
\enddata
\end{deluxetable}
\clearpage
\begin{deluxetable}{ccclllll}
\singlespace
\tablenum 4
\tablecolumns{8}
\tablecaption{\small\bf  Model Parameters }
\tabletypesize{\scriptsize}
\bf
\tablehead{
\colhead{Source}&\colhead{Epoch}&\colhead{Freq.}&\colhead{Comp.}&\colhead{Flux}&
\colhead{$R$}&\colhead{$\Theta$}&\colhead{$a$} \\
\colhead{}&\colhead{}&\colhead{[GHz]}&\colhead{}&\colhead{[Jy]}&
\colhead{[mas]}&\colhead{[deg.]}&\colhead{[mas]}} 
\startdata
0219$+$428& 1994.65 & 22.2 & A & 0.90$\pm$0.06 & 0.0$\pm$0.05  &\nodata&  0.08$\pm$0.04 \\*
&&&D3&0.09&0.56&$-$158&0.22 \\*
&&&D2&0.04&2.75&$-$166&0.03 \\*
&&&D1&0.05&4.62&$-$173&0.28 \\
0219$+$428& 1995.47 & 43.2 & A & 0.49$\pm$0.04 & 0.0$\pm$0.03  &\nodata&  0.05$\pm$0.02 \\*
&&&B3&0.41&0.11&$-$151&0.08 \\*
&&&C&0.05&0.51&$-$154&0.17 \\*
&&&B2&0.06&0.81&$-$158&0.13 \\
\enddata
\end{deluxetable}                                                 
\clearpage
\begin{deluxetable}{lllllll}
\singlespace
\tablenum 5
\tablecolumns{7}
\tablecaption{\small\bf Proper motion}
\tabletypesize{\footnotesize}
\bf
\tablehead{
\colhead{Source}&\colhead{z}&\colhead{Comp.}&\colhead{$\mu$[mas yr$^{-1}$]}&
\colhead{$\beta_{app}$[h$^{-1}$]}&\colhead{$T_\circ$}&\colhead{Figure} }
\startdata
0202$+$149&0.833&B&0.18$\pm$0.01 &5.2$\pm$0.3&1994.83$\pm$0.06&\nodata\tablenotemark{(a)}   \\
0219$+$428&0.444&B4&0.57$\pm$0.14 & 9.9$\pm$2.4&1995.67$\pm$0.33& 3c   \\*
&&B3&0.82$\pm$0.03 &14.2$\pm$0.5&1995.37$\pm$0.02&   \\*
&&B2&1.11$\pm$0.20 &19.3$\pm$3.5&1994.8$\pm$0.3&    \\
0235$+1$64&$\ge$0.94&B1&0.93$\pm$0.16 & 29.6$\pm$5.1&1995.1$\pm$0.2& 5b   \\*
&&B2&0.61$\pm$0.16 & 19.4$\pm$5.1&1994.8$\pm$0.2&   \\
0336$-$019&0.852&C4&0.18$\pm$0.07 & 5.3$\pm$2.1&1995.13$\pm$0.16&\nodata\tablenotemark{(b)}   \\*
&&C3&0.42$\pm$0.04 &12.4$\pm$1.2&1993.9$\pm$0.1&   \\
\enddata
\tablenotetext{(a)}{Fig. 15 in Pyatunina et al. 2000}
\tablenotetext{(b)}{Fig. 7 in Mattox et al. 2001}
\tablenotetext{(c)}{Fig. 4 in Yurchenko, Marchenko-Jorstad, \& Marscher 2000}
\tablenotetext{(d)}{Fig. 4 in Marscher 1999}
\tablenotetext{(e)}{Fig. 24 in Denn, Mutel, \& Marscher 2000}
\end{deluxetable}
\clearpage
\begin{deluxetable}{llllcr}
\singlespace
\tablenum 6
\tablecolumns{6}
\tablecaption{\bf Positions of Stationary Features}
\tabletypesize{\footnotesize}
\bf
\tablehead{
\colhead{Source}&\colhead{z}&\colhead{Freq.[GHz]}&\colhead{Comp.}&
\colhead{$\bar R$[mas]}&\colhead{$\bar \Theta$[deg.]} }
\startdata
0202$+$149\tablenotemark{(a)}&0.833&43.2&C&0.53$\pm$0.01&$-$79$\pm$3    \\*
&&&D&5.15$\pm$0.02&-57$\pm$2 \\
0219$+$428&0.444&43.2&C&0.53$\pm$0.02& $-$157$\pm$1    \\
0234$+$285&1.213&22.2&C1&0.32$\pm$0.02& $-$23$\pm$1   \\*
&&&C2&1.08$\pm$0.04& $-$20$\pm$1   \\*
&&&C3&3.20$\pm$0.03& $-$12$\pm$1   \\
0336$-$140\tablenotemark{(b)}&0.852&43.2&C2&1.2$\pm$0.2& $+$77$\pm$6   \\*
&&22.2&C2&1.18$\pm$0.08&$+$80$\pm$8 \\
\enddata
\tablenotetext{(a)}{in Pyatunina et al. 2000}
\tablenotetext{(b)}{in Mattox et al. 2001}
\tablenotetext{(c)}{in Yurchenko, Marchenko-Jorstad, \& Marscher 2000}
\end{deluxetable}

\clearpage
\begin{deluxetable}{ccccl}
\singlespace
\tablenum 7
\tablecolumns{5}
\tablecaption{\small\bf Spectral indices of VLBI-core and moving and stationary knots}
\tabletypesize{\footnotesize}
\bf
\tablehead{
\colhead{Name}&\colhead{Component}&\colhead{Type of Comp.}&\colhead{Range of Freq.[GHz] }&\colhead{$\alpha$} }
\startdata
0336$-$019&A&S&15$-$43&$+$0.6$\pm$0.1 \\ 
0336$-$019&C2&S&15$-$43&$-$0.73$\pm$0.09 \\ 
0336$-$019&C3&M&15$-$43&$-$0.68$\pm$0.14 \\
0336$-$019&C4&M&22$-$43&$-$0.31$\pm$0.09 \\
0528$+$134&A&S&22$-$43&$+$0.6$\pm$0.2  \\ 
0528$+$134&B1&M&22$-$43&$-$0.3$\pm$0.3 \\
0528$+$134&B2&M&22$-$43&$-$0.95$\pm$0.05 \\
0836$+$710&A&S&22$-$43&$+$0.05$\pm$0.15 \\ 
0836$+$710&B&M&22$-$43&$-$0.77$\pm$0.25 \\
0836$+$710&C2+C3&S&22$-$43&$-$1.5$\pm$0.4 \\
0851$+$202&A&S&15$-$43&$+$0.32$\pm$0.02 \\ 
0851$+$202&C2&S&15$-$43&$-$1.3$\pm$0.3 \\
1226$+$023&A&S&8$-$22&$-$0.33$\pm$0.05 \\
1226$+$023&A&S&22$-$43&$-$0.06$\pm$0.08 \\
1226$+$023&B1&M&22$-$43&$-$0.53$\pm$0.26 \\
1226$+$023&B4&M&22$-$43&$-$0.73$\pm$0.35 \\
1226$+$023&B5&M&22$-$43&$-$1.1$\pm$0.4 \\
1253$-$055&A&S&8$-$22&$+$0.24$\pm$0.02 \\
1253$-$055&A&S&22$-$43&$+$0.6$\pm$0.5 \\
1253$-$055&C&S&22$-$43&$-$1.1$\pm$0.3 \\
1253$-$055&D&M&22$-$43&$-$0.87$\pm$0.15 \\
1611$+$343&A&S&15$-$43&$-$0.6$\pm$0.2 \\
1611$+$343&C2&S&15$-$43&$-$1.5$\pm$0.3 \\
1622$-$297&A&S&15$-$43&$+$0.5$\pm$0.2 \\
1622$-$297&B1&M&15$-$43&$-$0.91$\pm$0.05 \\
1692$-$297&B2&M&15$-$43&$-$0.86$\pm$0.08 \\
1730$-$130&A&S&22$-$43&$+$0.67$\pm$0.06 \\
1730$-$130&B1&M&22$-$43&$-$0.24$\pm$0.15 \\
\enddata
\end{deluxetable}

\clearpage
\begin{deluxetable}{clccc}
\singlespace
\tablenum 8
\tablecolumns{5}
\tablecaption{\small\bf Jet Properties}
\tabletypesize{\footnotesize}
\tablehead{
\colhead{Name}&\colhead{$\Theta$}&\colhead{$\Delta\Theta_{max}$}&
\colhead{$R_{\Delta\Theta}$}&\colhead{S$_{jet}$/S$_{core}$}\\
\colhead{}&\colhead{[deg]}&\colhead{[deg]}&\colhead{[mas]}&\colhead{}
}
\startdata
0202$+$149&$-$32&73&0.4&1.78  \\
0219$+$428&$-$158&29&1.0&1.06 \\
0234$+$285&$-$20&0&\nodata&0.66   \\
0235$+$164&$+$0&59&0.15&0.14   \\
0336$-$019&$+$75&28&0.6&0.34   \\
0420$-$014&$-$166&16&0.6&0.14   \\
0440$-$003&$-$123&0&\nodata&1.49  \\
0446$+$112&$+$143&0&\nodata&0.84   \\
0454$-$234&\nodata&\nodata&\nodata\nodata   \\
0458$-$020&$-$7&70&1.0&0.64  \\
0528$+$134&$+$81&45&0.75&1.44   \\
0716$+$714&$+$10&0&\nodata&0.07   \\
0804$+$499&$+$136&85&1.0&0.19   \\
0827$+$243&$+$134&17&1.2&1.24   \\
0829$+$046&$+$72&0&\nodata&0.68   \\
0836$+$710&$-$130&20&1.1&0.65   \\
0851$+$202&$-$86&0&\nodata&0.52   \\
0917$+$449&$+$180&0&\nodata&0.88   \\
0954$+$658&$-$22&53&1.65&0.34   \\
1101$+$384&$-$42&13&1.9&0.14   \\
1127$-$145&$+$96&15&1.35&1.12  \\
1156$+$296&$+$0&31&0.30&0.79   \\
1219$+$285&$+$102&30&2.35&0.39   \\
1222$+$216&$-$14&18&1.0&0.75  \\
1226$+$023&$-$108&16&3.2&1.93 \\
1253$-$055&$-$113&18&1.1&1.32   \\
1406$-$076&$-$81&0&\nodata&0.46   \\
1510$-$089&$-$32&0&\nodata&0.16   \\
1606$+$106&$-$62&31&1.2&0.37   \\
1611$+$343&$+$173&105&2.8&1.42   \\
1622$-$253&$-$57&0&\nodata&0.30 \\
1622$-$297&$-$83&27&2.0&0.80 \\
1633$+$382&$-$85&0&\nodata&1.86  \\
1652$+$398&$+$154&90&3.0&\nodata   \\
1730$-$130&$+$17&123&0.22&6.13   \\
1739$+$522&$+$61&0&\nodata&0.66   \\
1741$-$038&$-$140&52&0.7&0.41   \\ 
1908$-$201&$-$25&104&1.9&0.25  \\  
2200$+$420&$+$167&0&\nodata&\nodata   \\
2209$+$236&$+$12&0&\nodata&\nodata\\
2230$+$114&$+$132&55&1.8&0.69  \\
2251$+$158&$-$66&54&0.6&9.65   \\
\enddata
\end{deluxetable}

\clearpage
\begin{deluxetable}{llll}
\singlespace
\tablenum 9
\tablecolumns{4}
\tablecaption{\small\bf Results of Partial Correlation Analysis  }
\tabletypesize{\footnotesize}
\bf
\tablehead{
\colhead{Coeff.}&\colhead{BL Lacs}&\colhead{Quasars}&\colhead{All} }
\startdata
$r_{rz}$&0.841&0.716&0.771 \\
$r_{gz}$&0.892&0.875&0.867 \\
$r_{\beta z}$&0.830&0.366&0.427 \\
$r_{\beta r}$&0.840&0.070&0.415 \\
$r_{\beta g}$&0.906&0.399&0.512 \\
$r_{\beta r,z}$&0.471&$-$0.13&0.149 \\
$r_{\beta g,z}$&0.659&0.174&0.314 \\
\enddata
\end{deluxetable}

\clearpage
\begin{figure}
\figurenum{1}
\plotone{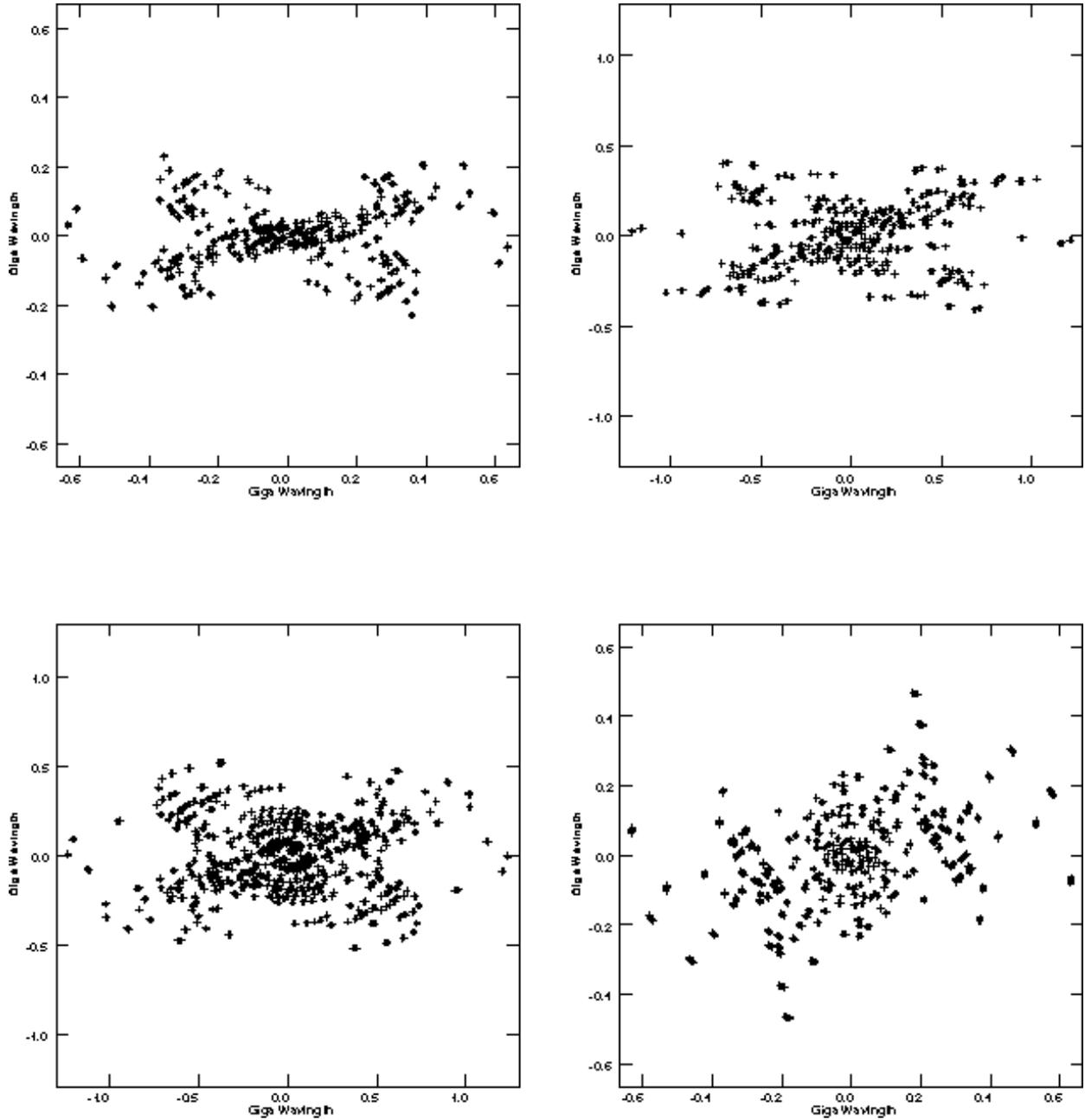}
\caption{The typical {\it uv} - coverage for 
sources over a range of declinations at epoch 1996.60: 
top left panel - 1908$-$201; top right panel - 0420$-$014; 
bottom left panel - 2230$+$114; bottom right panel - 0917$+$449.}
\end{figure}
\begin{figure}
\figurenum{2}
\plotone{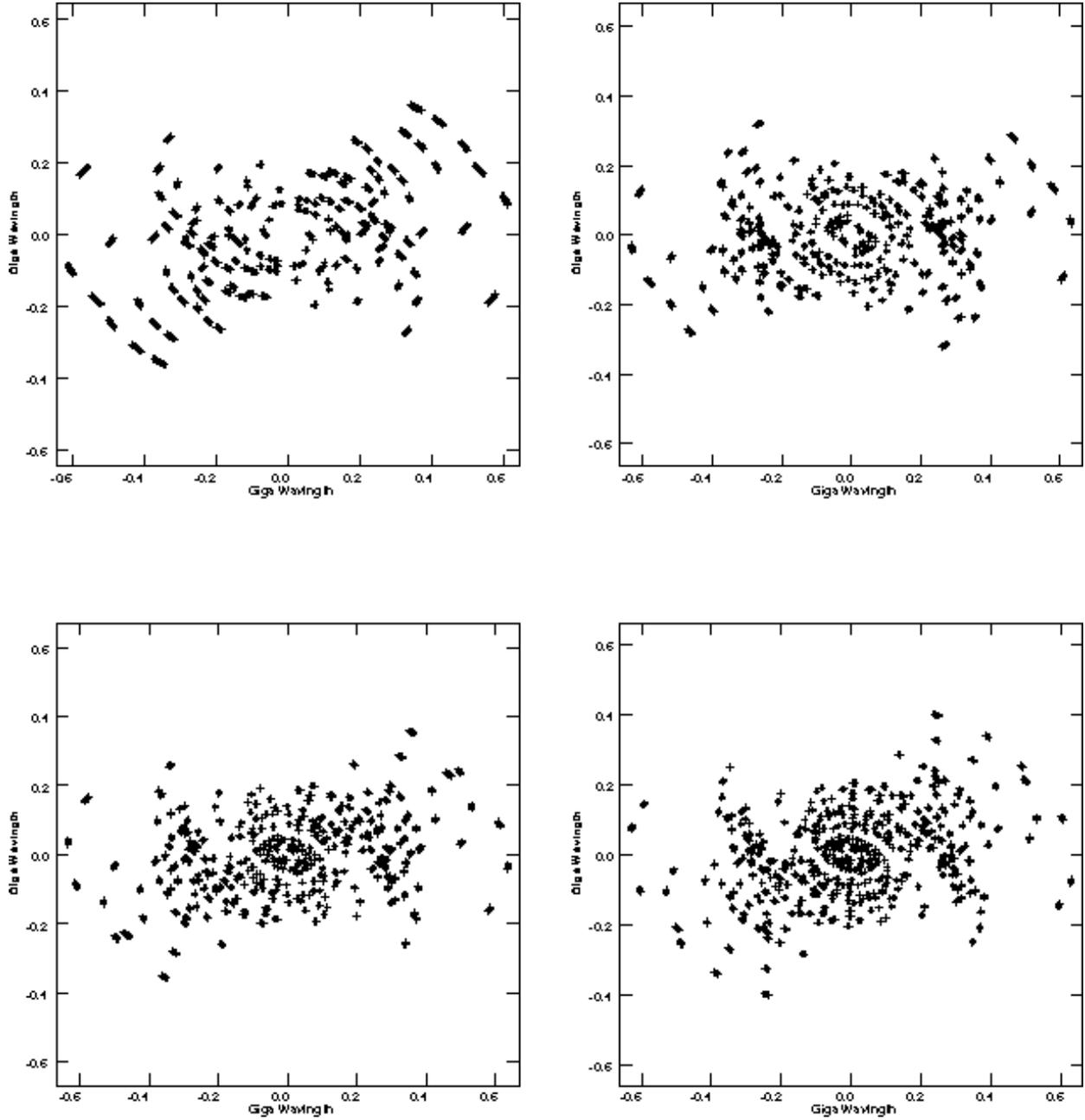}
\caption{The typical {\it uv} - coverage for source 1611$+$343  over a range of epochs:
top left panel - 1994.76; top right panel - 1995.31; bottom left panel - 1996.60; 
bottom right panel - 1997.58.}
\end{figure}
\begin{figure}
\figurenum{3a}
\plotone{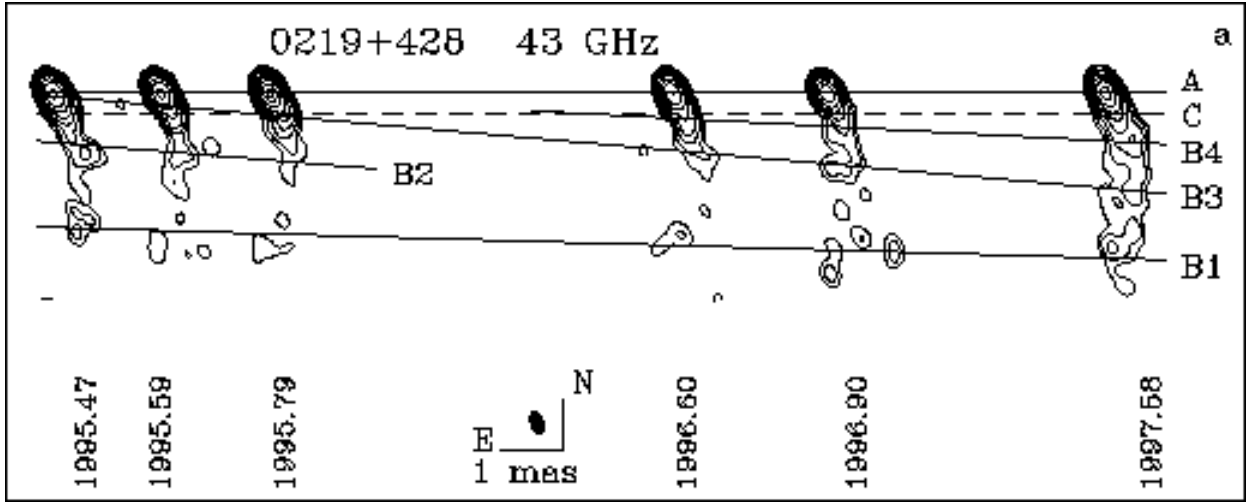}
\caption{Hybrid maps of 3C~66A at 43~GHz.}
\end{figure}
\begin{figure}
\figurenum{3b}
\plotone{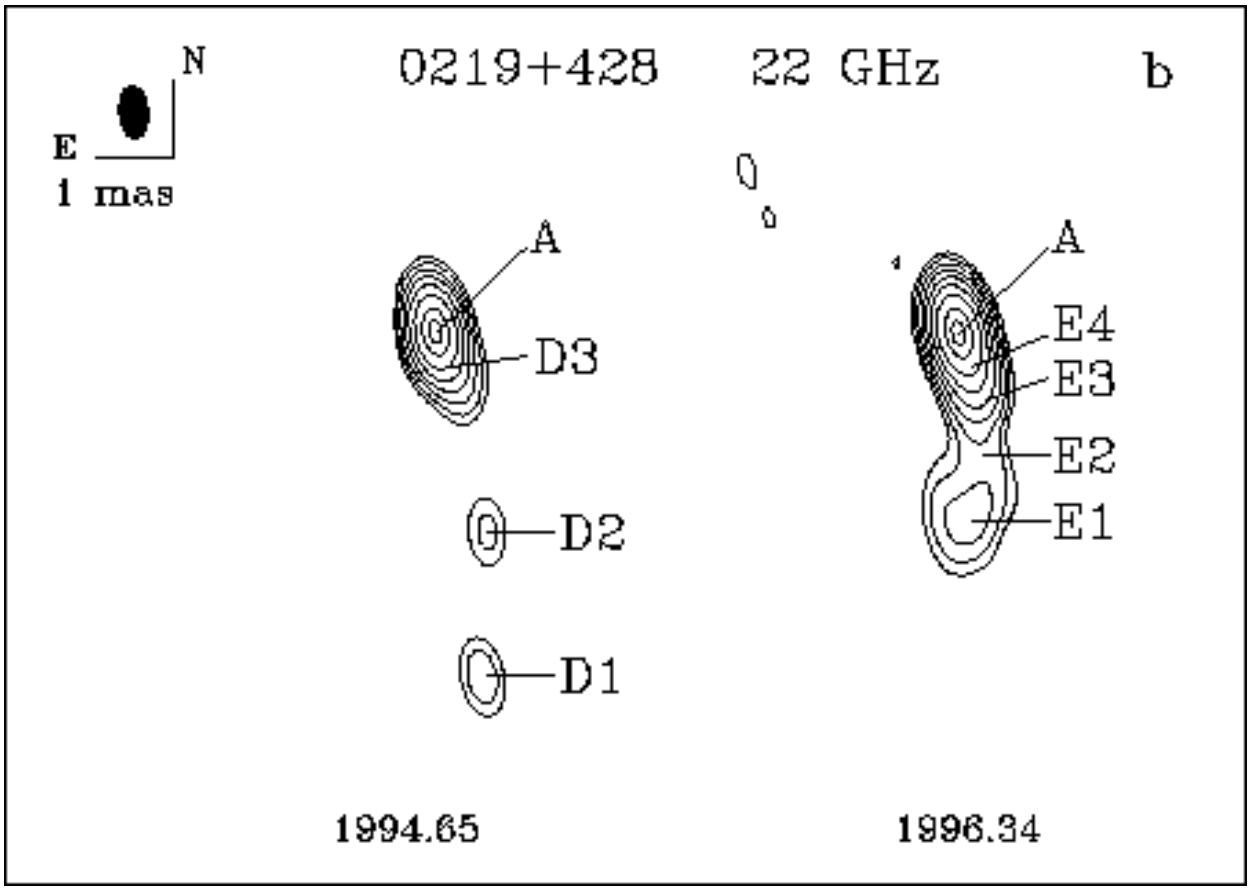}
\caption{Hybrid maps of 3C~66A at 22~GHz.}
\end{figure}
\begin{figure}
\figurenum{3c}
\plotone{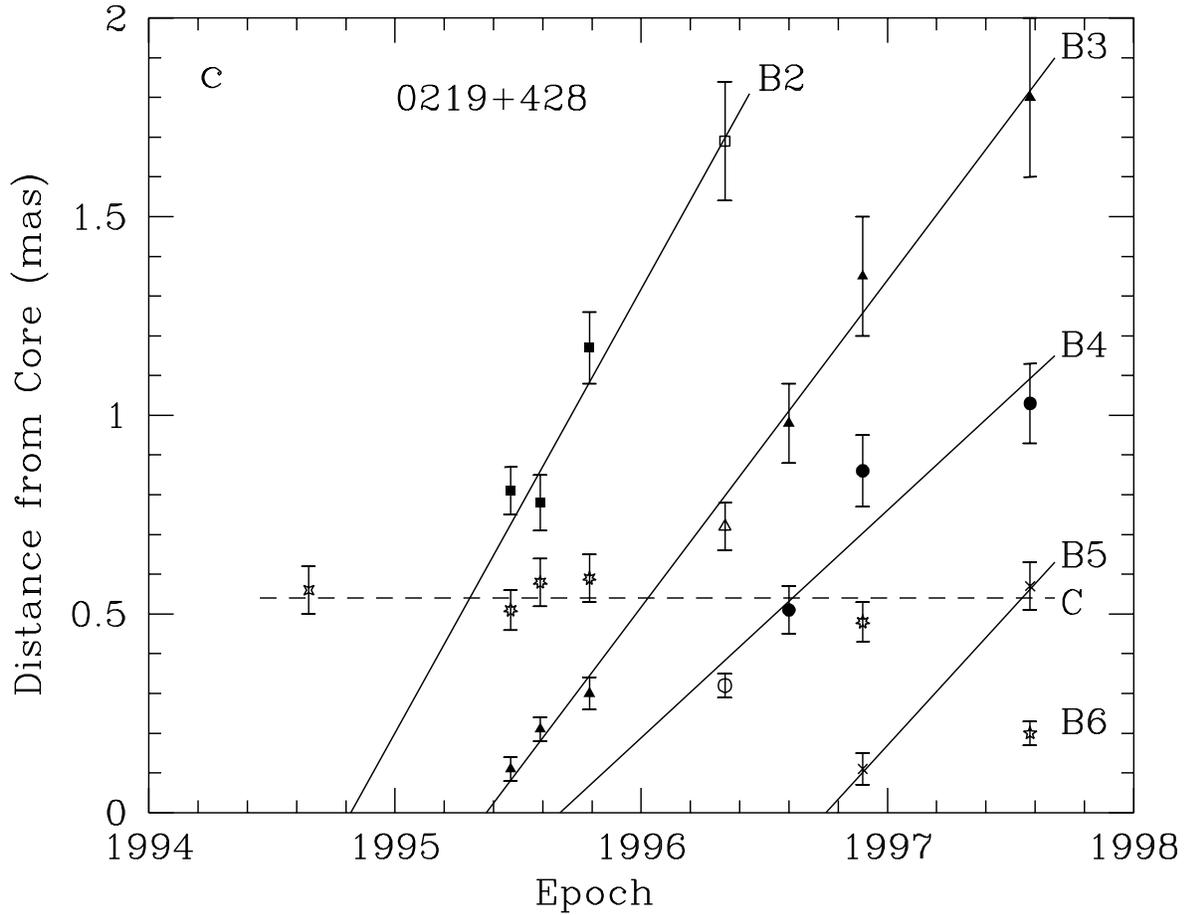}
\caption{Positions of components with respect to the core at different
epochs from model fitting for 3C~66A; designations of components are as follows:
7-point stars - component $C$ at 43~GHz, 4-point stars - component $D3$ 
at 22~GHz, filled squares -  $B2$ at 43~GHz,  open squares -
$E2$ at 22~GHz, filled triangles - $B3$ at 43~GHz,
open triangles - $E3$ at 22~GHz, filled circles - $B4$ at 43~GHz, open circles -
$E4$ at 22~GHz, crosses - $B5$ at 43~GHz, 5-point stars - $B6$ at 43~GHz.}
\end{figure}
\begin{figure}
\figurenum{4}
\plotone{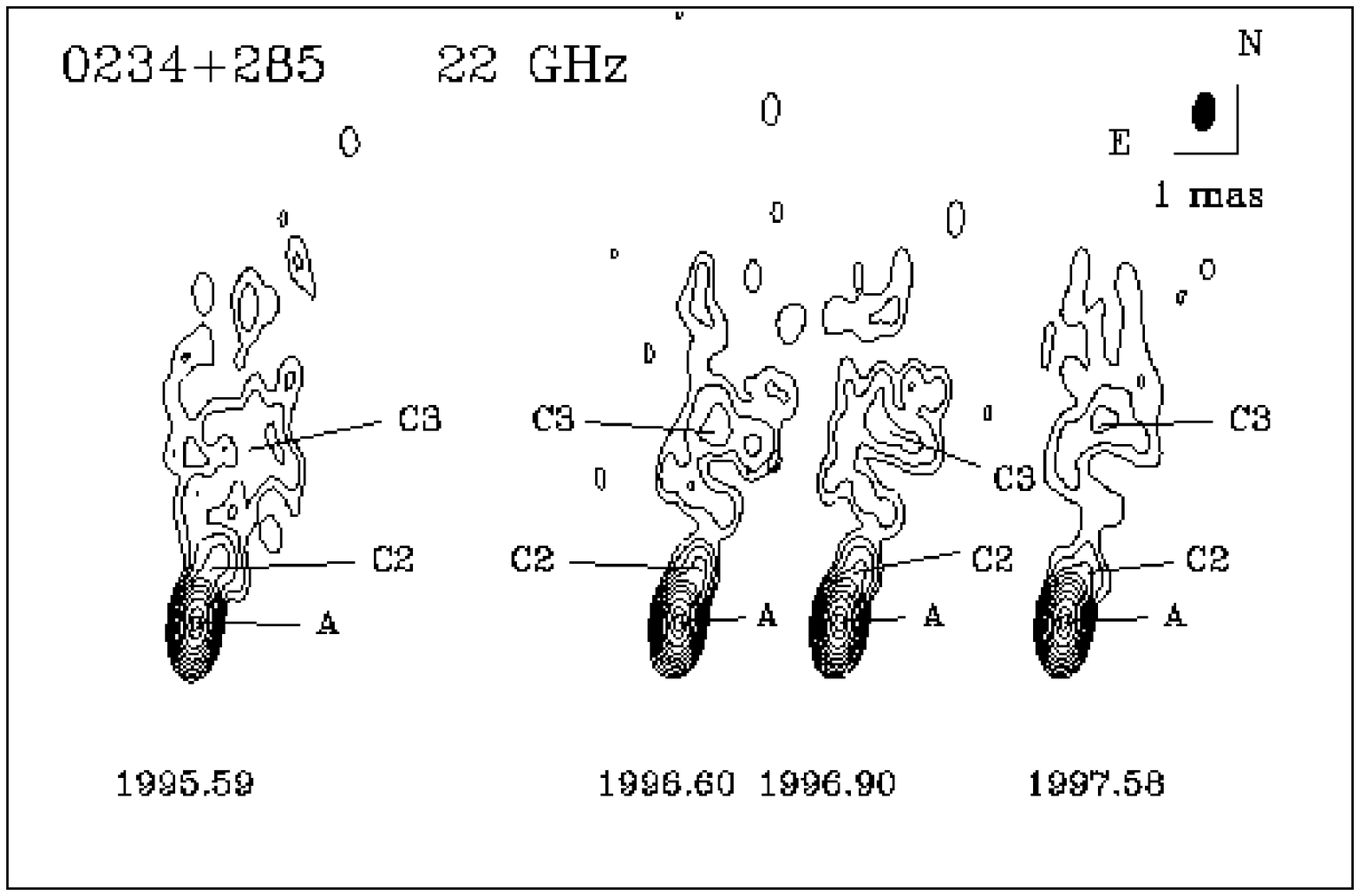}
\caption{Hybrid maps of 0234$+$285 at 22~GHz.}
\end{figure}
\begin{figure}
\figurenum{5a}
\plotone{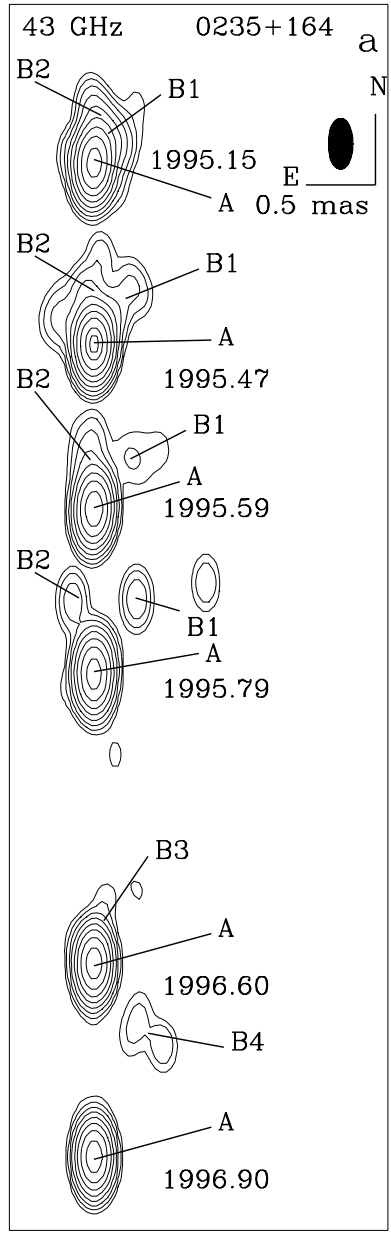}
\caption{Hybrid maps of 0235$+$164 at 43~GHz (time increases downward).}
\end{figure}
\begin{figure}
\figurenum{5b}
\plotone{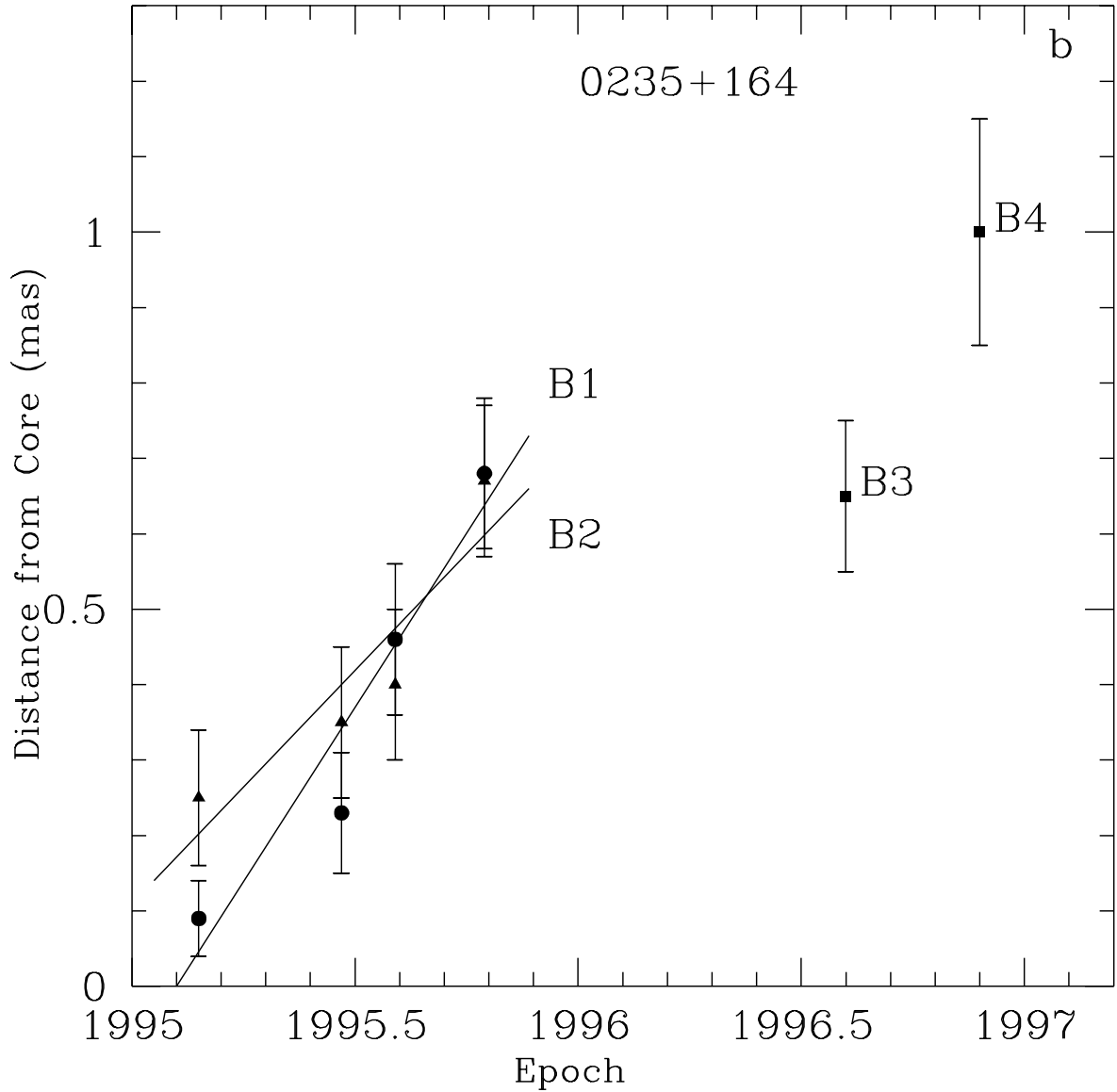}
\vspace{-1cm}
\caption{Positions of components with respect to the core at different
epochs from model fitting for 0235$+$164; designations of components are as follows:
filled triangles - component $B2$,
filled circles - $B1$, filled squares - components $B3$ and $B4$.}
\end{figure}
\begin{figure}
\figurenum{6a}
\plotone{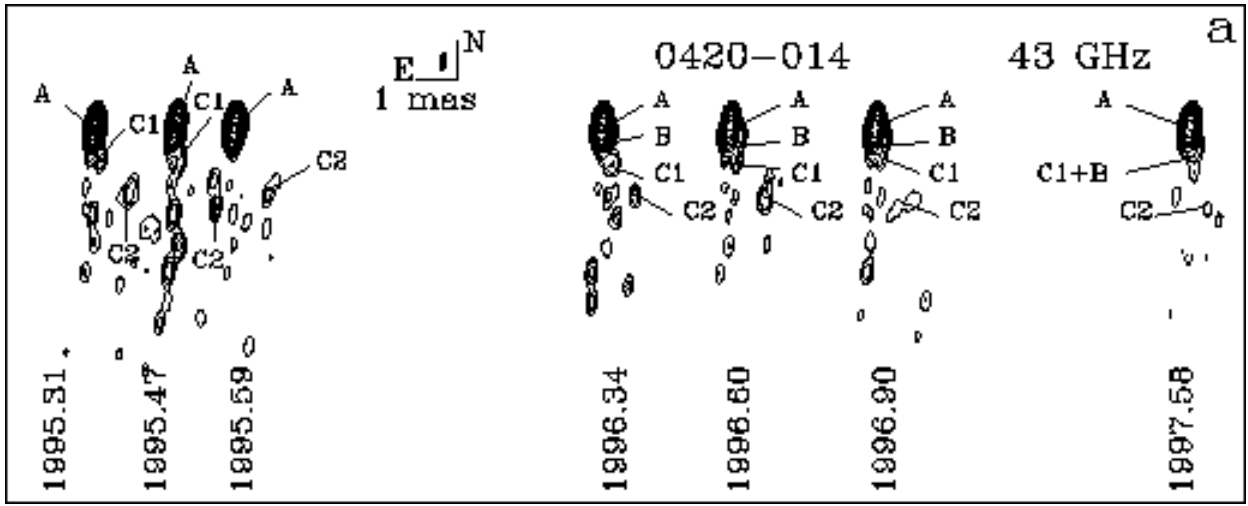}
\caption{Hybrid maps of 0420$-$014 at 43~GHz.}
\vspace{-1cm}
\end{figure}
\begin{figure}
\figurenum{6b}
\plotone{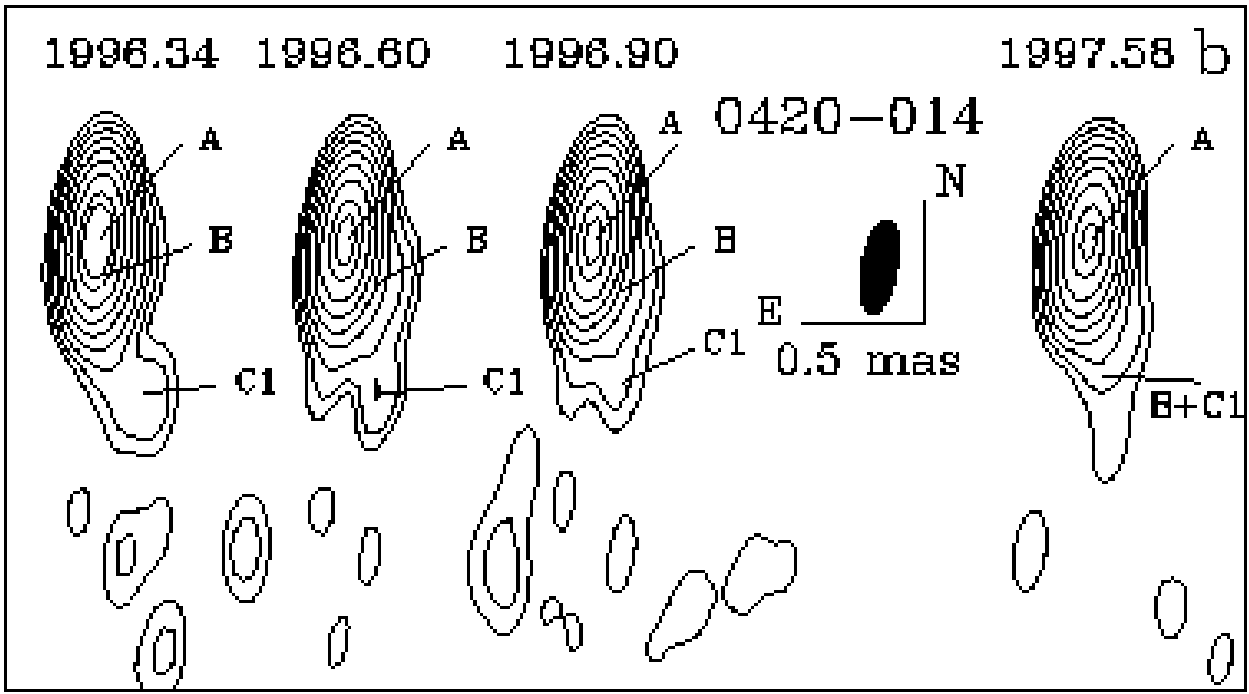}
\caption{Hybrid maps of 0420$-$014 at 43~GHz at the last four epochs
on a scale that accentuates the region near the core.}
\vspace{-1cm}
\end{figure}
\begin{figure}
\figurenum{6c}
\plotone{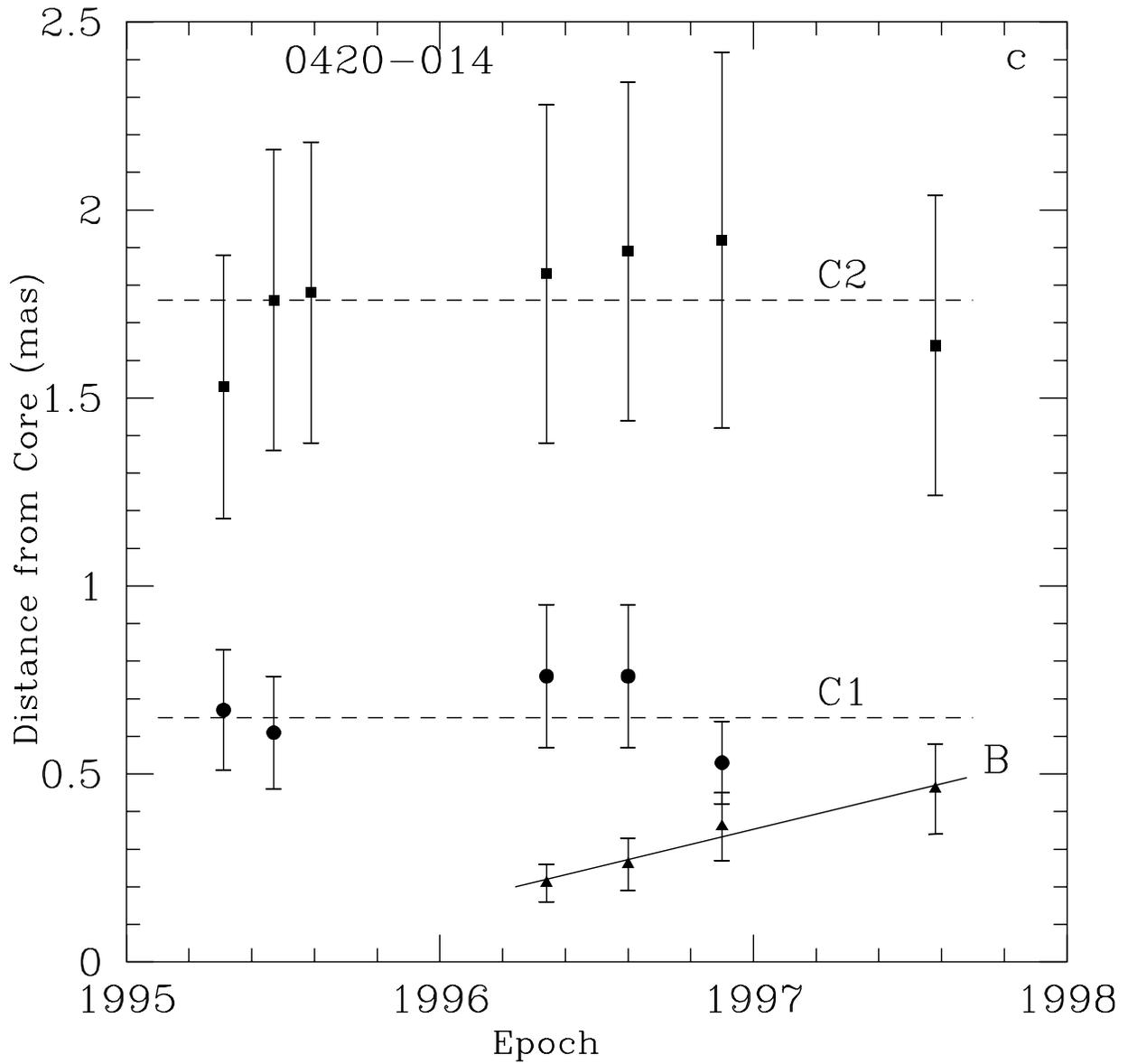}
\caption{Positions of components with respect to the core at different
epochs from model fitting for 0420$-$014; designations of components are as follows:
filled triangles - component $B$,
filled circles - $C1$, filled squares - $C2$. }
\end{figure}
\begin{figure}
\figurenum{7}
\plotone{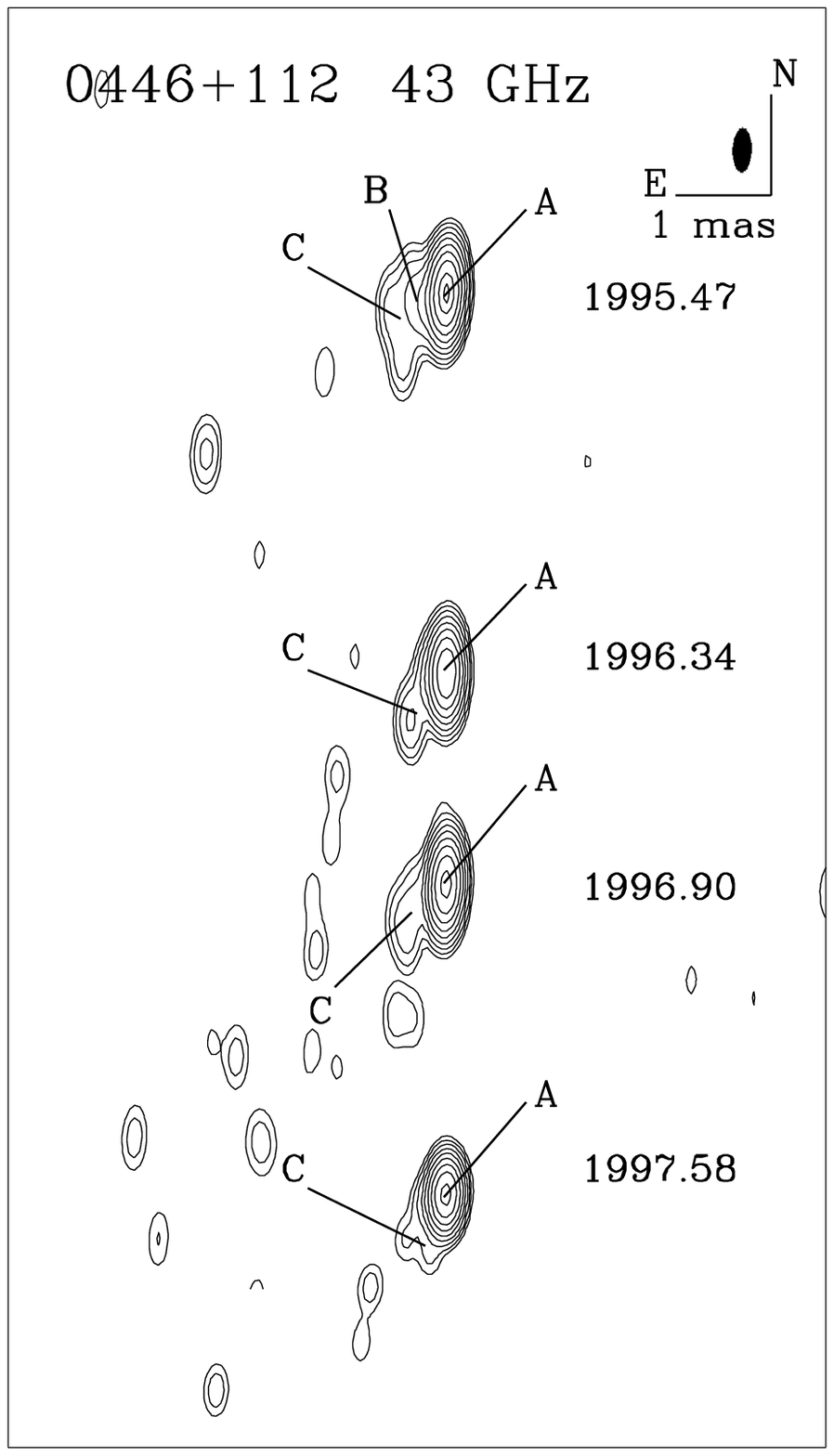}
\caption{Hybrid maps of 0446$+$112 at 43~GHz.}
\end{figure}
\begin{figure}
\figurenum{8}
\plotone{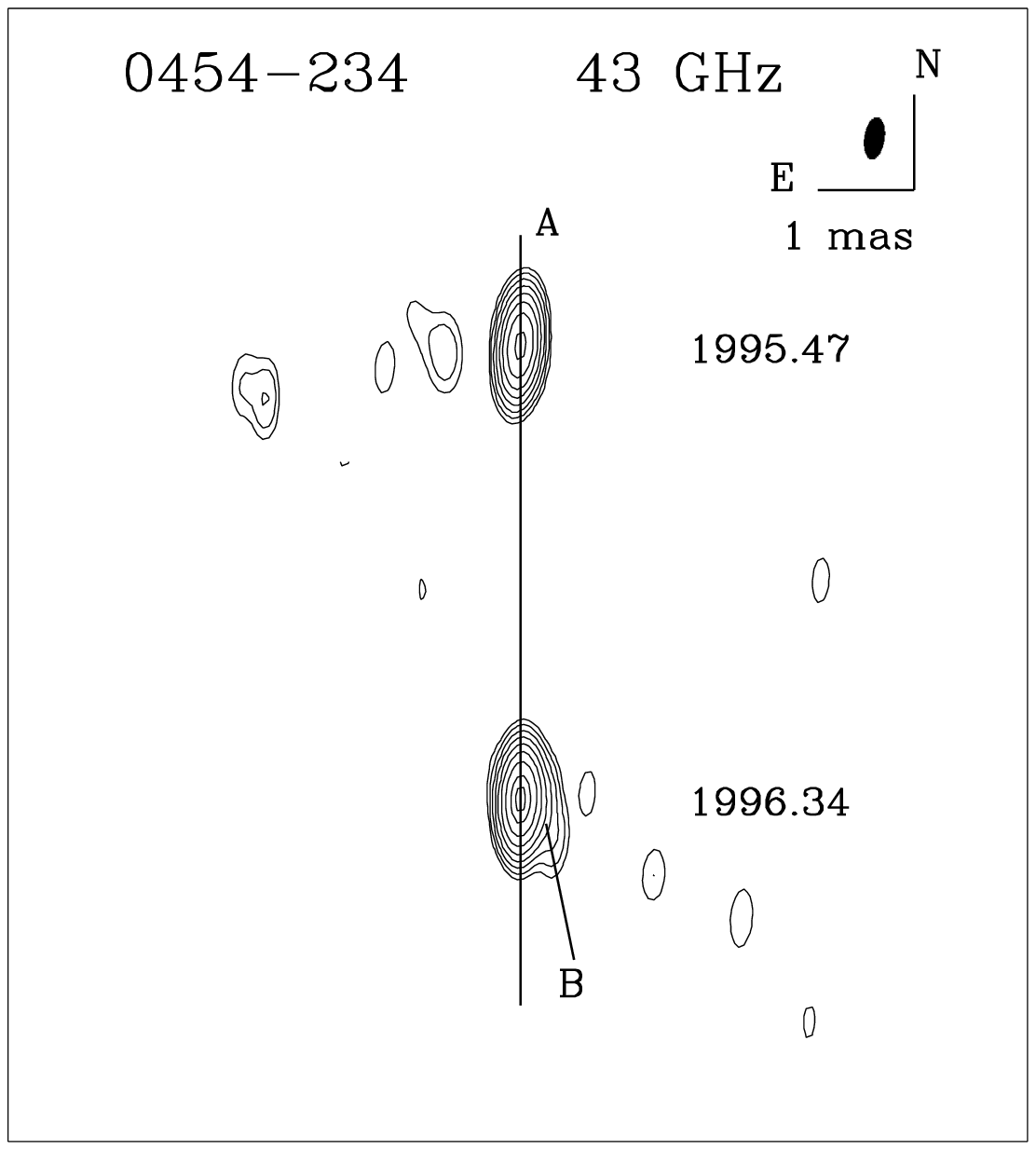}
\caption{Hybrid maps of 0454$-$234 at 43~GHz.}
\end{figure}
\begin{figure}
\figurenum{9a}
\plotone{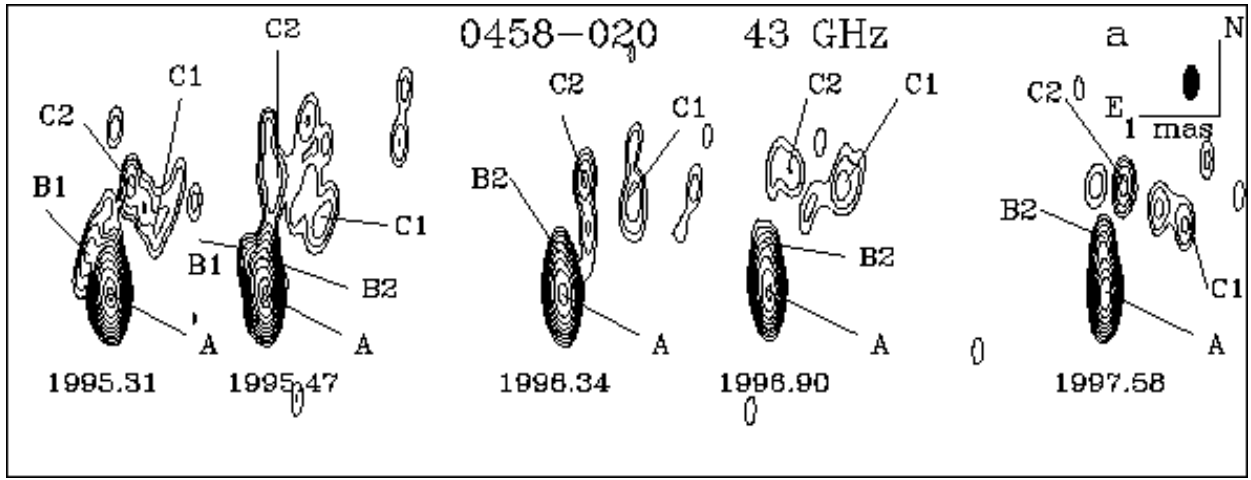}
\figcaption{Hybrid maps of 0458$-$020 at 43~GHz.}
\end{figure}
\begin{figure}
\figurenum{9b}
\plotone{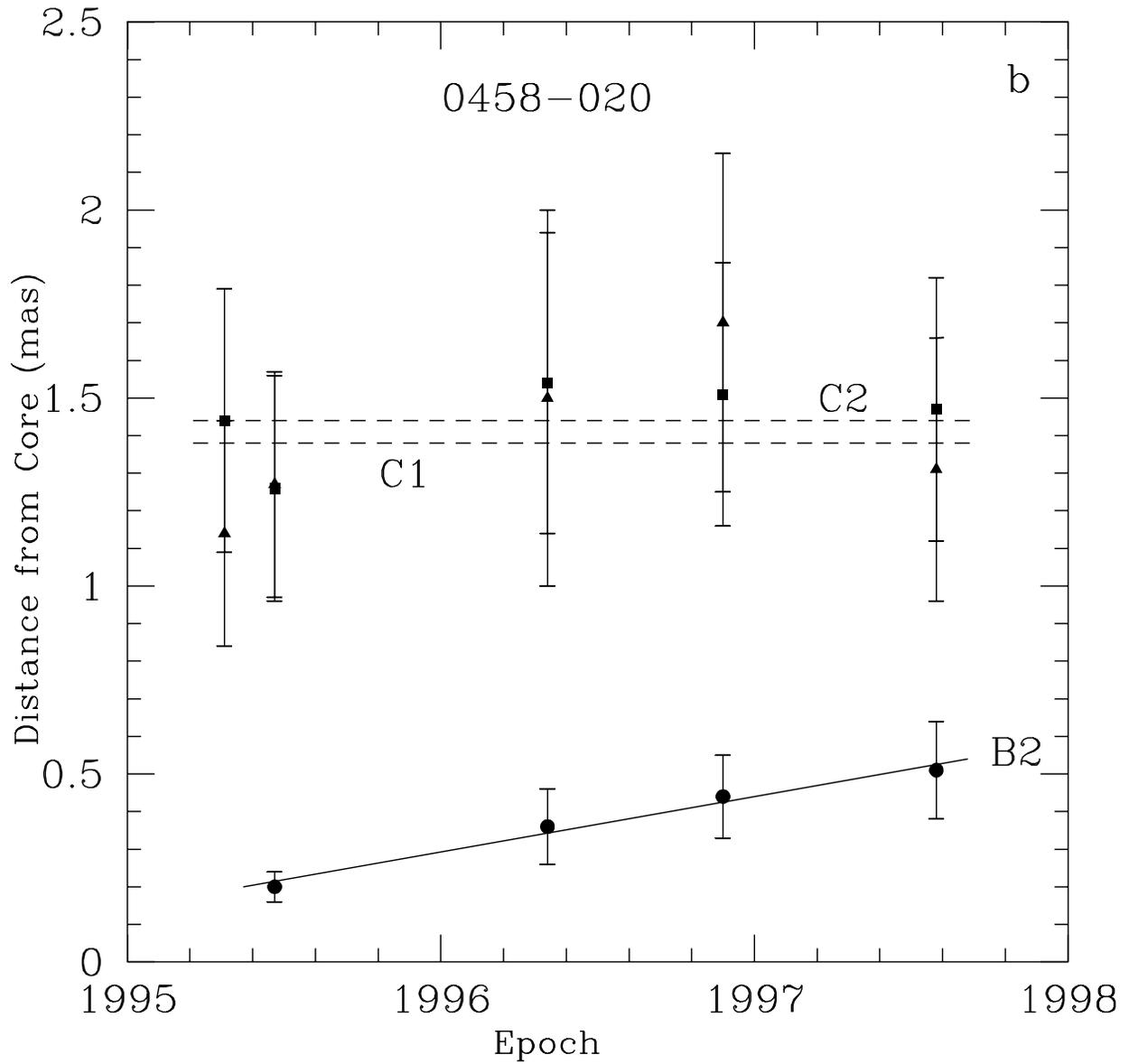}
\caption{Positions of components with respect to the core at different
epochs from model fitting for 0458$-$020; designations of components are as follows:
filled circles - component $B2$,
filled triangles - $C1$, filled squares - $C2$.}
\end{figure}
\begin{figure}
\figurenum{10a}
\plotone{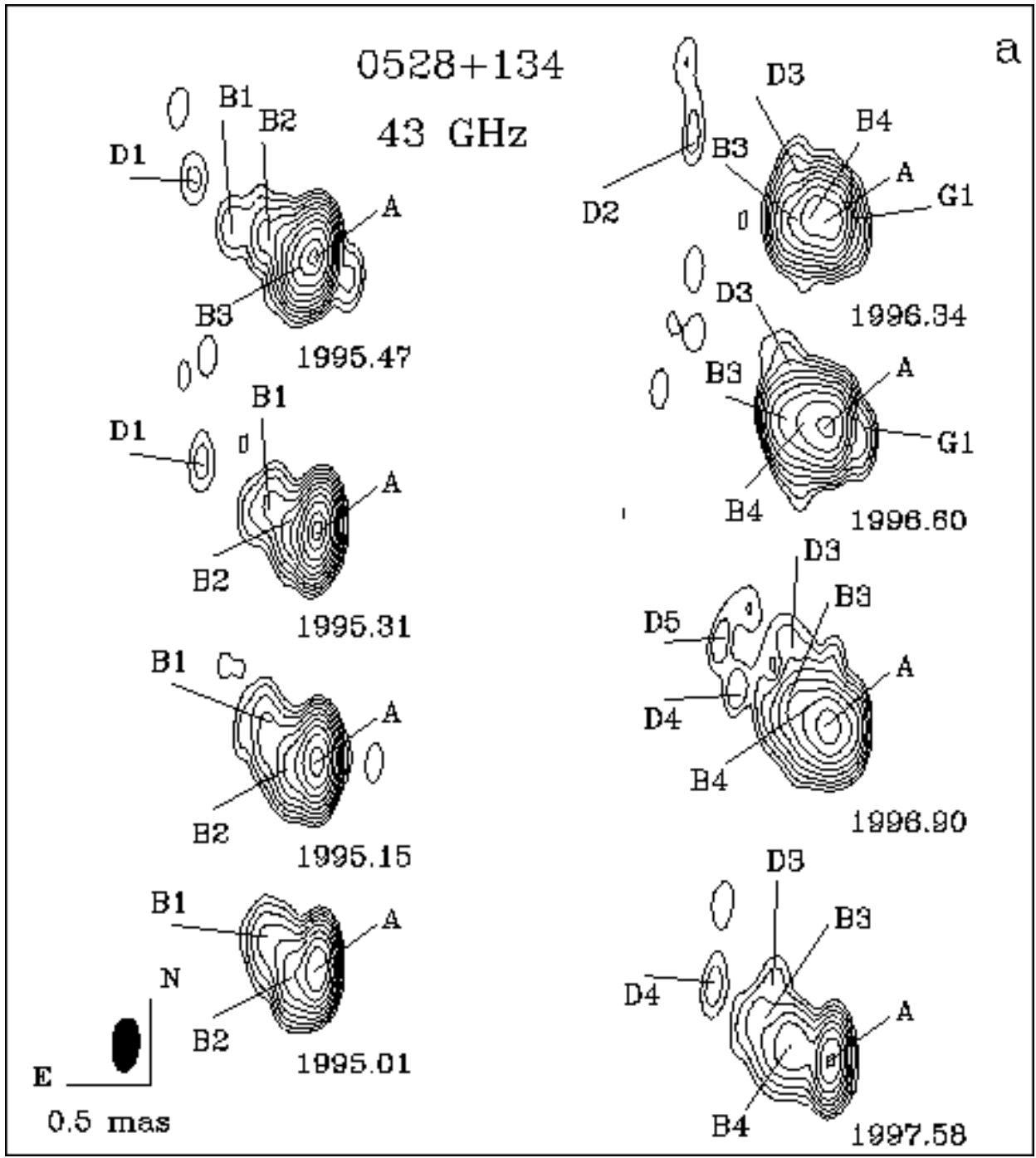}
\caption{Hybrid maps of 0528$+$134 at 43~GHz.}
\end{figure}
\begin{figure}
\figurenum{10b}
\plotone{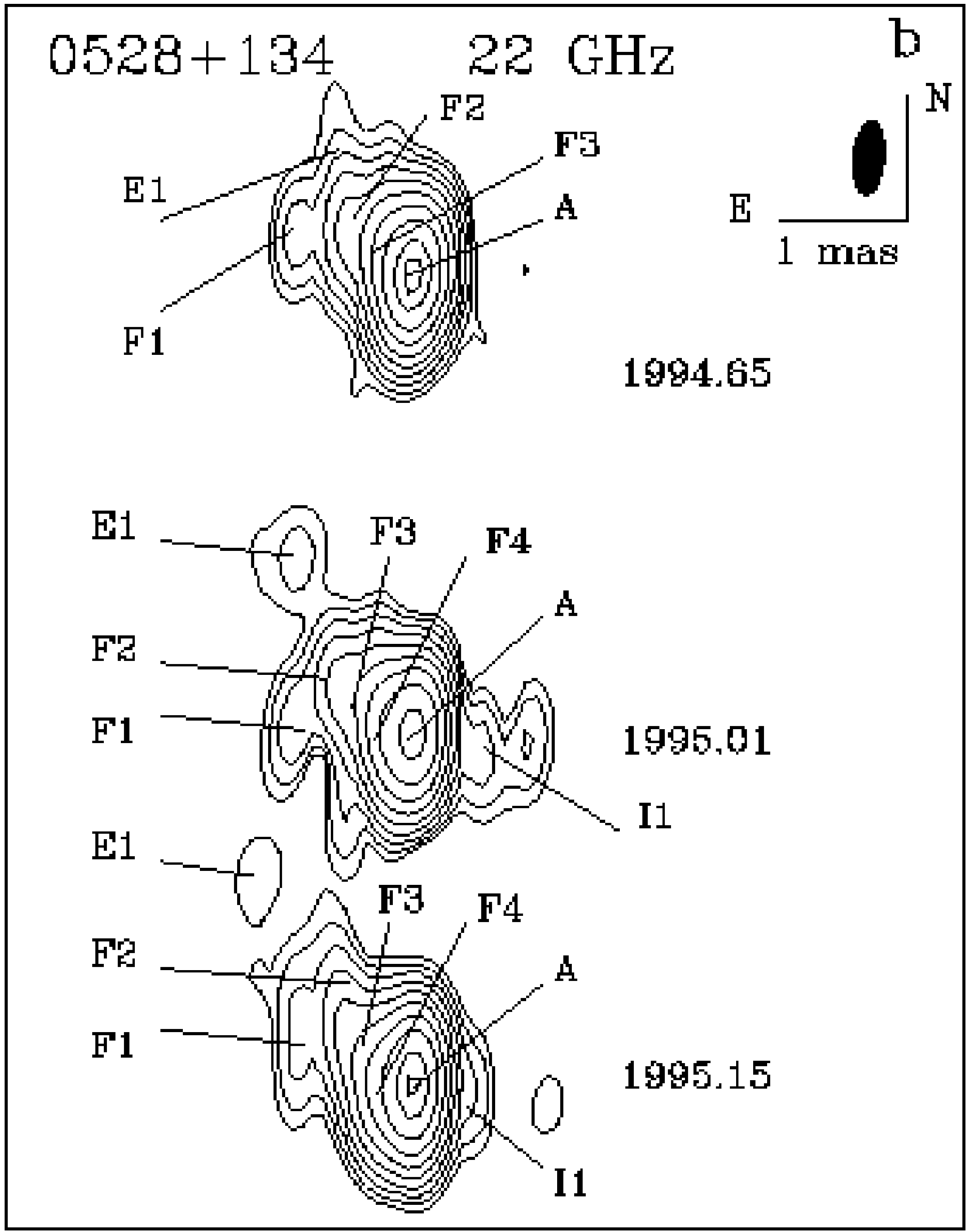}
\caption{Hybrid maps of 0528$+$134 at 22~GHz.}
\end{figure}
\begin{figure}
\figurenum{10c}
\plotone{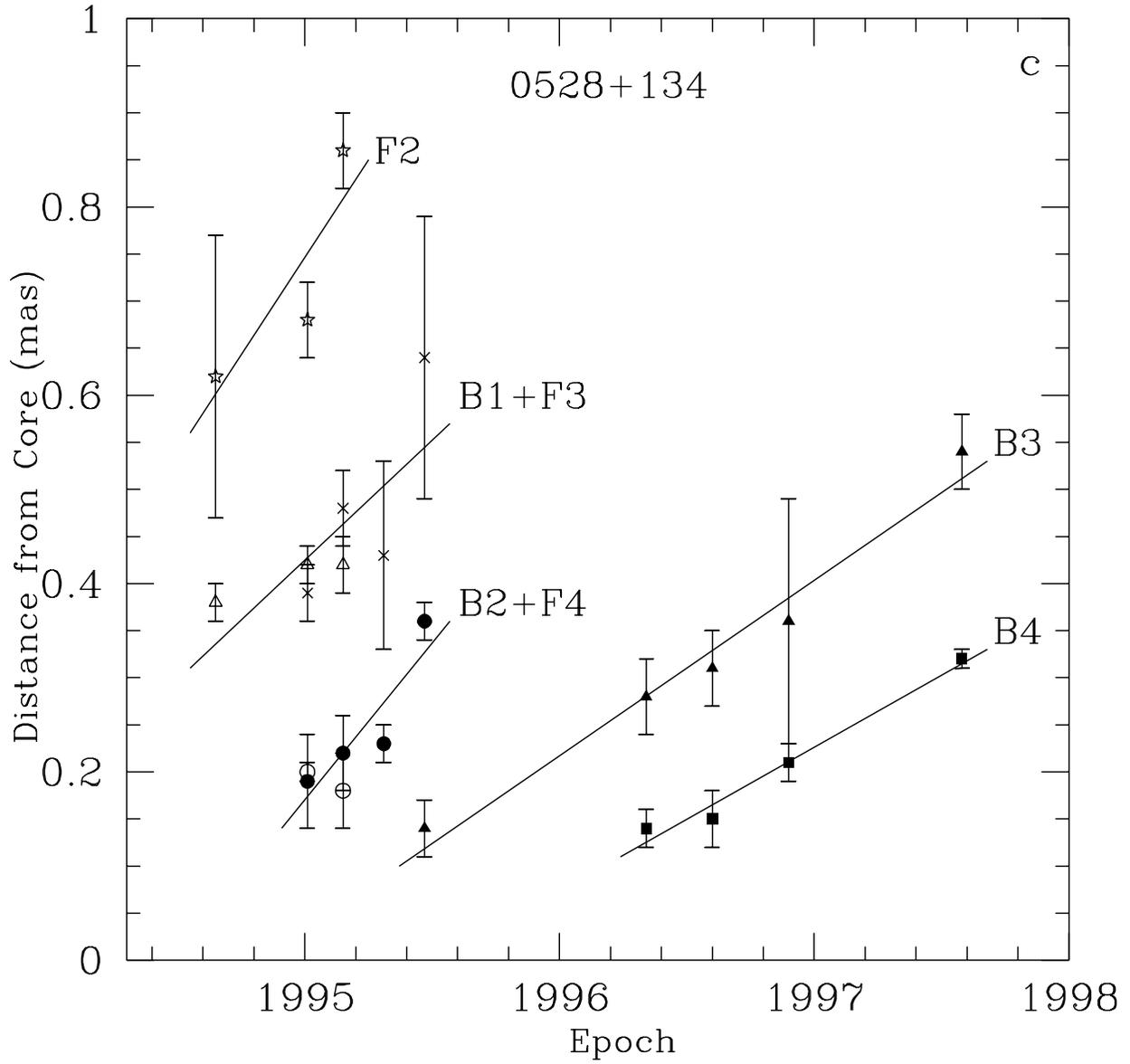}
\caption{Positions of components with respect to the core at different
epochs from model fitting for 0528$+$134; designations of components are as follows:
filled squars - component $B4$ at 43~GHz,
filled triangles - $B3$ at 43~GHz, filled circles - $B2$ at 43~GHz and
open circles - $F4$ at 22~GHz, crosses - $B1$ at 43~GHz and open triangles -
$F3$ at 22~GHz,  stars - $F2$ at 22~GHz.}
\end{figure}
\clearpage
\begin{figure}
\figurenum{11a}
\plotone{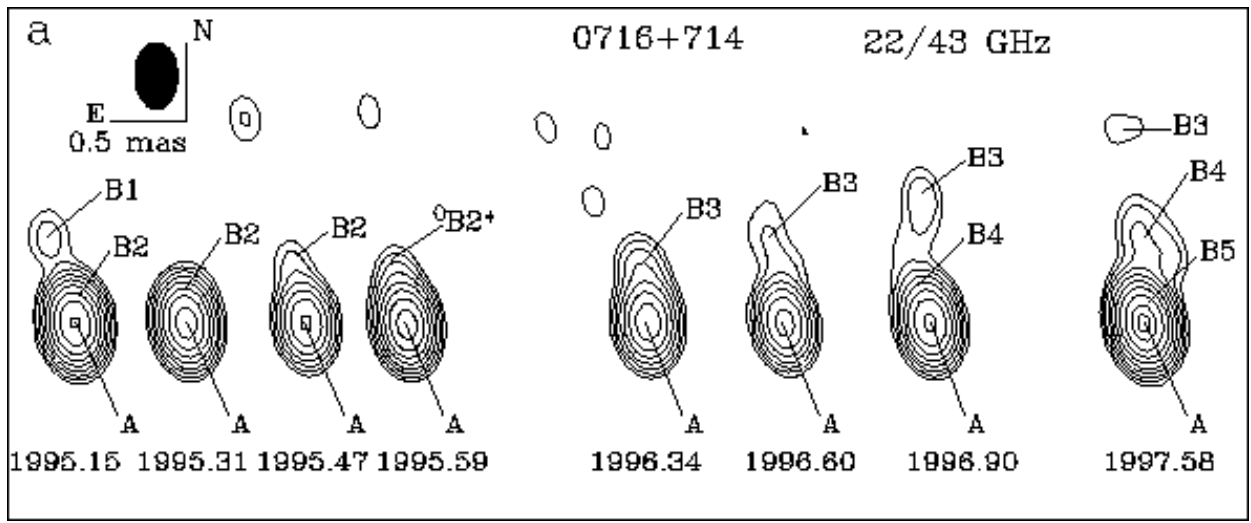}
\caption{Hybrid maps of 0716$+$714 at 22~GHz.
The map at epoch 1995.59 is represented by the image at 43~GHz convolved with 
a beam (dark ellipse) corresponding to the beam at 22~GHz.}
\end{figure}
\begin{figure}
\figurenum{11b}
\plotone{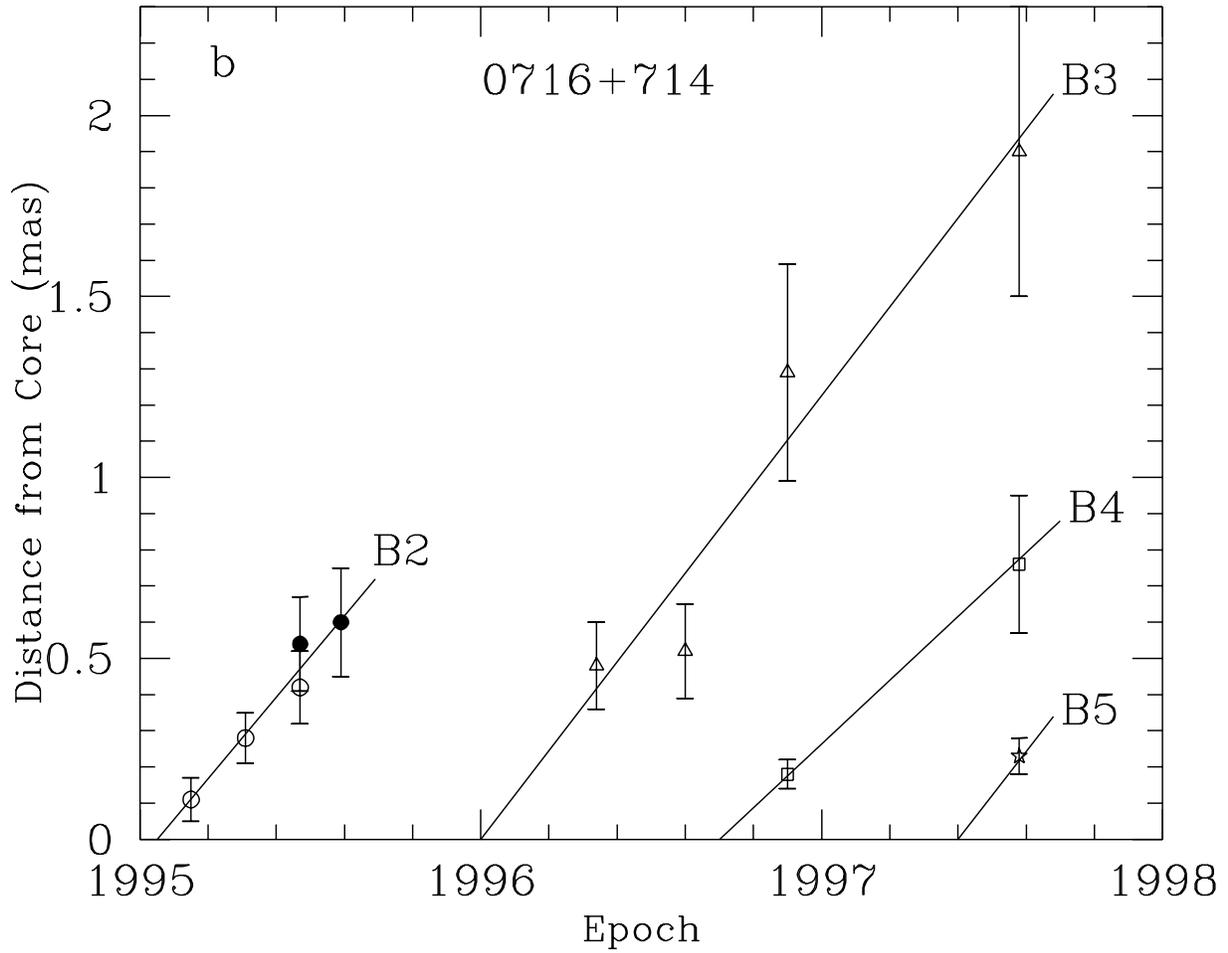}
\figcaption{Positions of components with respect to the core at different
epochs from model fitting for 0716$+$714; designations of components are as follows:
star - component $B5$ at 22~GHz,
open squares - $B4$ at 22~GHz, open triangles -  $B3$ at 22~GHz
open circles - $B2$ at 22~GHz and filled circles - $B2^*$ at 43~GHz.}
\end{figure}
\begin{figure}
\figurenum{12}
\plotone{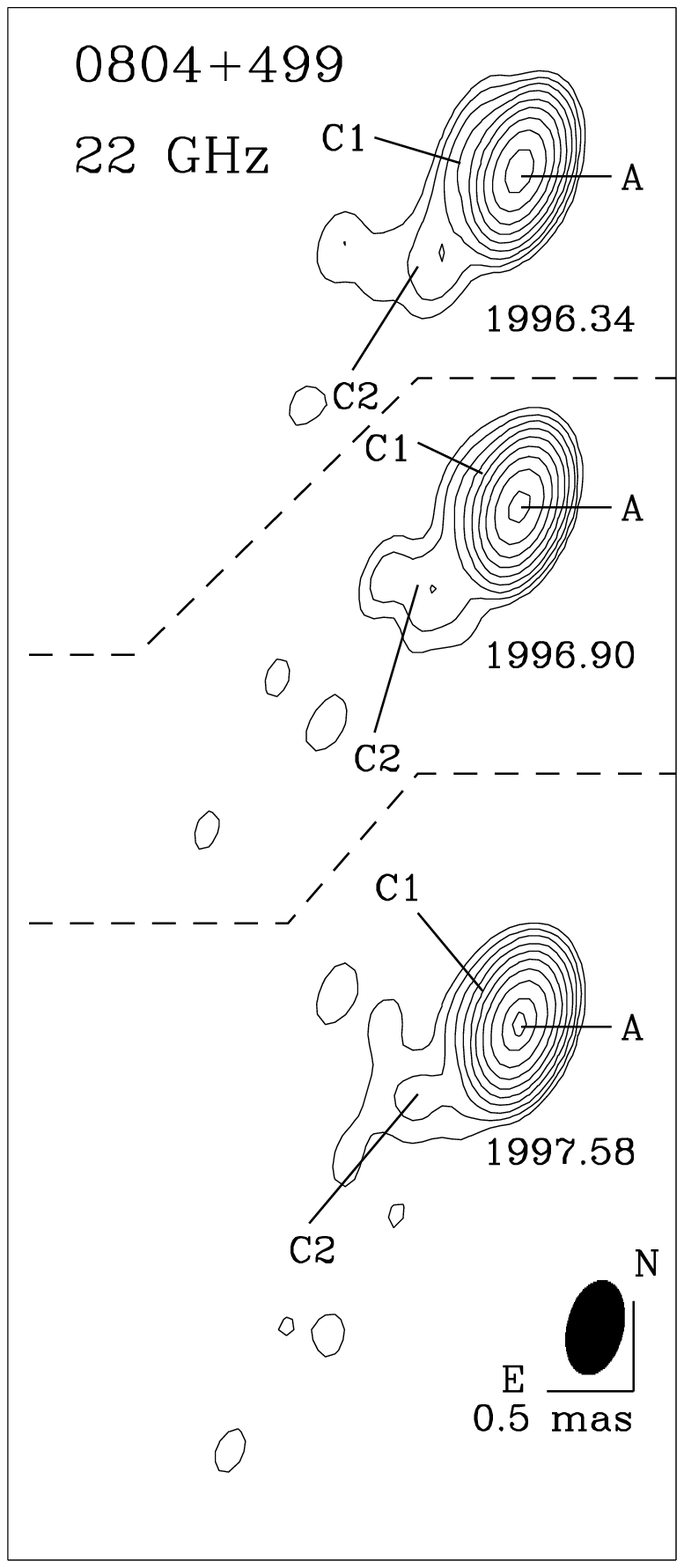}
\caption{Hybrid maps of 0804$+$499 at 22~GHz.}
\end{figure}
\begin{figure}
\figurenum{13a}
\plotone{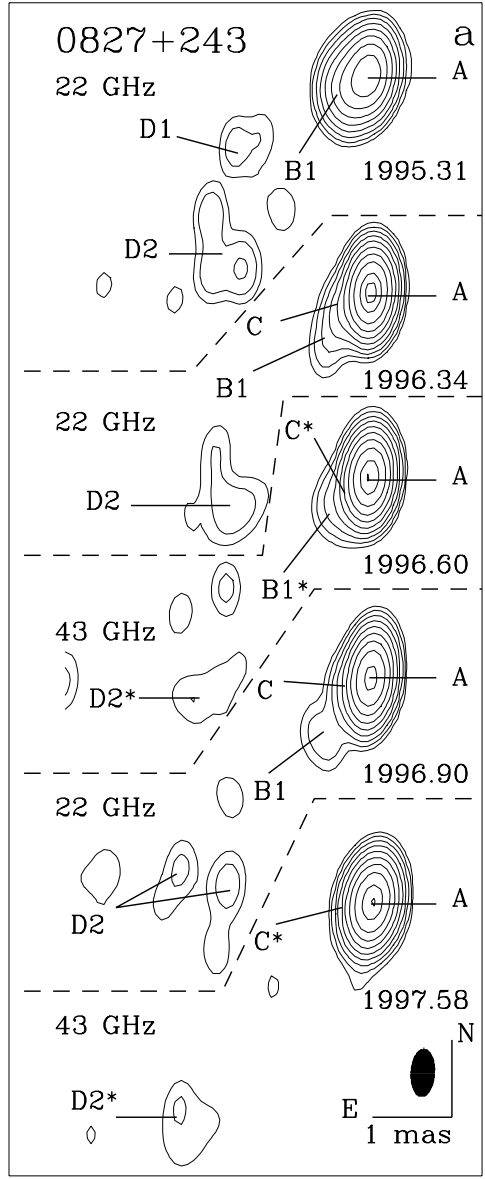}
\caption{Hybrid maps of 0827$+$243 at 43 and 22~GHz.
The images at 43~GHz are convolved with 
a beam (dark ellipse) corresponding to the beam at 22~GHz.
At epoch 1996.90 there are observations at 22 and 43~GHz, but only the data at 
22~GHz are shown.}
\end{figure}
\begin{figure}
\figurenum{13b}
\plotone{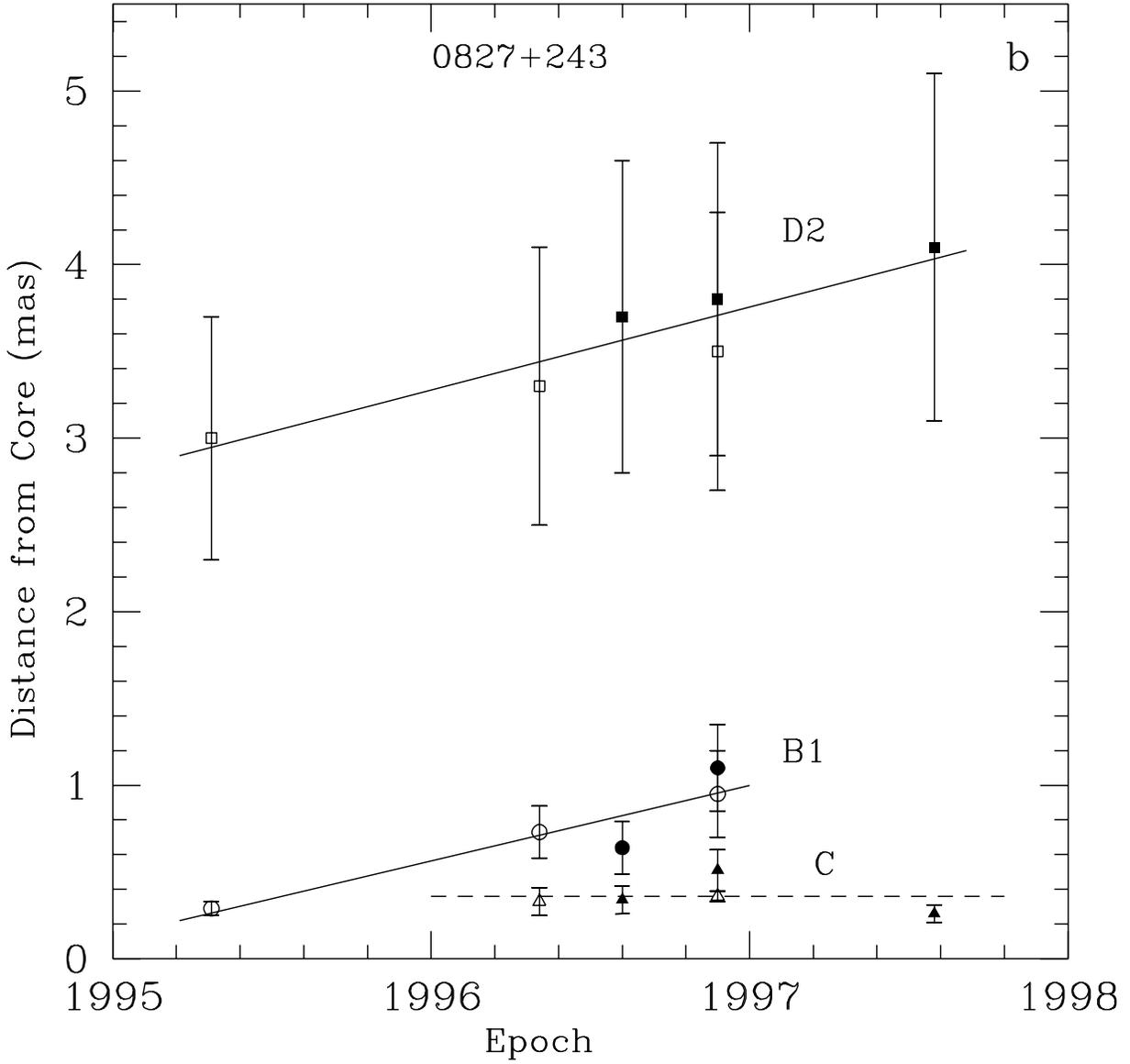}
\caption{Positions of components with respect to the core at different
epochs from model fitting for 0827$+$243; designations of components are as follows:
filled triangles - component $C^*$ at 43~GHz and open triangles -
$C$ at 22~GHz,
filled circles - $B1^*$ at 43~GHz and open circles - $B1$ at 22~GHz,
filled squares - $D2^*$ at 43~GHz and  open squares - $D2$ at 22~GHz.}
\end{figure}
\begin{figure}
\figurenum{14a}
\plotone{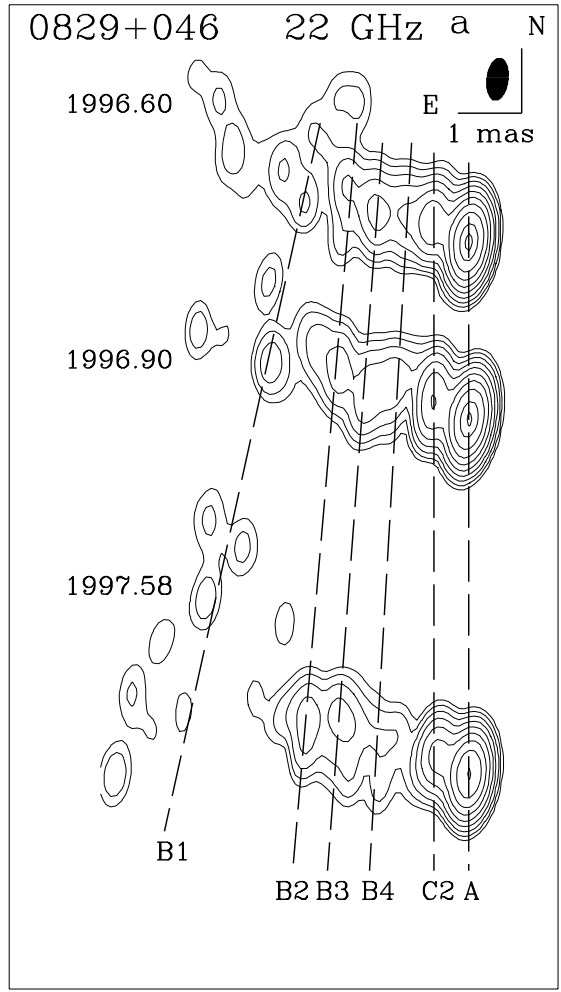}
\caption{Hybrid maps of 0829$+$046 at 22~GHz.}
\end{figure}
\begin{figure}
\figurenum{14b}
\plotone{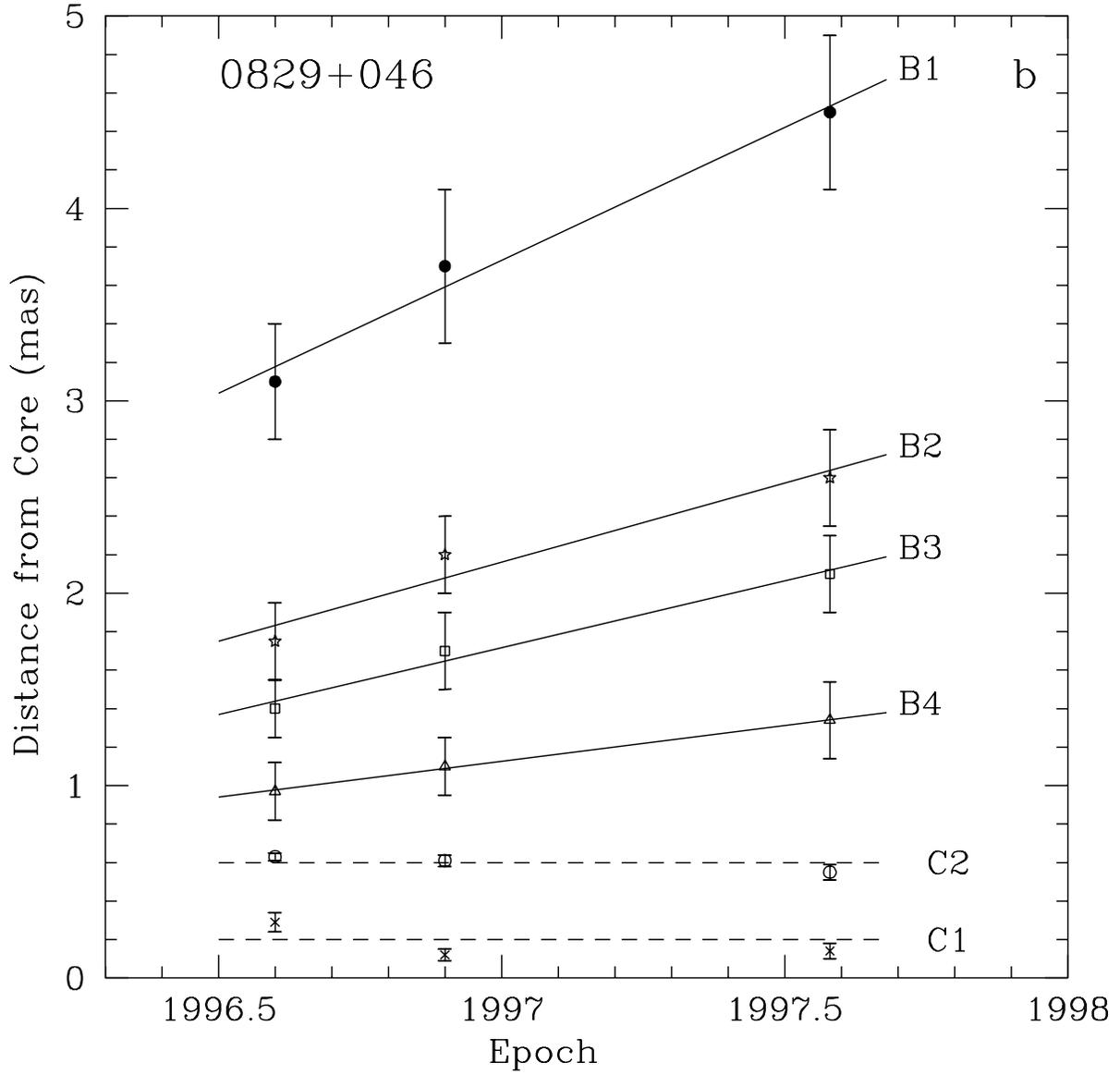}
\caption{Positions of components with respect to the core at different
epochs from model fitting for 0829$+$046; designations of components are as follows:
crosses - component $C1$, open circles - $C2$,
open triangles - $B4$, open squares - $B3$,
stars - $B2$, and  filled circles - $B1$. }
\end{figure}
\begin{figure}
\figurenum{15a}
\plotone{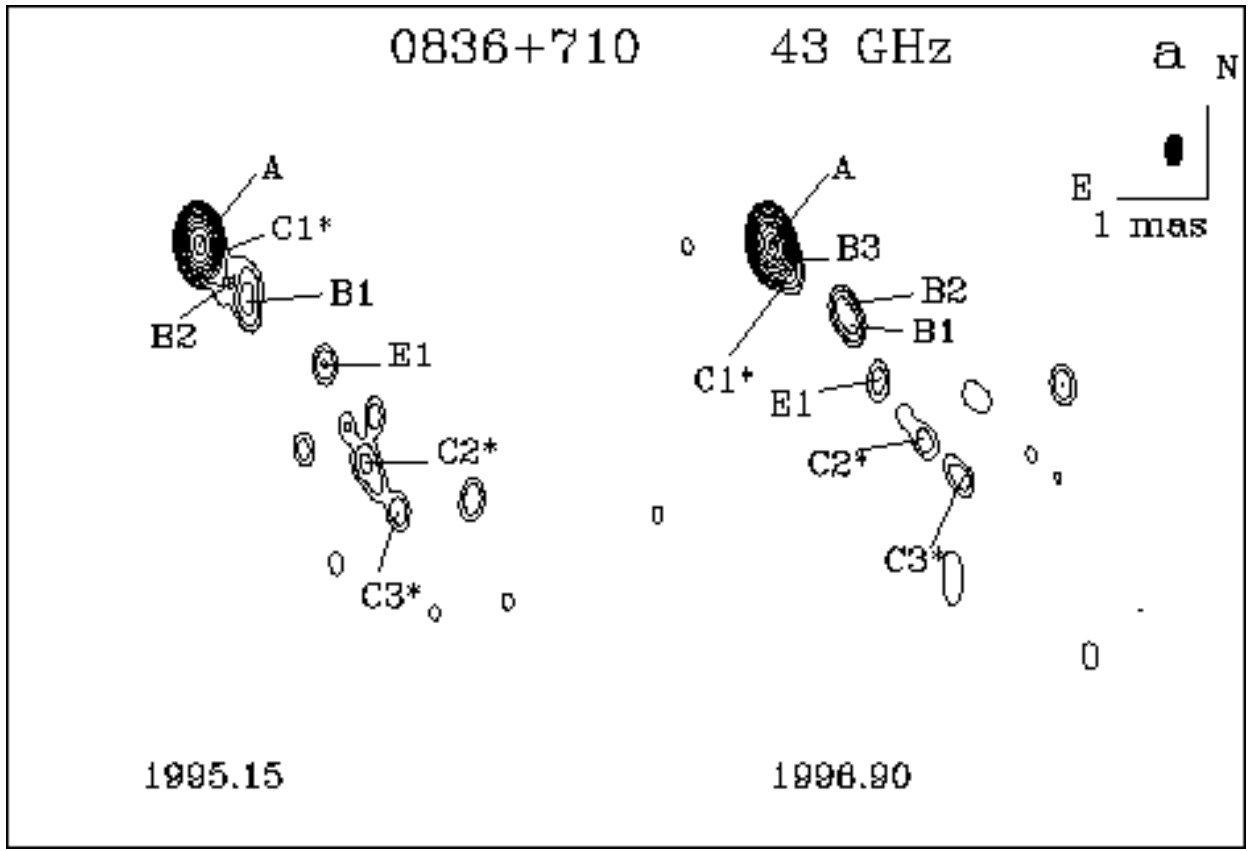}
\caption{Hybrid maps of 0836$+$710 at 43~GHz.}
\end{figure}
\begin{figure}
\figurenum{15b}
\plotone{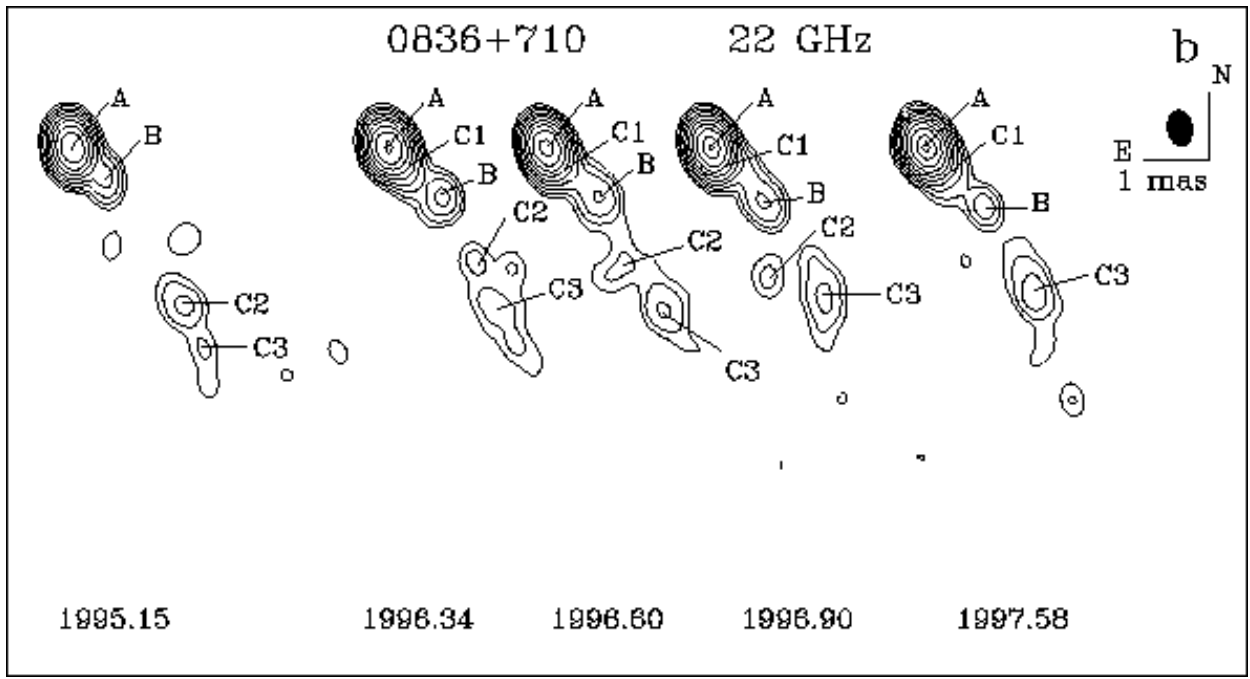}
\caption{Hybrid maps of 0836$+$710 at 22~GHz.}
\end{figure}
\begin{figure}
\figurenum{15c}
\plotone{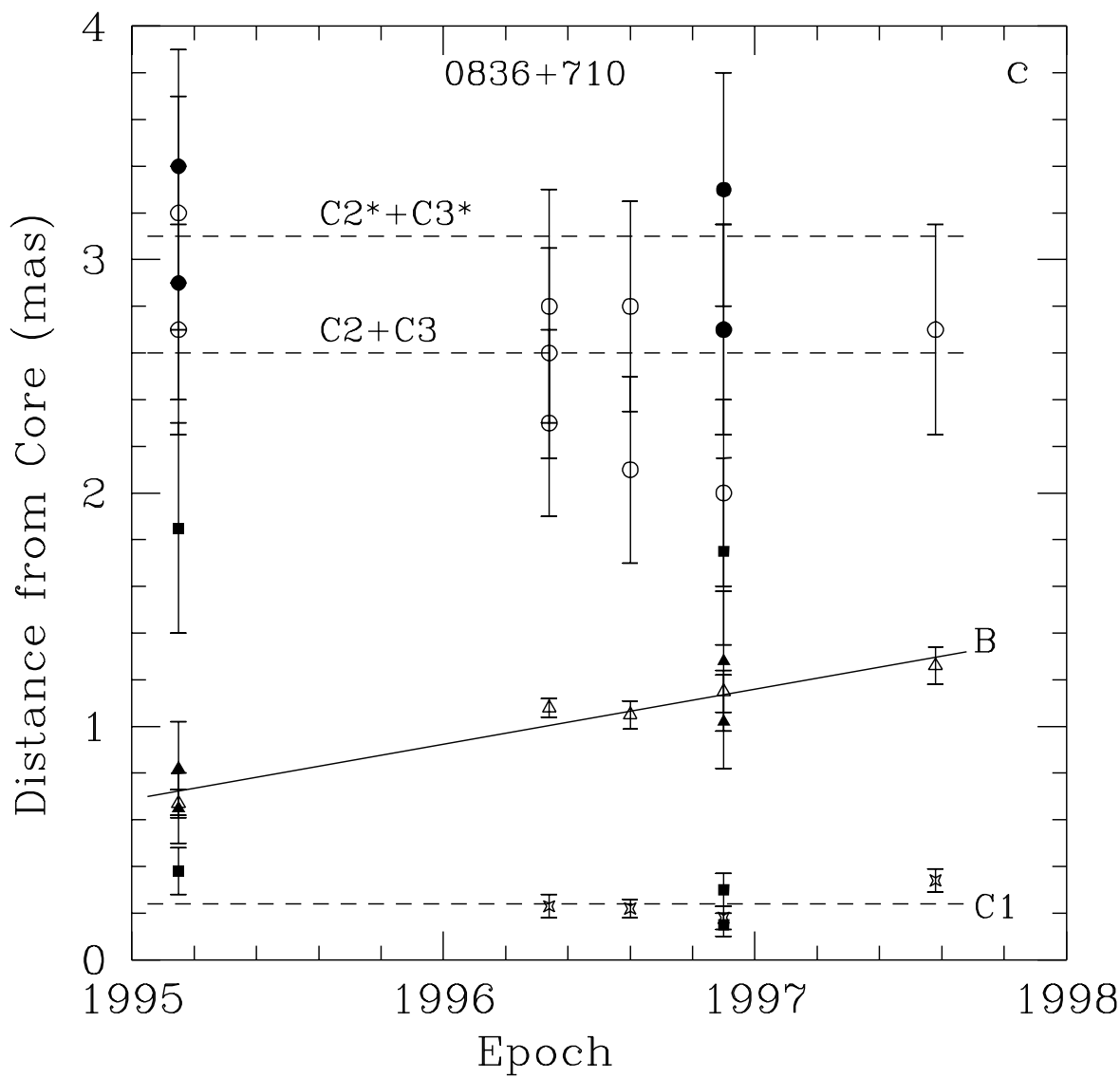}
\caption{Positions of components with respect to the core at different
epochs from model fitting for 0836$+$710; designations of components are as follows:
4-point stars - component $C1$ at 22~GHz, filled squares - $C1^*$ at 43~GHz, open triangles - 
component $B$ at 22~GHz, filled triangles -  components $B1$ and $B2$ at 43~GHz,   
open circles - components $C2$ and $C3$ at 22~GHz, filled circles - components $C2^*$ and $C3^*$ at 43~GHz.}
\end{figure}
\clearpage
\begin{figure}
\figurenum{16a}
\plotone{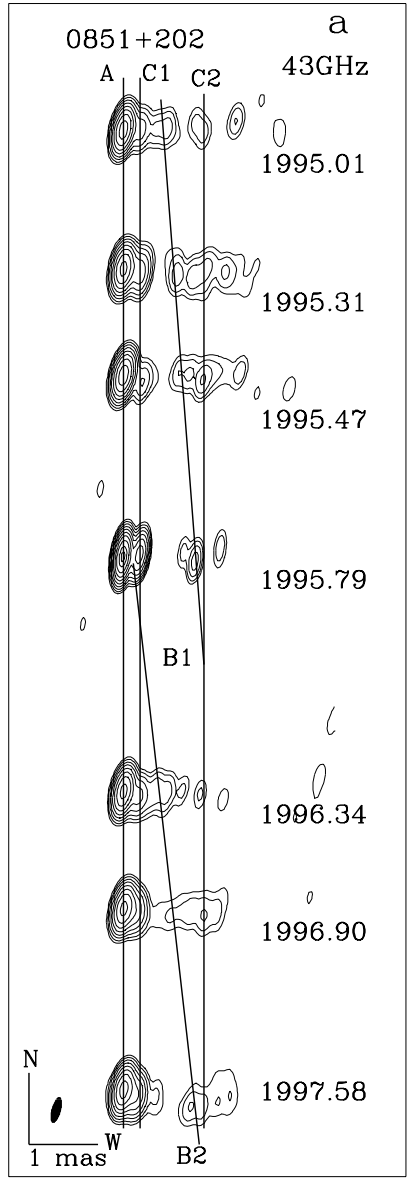}
\caption{Hybrid maps of OJ~287 at 43~GHz.}
\end{figure}
\begin{figure}
\figurenum{16b}
\plotone{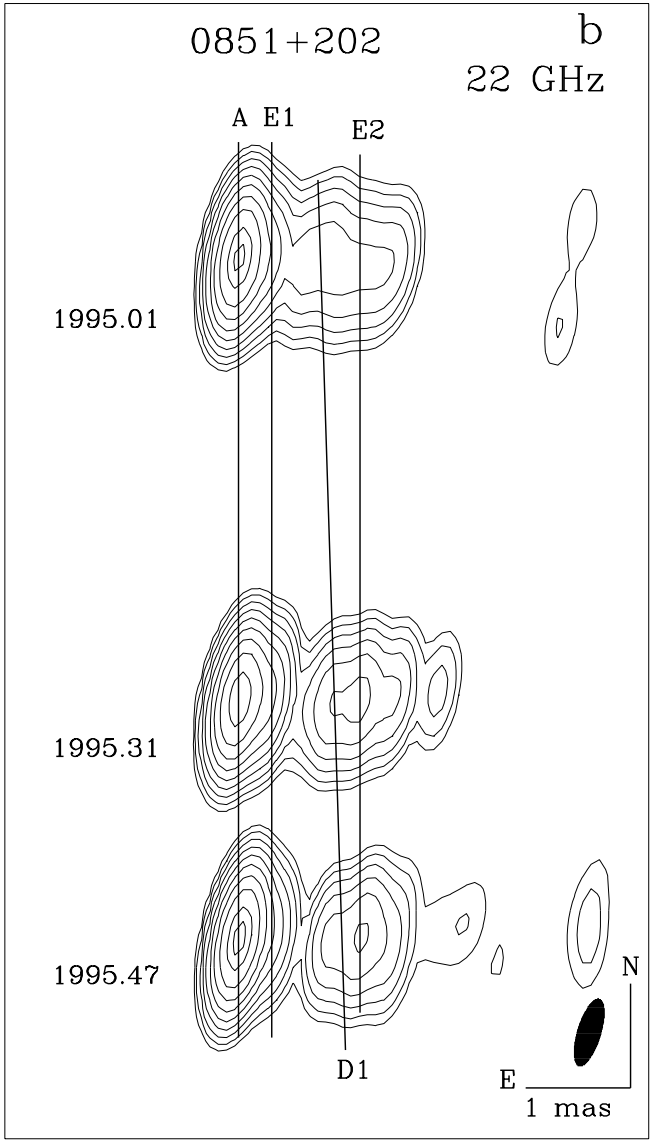}
\caption{Hybrid maps of OJ~287 at 22~GHz.}
\end{figure}
\begin{figure}
\figurenum{16c}
\plotone{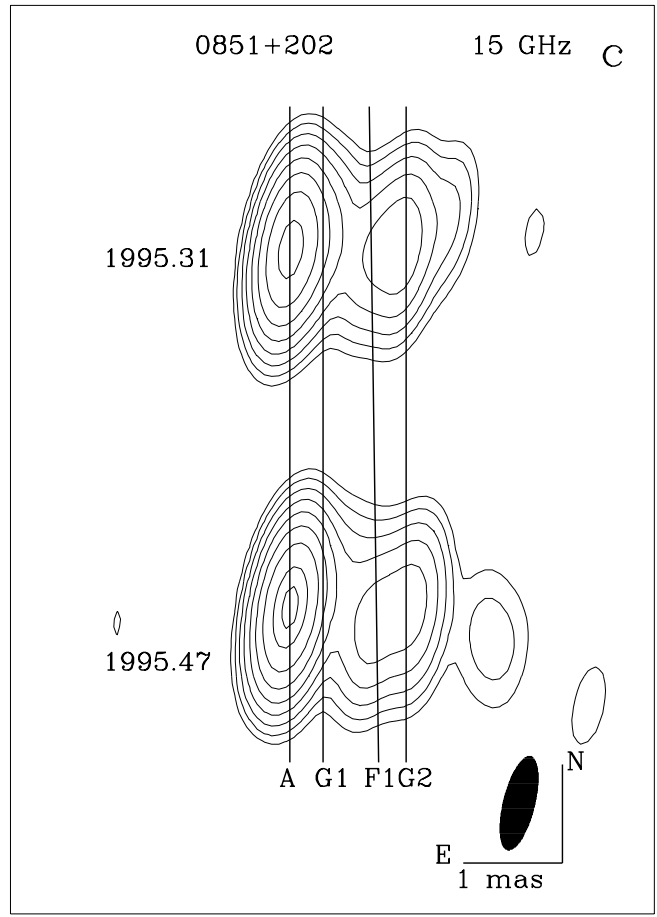}
\caption{Hybrid maps of OJ~287 at 15~GHz.}
\end{figure}
\begin{figure}
\figurenum{16d}
\plotone{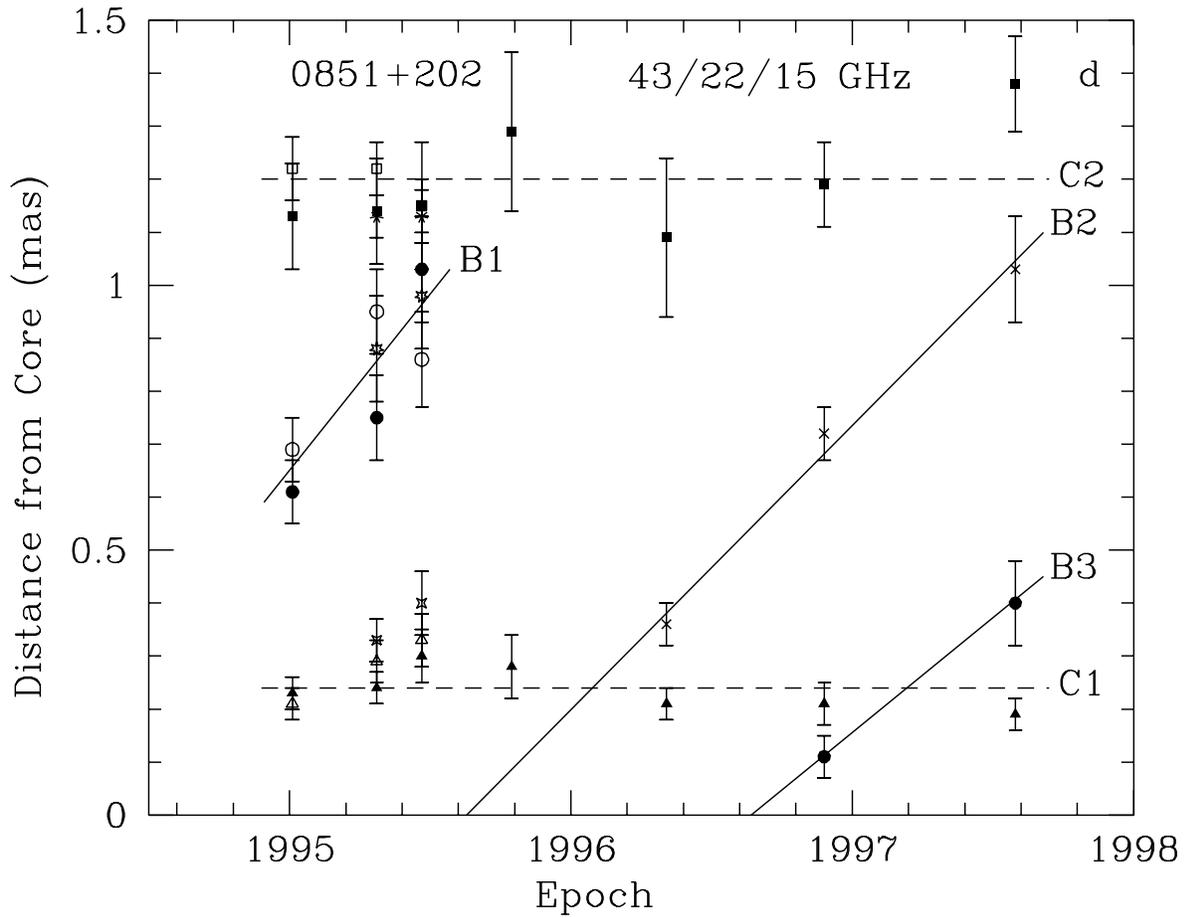}
\caption{Positions of components with respect to the core at different
epochs from model fitting for OJ~287; designations of components are as follows:
at 43~GHz - components $B2$ (crosses), B1 (filled circles),
C1 (filled triangles), and C2 (filled squares);
at 22~GHz - components $D1$ (open circles), $E1$ (open triangles),
and $E2$ (open squares); at 15~GHz - components  $F1$ (7-point stars),
$G1$ (4-point stars),  and  $G2$ (asterisks).}
\end{figure}
\begin{figure}
\figurenum{17a}
\plotone{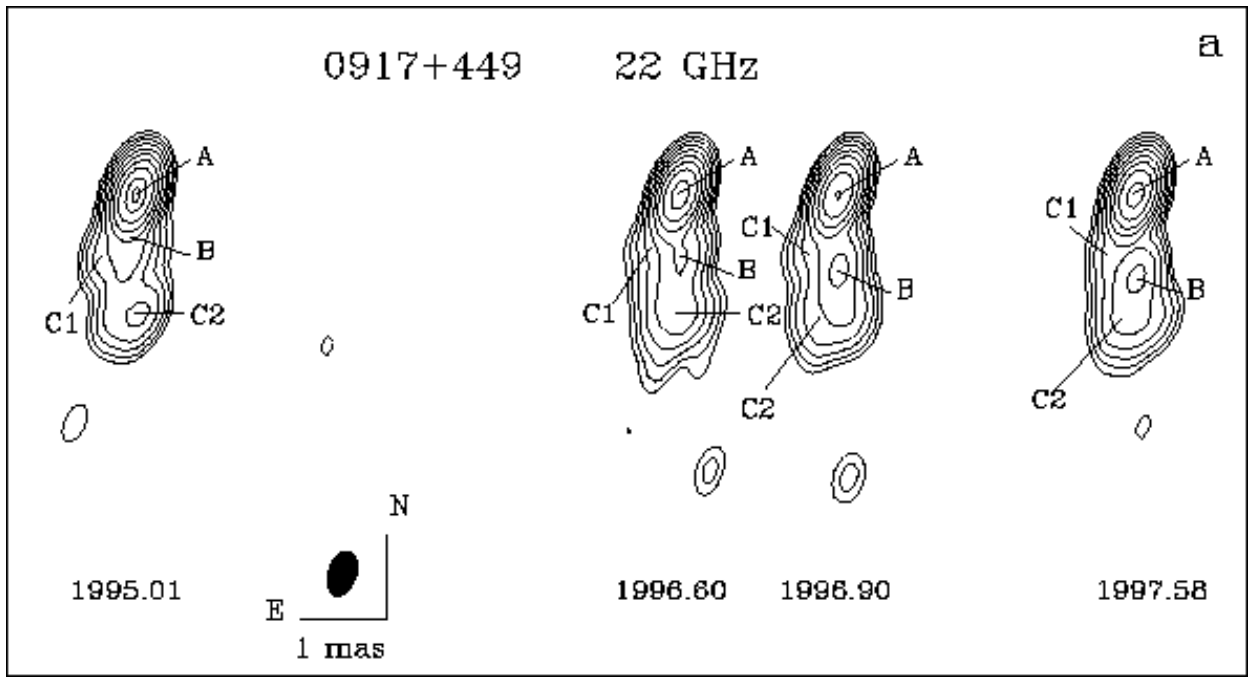}
\caption{Hybrid maps of 0917$+$449 at 22~GHz.}
\end{figure}
\begin{figure}
\figurenum{17b}
\plotone{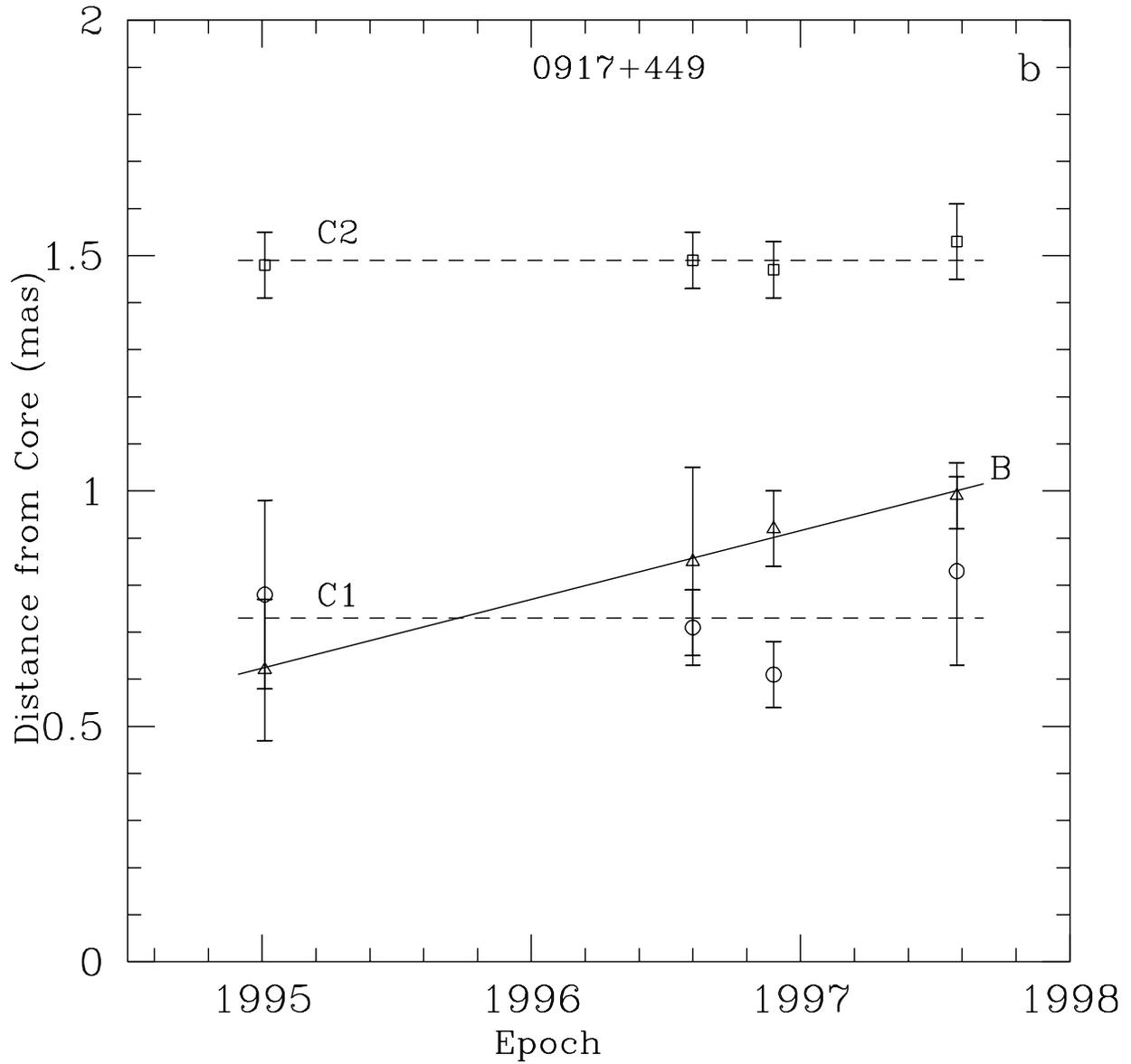}
\caption{Positions of components with respect to the core at different
epochs from model fitting for 0917$+$449; designations of components are as follows:
open triangles - component $B$,
open circles - $C1$, open squares - $C2$. }
\end{figure}
\begin{figure}
\figurenum{18}
\plotone{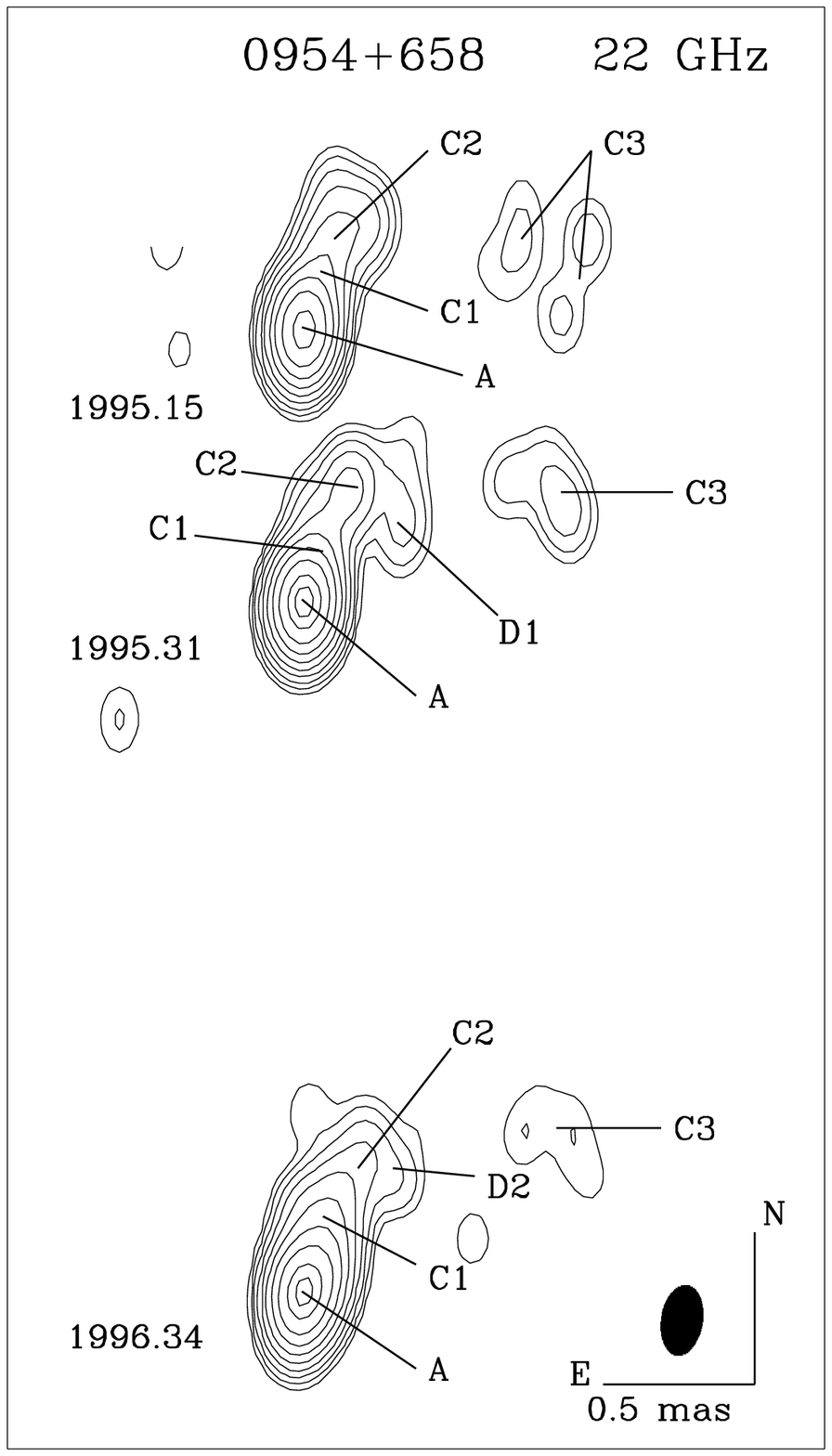}
\caption{Hybrid maps of 0954$+$658 at 22~GHz.}
\end{figure}
\clearpage
\begin{figure}
\figurenum{19a}
\plotone{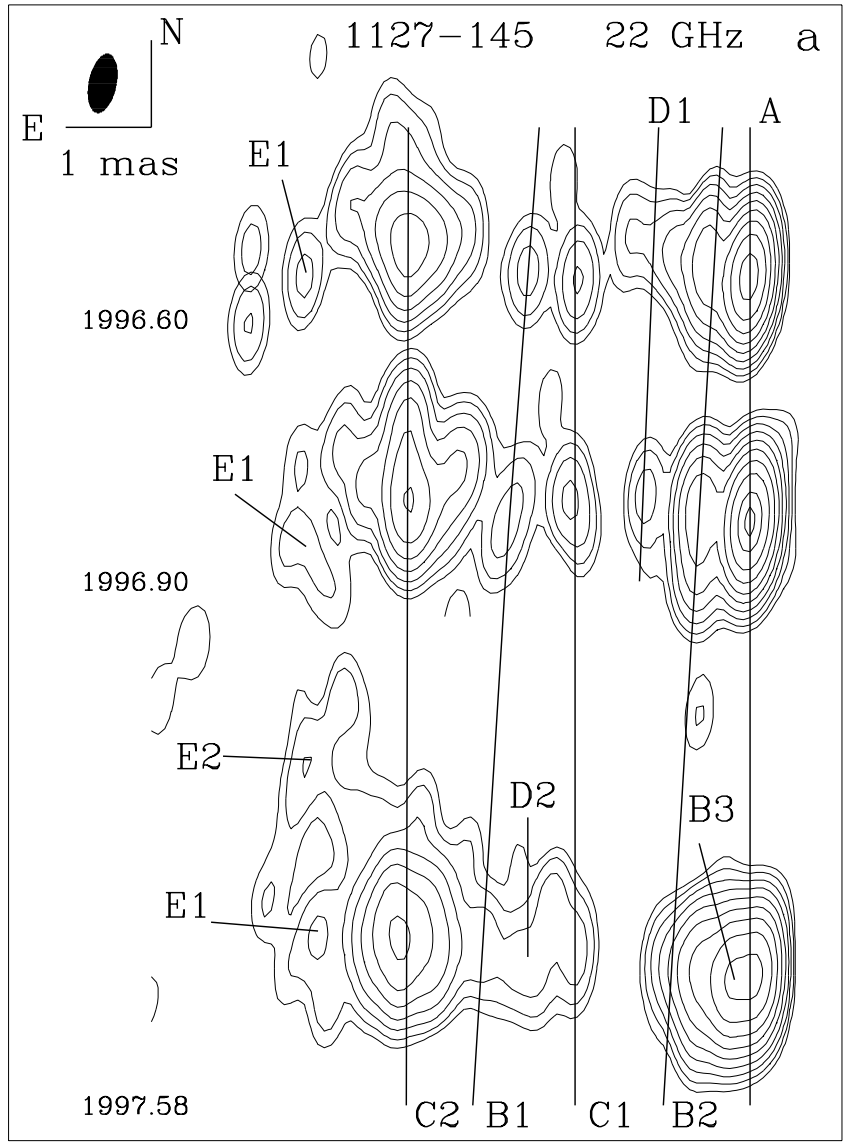}
\caption{Hybrid maps of 1127$-$145 at 22~GHz.}
\end{figure}
\begin{figure}
\figurenum{19b}
\plotone{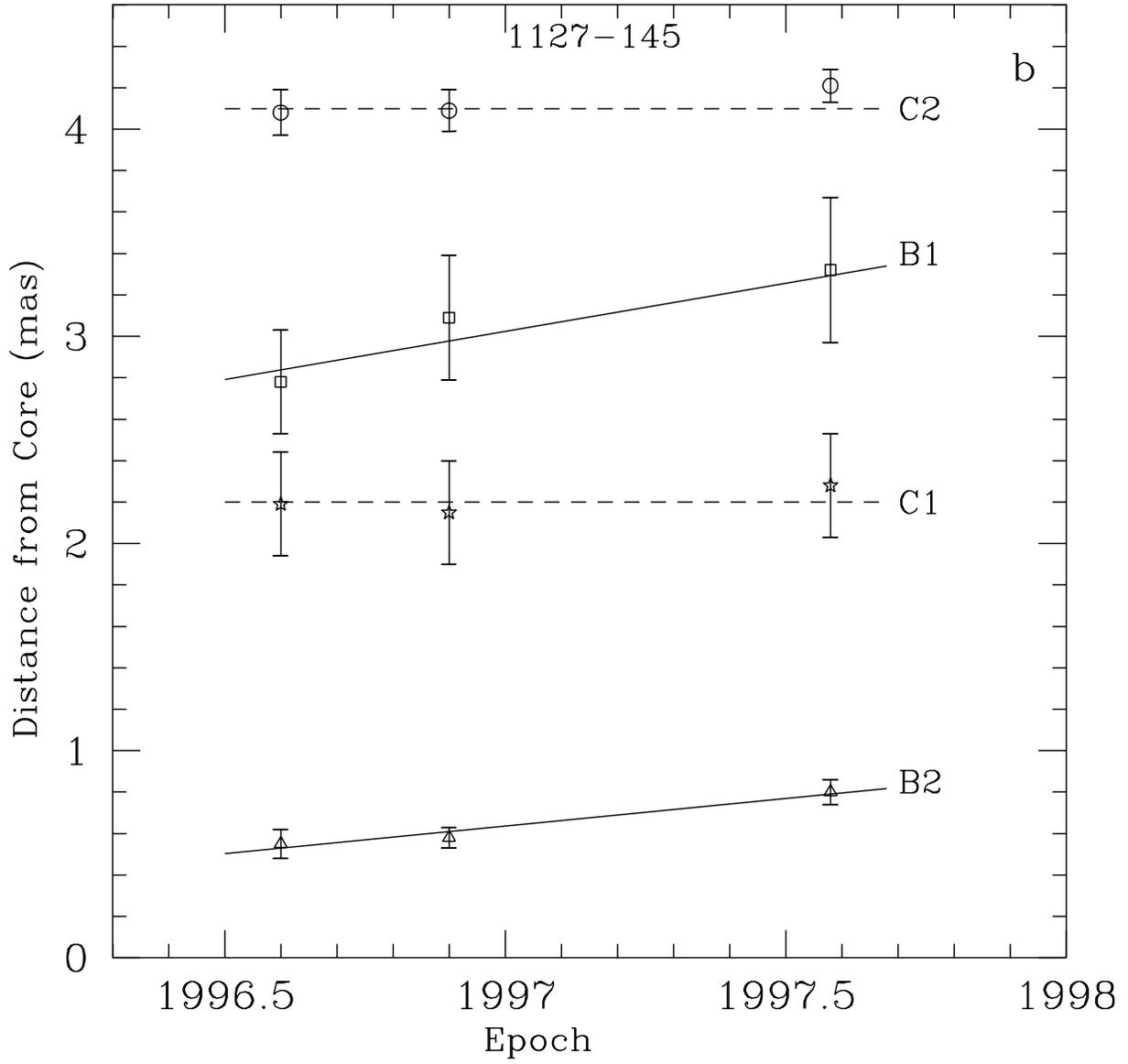}
\caption{Positions of components with respect to the core at different
epochs from model fitting for 1127$-$145; designations of components are as follows:
open triangles - component $B2$, stars - component $C1$,
open squares - $B1$, open circles - component $C2$.}
\end{figure}
\begin{figure}
\figurenum{20a}
\plotone{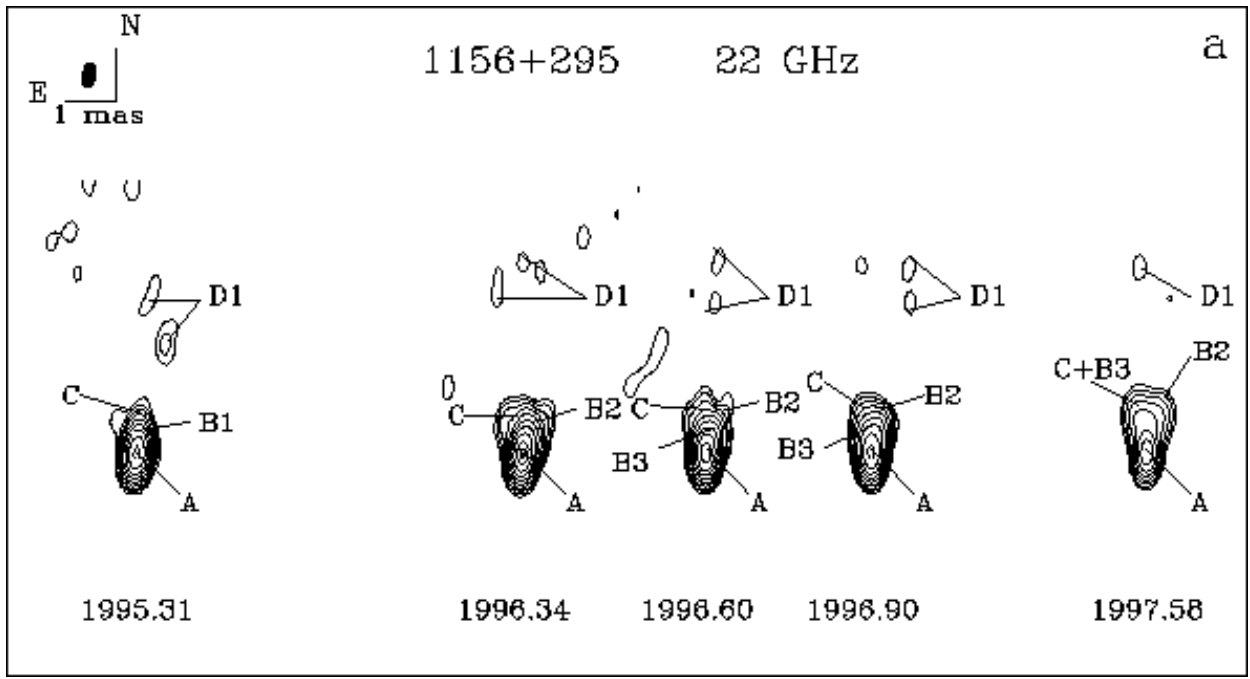}
\caption{Hybrid maps of 1156$+$295 at 22~GHz.}
\end{figure}
\begin{figure}
\figurenum{20b}
\plotone{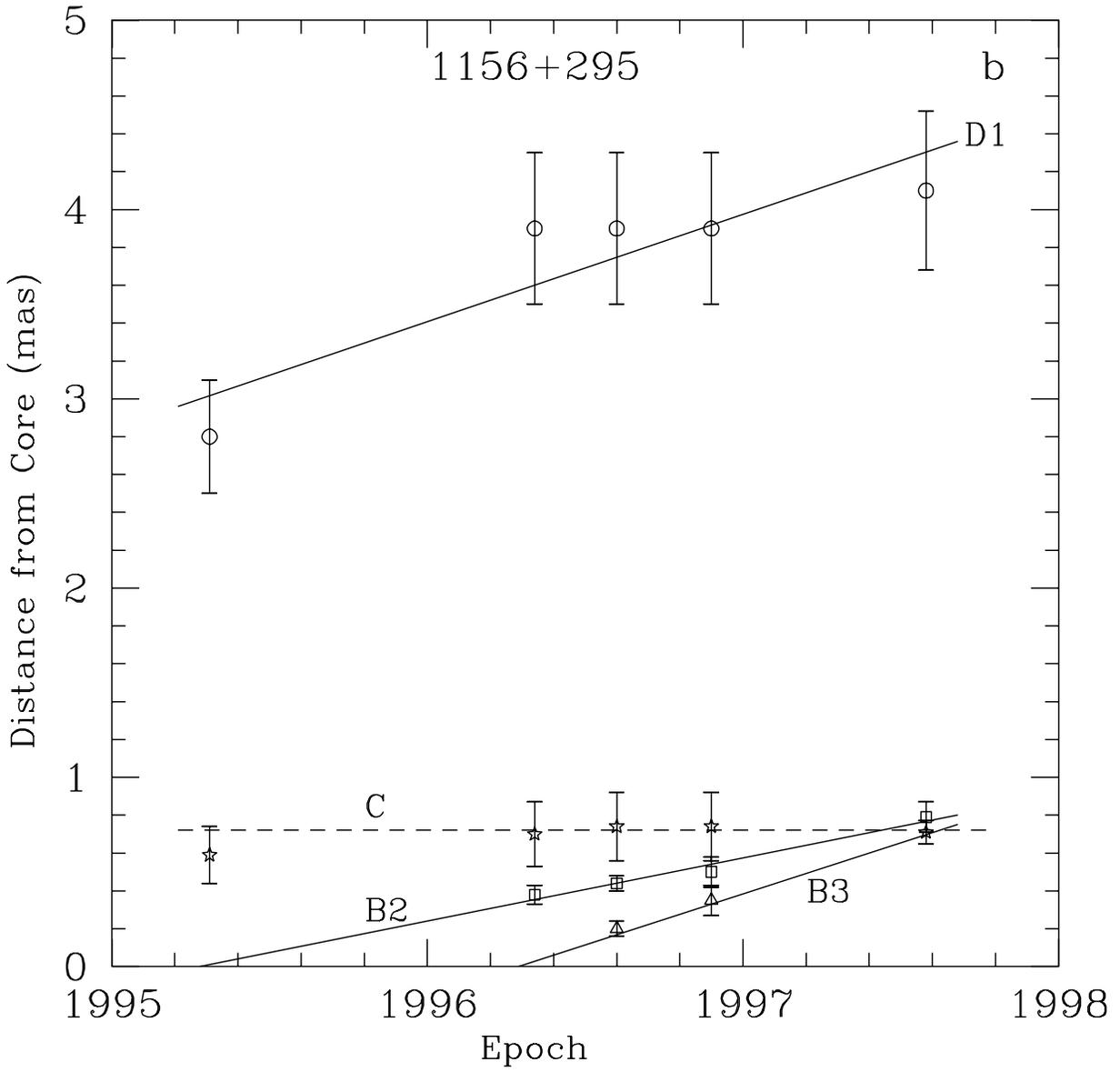}
\caption{Positions of components with respect to the core at different
epochs from model fitting for 1156$+$295; designations of components are as follows:
open triangles - component $B3$,
open squares - $B2$, stars - $C$, open circles - $D1$. }
\end{figure}
\begin{figure}
\figurenum{21a}
\plotone{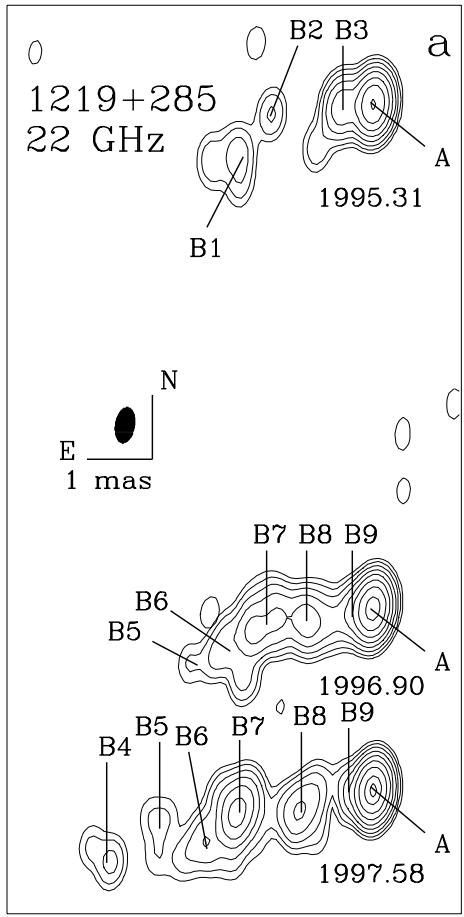}
\caption{Hybrid maps of 1219$+$285 at 22~GHz.}
\end{figure}
\begin{figure}
\figurenum{21b}
\plotone{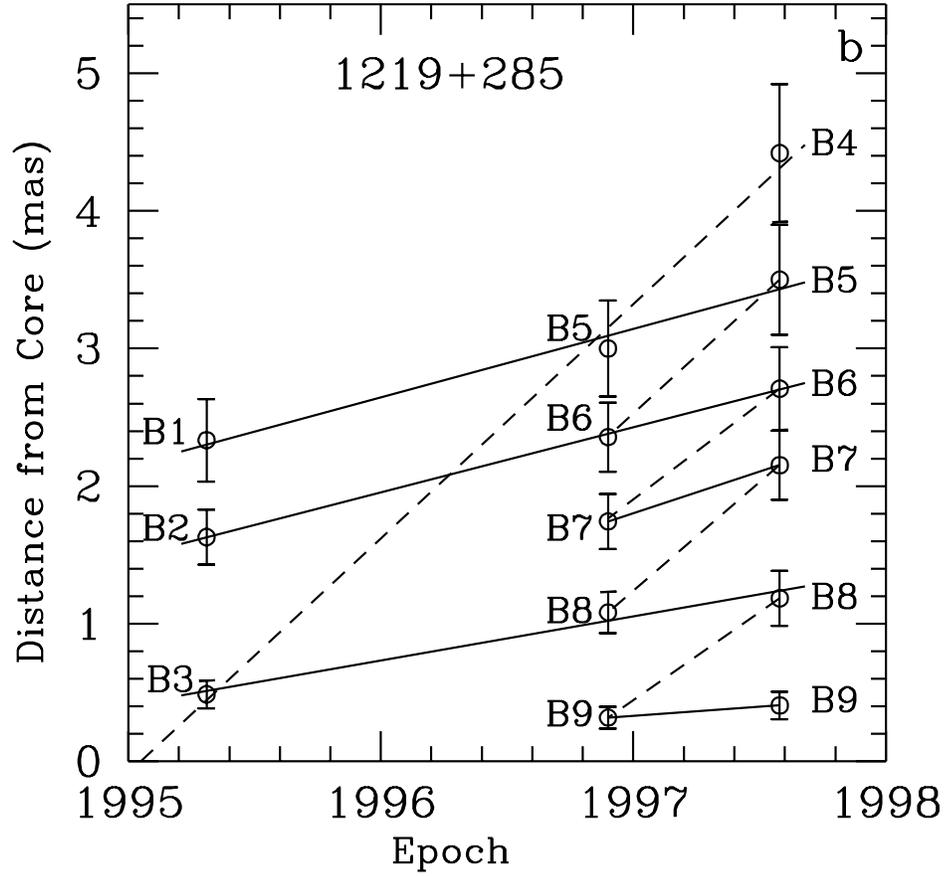}
\caption{Positions of components with respect to the core at different
epochs from model fitting for 1219$+$285, the solid lines show linear fits
corresponding to proper motion listed in Table 5, the dashed lines
indicate another possible identification of components through epochs with higher
proper motion.}
\end{figure}
\begin{figure}
\figurenum{22a}
\plotone{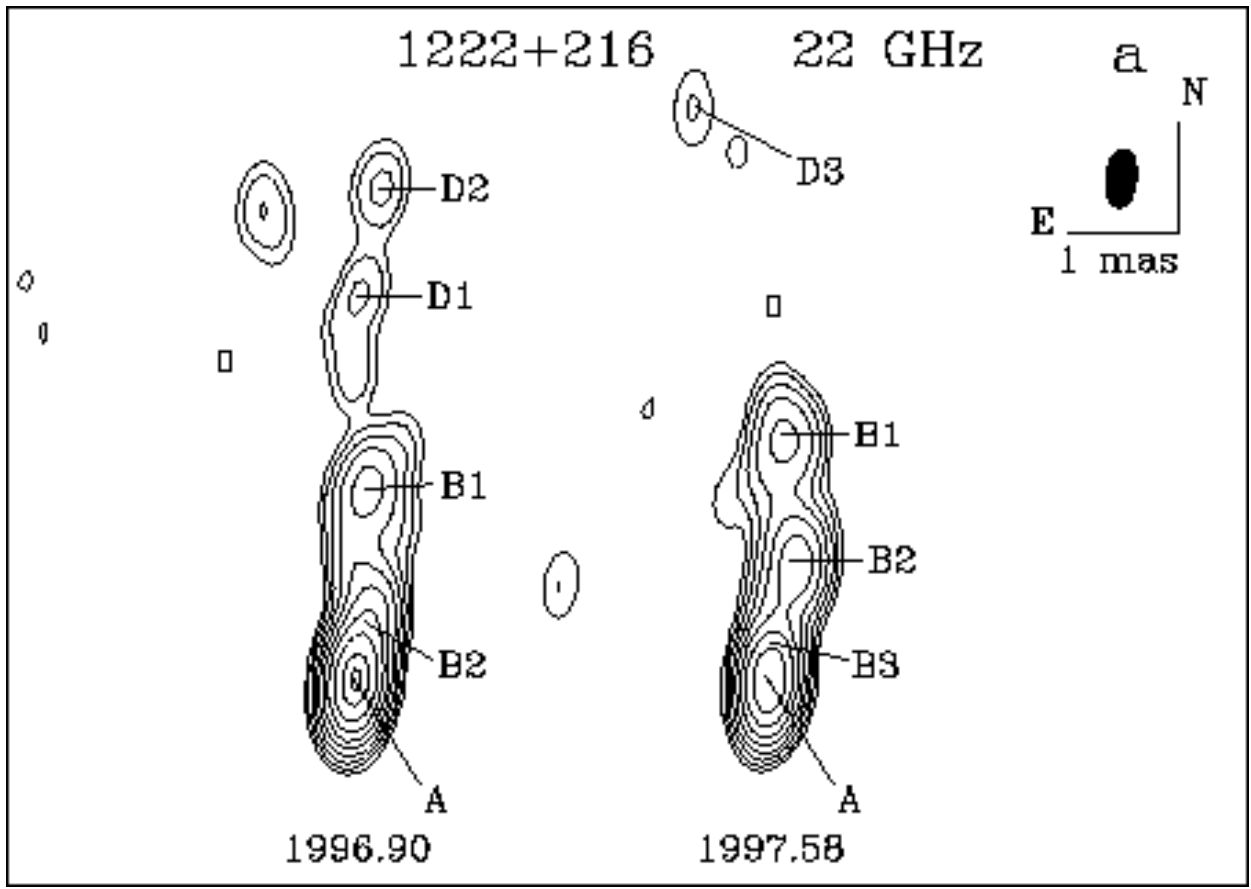}
\caption{Hybrid maps of 1222$+$216 at 22~GHz.}
\end{figure}
\begin{figure}
\figurenum{22b}
\plotone{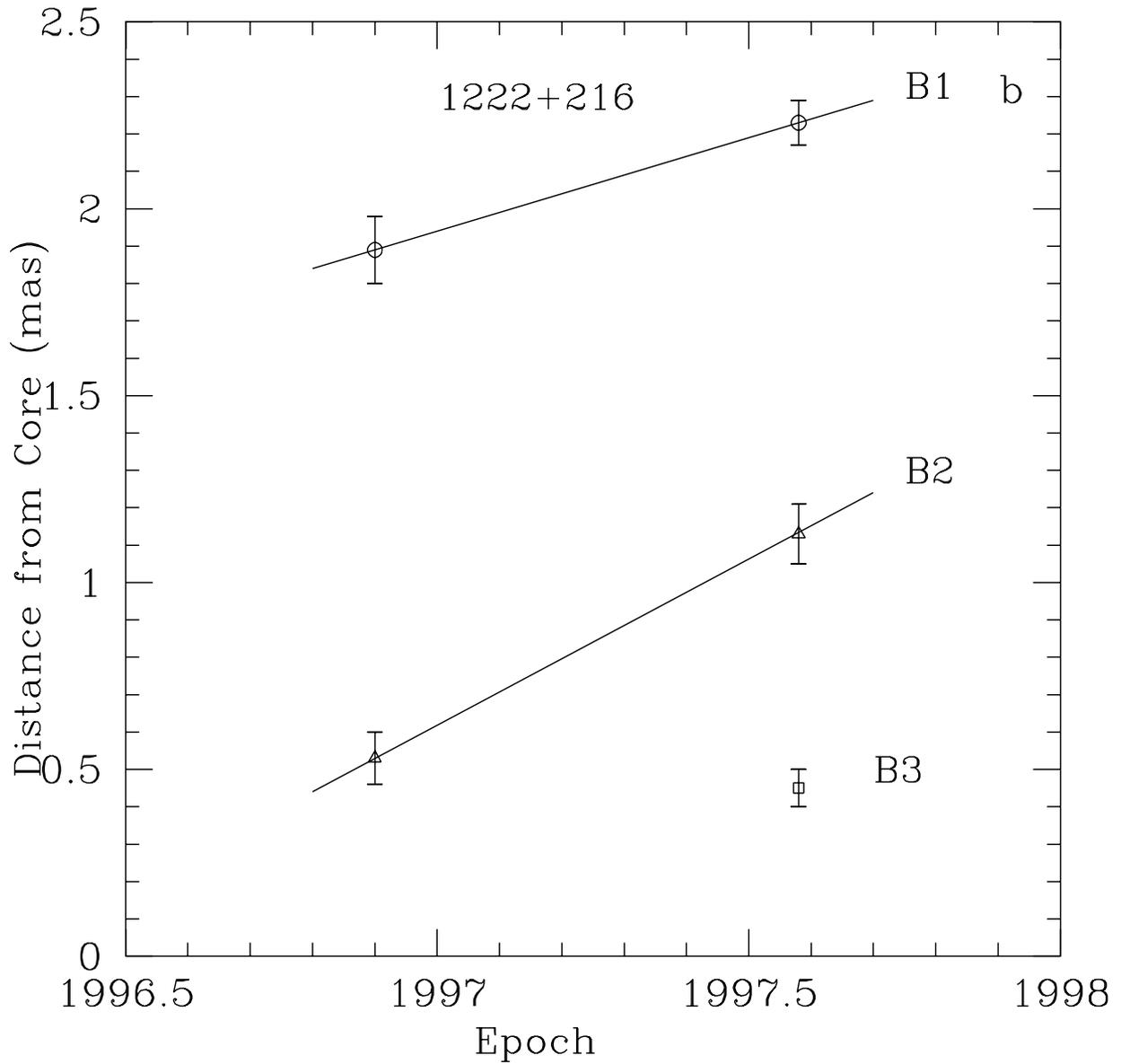}
\caption{Positions of components with respect to the core at 
different epochs from model fitting for 1222$+$216, designations of components are as follows:
open square - $B3$, open triangles - component $B2$, open circles - $B1$. }
\end{figure}
\clearpage
\begin{figure}
\figurenum{23a}
\plotone{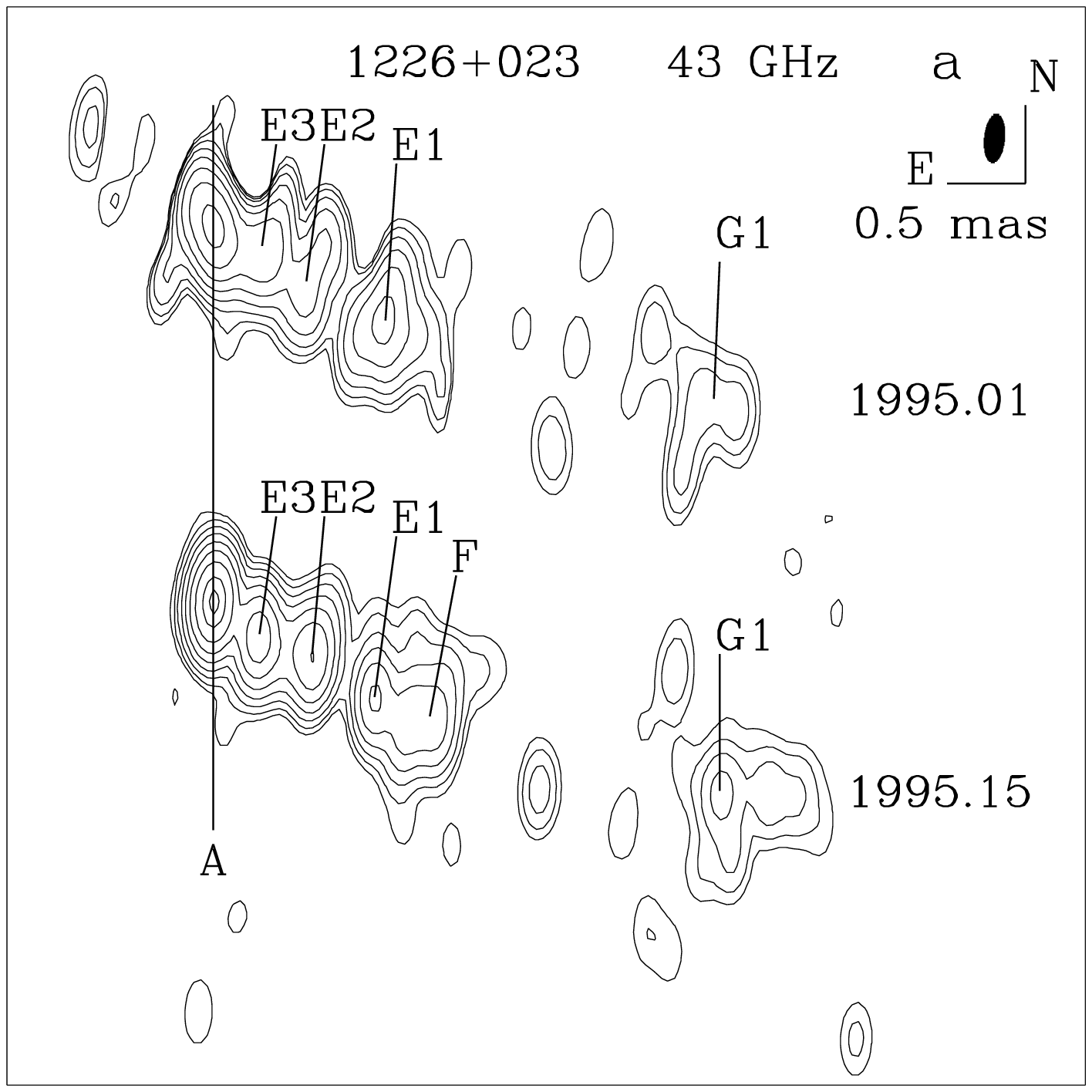}
\caption{Hybrid maps of 3C~273 at 43~GHz.}
\end{figure}
\begin{figure}
\figurenum{23b}
\plotone{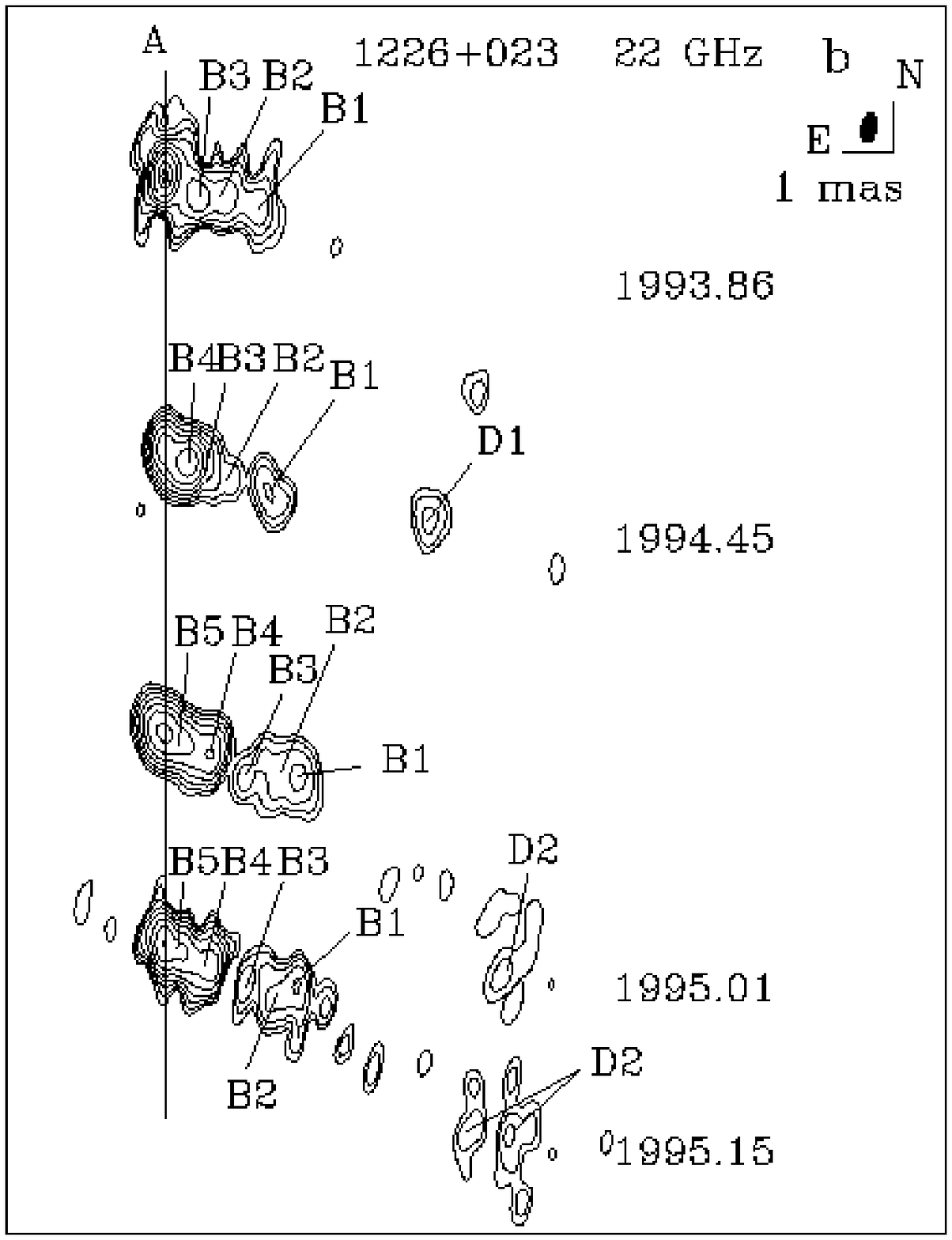}
\caption{Hybrid maps of 3C~273 at 22~GHz.}
\end{figure}
\begin{figure}
\figurenum{23c}
\plotone{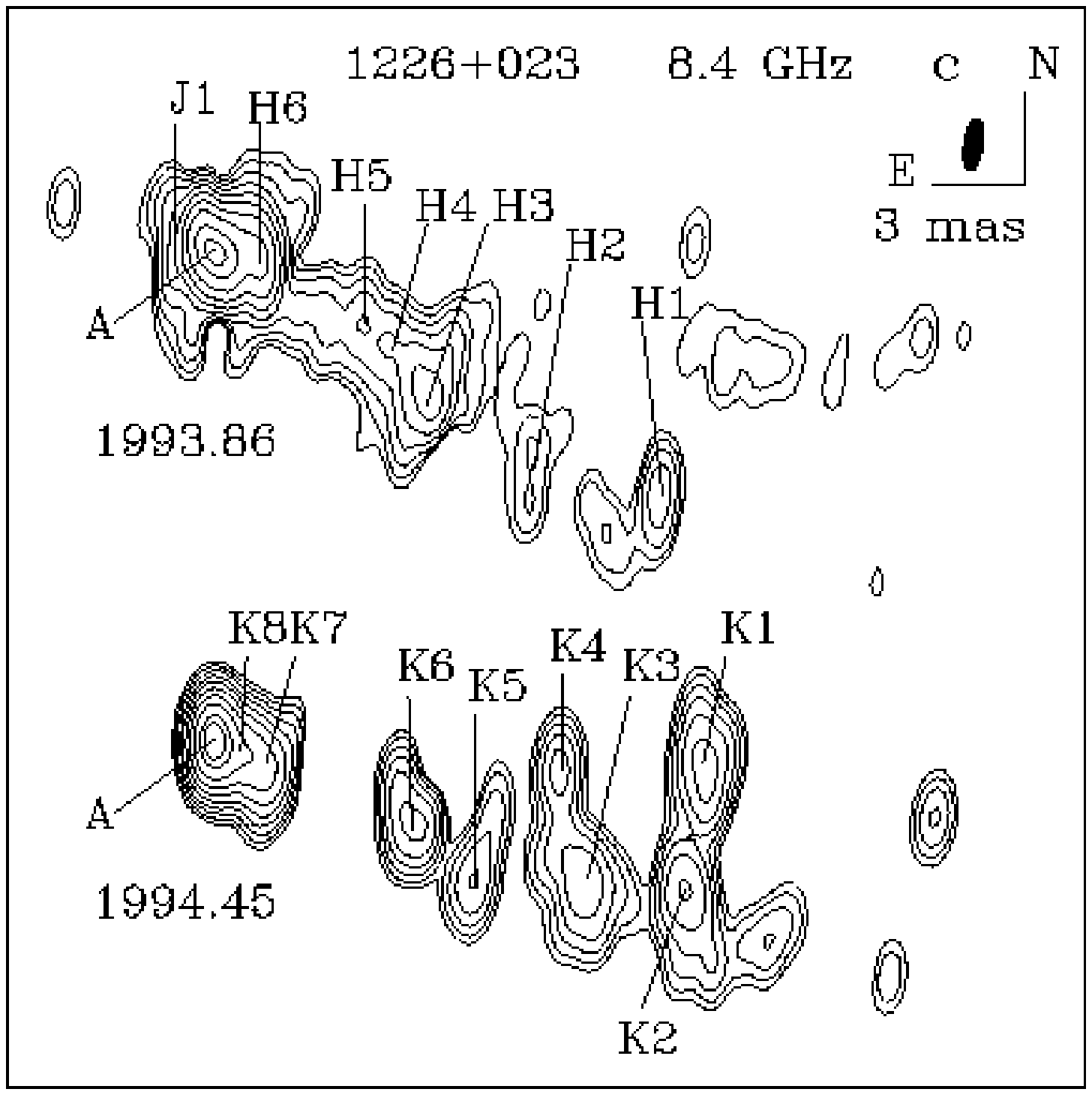}
\caption{Hybrid maps of 3C~273 at 8~GHz.}
\end{figure}
\begin{figure}
\figurenum{23d}
\plotone{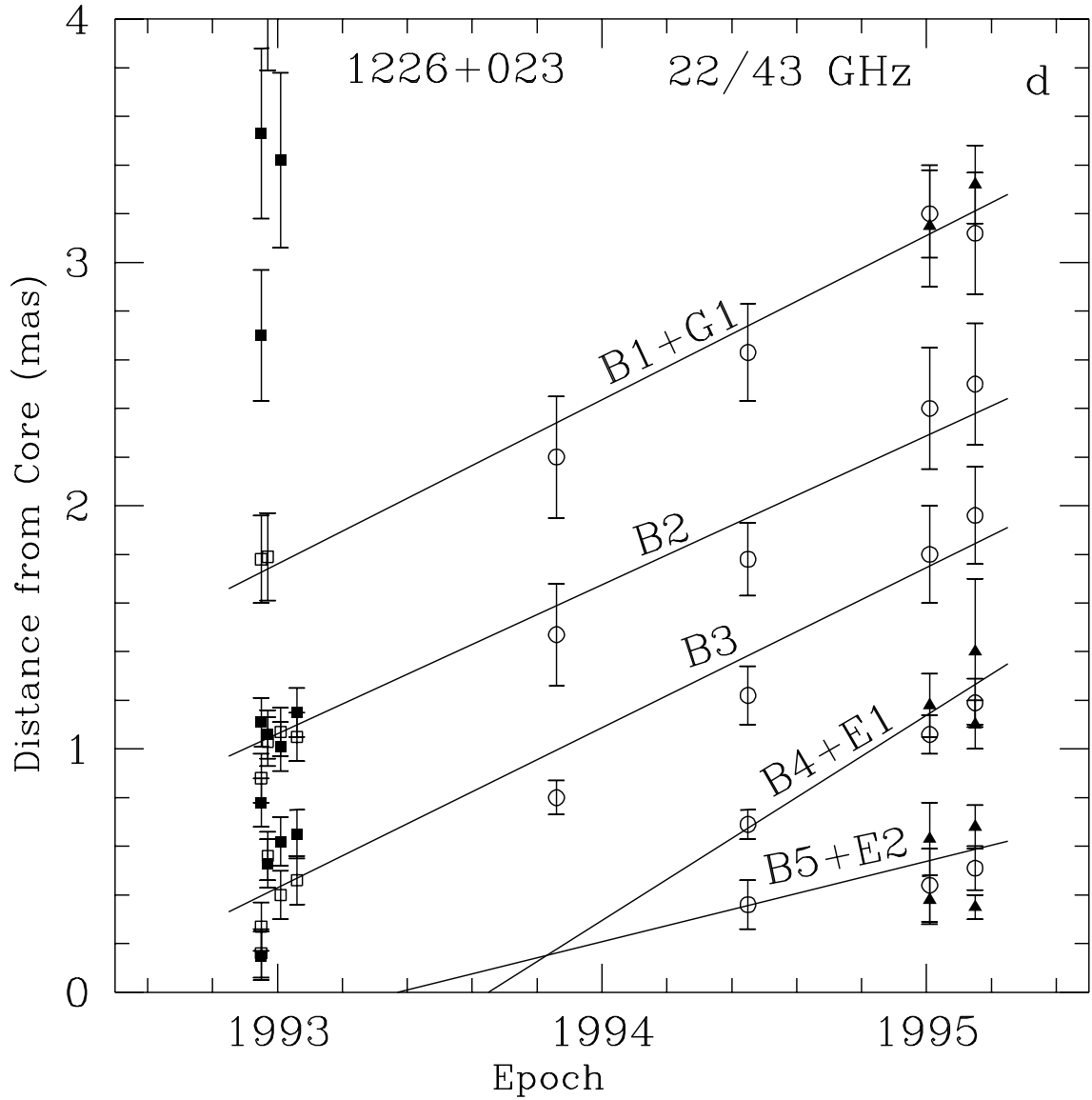}
\caption{Positions of components over a range from 0 to 4~mas with
respect to the core at different
epochs from model fitting for 3C~273, the plot shows our results
and the model fitting parameters published by Mantovani et al.(1999),
designations of components are as follows:
our results at 43 and 22~GHz
are filled triangles and open circles, respectively, the results of
Mantovani et al. (1999) at 43 and 22~GHz are filled and open squares, 
repectively.}
\end{figure}
\begin{figure}
\figurenum{23e}
\plotone{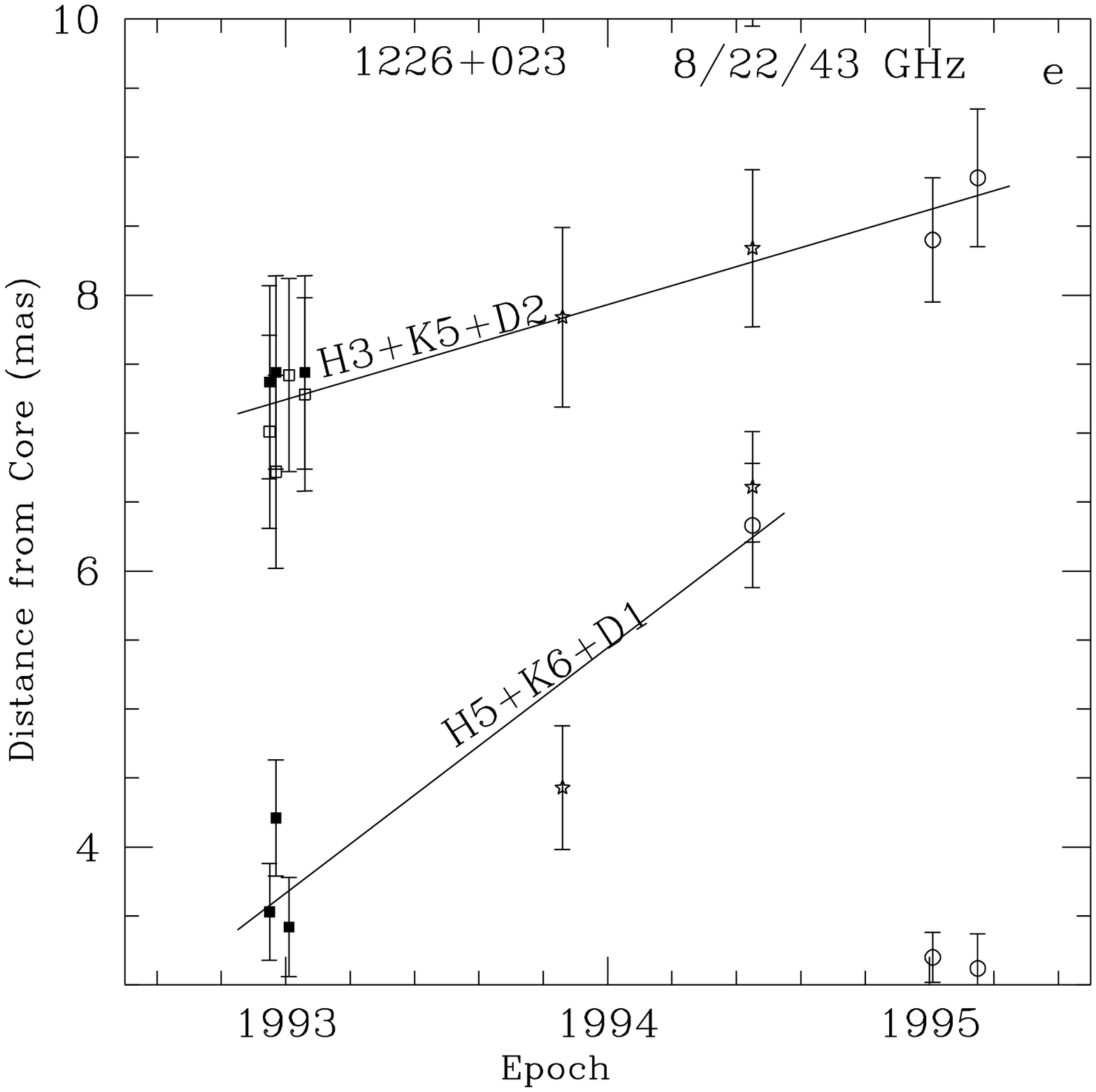}
\caption{The same as plot (d) over a range of a distance with respect
to the core from 3 to 10~mas; designations of components at 22 and 
43~GHz are the same as in plot (d),  in addition,  the data 
at 8.4~GHz are shown by 5-point stars.}
\end{figure}
\begin{figure}
\figurenum{24a}
\plotone{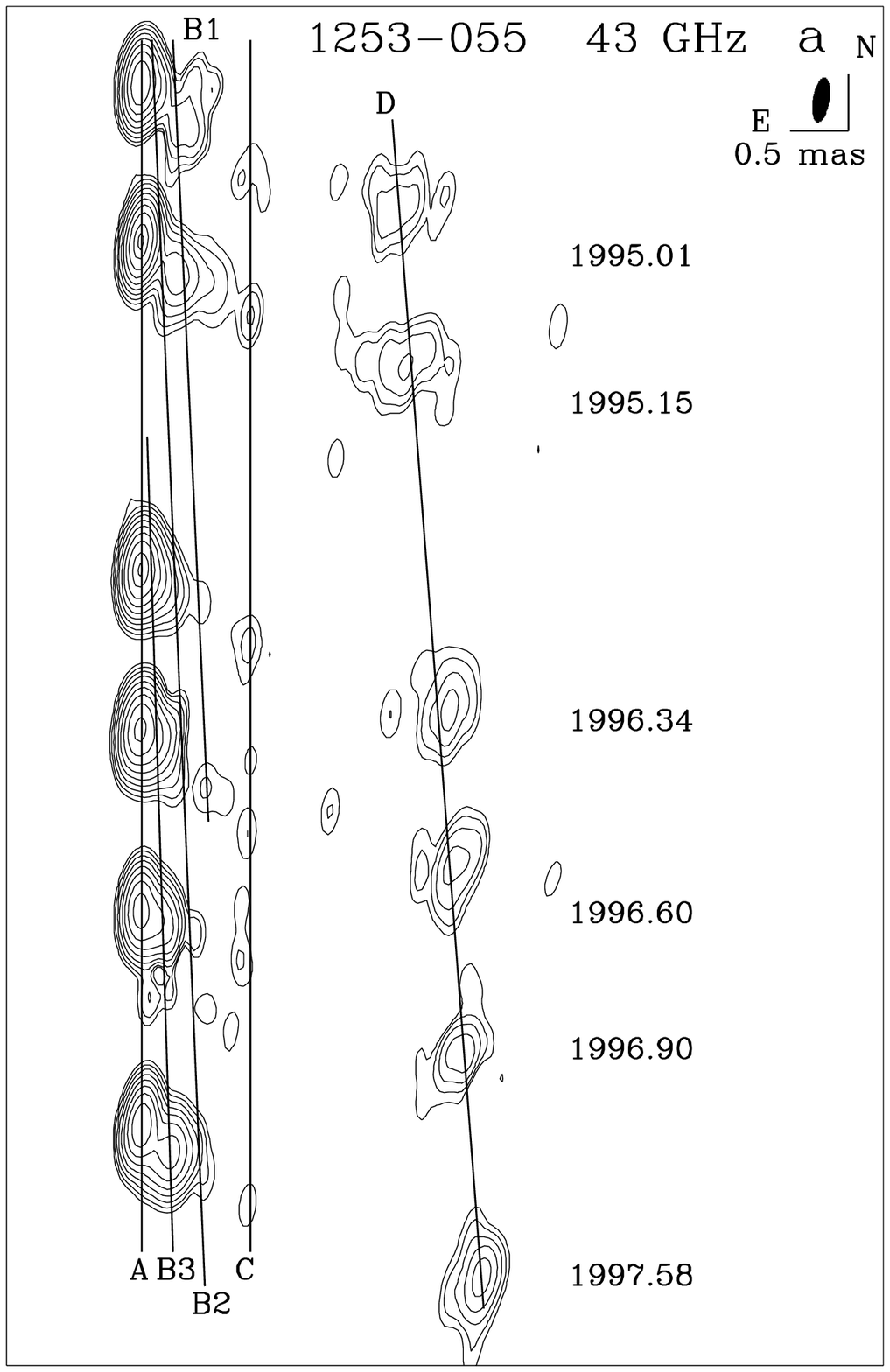}
\caption{Hybrid maps of 3C~279 at 43~GHz.}
\end{figure}
\begin{figure}
\figurenum{24b}
\plotone{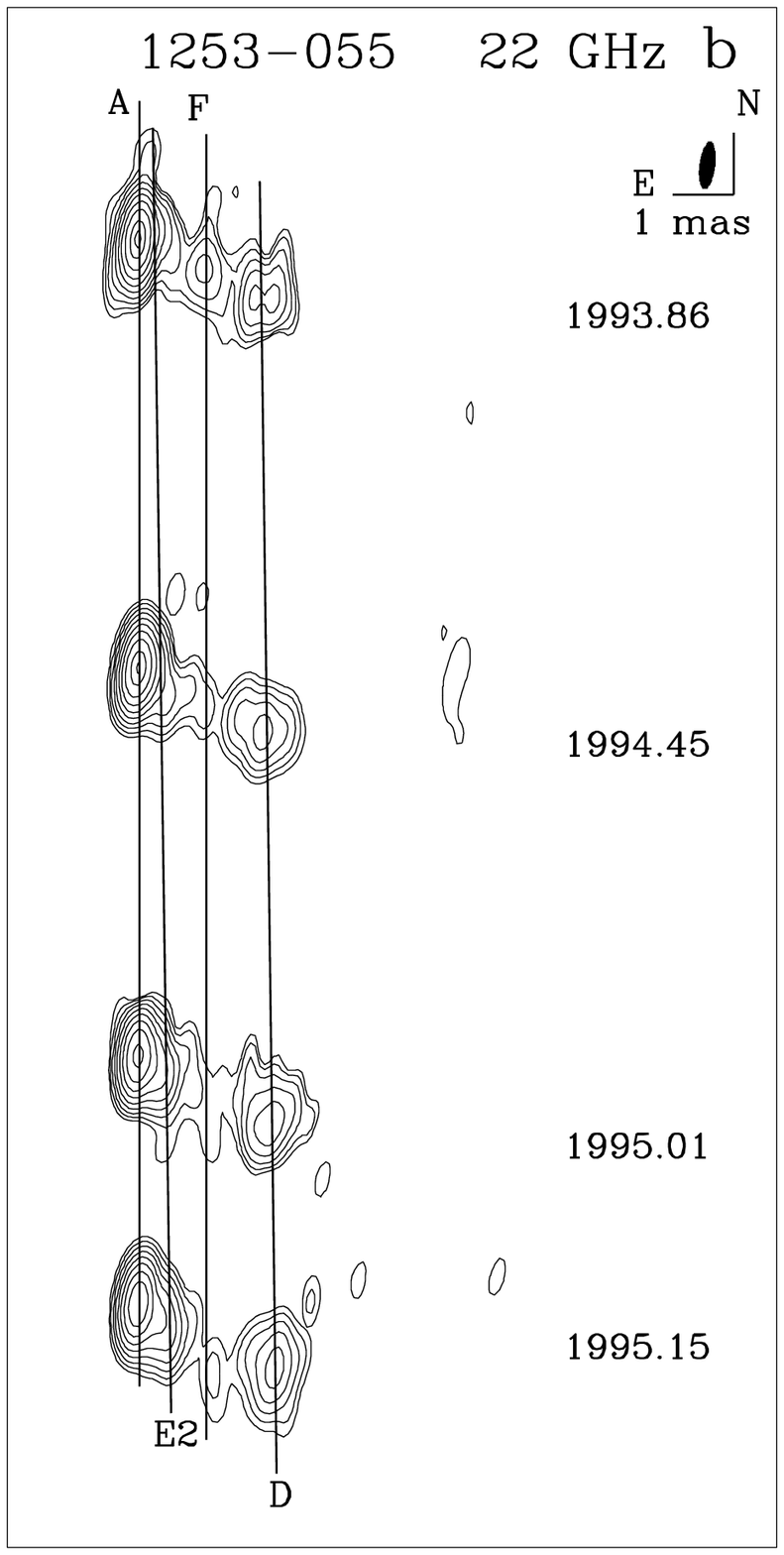}
\caption{Hybrid maps of 3C~279 at 22~GHz.}
\end{figure}
\begin{figure}
\figurenum{24c}
\plotone{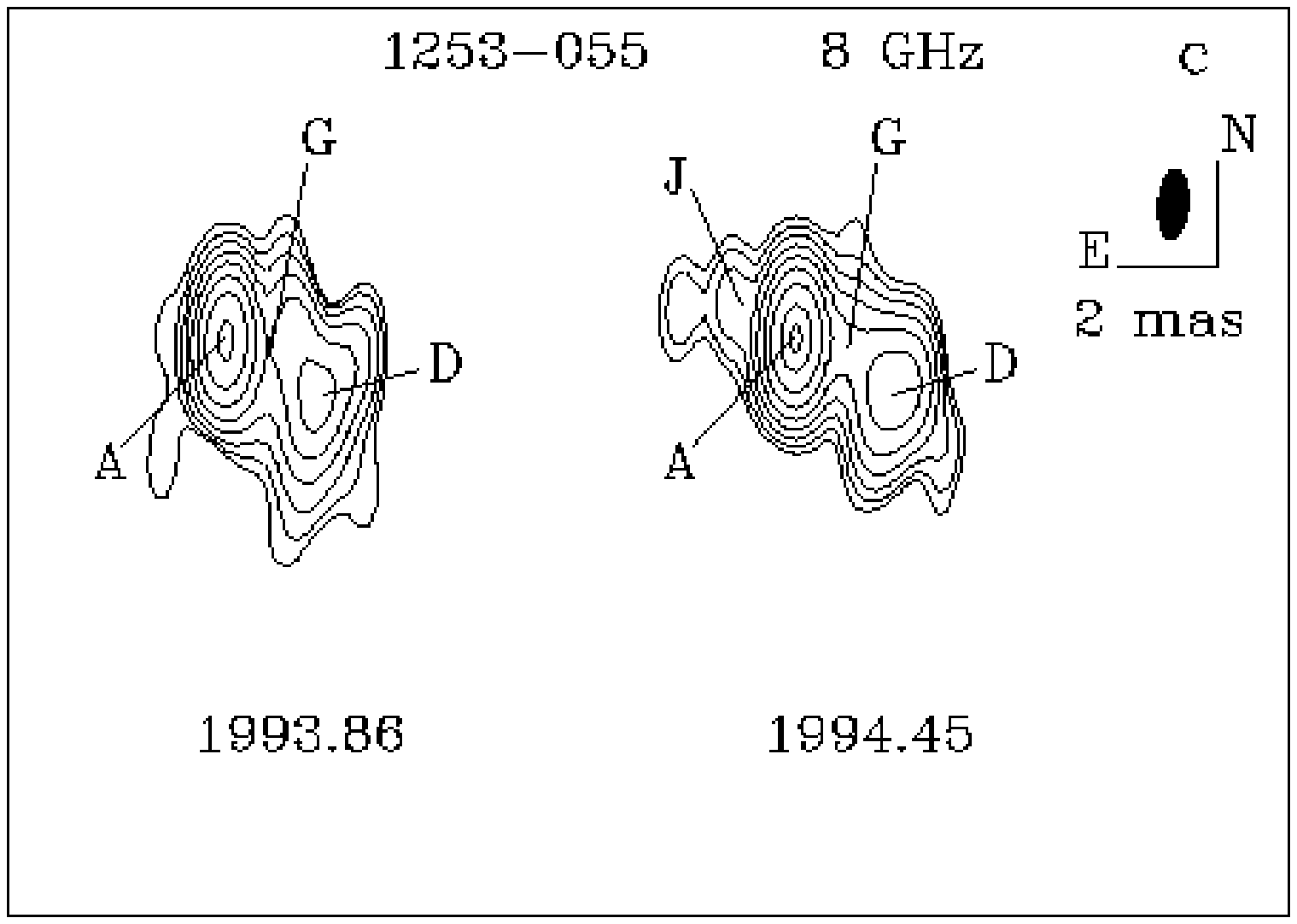}
\caption{Hybrid maps of 3C~279 at 8~GHz.}
\end{figure}
\begin{figure}
\figurenum{24d}
\plotone{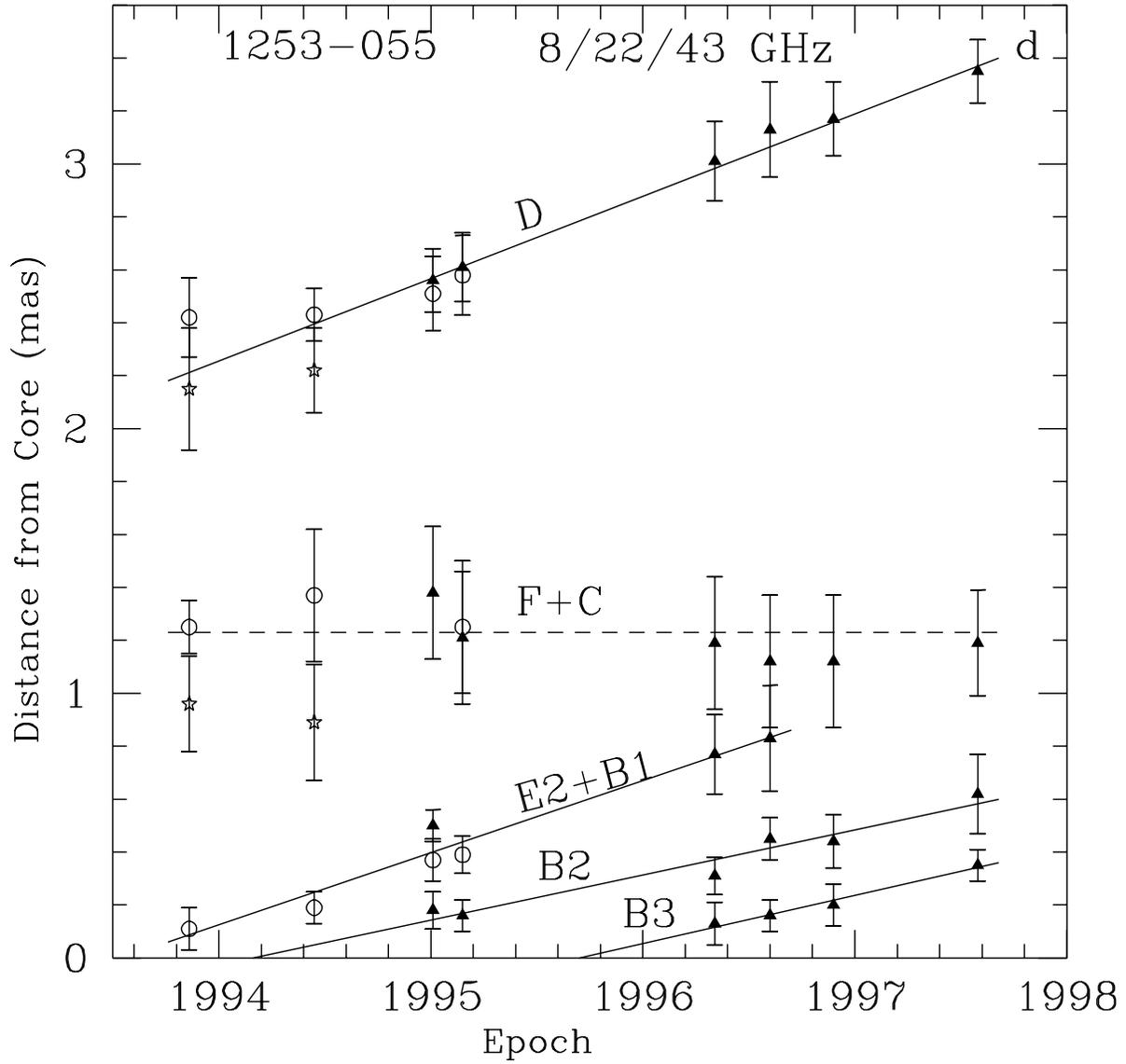}
\caption{Positions of components with  
respect to the core at different
epochs from model fitting for 3C~279, 
designations of components are as follows:
components at 8.4 are labeled by 5-point stars, 
at 22~GHz by open circles, and  at 43~GHz by filled triangles.}
\end{figure}
\begin{figure}
\figurenum{25a}
\plotone{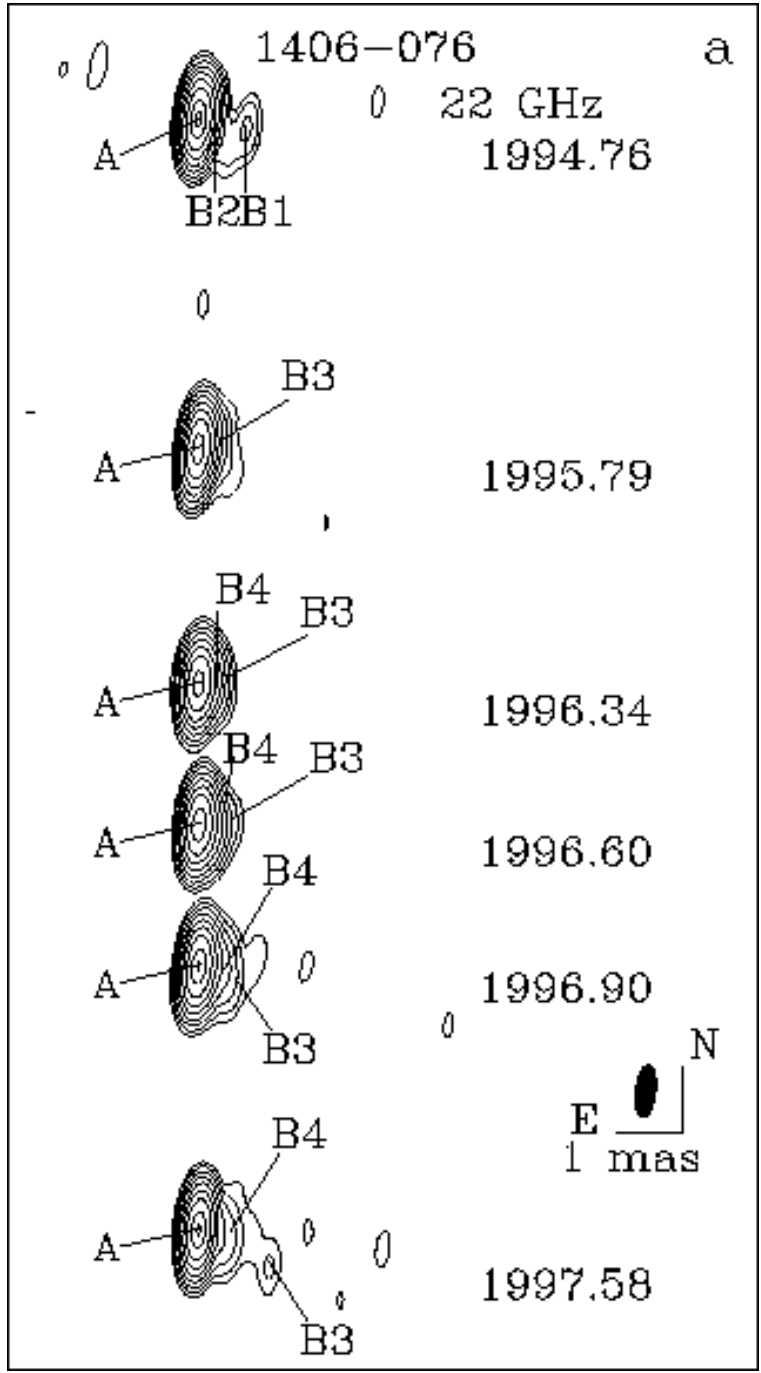}
\caption{Hybrid maps of 1406$-$076 at 22~GHz.}
\end{figure}
\begin{figure}
\figurenum{25b}
\plotone{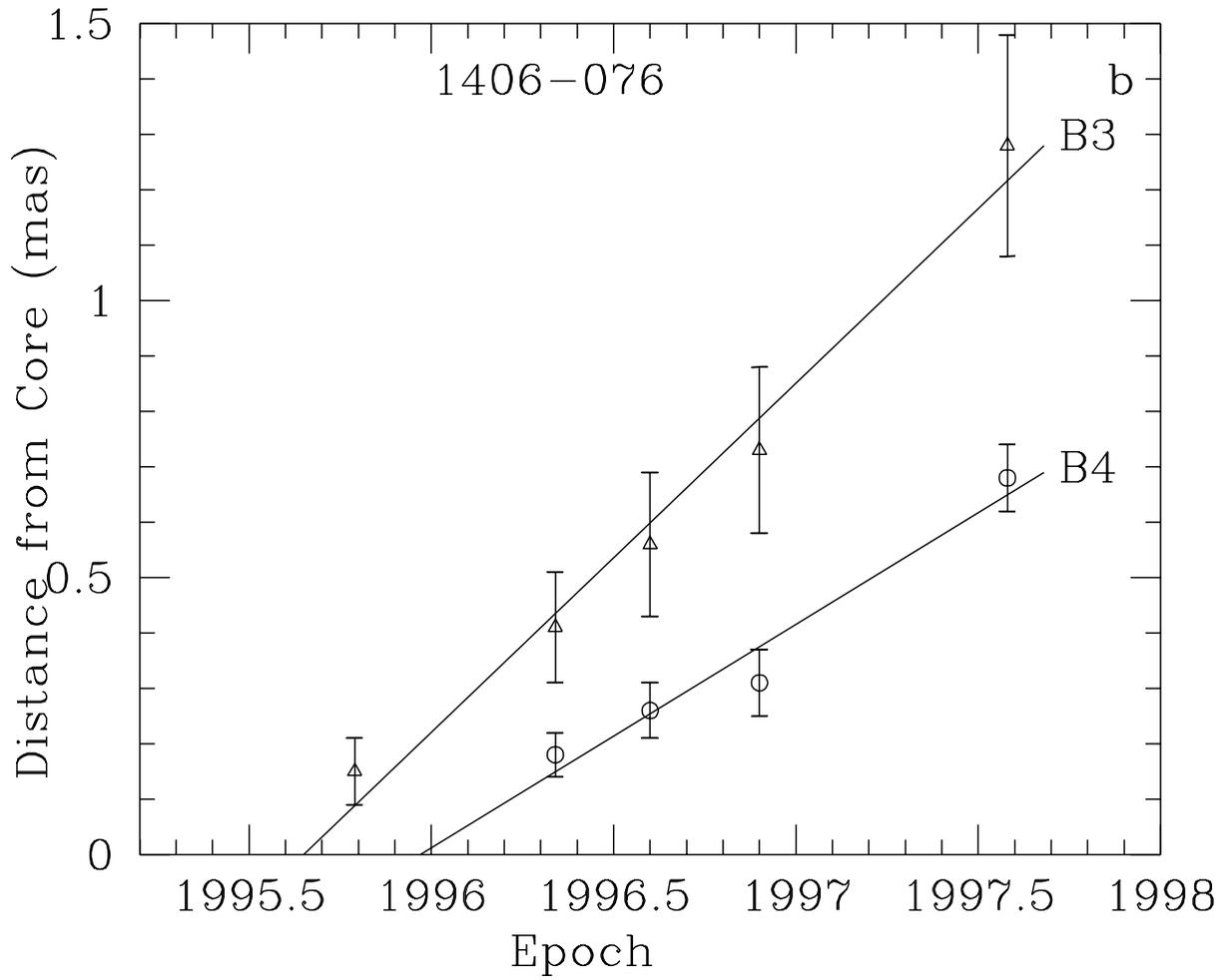}
\caption{Positions of components with respect to the core at 
different epochs from model fitting for 1406$-$076; designations of components are as follows:
open circles - component $B4$, open triangles - component $B3$.}
\end{figure}
\begin{figure}
\figurenum{26a}
\plotone{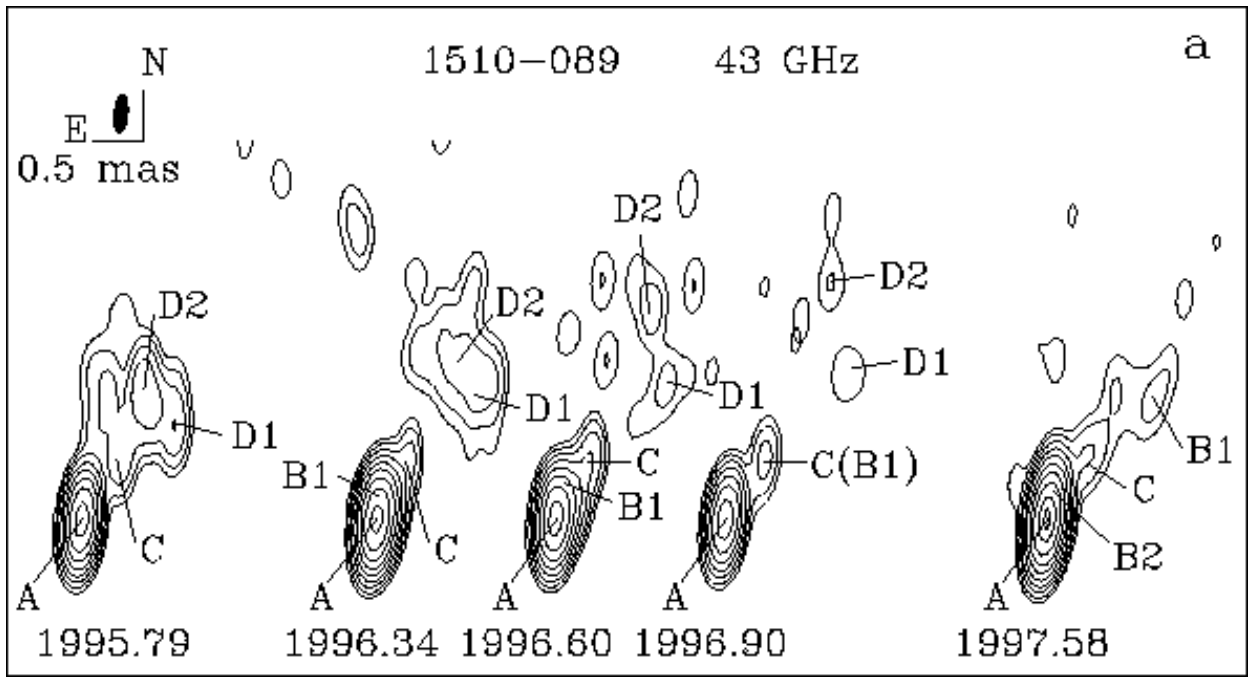}
\caption{Hybrid maps of 1510$-$089 at 43~GHz.}
\end{figure}
\begin{figure}
\figurenum{26b}
\plotone{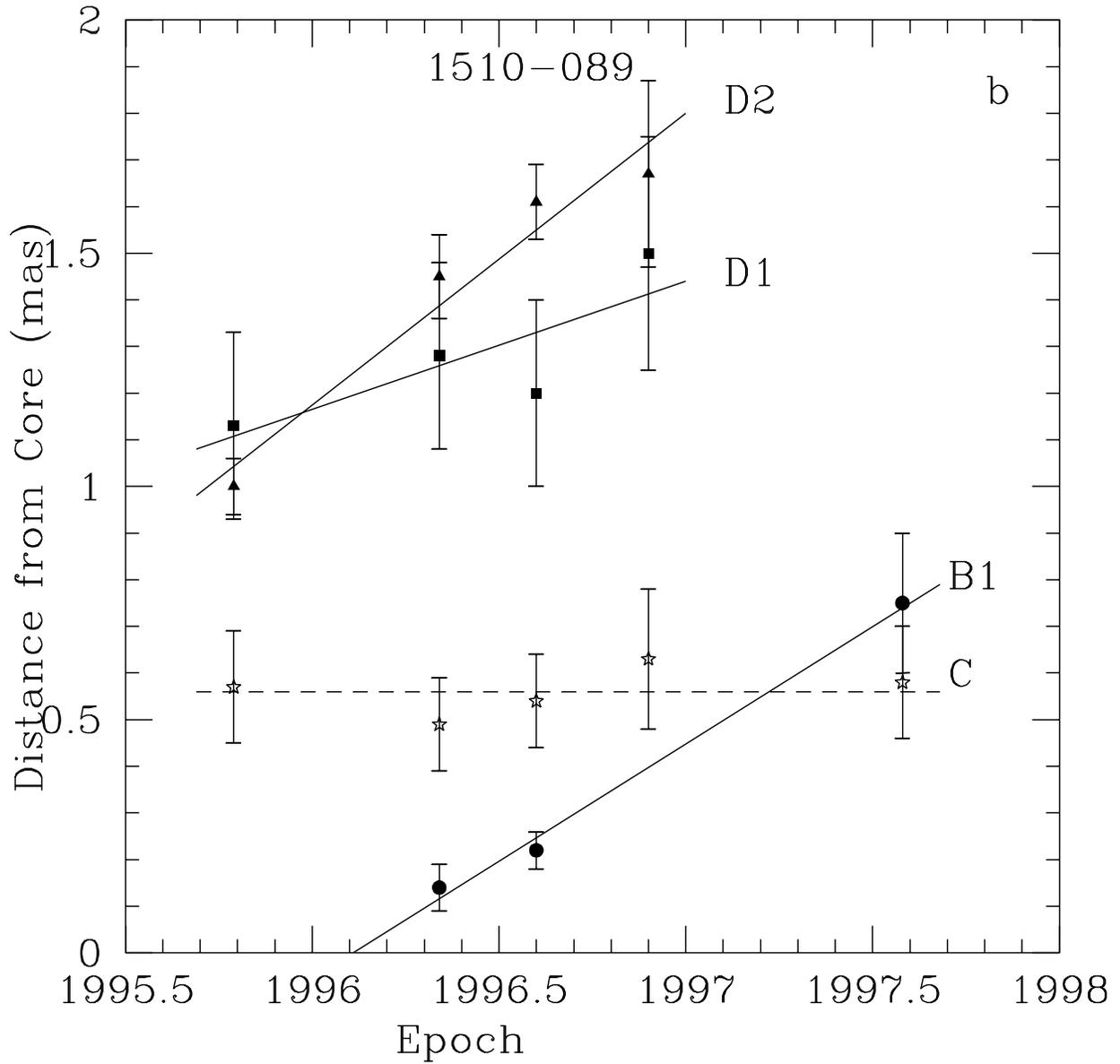}
\caption{Positions of components with respect to the core at different
epochs from model fitting for 1510$-$089; designations of components are as follows:
filled circles - component $B1$, stars - component $C$,
filled squares - $D1$, filled triangles - component $D2$.}
\end{figure}
\begin{figure}
\figurenum{27}
\plotone{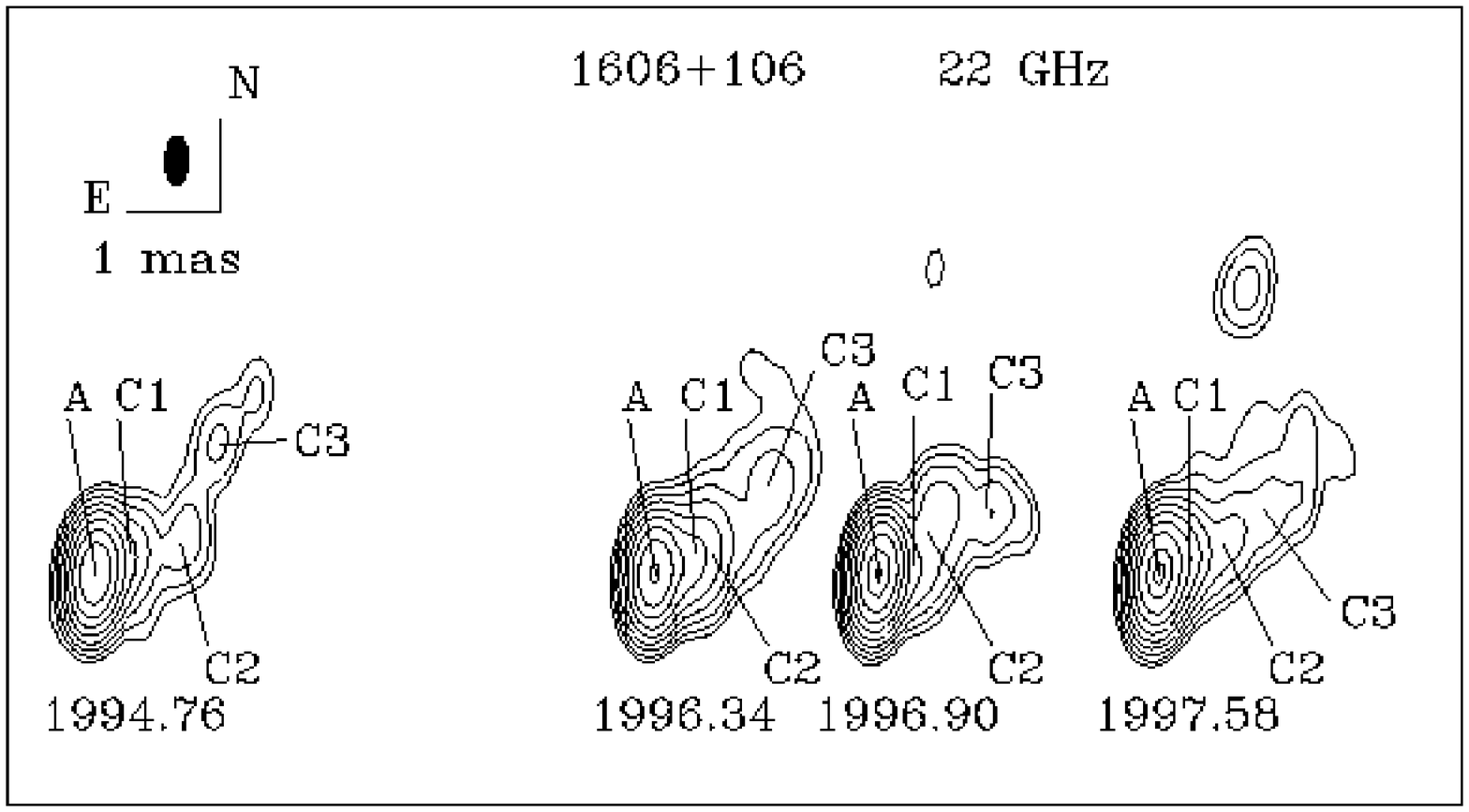}
\caption{Hybrid maps of 1606$+$106 at 22~GHz.}
\end{figure}
\begin{figure}
\figurenum{28a}
\plotone{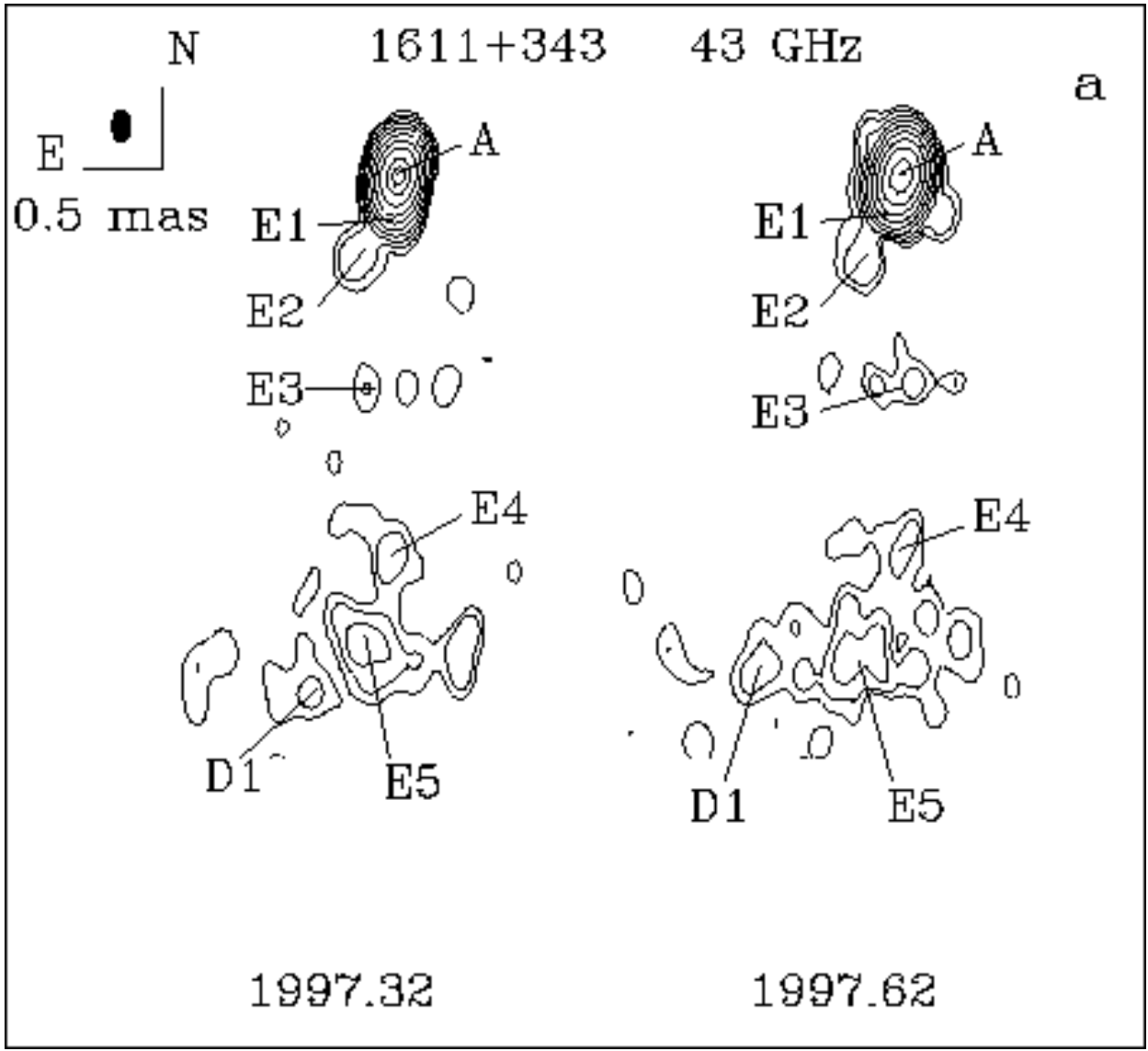}
\caption{Hybrid maps of 1611$+$343 at 43~GHz.}
\end{figure}
\begin{figure}
\figurenum{28b}
\plotone{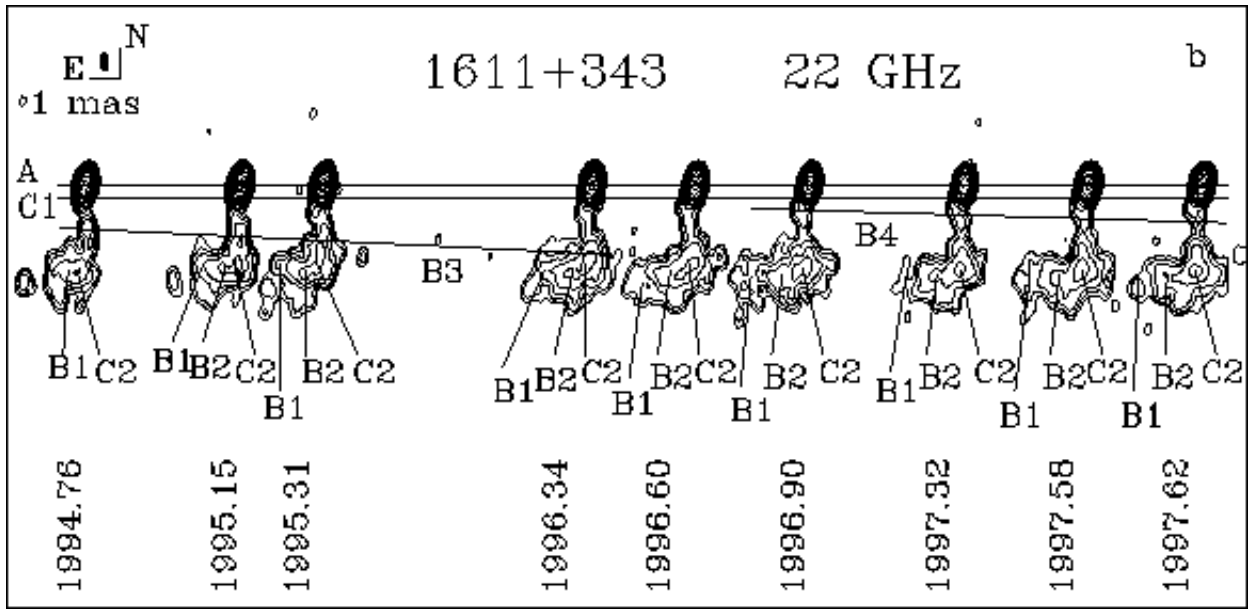}
\caption{Hybrid maps of 1611$+$343 at 22~GHz.}
\end{figure}
\begin{figure}
\figurenum{28c}
\plotone{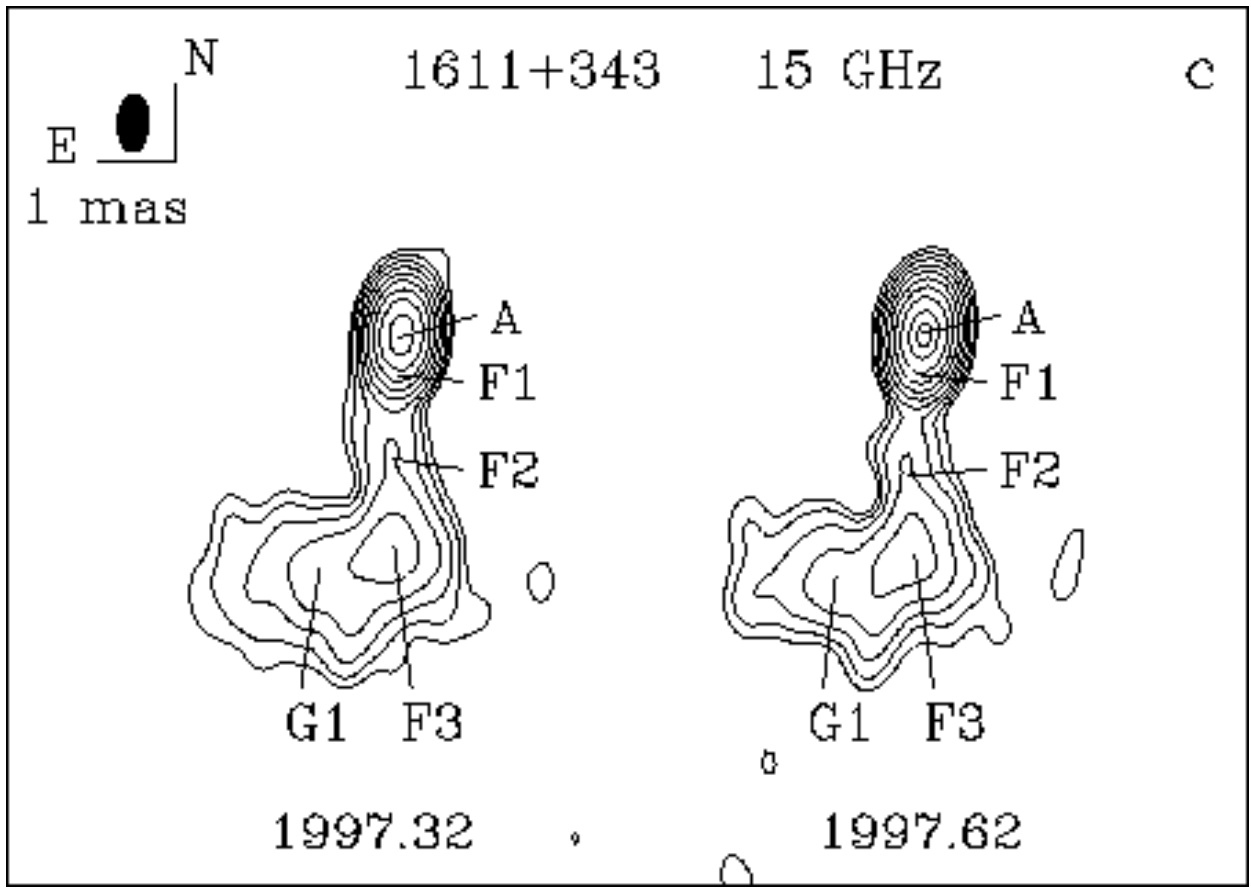}
\caption{Hybrid maps of 1611$+$343 at 15~GHz.}
\end{figure}
\begin{figure}
\figurenum{28d}
\plotone{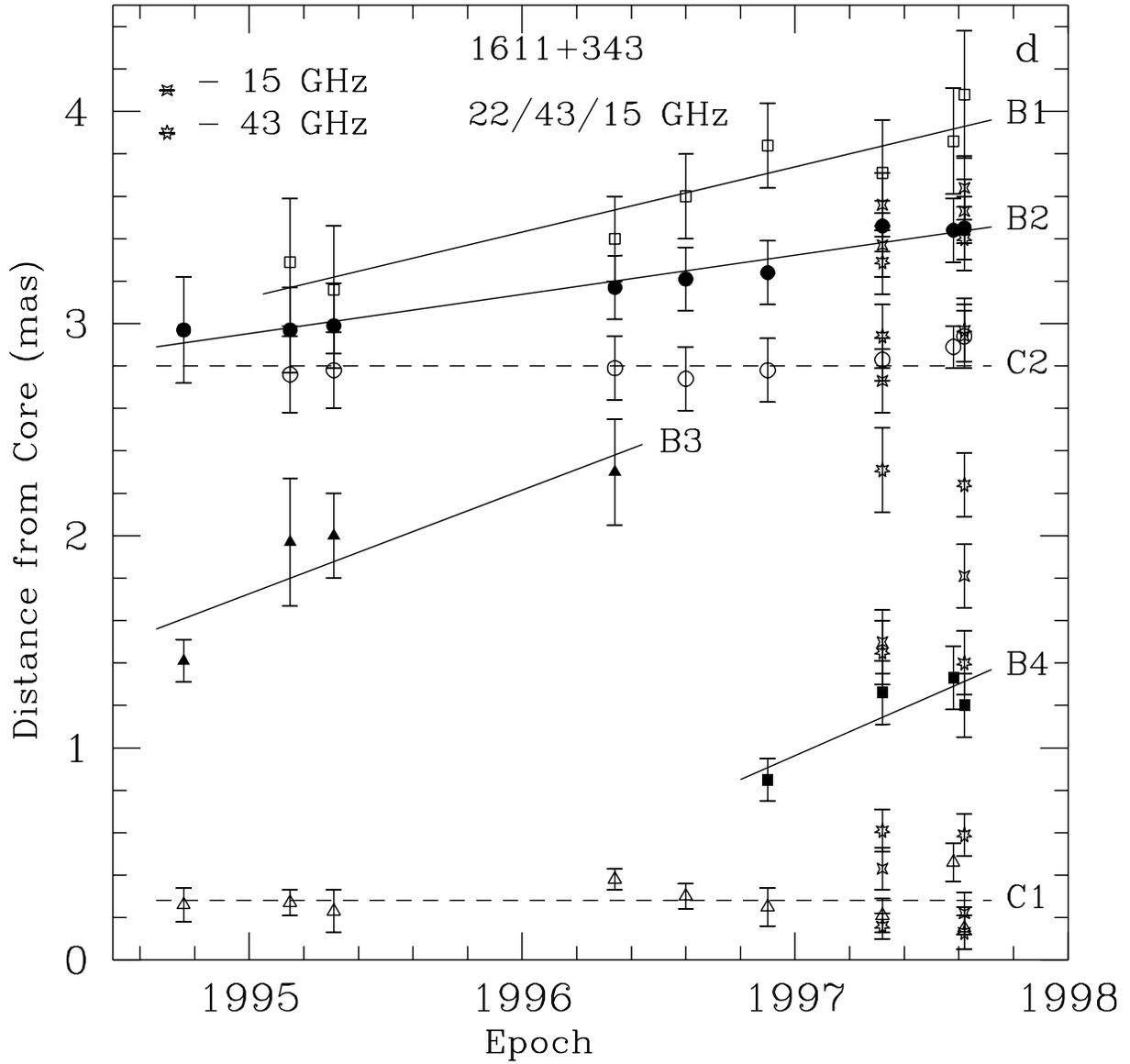}
\caption{Positions of components with respect to the core 
at different epochs from model fitting for 1611$+$343; 
designations of components are as follows:
at 22~GHz - components $C1$ (open triangles), $C2$ (open circles),
$B4$ (filled squares), $B3$ (filled triangles), B2 (filled circles),
$B1$ (open squares); all components at 43~GHz are denoted by
7-point stars; all components at 15~GHz are designated 
by 4-point stars.}
\end{figure}
\clearpage
\begin{figure}
\figurenum{29a}
\plotone{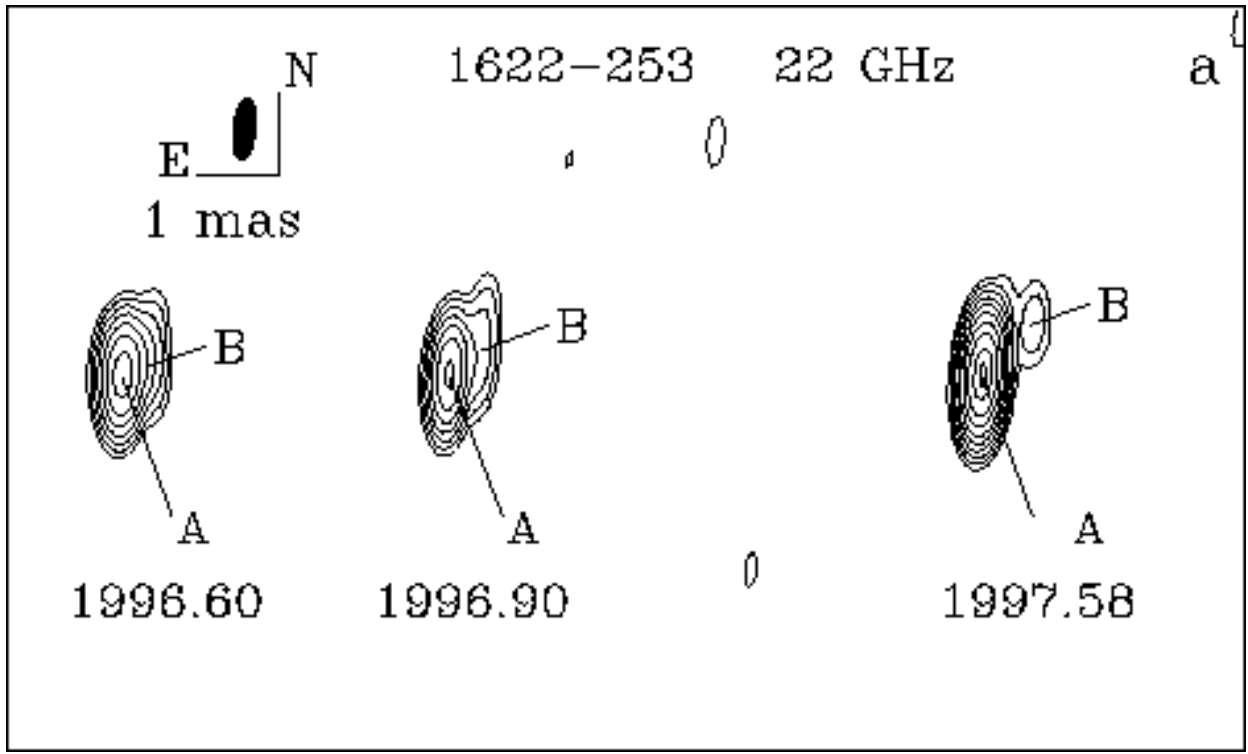}
\caption{Hybrid maps of 1622$-$253 at 22~GHz.}
\end{figure}
\begin{figure}
\figurenum{29b}
\plotone{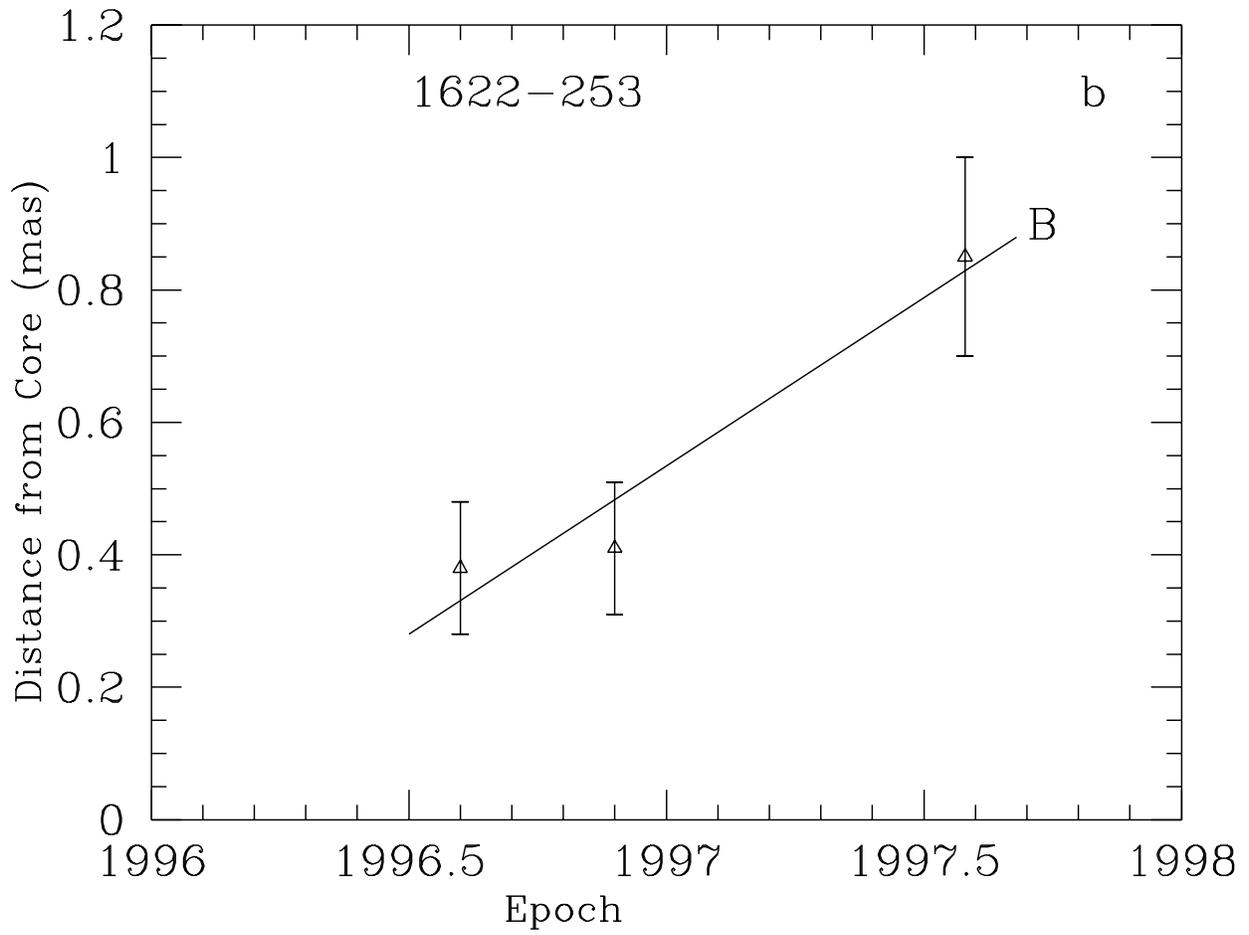}
\caption{Positions of componet $B$ with respect to the core at different
epochs from model fitting for 1622$-$253.}
\end{figure}
\begin{figure}
\figurenum{30a}
\plotone{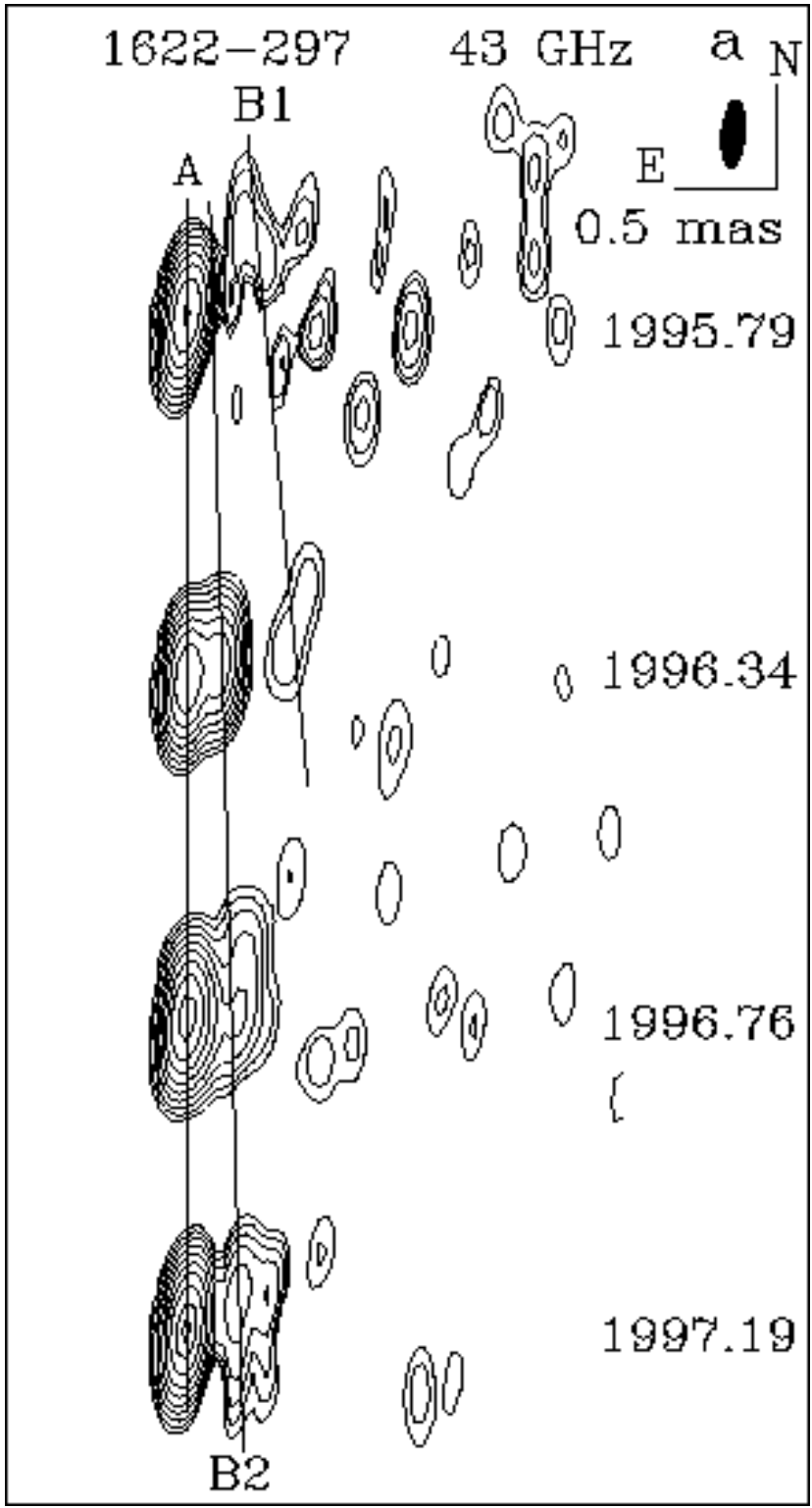}
\caption{Hybrid maps of 1622$-$297 at 43~GHz.}
\end{figure}
\begin{figure}
\figurenum{30b}
\plotone{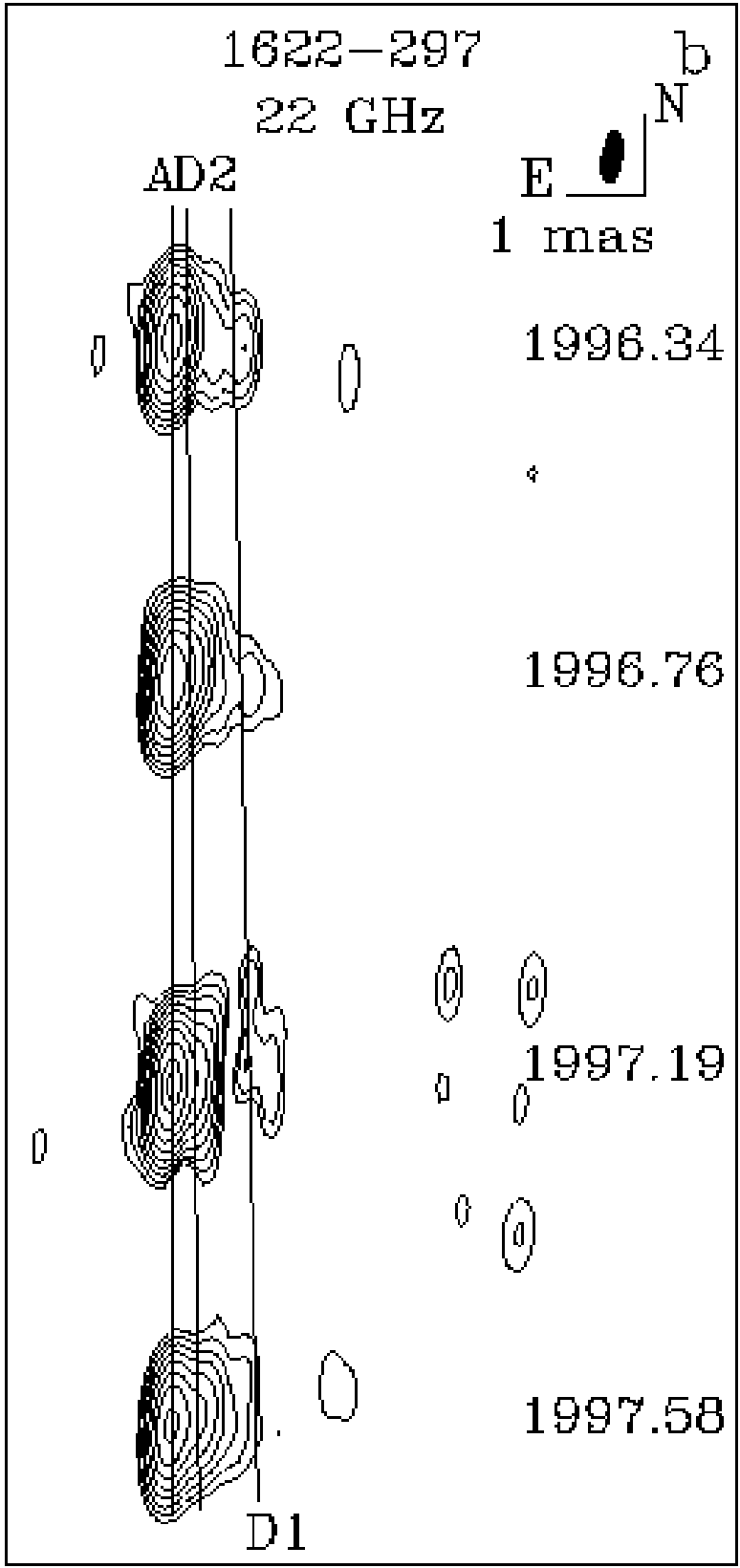}
\caption{Hybrid maps of 1622$-$297 at 22~GHz.}
\end{figure}
\begin{figure}
\figurenum{30c}
\plotone{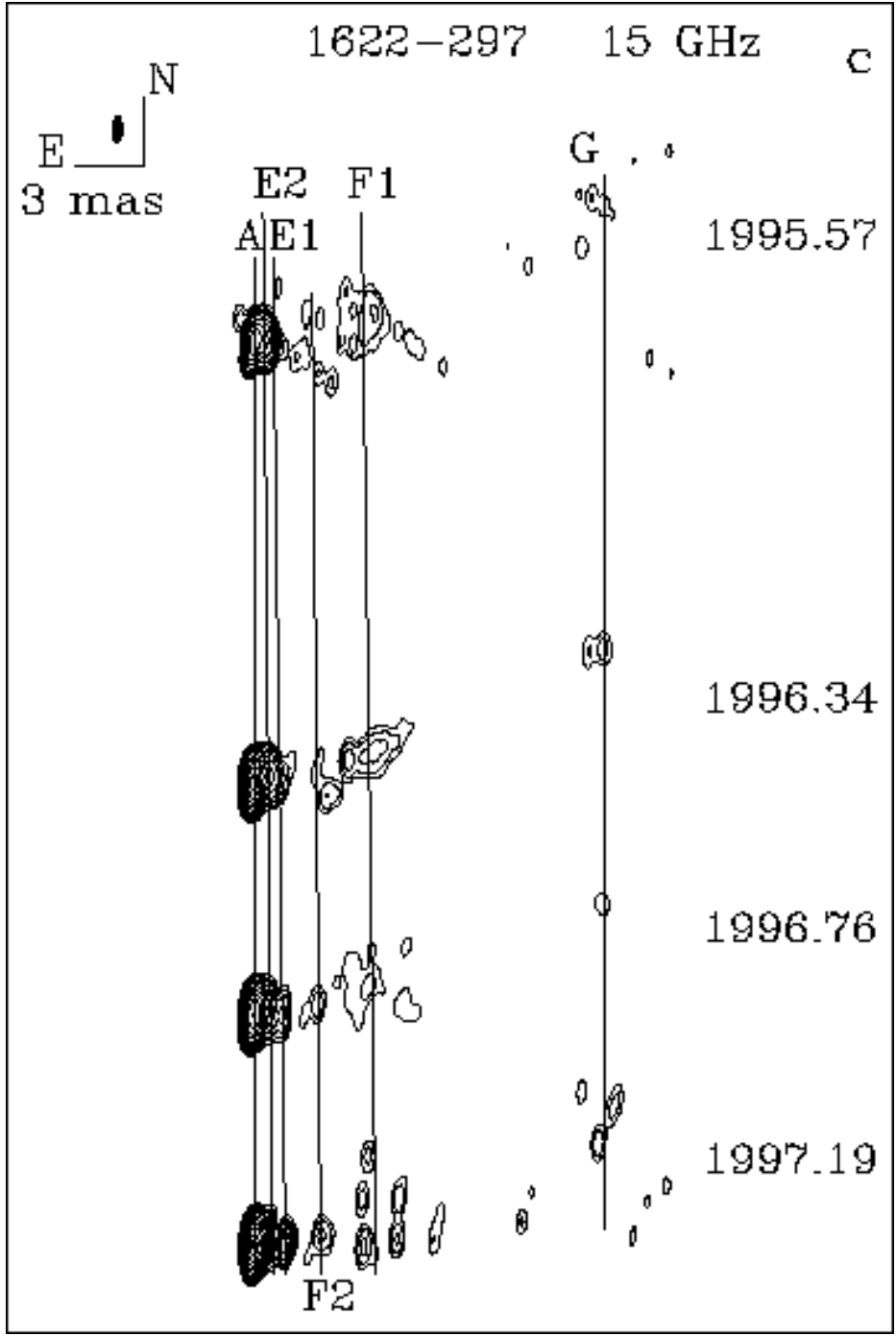}
\caption{Hybrid maps of 1622$-$297 at 15~GHz.}
\end{figure}
\begin{figure}
\figurenum{30d}
\plotone{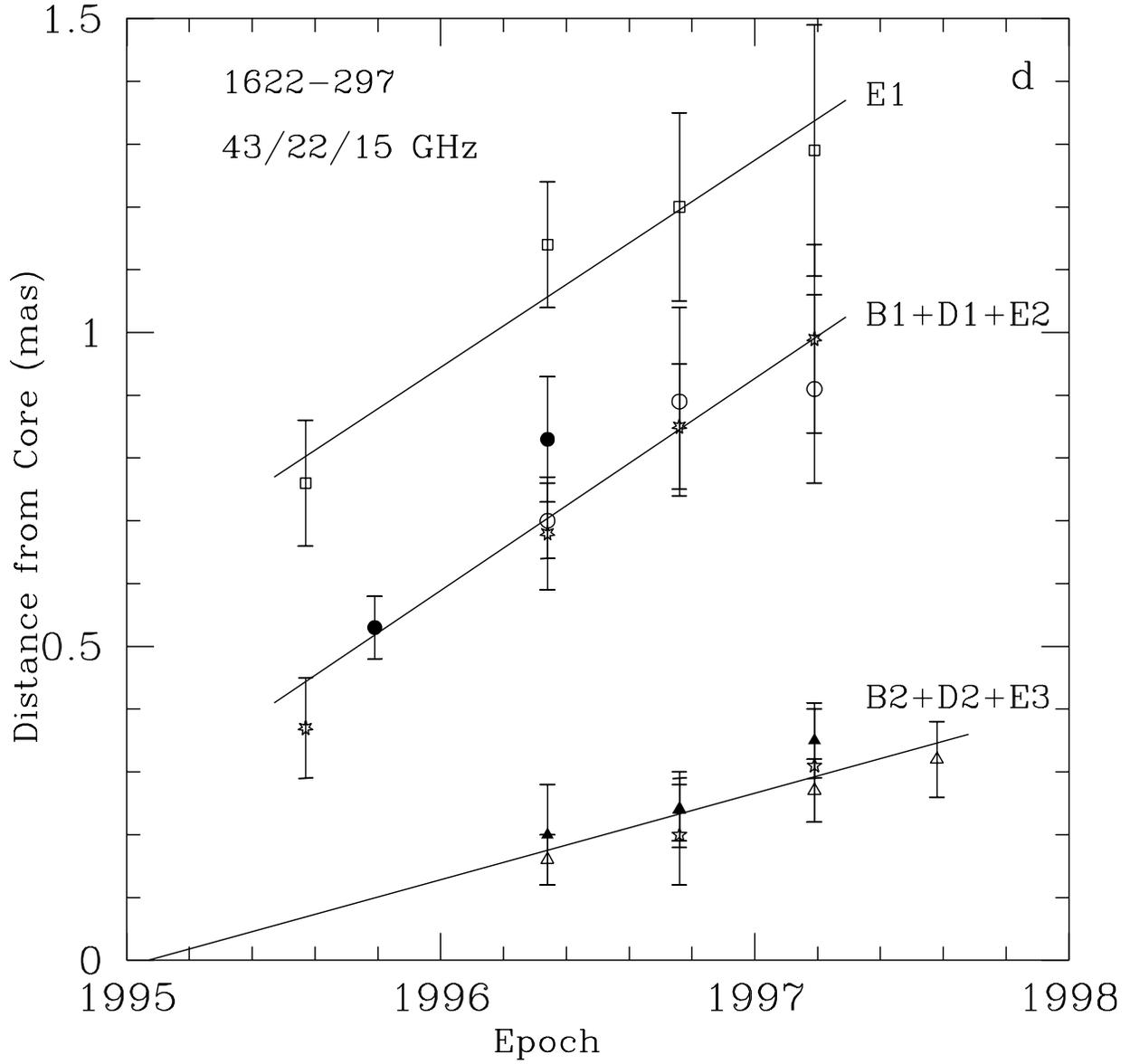}
\caption{Positions of components over a range from 0 to 1.5~mas
with respect to the core at different
epochs from model fitting for 1622$-$297 at 15, 22, and 43~GHz,
designations of components are as follows:
component $B2$ is designated by filled triangles, $B1$ - filled circles,
$D2$ - open triangles, $D1$ - open circles, $E3$ - 5-point stars; 
$E2$ - 7-point stars; $E1$ - open squares.}
\end{figure}
\begin{figure}
\figurenum{30e}
\plotone{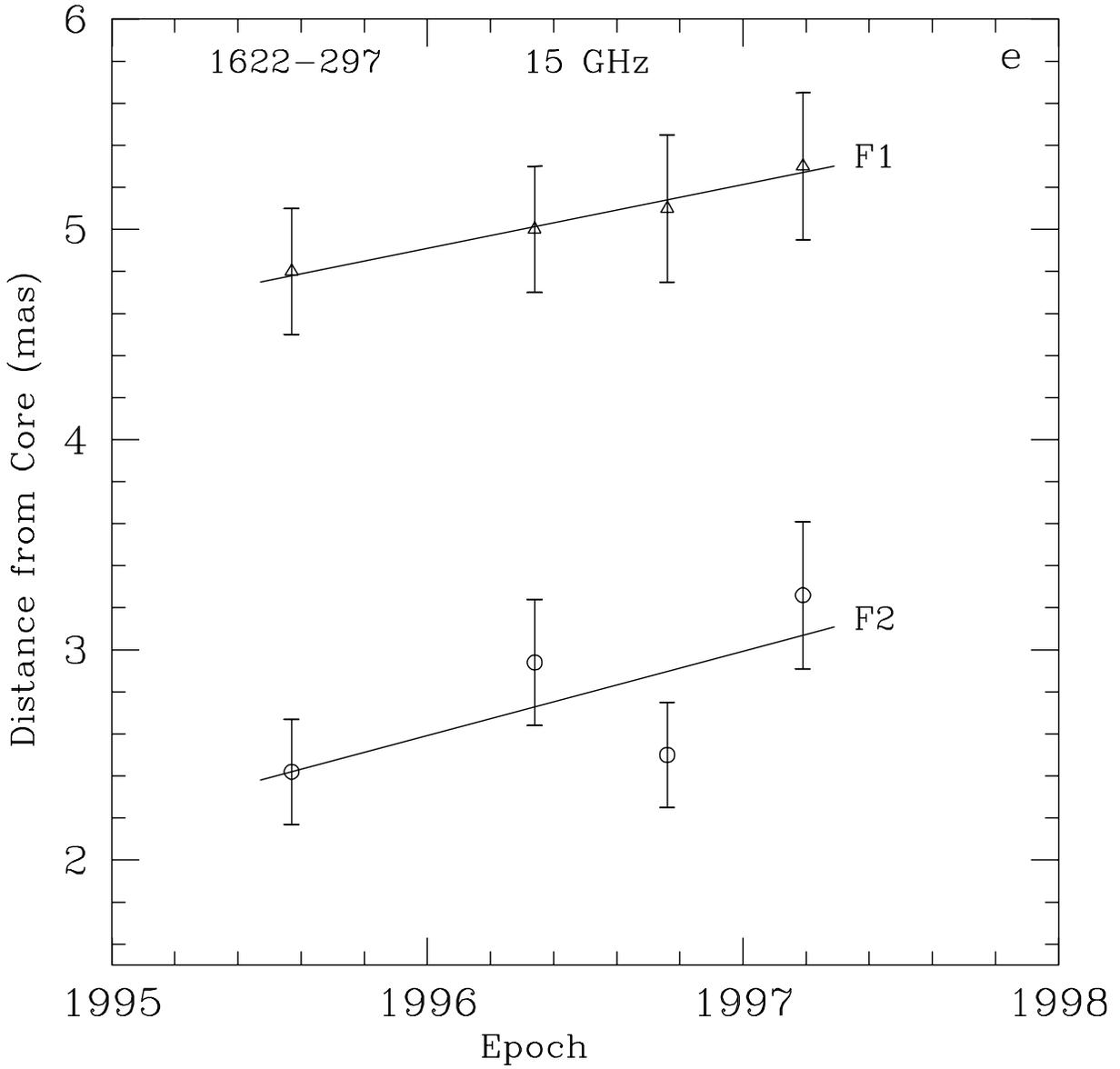}
\caption{The same as plot (d) at 15~GHz over a range of a distance with respect
to the core from 2 to 6~mas; component $F2$ is denoted by 
open circles, component $F1$ is shown by open triangles.}
\end{figure}
\begin{figure}
\figurenum{31a}
\plotone{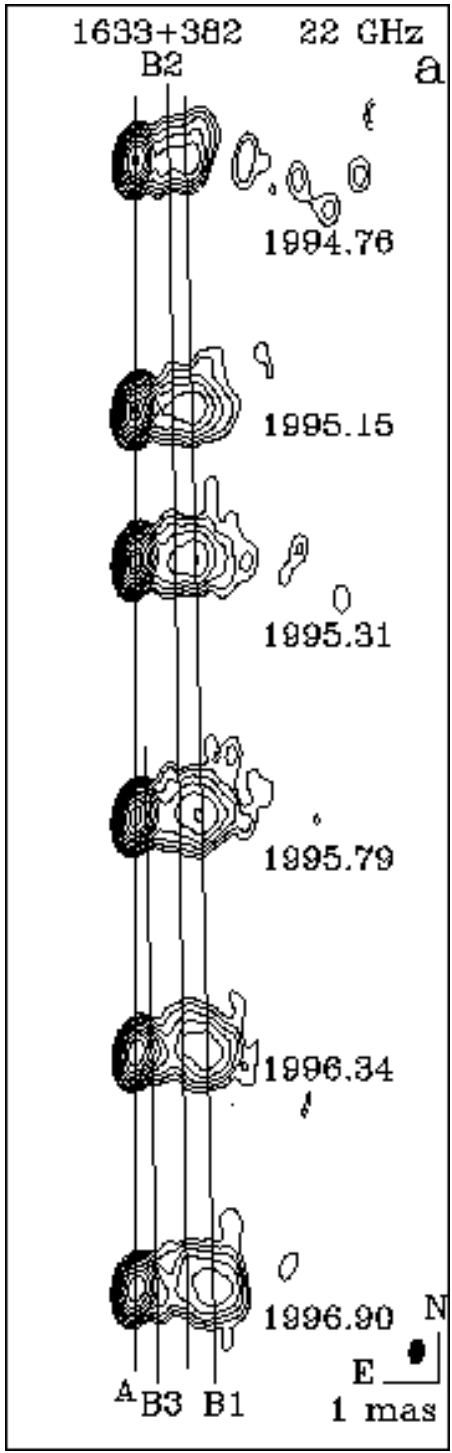}
\caption{Hybrid maps of 1633$+$382 at 22~GHz.}
\end{figure}
\begin{figure}
\figurenum{31b}
\plotone{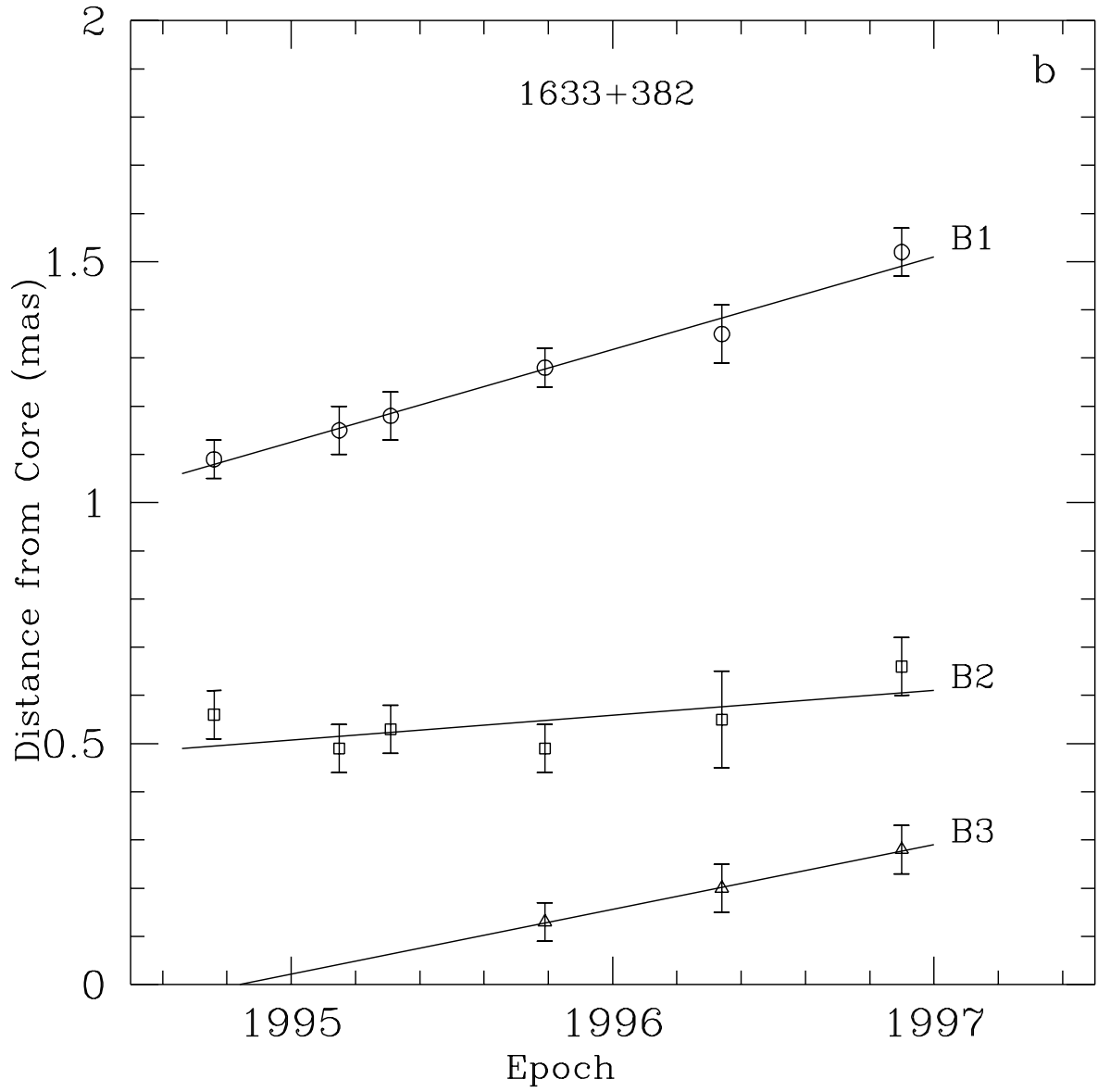}
\caption{Positions of components with respect to the core at 
different epochs from model fitting for 1633$+$382; designations of components are as follows:
open triangles - component $B3$, 
open squares - $B2$, open circles - component $B1$.}
\end{figure}
\begin{figure}
\figurenum{32a}
\plotone{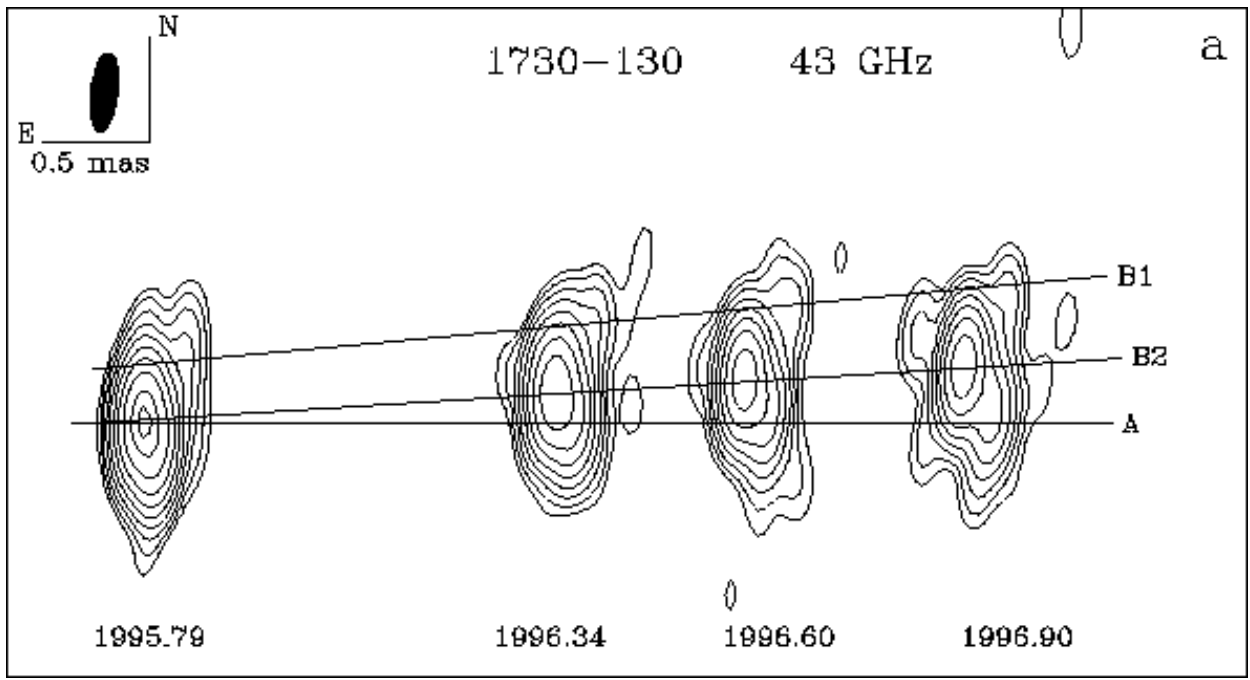}
\caption{Hybrid maps of 1730$-$130 at 43~GHz.}
\end{figure}
\begin{figure}
\figurenum{32b}
\plotone{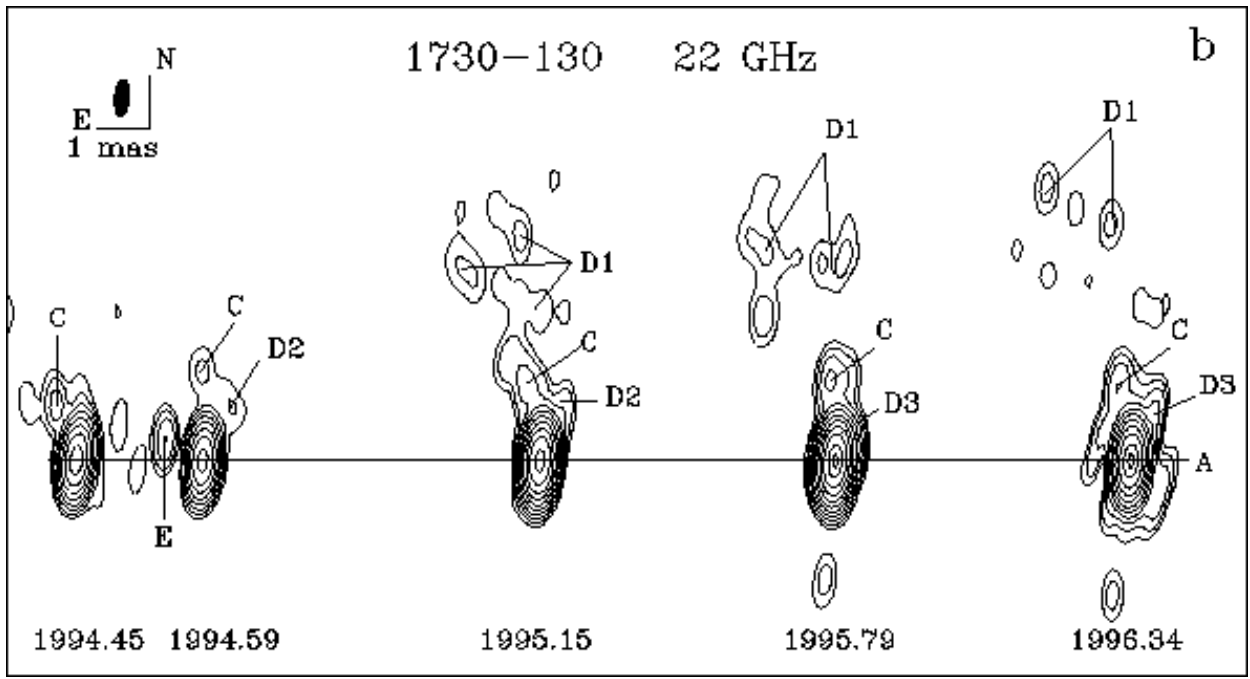}
\caption{Hybrid maps of 1730$-$130 at 22~GHz.}
\end{figure}
\begin{figure}
\figurenum{32c}
\plotone{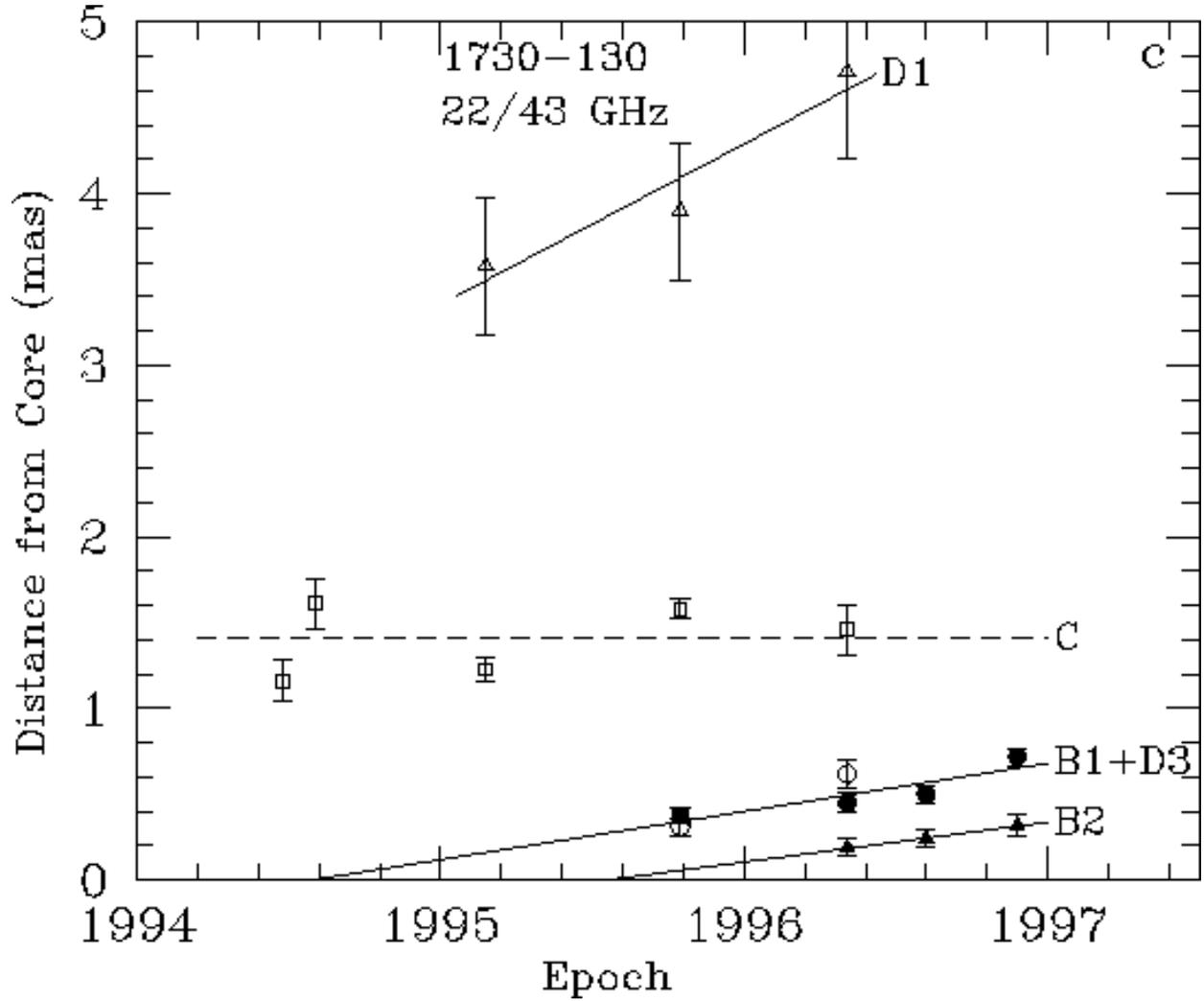}
\caption{Positions of components with respect to the core at different
epochs from model fitting for 1730$-$130; designations of components are as follows:
filled triangles - component $B2$ and 
filled circles - $B1$ at 43~GHz; open circles - component $D3$
open squares - component $C$, and open triangles - component $D1$
at 22~GHz.}
\end{figure}
\clearpage
\begin{figure}
\figurenum{33}
\plotone{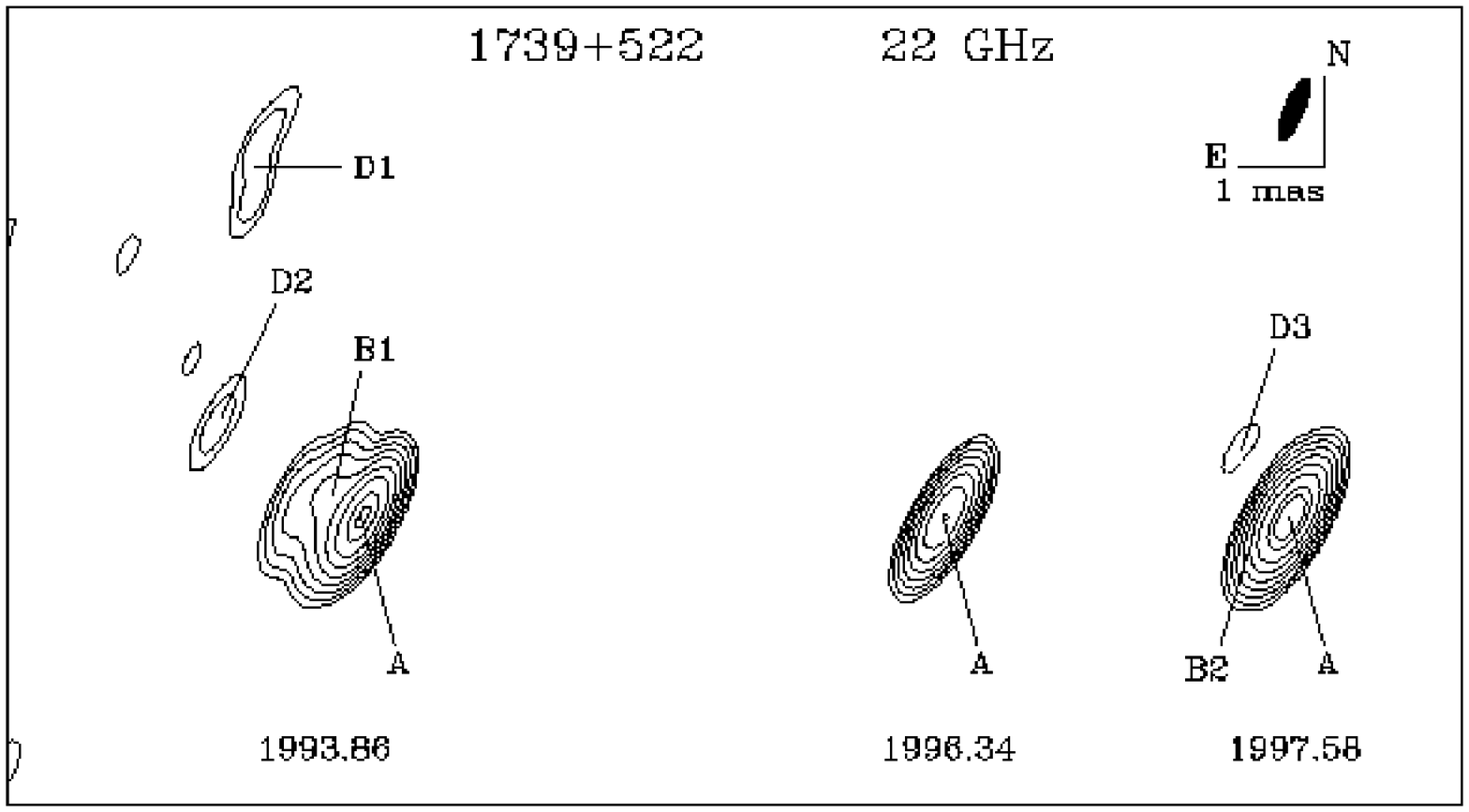}
\caption{Hybrid maps of 1739$+$522 at 22~GHz.}
\end{figure}
\begin{figure}
\figurenum{34}
\plotone{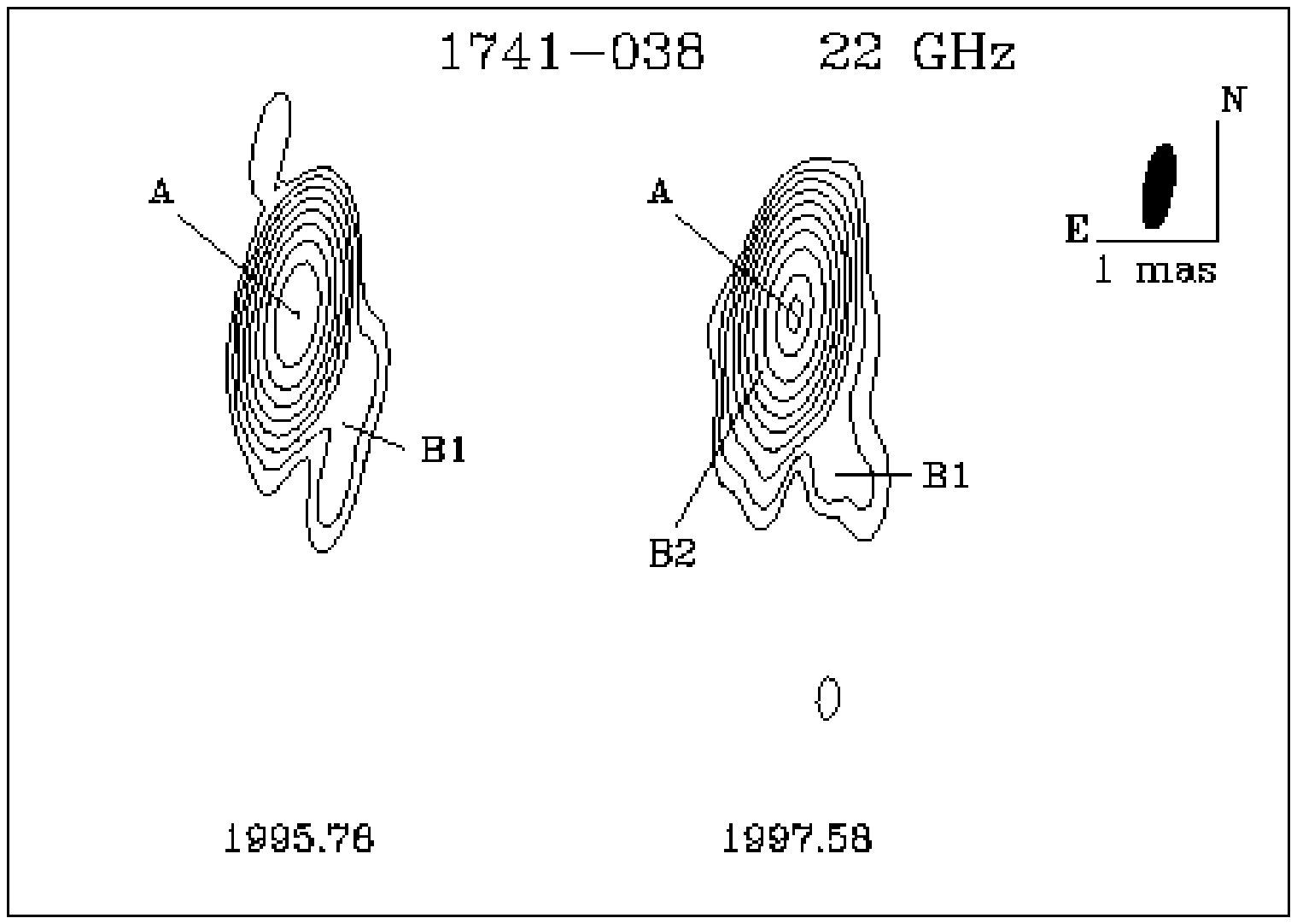}
\caption{Hybrid maps of 1741$-$038 at 22~GHz.}
\end{figure}
\begin{figure}
\figurenum{35a}
\plotone{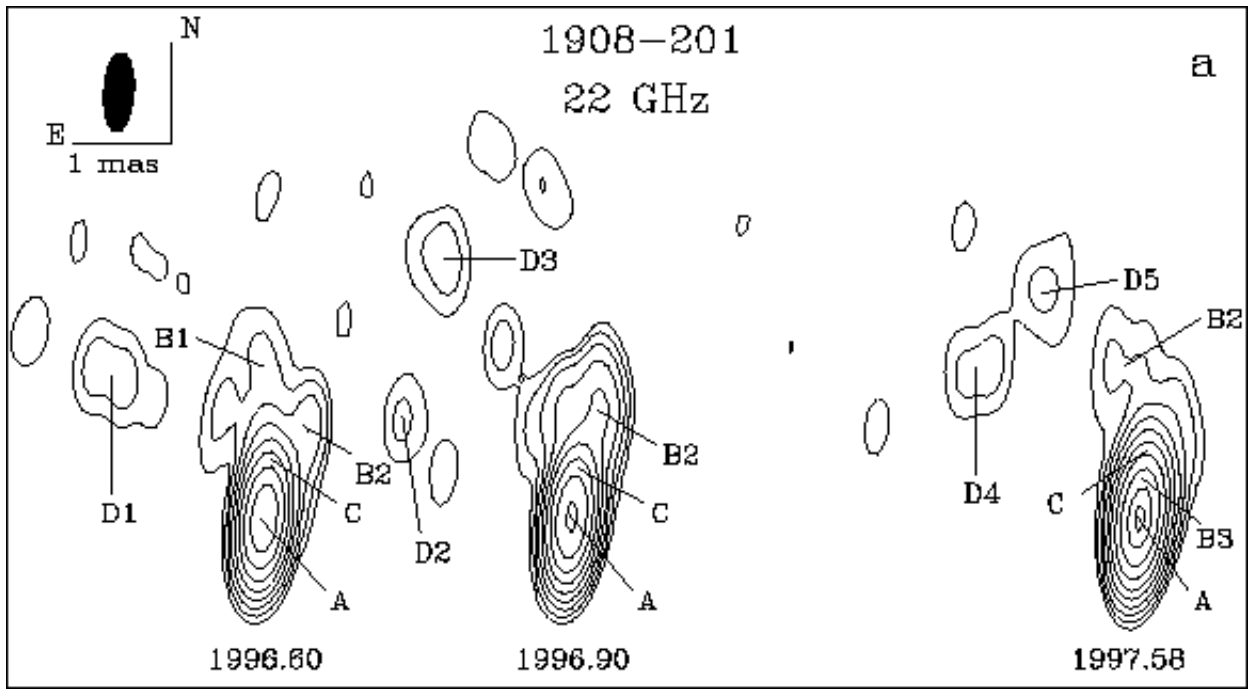}
\caption{Hybrid maps of 1908$-$201 at 22~GHz.}
\end{figure}
\begin{figure}
\figurenum{35b}
\plotone{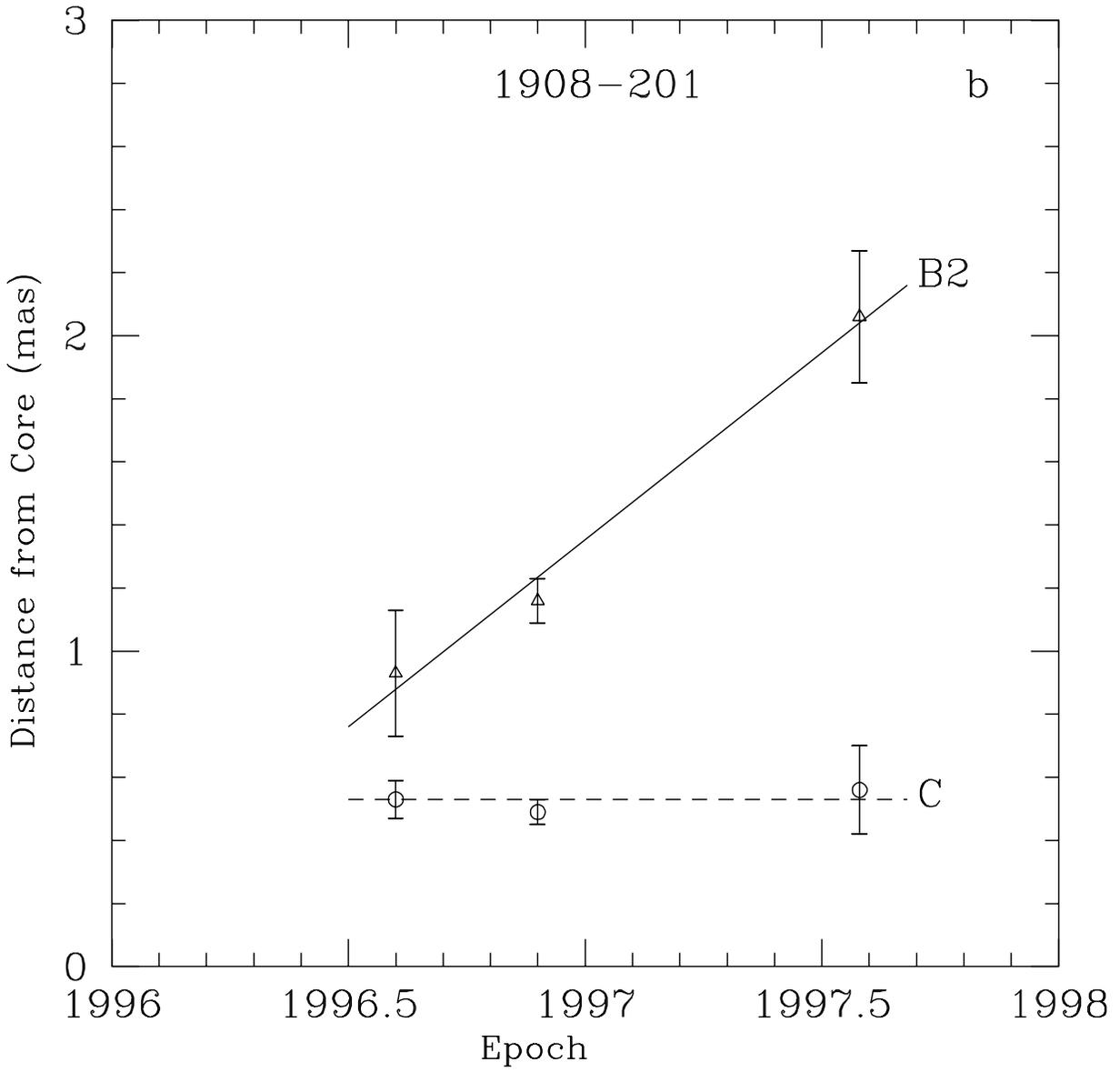}
\caption{Positions of components with respect to the core 
at different epochs from model fitting for 1908$-$201; designations of components are as follows:
open triangles - component $B2$, 
open circles - component $C$.}
\end{figure}
\begin{figure}
\figurenum{36}
\plotone{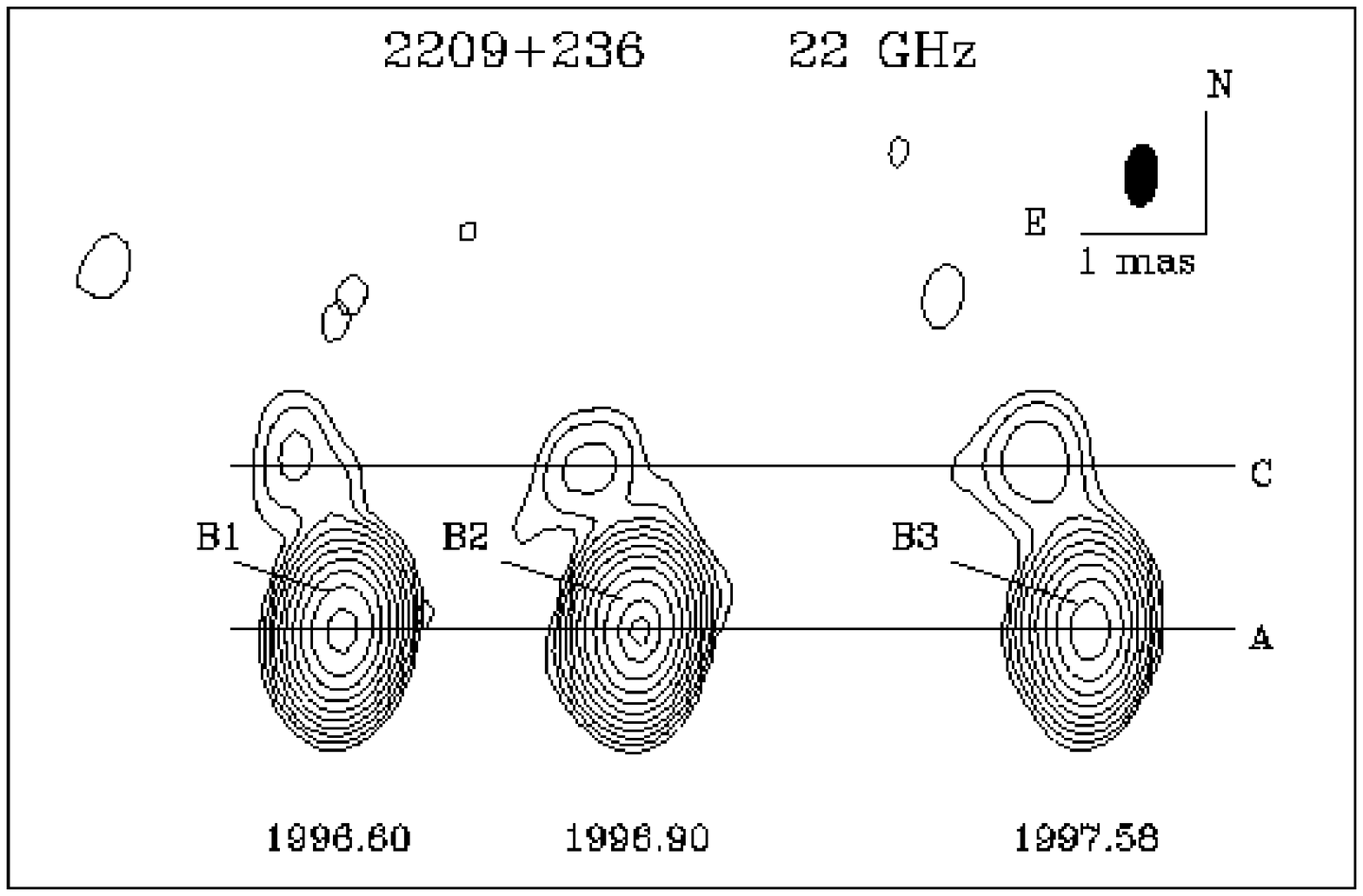}
\caption{Hybrid maps of 2209$+$236 at 22~GHz.}
\end{figure}
\begin{figure}
\figurenum{37a}
\plotone{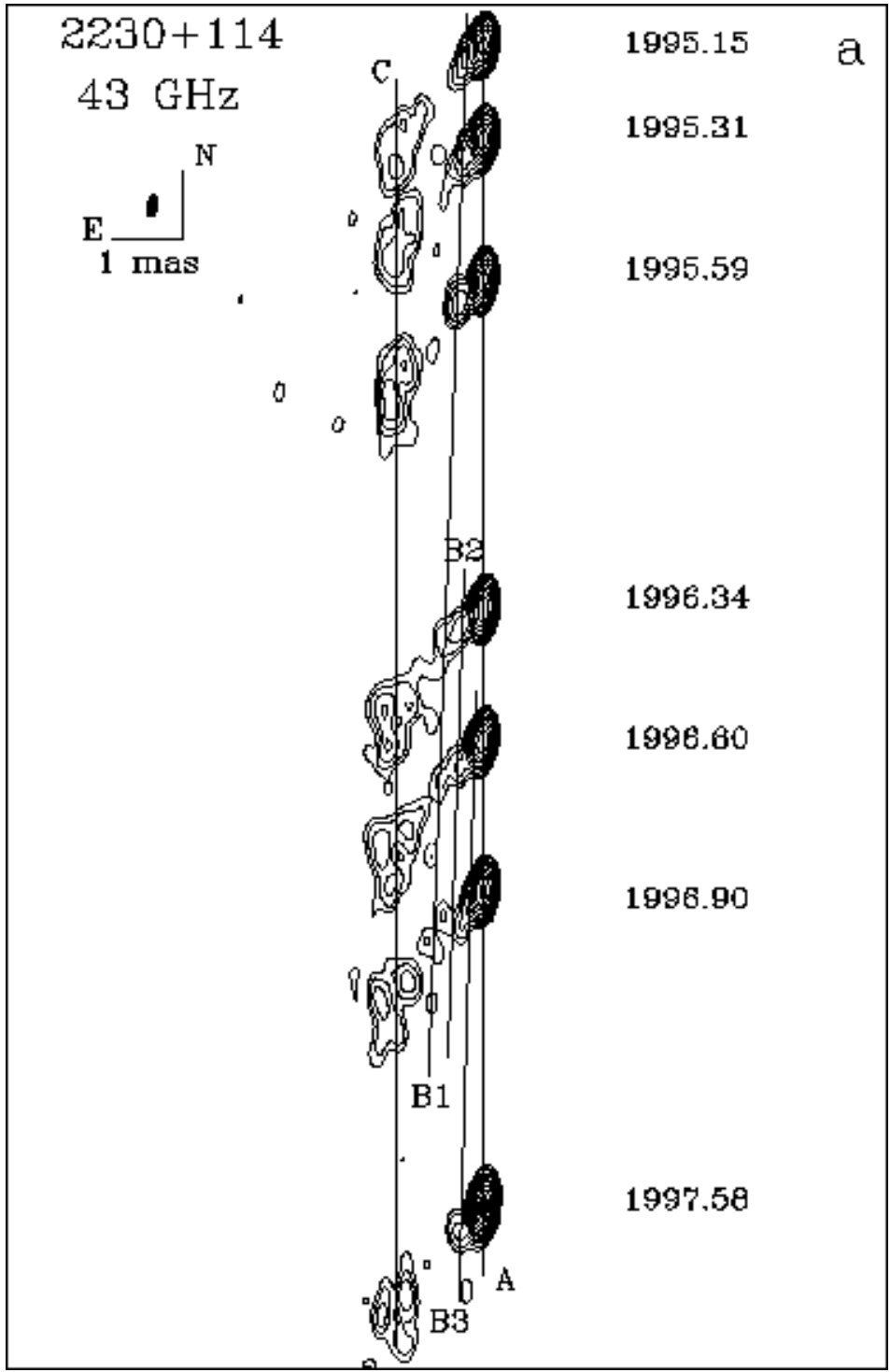}
\caption{Hybrid maps of CTA~102 at 43~GHz.}
\end{figure}
\begin{figure}
\figurenum{37b}
\plotone{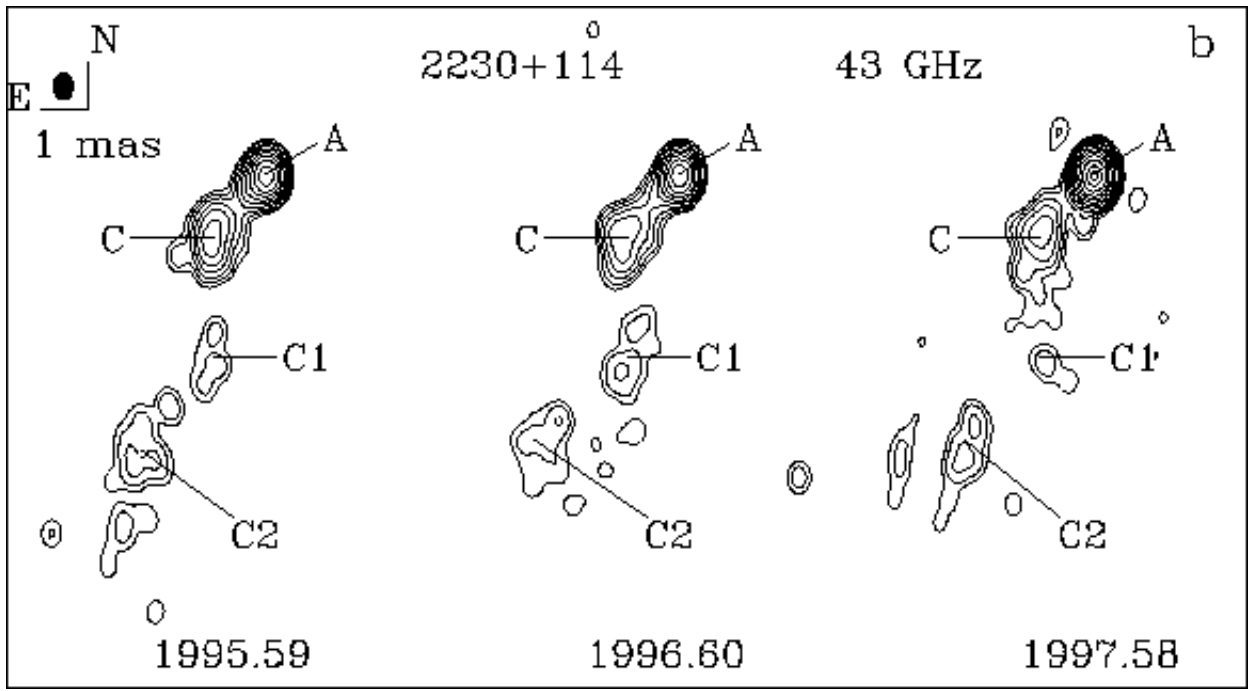}
\caption{Tapered maps of CTA~102 at 43~GHz. 
The restoring beam is 0.6$\times$0.4~mas$^2$ at PA=0$^\circ$, 
the peak flux density is 6.31~Jy/Beam in 1997.58, and
the lowest contour is 0.0625\% of the peak flux density.}
\end{figure}
\begin{figure}
\figurenum{37c}
\plotone{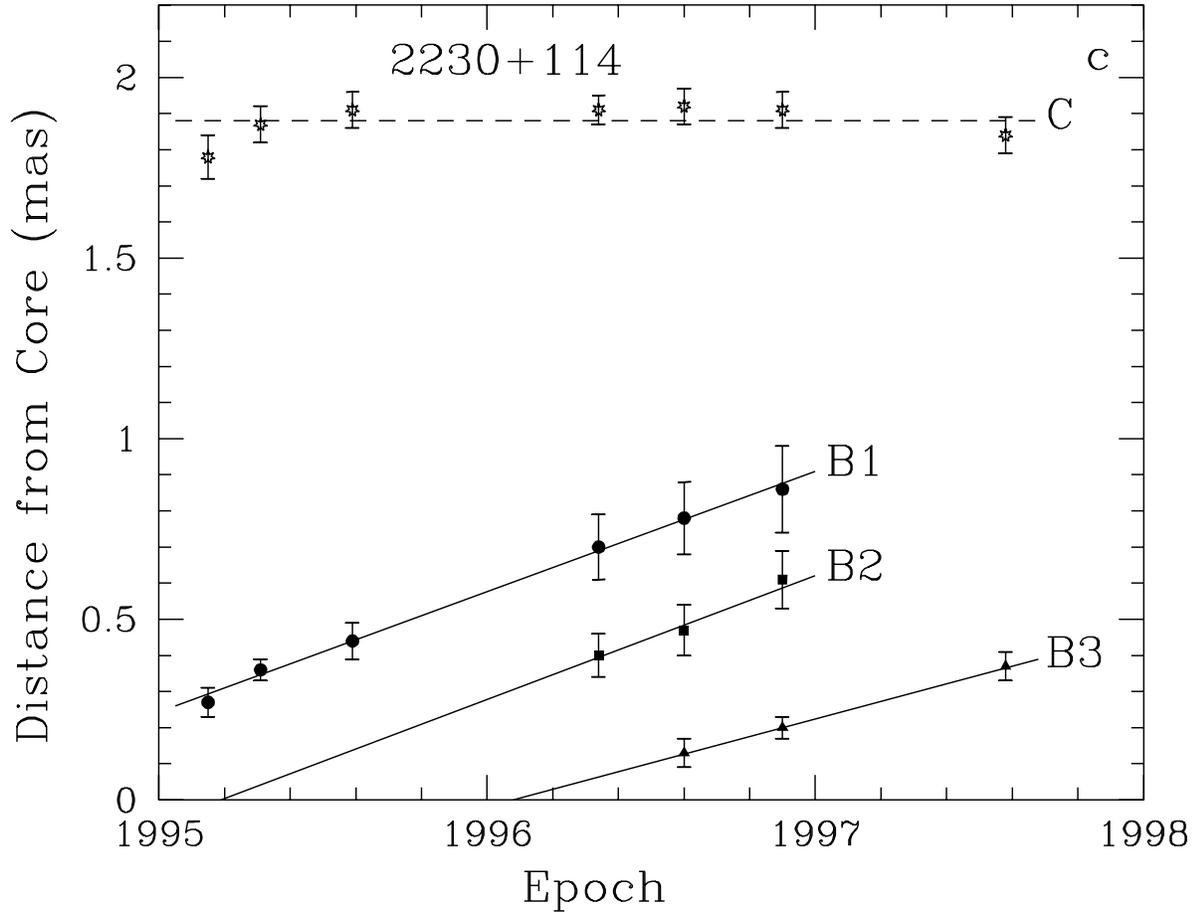}
\caption{Positions of components with respect to the core  
at different epochs from model fitting for CTA~102; designations of components are as follows:
filled triangles - component $B3$, filled squares - $B2$,
filled circles - component $B1$, stars - component $C$.}
\end{figure}
\begin{figure}
\figurenum{38a}
\plotone{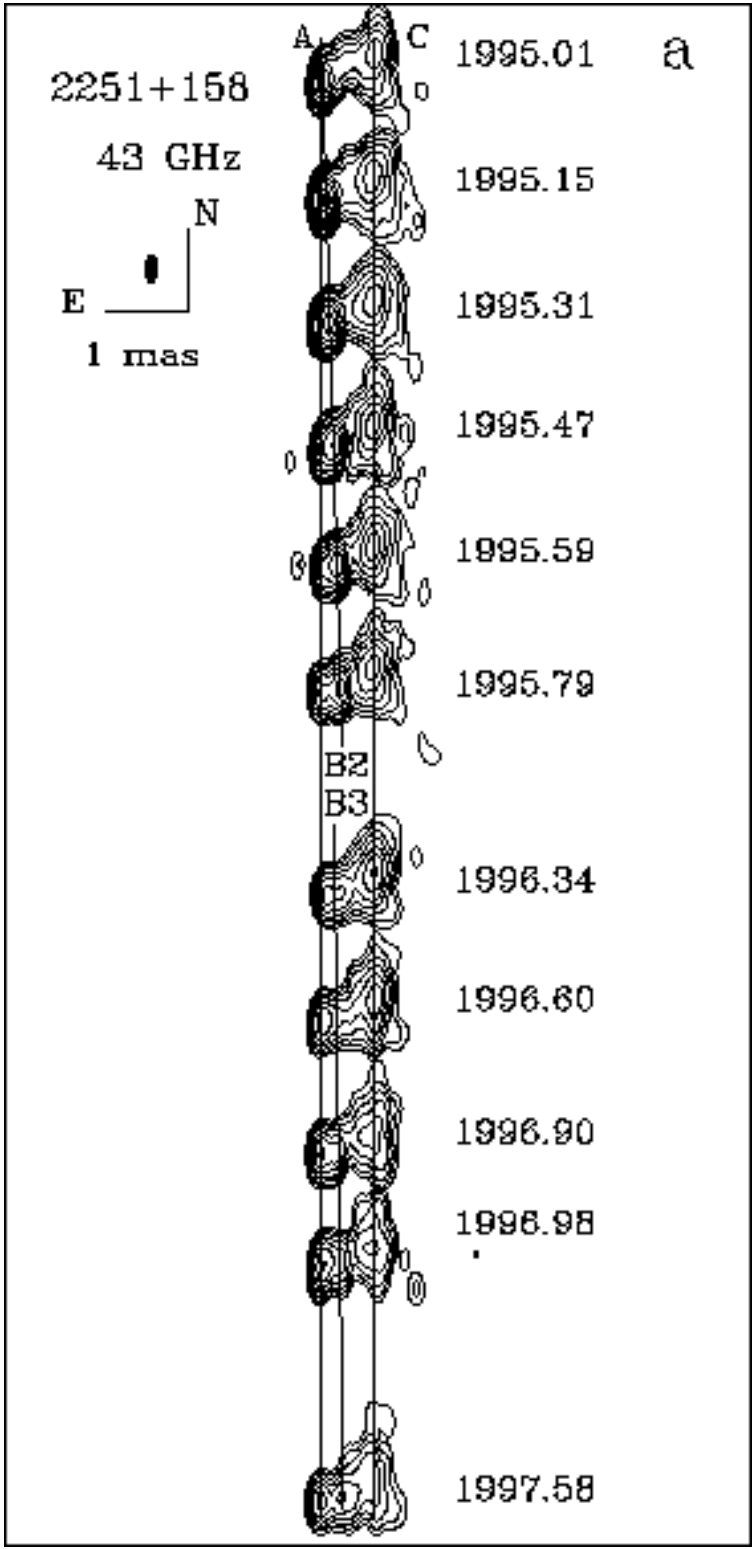}
\caption{Hybrid maps of 3C~454.3 at 43~GHz.}
\end{figure}
\begin{figure}
\figurenum{38b}
\plotone{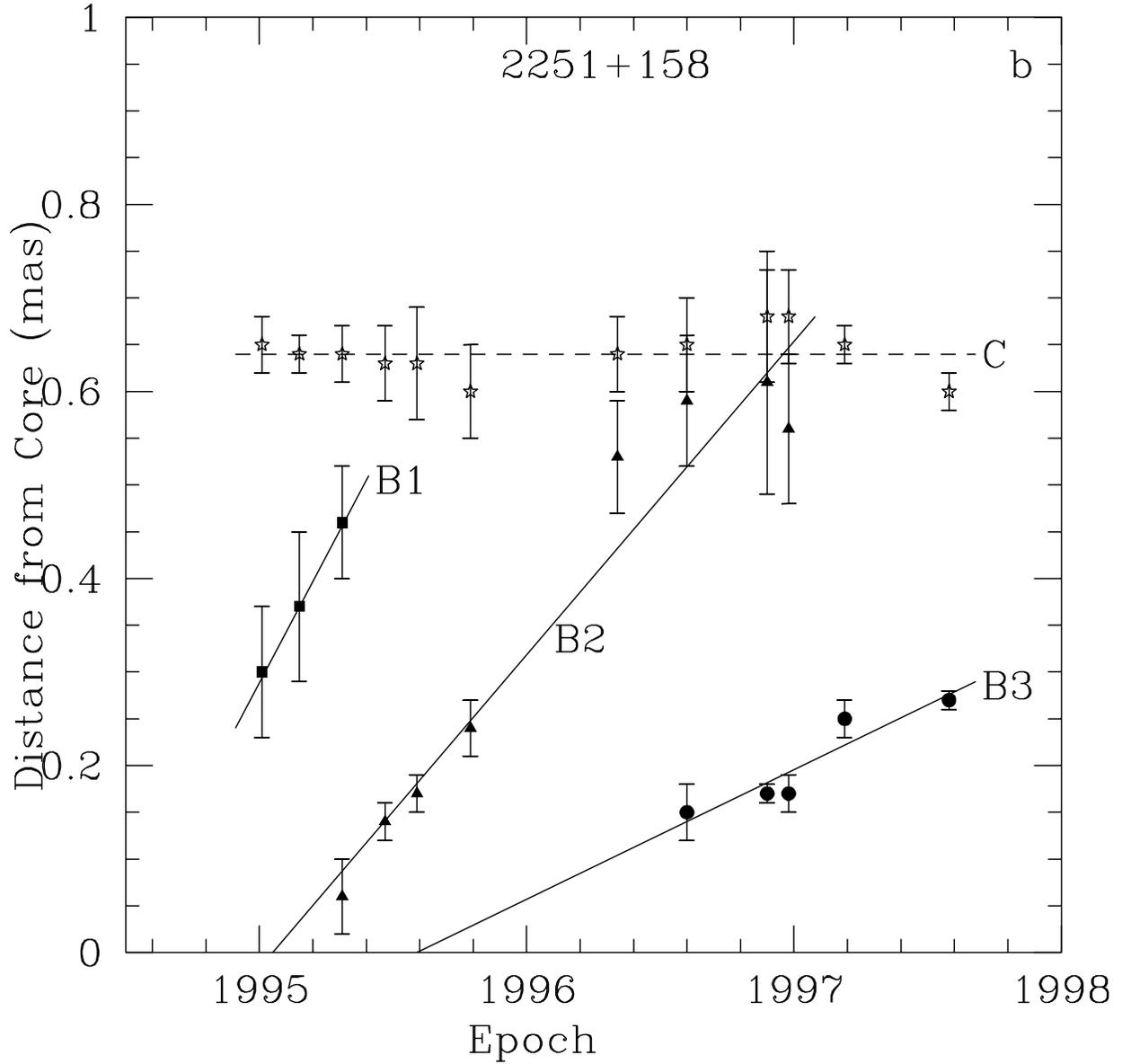}
\caption{Positions of components with respect to the core 
at different epochs from model fitting for 3C~454.3; designations of components are as follows:
filled circles - component $B3$, filled triangles - $B2$,
filled squares - component $B1$, stars - component $C$.}
\end{figure}
\begin{figure}
\figurenum{39}
\plotone{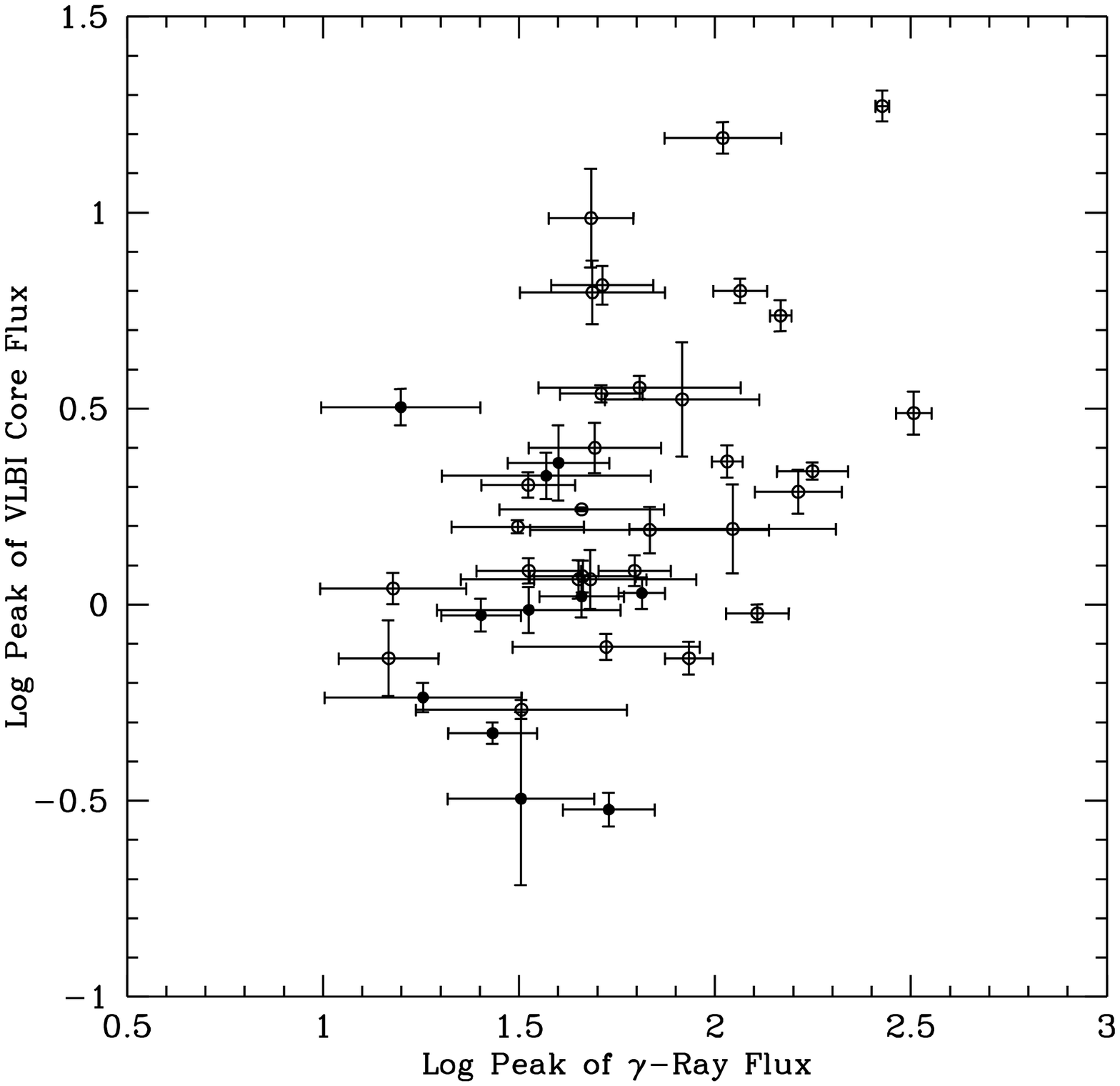}
\caption{Maximum of VLBI core flux at 22/43~GHz vs.
maximum of $\gamma$-ray flux; filled circles correspond to BL~Lac objects,
open circles to quasars.}
\end{figure}
\begin{figure}
\figurenum{40}
\plotone{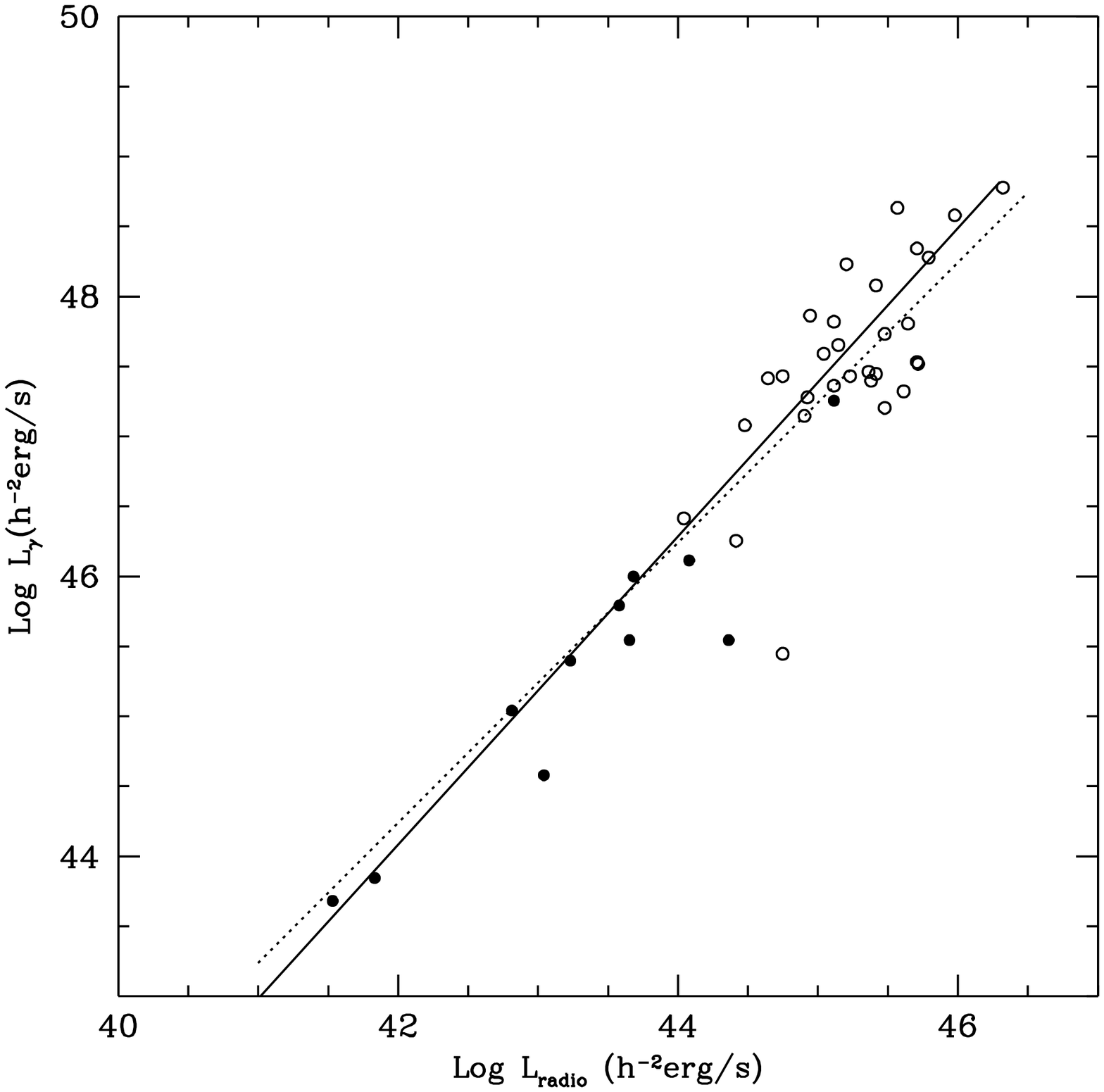}
\caption{Correlation of the $\gamma$-ray luminosity
with radio luminosity; filled circles correspond to BL~Lac objects,
open circles to quasars. The solid line shows the fit of the data
by the dependence L$_\gamma\propto$L$_{radio}^{1.10\pm0.04}$, while
the dotted line corresponds to a slope of 1.00.}
\end{figure}
\begin{figure}
\figurenum{41}
\plotone{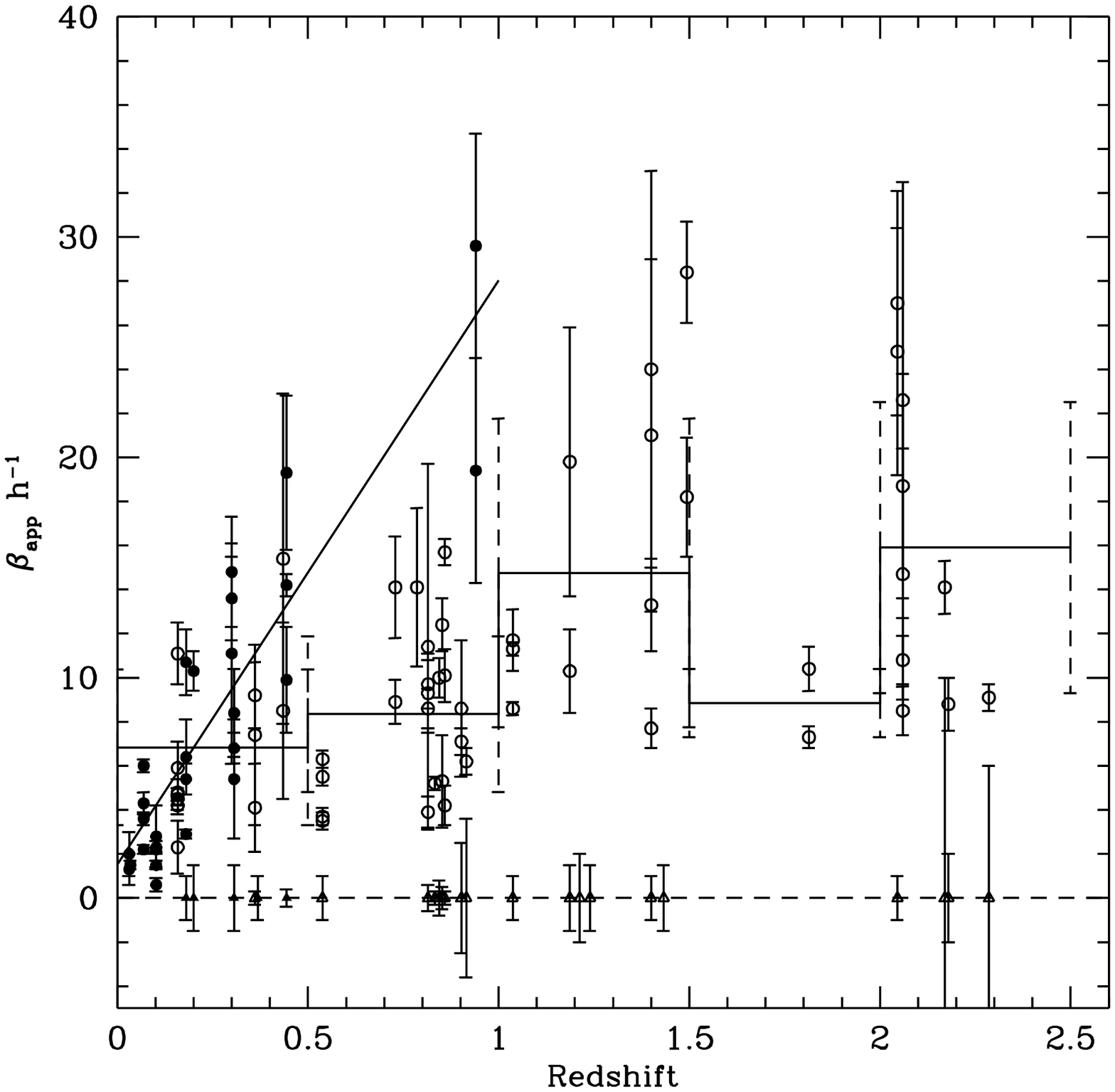}
\caption{Dependence of apparent velocities of jet components
on redshift. Moving components for quasars are designated by open circles, 
for BL~Lacs by filled circles; triangles correspond to components with proper 
motion equal to 0. The solid broken line is the arithmetic average of apparent
velocities of moving components for quasars calculated in redshift intervals of 0.5; the vertical dashed
line is the standard deviation. The solid straight line corresponds to the linear fit
$\beta_{app}$(z)=[(1.5$\pm$0.7)+(26.5$\pm$9.3)z]~$h^{-1}c$ for BL~Lac objects.} 
\end{figure}
\begin{figure}
\figurenum{42a}
\plotone{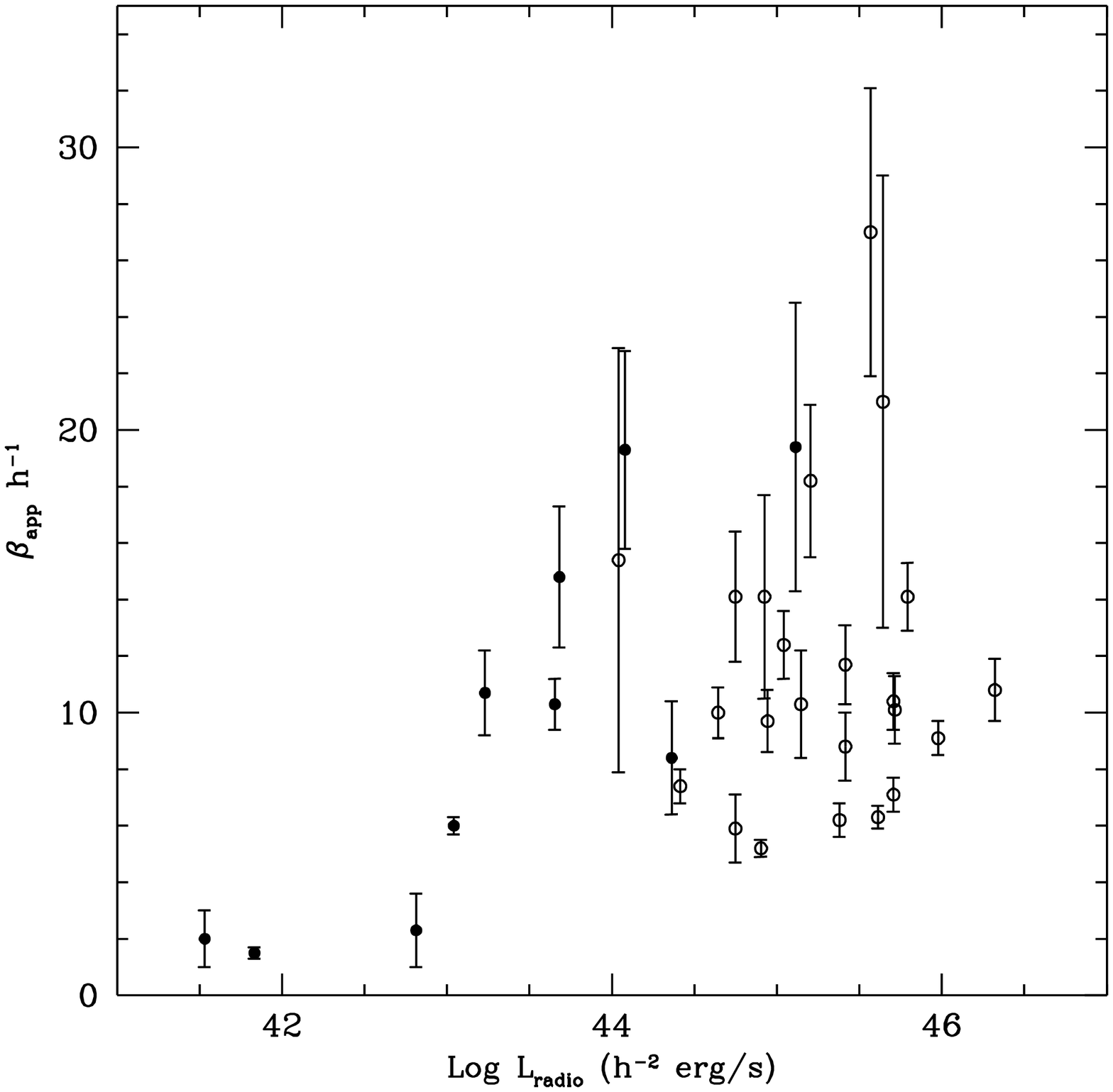}
\caption{Dependence of apparent speed of jet components 
on the radio luminosity; open circles correspond to quasars, filled circles to BL~Lac objects.
In the case of multiple moving components the highest speed obtained at 3 or more epochs is
used.}
\end{figure}
\begin{figure}
\figurenum{42b}
\plotone{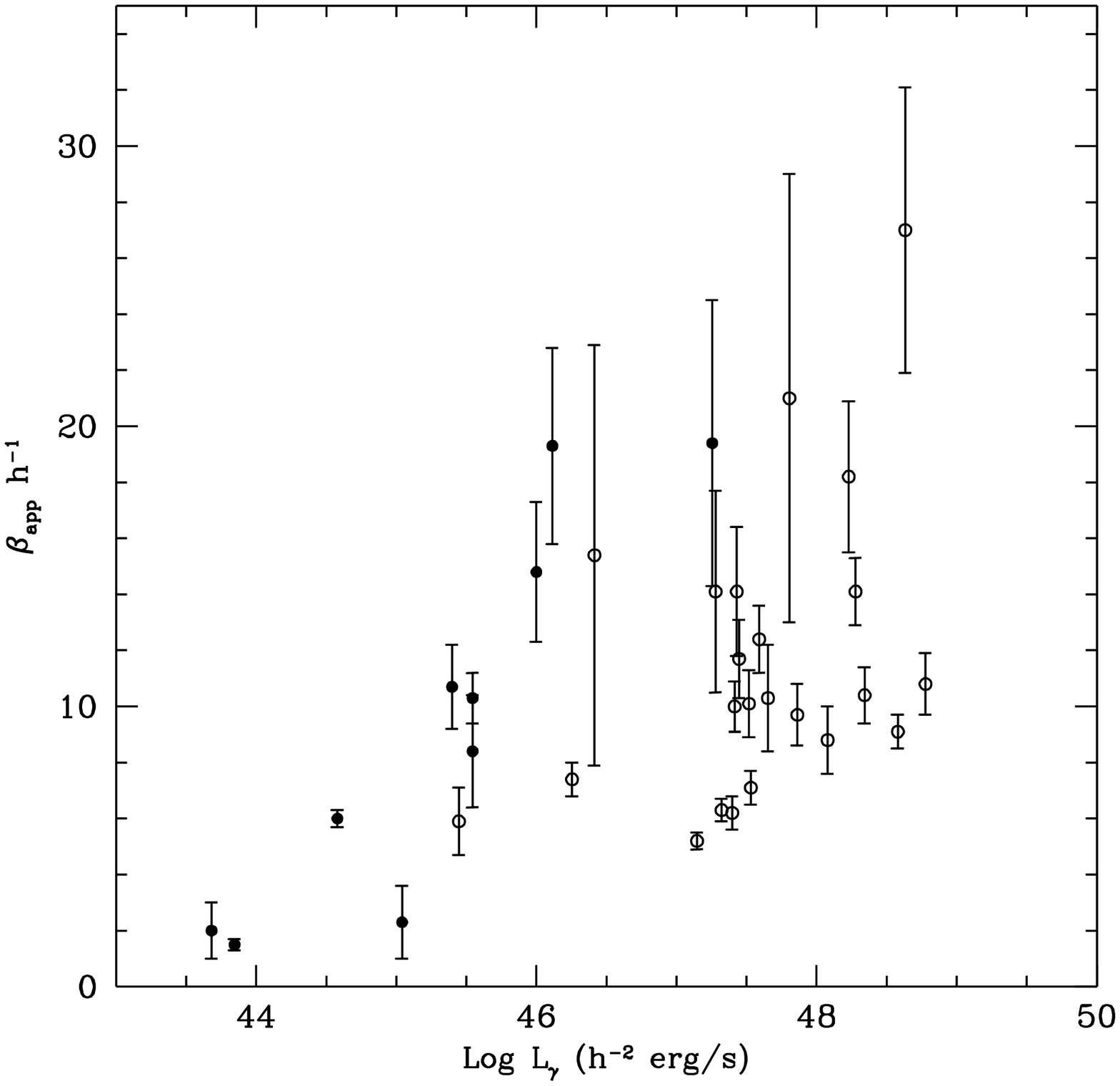}
\caption{Dependence of apparent speed of jet components 
on the $\gamma$-ray luminosity; open circles correspond to quasars, filled circles to BL~Lac objects.
In the case of multiple moving components the highest speed obtained at 3 or more epochs is
used.}
\end{figure}
\begin{figure}
\figurenum{43a}
\plotone{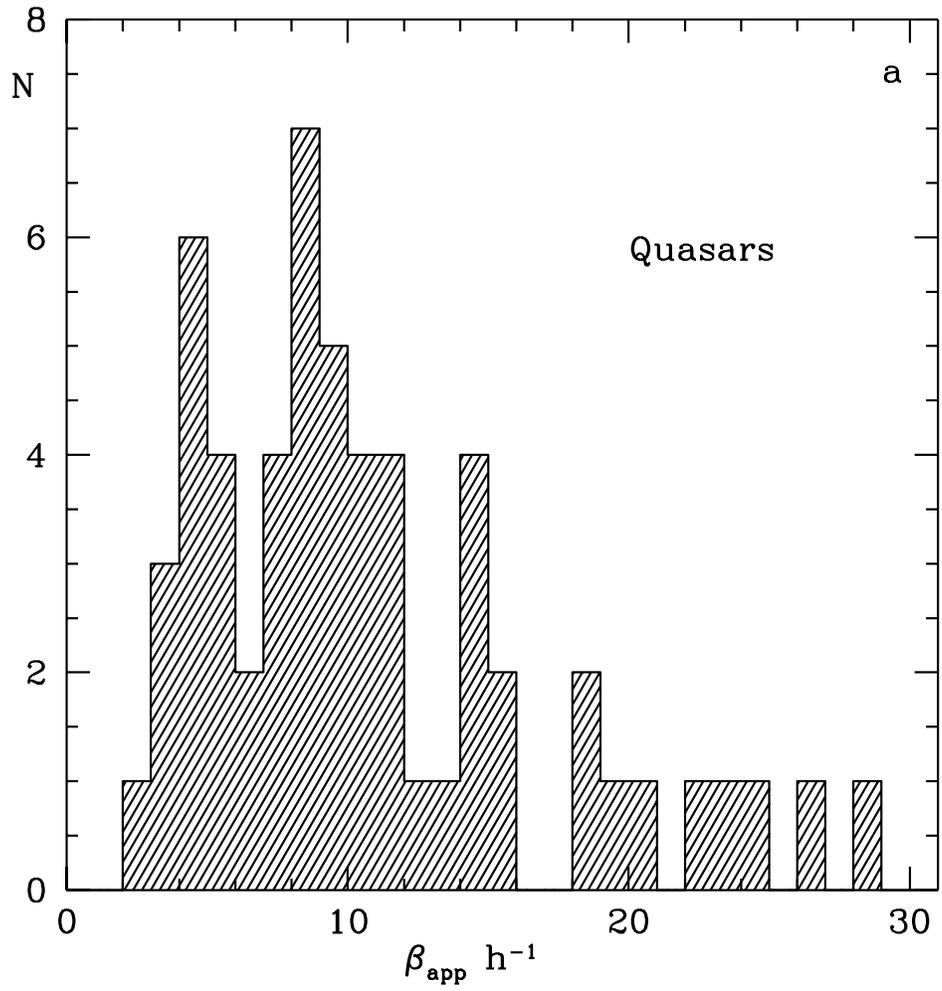}
\caption{Distribution of apparent speeds of jet components in quasars.}
\end{figure}
\begin{figure}
\figurenum{43b}
\plotone{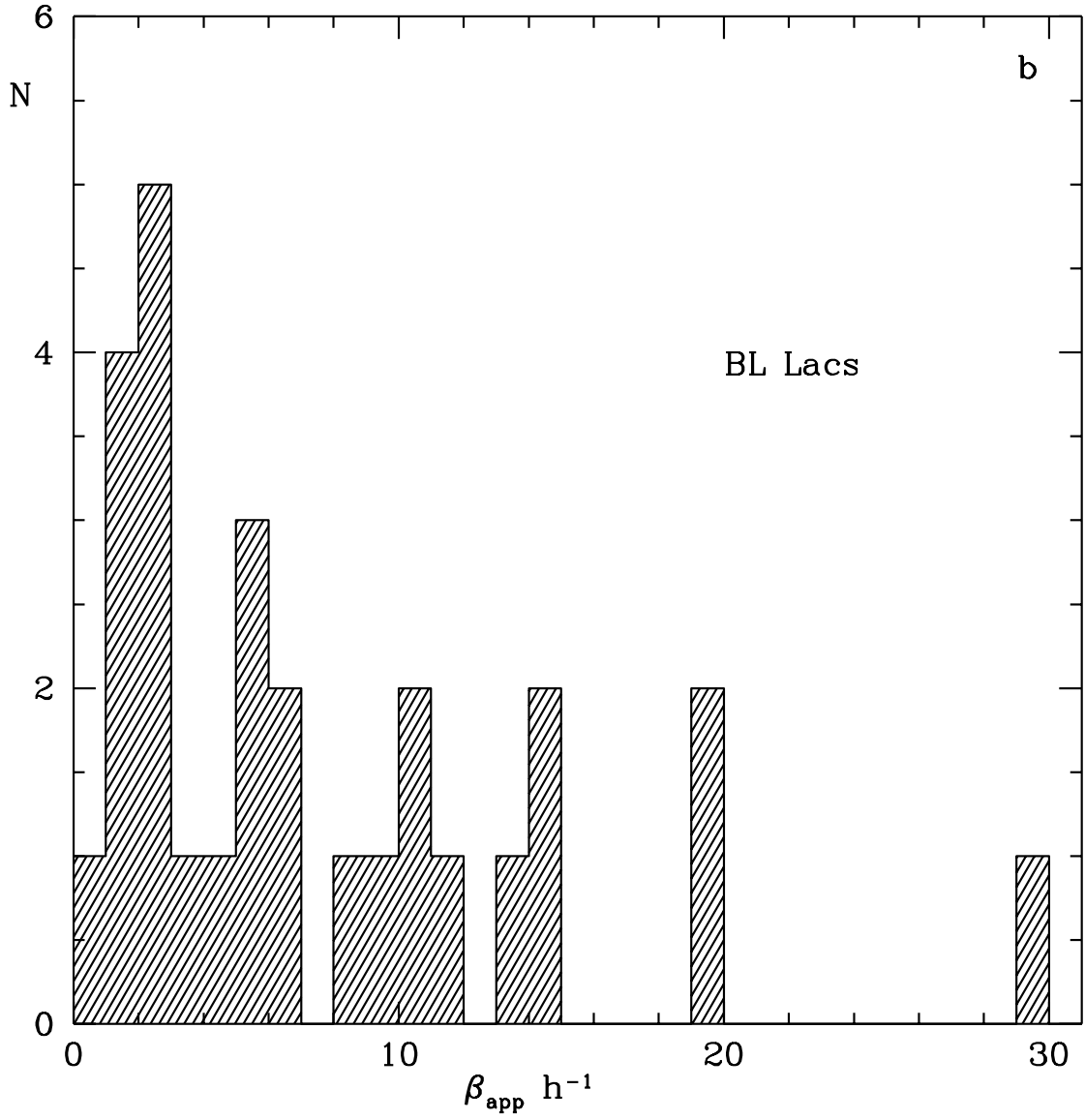}
\caption{Distribution of apparent speeds of jet components in BL~Lac objects.}
\end{figure}
\begin{figure}
\figurenum{44}
\plotone{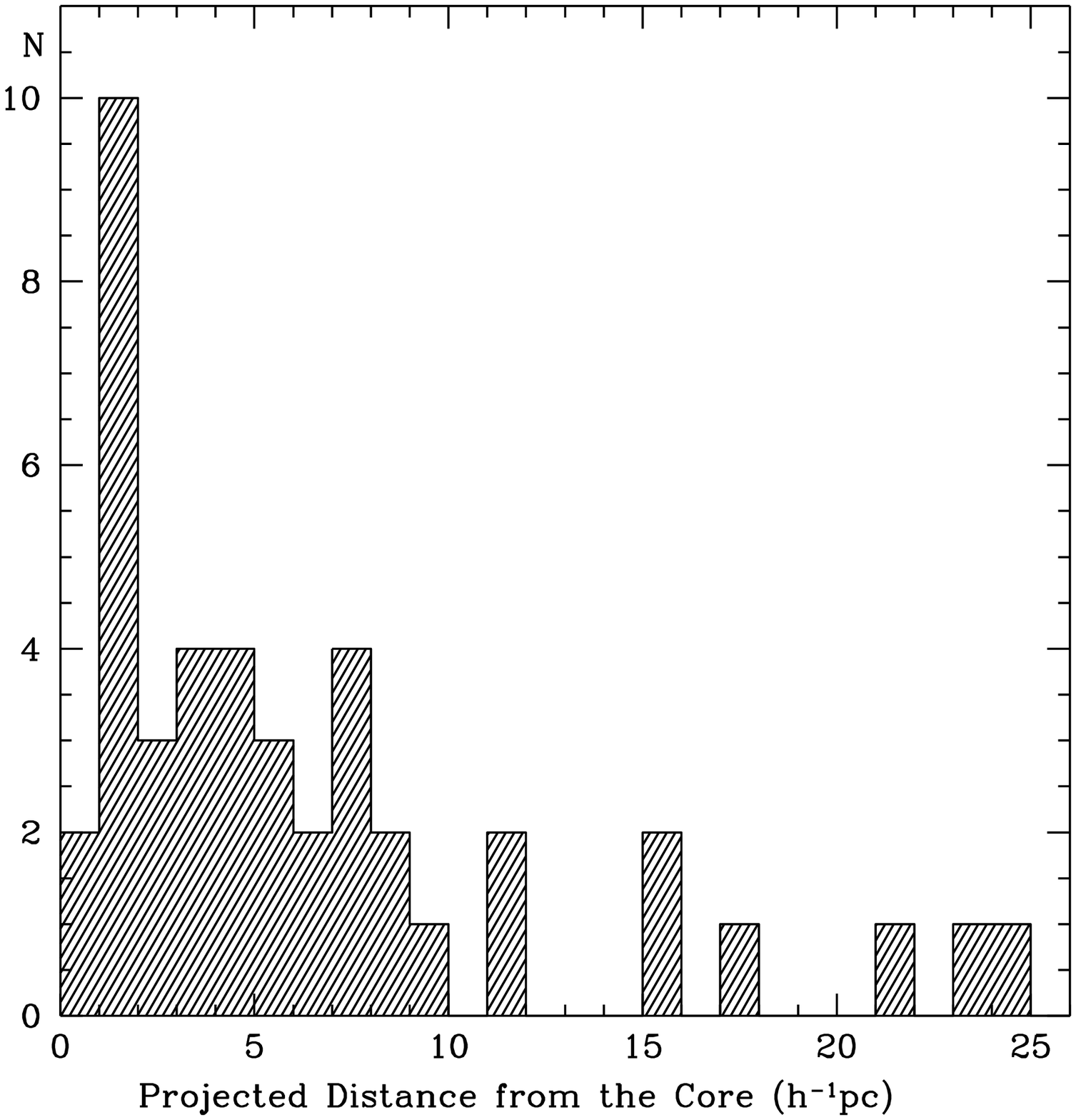}
\caption{Distribution of location of stationary components with respect to the core,
in projection on the sky plane.}
\end{figure}
\clearpage
\begin{figure}
\figurenum{45}
\plotone{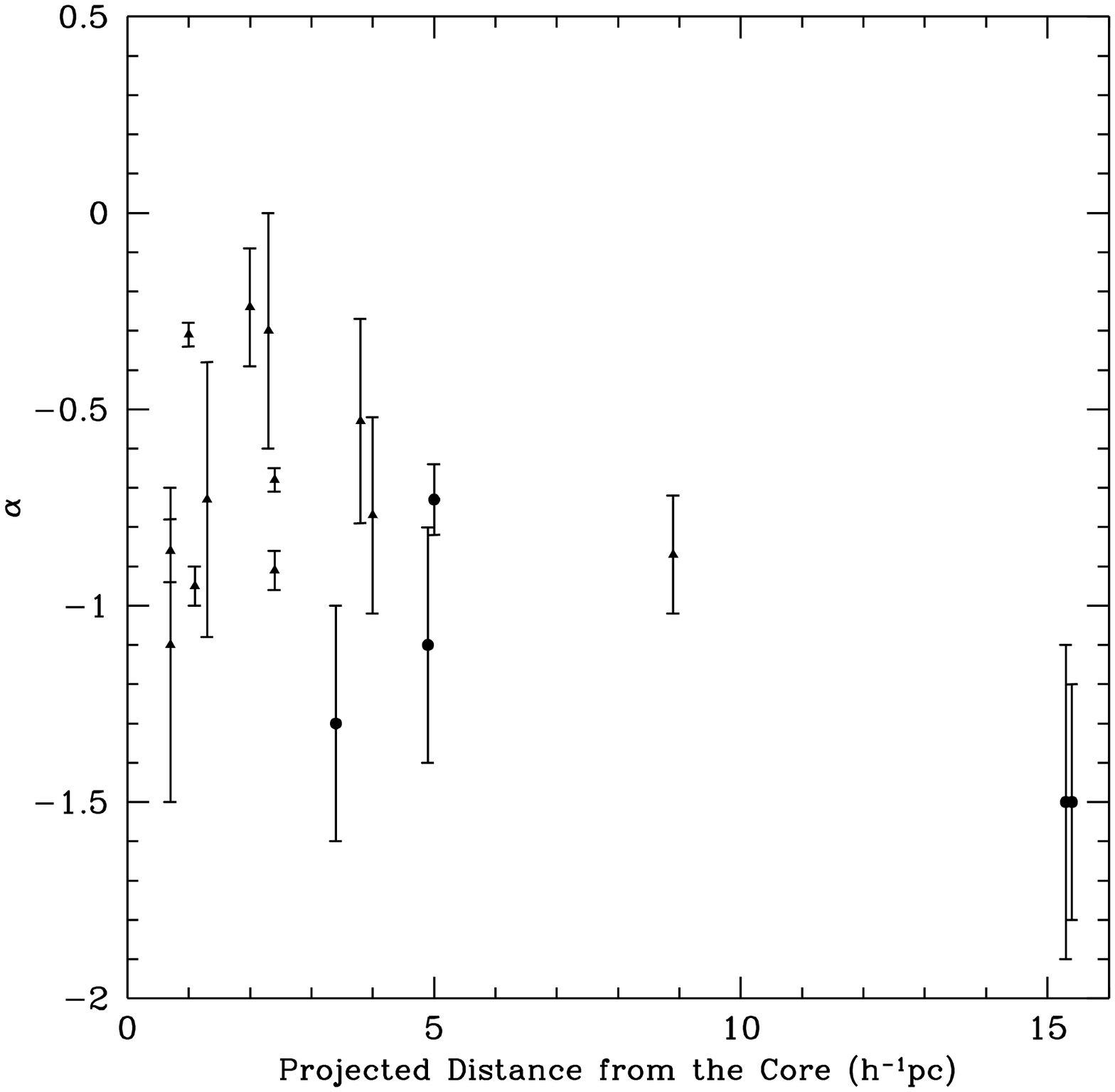}
\caption{Dependence of spectral indices of moving (triangles)
and stationary (circles) components with distance from the core.}
\end{figure}
\begin{figure}
\figurenum{46}
\plotone{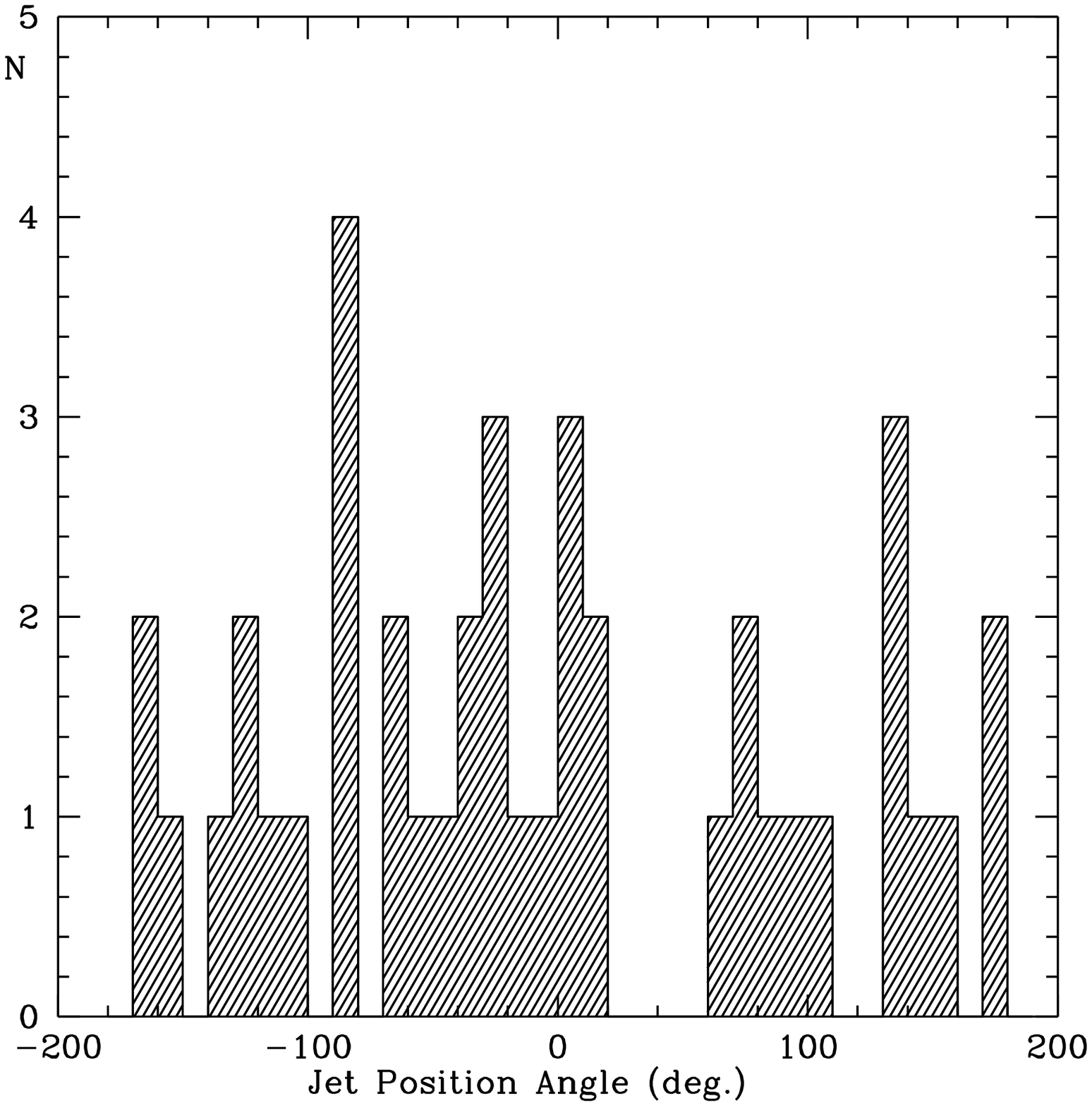}
\figcaption{Distribution of jet position angles in projection on the sky plane.}
\end{figure}
\begin{figure}
\figurenum{47}
\plotone{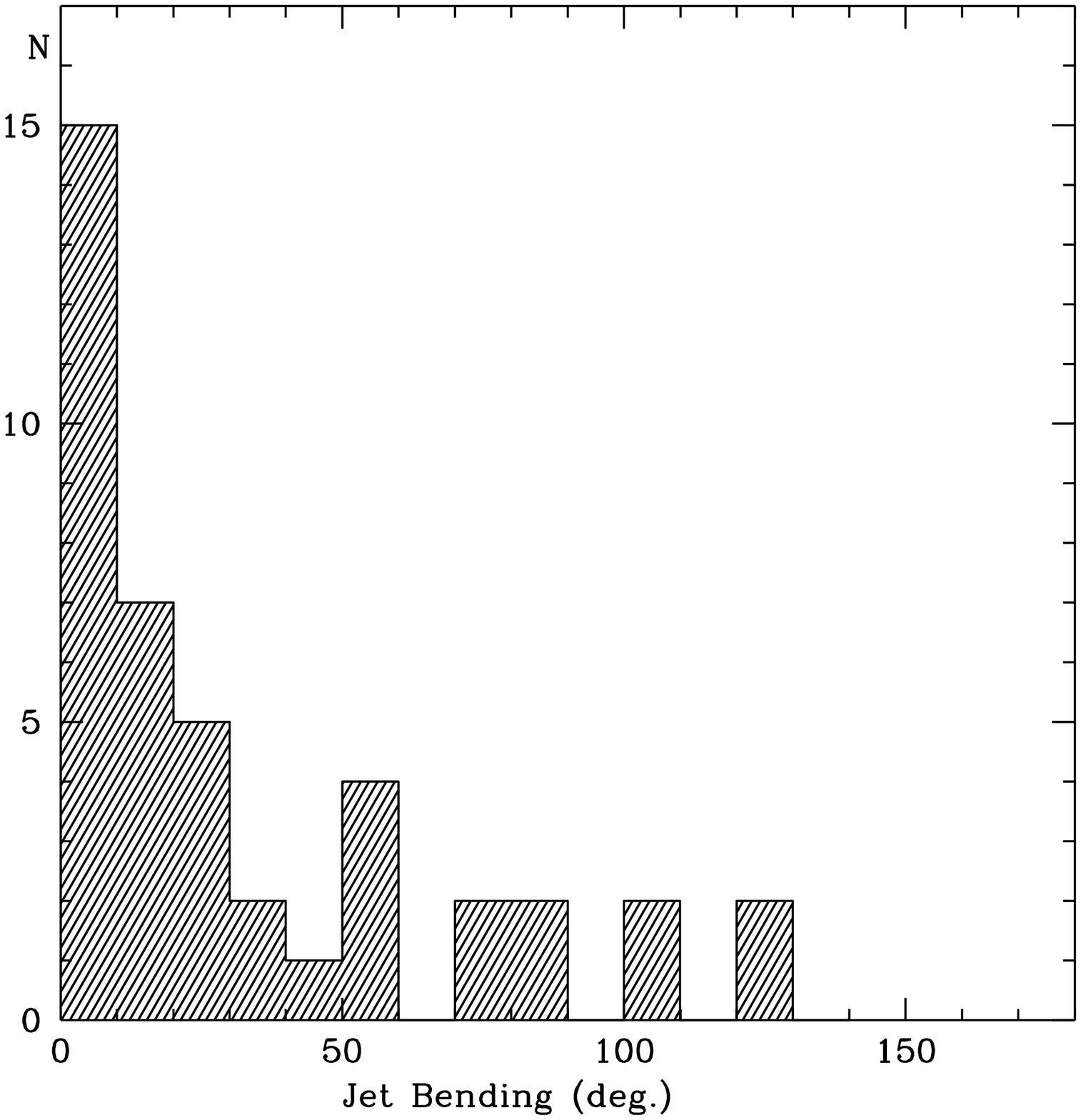}
\caption{Distribution of the angles of maximum local bends in the jets.}
\end{figure}
\begin{figure}
\figurenum{48}
\plotone{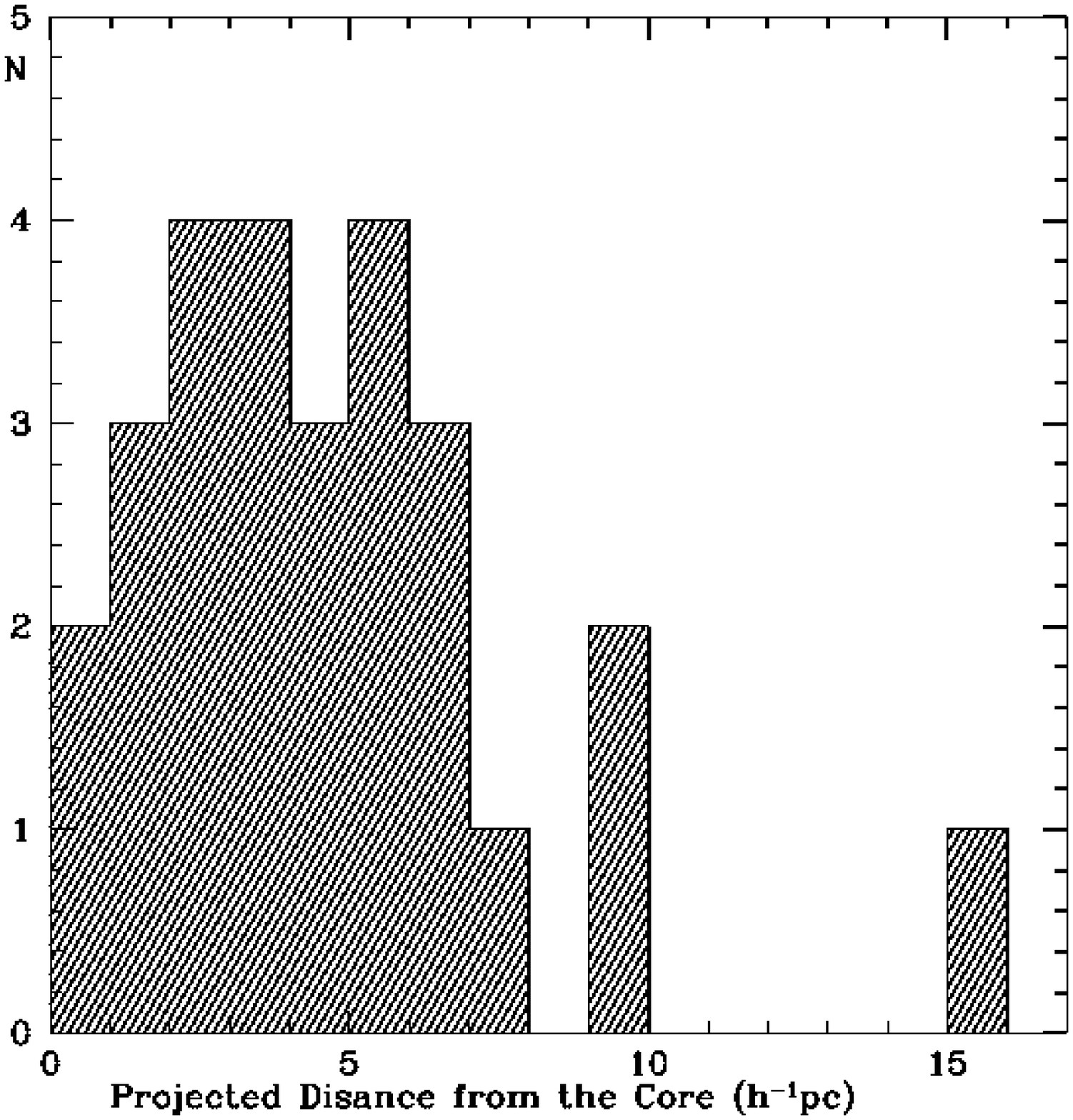}
\caption{Distribution of projected distance from the core at which the
maximum jet bending occurs in the innermost part of the jets.}
\end{figure}


\begin{references}
\reference{} Alberdi, A., et al. 1993, \apj, 402, 160
\reference{} Aller, M.F, Aller, H.D., \& Hughes, P.A. 1999, BAAS,  195, 1609
\reference{} Altschuler, D. R., et al. 1984, \aj, 89, 1784
\reference{} Antonucci, R.R.J. \& Ulvestad, J.S. 1985, \apj, 294, 158
\reference{} B{\aa}{\aa}th, L.B. 1987, in Superluminal Radio Sources, eds. Zensus J.A. \& 
Pearson T.J., Cambridge: Cambridge Univ. Press, 206
\reference{} Barthel P.D., C.E., Conway, J.E., Meyers, S.T., Pearson, T.J., \& Readhead, A.C.S. 1995, \apj, 444, 21
\reference{} Begelman, M.C., Rees, M.J., \& Sikora, M. 1994,  ApJ, 429, L57
\reference{} Bondi, M., et al. 1996, \aap, 308, 415
\reference{} Bower, G.C., Backer, D.C., Wright, M., \& Forster, J.R. 1997, \apj, 484, 118
\reference{} Britzen, S., et al.  1999, in BL~Lac Phenomenon, 
eds. Takalo, L.O. \&  Sillanp\"a\"a A., ASP Conf. Ser., 159, 431
\reference{} Britzen, S., Witzel, A., Krichbaum, T.P., Roland, J, \& Wagner, S.J. 1998, 
in IAU Colloq. 164,
Radio Emission from Galactic and Extragalactic Compact Sources, eds. Zensus, J.A., Taylor, G.B.,
\& Wrobel J.M., ASP Conf. Ser., 144, 43
\reference{} Cawthorne, T.V. \& Gabuzda, D.C. 1996, \mnras, 278, 861
\reference{} Chu, H.S., B{\aa}{\aa}th, L.B., Rantakyr\"o, F.T., Zhang, H.S., \& Nicholson, G. 1996, \aap, 307, 15
\reference{} Cotton, W. D., \& Spangler, S. R. 1979, \apjl, 228, L63
\reference{} Daly, R.A., \& Marscher, A.P. 1988, \apj, 334, 539
\reference{} Denn, G.R., Mutel, R.L., \& Marscher, A.P.  2000, \apjs, 129, 61
\reference{} Dondi, L. \& Ghisellini G. 1995, \mnras, 273, 583
\reference{} Fey, A.L. \& Charlot, P. 1997, \apjs, 111, 95
\reference{} Fey, A.L., Clegg, A.W., \& Fomalont, E.B. 1996, \apjs, 105, 299
\reference{} Fichtel, C.E., et al. 1993, A\&AS, 97, 13
\reference{} Flatters, C. 1996, SLIME User's Guide and Reference Manual, NRAO
\reference{} Gabuzda, D.C., et al. 1998, \aap, 333, 445
\reference{} Gabuzda, D.C. \& Cawthorne, T.V. 1996, \mnras, 283, 759
\reference{} Gabuzda, D.C., Mullan C.M., Cawthorne, T.V., Wardle, J.F.C., \& Roberts, D.H. 1994, \apj, 435, 140
\reference{} Gabuzda, D.C., Cawthorne, T.V., Roberts, D.H., \& Wardle, J.F.C. 1992, \apj, 388, 40
\reference{} Gabuzda, D.C., Wardle, J.F.C., \& Roberts, D.H.  1989, \apj, 336, L59
\reference{} G\'omez, J.L.,  Marscher A.P., \& Alberdi, A. 1999, \apj, 522, 74
\reference{} G\'omez, J.L., Marti, J.M., Marscher A.P., Ib\'a\~nez, J.M., \& Alberdi, A.  1997, \apj, 482, L33
\reference{} G\'omez, J.L., Marti, J.M., Marscher A.P., Ib\'a\~nez, J.M., \& Marcaide, J.M.  1995, \apj, 449, L19
\reference{} Grandi, P., et al.  1996, \apj, 459, 73
\reference{} Hartman, R.C., et al. 1999, \apjs, 123, 79
\reference{} Homan, D.C. \&  Wardle, J.F.C. 1999, \aj, 118, 1942
\reference{} Hooimeyer, J.R., Schilizzi, R.T., Miley, G.K., \& Barthel, P.D. 1992, \aap, 261, 5
\reference{} Hummel, C.A., et al. 1992, \aap, 266, 93
\reference{} Jorstad, S.G., et al. 2001, in preparation
\reference{} Kellermann, K.I., Vermeulen, R.C., Zensus, J.A., \& Cohen, M.H. 1998, \aj, 115, 1295
\reference{} Kemball, A.J., Diamond, P.J., \& Pauliny-Toth, I.I.K. 1996, \apj, 464, L55
\reference{} Krichbaum, T.P., Kraus, A., Otterbein, K., Britzen, S., Witzel, A., \& Zensus, J.A. 1998, in IAU Colloq. 164,
Radio Emission from Galactic and Extragalactic Compact Sources, eds. Zensus, J.A., Taylor, G.B.,
\& Wrobel J.M., ASP Conf. Ser., 144, 37
\reference{} Krichbaum, T.P., et al. 1990, \aap, 230, 271
\reference{} Kollgaard, R.I., Wardle, J.F.C., Roberts, D.H., \& Gabuzda, D.C. 1992, \aj, 104, 1687
\reference{} Lainela, M., et al. 1999, \apj, 521, 561
\reference{} Lister, M.L. \& Marscher, A.P. 1999, Astroparticle Physics, 11, 65
\reference{} Lister, M.L., Marscher, A.P., \&  Gear, W.K.  1998, \apj, 504, 702
\reference{} Lister, M.L. \& Marscher, A.P. 1997, \apj, 476, 572
\reference{} Mantovani, F., Junor, W., Valerio, C., \& McHardy, I. 1999, \aap, 346, 397
\reference{} Marscher, A.P., Jorstad, S.G., Mattox, J.R., \& Wehrle, A.E.  2001, in preparation
\reference{} Marscher, A.P. 1999, Astroparticle Physics, 11, 19
\reference{} Marscher, A.P. 1998, in IAU Colloq. 164,
Radio Emission from Galactic and Extragalactic Compact Sources, eds. Zensus, J.A., Taylor, G.B.,
\& Wrobel J.M., ASP Conf. Ser., 144, 25 
\reference{} Mattox, J.R., et al. 2001, submitted to \apj
\reference{} Mattox, J.R., Schachter, J., Molnar, L., Hartman, R.C., \& Patnail, A.R. 1997a, \apj, 481, 95
\reference{} Mattox, J.R., et al. 1997b, \apj, 476, 692
\reference{} McHardy, I., Lawson, A., Newsam, A., Marscher, A., Robson, I., \& Stevens, J. 1999, \mnras, 310, 571
\reference{} McHardy, I.M., Marscher A.P., Gear, W.K., Muxlow, T., Lehto, H.J., \&
Abraham, R.G. 1990, \mnras, 246, 305
\reference{} Michelson, P., et al. 1993, BAAS, 182, 4409
\reference{} M\"ucke, A, et al. 1997, \aap, 320, 33
\reference{} Mukherjee, R, et al. 1997, \apj, 490, 116
\reference{} Otterbein, K., et al. 1998, \aap, 334, 489
\reference{} Padovani, P. 1992, \aap, 256, 399
\reference{} Pauliny-Toth, I.I.K 1998, in IAU Colloq. 164,
Radio Emission from Galactic and Extragalactic Compact Sources, ed. Zensus, J.A., Taylor, G.B.,
\& Wrobel J.M., ASP Conf. Ser., 144, 75
\reference{} Pauliny-Toth, I.I.K., et al. 1987, \nat, 328, 778
\reference{} Pearson, T.J., et al. 1998, in IAU Colloq. 164,
Radio Emission from Galactic and Extragalactic Compact Sources, eds. Zensus, J.A., Taylor, G.B.,
\& Wrobel J.M., ASP Conf. Ser., 144, 17
\reference{} Pearson, T.J. \& Readhead, A.C.S. 1988, \apj, 328, 114
\reference{} Pearson, T.J. \& Zensus, J.A.  1987, in Superluminal Radio Sources, eds. Zensus J.A. \& 
Pearson T.J., Cambridge: Cambridge Univ. Press, 1
\reference{} Perley, R.A. 1982, \aj, 87, 859
\reference{} Piner, B.G., et al. 1999, BAAS, 195, 1504
\reference{} Piner, B.G. \& Kingham, K.A. 1998, \apj, 507, 706
\reference{} Piner, B.G. \& Kingham, K.A. 1997a, \apj, 479, 684
\reference{} Piner, B.G. \& Kingham, K.A. 1997b, \apj, 485, L61
\reference{} Pohl, M., et al. 1995, \aap, 303, 383
\reference{} Polatidis, A.G., et al. 1995, \apjs, 98, 1
\reference{} Price, R., Gower, A.C., Hutchings, J.B., Talon, S., Duncan, D., \& Ross, G. 1993, \apjs, 86, 365
\reference{} Pyatunina, T.B., et al. 2000, \aap, 358, 451
\reference{} Rantakyr\"o, F.T., B{\aa}{\aa}th, L.B., Dallacassa, D., Jones, D.L., \& Wehrle, A.E. 1996, \aap, 310, 66
\reference{} Shepherd, M.C., Pearson, T.J., \& Taylor, G.B. 1994, BAAS, 26, 987
\reference{} Sillanp\"a\"a, A., et al. 1996, \aap, 315, 13
\reference{} Singh, K.P., Shrader, C.R., \& George, I.M. 1997, \apj, 491, 515
\reference{} Sitko, M.L., Schmidt, G.D., \& Stein, W.A. 1985, \apjs, 59, 323
\reference{} Smith, P.S., Balonek, T.J., Elston, R., \& Heckert, P.A. 1987, \apjs, 64, 459
\reference{} Tateyama, C.E., Kingham, K.A., Kaufmann, P., Piner, B.G., Botti, L.C.L., \& de Lucena, A.M.P 1999, \apj, 520, 627
\reference{} Tingay, S.J., Murphy, D.W., \& Edwards, P.G. 1998, \apj, 500, 673
\reference{} Thompson, D.J., et al. 1993, \apjl, 415, 13
\reference{} Unwin, S.C., Wehrle, A.E., Xu, W., Zook, A.C., \& Marscher, A.P. 1998, in IAU Colloq. 164,
Radio Emission from Galactic and Extragalactic Compact Sources, ed. Zensus, J.A., Taylor, G.B.,
\& Wrobel J.M., ASP Conf. Ser., 144, 69 
\reference{} Valtaoja, E. \& Teraesr\"anta, H.  1996, A\&AS, 120, 491
\reference{} Valtaoja, E., Lahteenm\"aki, A., \& Ter\"aesranta, H.  1992, A\&AS, 95, 73
\reference{} Vermeulen, R.C. \& Cohen, M.H. 1994, \apj, 430, 467
\reference{} von Montigny, C., et al. 1997, \apj, 483, 161
\reference{} von Montigny, C., et al. 1995a, \apj, 440, 525
\reference{} von Montigny, C., et al. 1995b, \aap, 299, 680
\reference{} Wagner, S.J., et al. 1996, \aj, 111, 2187
\reference{} Wagner, S.J., et al. 1995, \aap, 298, 688
\reference{} Wajima, K., Lovell, J.E.J., Kobayashi, H., Hirabayashi, H., Fujisawa, K., \& Tsuboi, M. 2000, PASJ, 52, 329
\reference{} Wardle, J.F.C., Homan, D.C., Ojha, R., \& Roberts, D.H. 1998, Nature, 395, 457
\reference{} Wehrle, A. E., et al. 2001, \apj, in press
\reference{} Wehrle, A. E., et al. 1998, \apj, 497, 178
\reference{} Wehrle, A. E., et al. 1992, \apj, 391, 589
\reference{} Wehrle, A. E. \& Cohen, M.  1989, \apj, 346, L69
\reference{} Whitney, A. R., et al. 1971, Science, 173, 225
\reference{} Witzel, A., et al. 1988, \aap, 206, 245
\reference{} Xie, G. Z., et al. 1994, A\&AS, 106, 361
\reference{} Xu, W., Readhead, A. C. S., Pearson, T. J., Polatidis, A. G., \& Wilkinson,
P. N. 1995, \apjs, 99, 297
\reference{} Yurchenko, A. V., Marchenko-Jorstad, S. G., \& Marscher, A. P. 2000,
\aap, 358, 428
\reference{} Zhou, Y. Y., Lu, Y. J., Wang, T. G., Yu, K. N., \& Young, E. C. M. 1997,
\apjl, 484, 47
\end{references}
\end{document}